\documentclass[upright, contnum]{umemoria}

\makeatletter
\g@addto@macro\titlepage{\pagenumbering{Alph}}
\g@addto@macro\endtitlepage{\pagenumbering{roman}}
\makeatother

\depto{Departamento de Física}
\author{Bruno Sebastián Scheihing Hitschfeld}
\title{Revelando la Estructura del Paisaje Primordial a través de no-Gaussianidad Primordial}
\auspicio{}
\date{2019}
\guia{Gonzalo Palma Quilodrán}
\carrera{Magíster en Ciencias, Mención Física}
\memoria{Tesis para optar al Grado de \break  Magíster en Ciencias, Mención Física}
\comision{Jorge Noreña Sánchez, Nelson Padilla, Domenico Sapone}

\deptoa{Physics Department}
\authora{Bruno Sebastián Scheihing Hitschfeld}
\titlea{Revealing the Structure of the Inflationary Landscape through Primordial non-Gaussianity}
\auspicioa{}
\datea{August 8, 2019}
\dateaa{8 de Agosto, 2019}
\guiaa{Gonzalo Palma Quilodrán}
\carreraa{Master of Science in Physics}
\memoriaa{Tesis para optar al Grado de \break  Magíster en Ciencias, Mención Física}
\comisiona{Jorge Noreña Sánchez, Nelson Padilla, Domenico Sapone}

\usepackage{lipsum}
\usepackage{color}
\usepackage[toc,page]{appendix}
\usepackage{braket}
\usepackage{hyperref}

\usepackage[utf8]{inputenc}
\usepackage[T1]{fontenc}

\usepackage{scalerel}
\usepackage{stackengine,wasysym}

\newcommand\reallywidetilde[1]{\ThisStyle{%
  \setbox0=\hbox{$\SavedStyle#1$}%
  \stackengine{-.1\LMpt}{$\SavedStyle#1$}{%
    \stretchto{\scaleto{\SavedStyle\mkern.2mu\AC}{.5150\wd0}}{.3\ht0}%
  }{O}{c}{F}{T}{S}%
}}

\newcommand{\be}{\begin{equation}}
\newcommand{\ee}{\end{equation}}
\newcommand{\bea}{\begin{eqnarray}}
\newcommand{\eea}{\end{eqnarray}}
\newcommand{\nn}{\nonumber}

\def\p{{\bf p}}
\def\q{{\bf q}}
\def\k{{\bf k}}
\def\r{{\bf r}}
\def\x{{\bf x}}
\def\y{{\bf y}}
\def\z{{\bf z}}
\def\w{{\bf w}}
\def\H{{\mathcal{H}}}
\def\L{{\mathcal L}}
\newcommand{\n}{ \hat{n} }

\begin{document}

\frontmatter
\maketitle

\begin{abstract2}
{ 

In this thesis, we show how the structure of the landscape potential of the primordial Universe may be probed through the properties of the primordial density perturbations responsible for the origin of the cosmic microwave background anisotropies and the large-scale structure of our Universe. Isocurvature fields ---fields orthogonal to the inflationary trajectory--- may have fluctuated across the barriers separating local minima of the landscape potential during inflation. We analyze how this process could have impacted the evolution of the primordial curvature perturbations. If the typical distance separating consecutive minima of the landscape potential and the height of the potential barriers are smaller than the Hubble expansion rate parametrizing inflation, the probability distribution function of isocurvature fields becomes non-Gaussian due to the appearance of bumps and dips associated with the structure of the potential. We show that this non-Gaussianity can be transferred to the statistics of primordial curvature perturbations if the isocurvature fields are coupled to the curvature perturbations. The type of non-Gaussian structure that emerges in the distribution of curvature perturbations cannot be fully probed with the standard methods of polyspectra; instead, the probability distribution function is needed. The latter is obtained by summing all the $n$-point correlation functions, which are of the local type. 

To substantiate our claims, we offer a concrete model consisting of an axionlike isocurvature perturbation with a sinusoidal potential and a linear derivative coupling between the isocurvature and curvature field. This result is generalized to arbitrary potentials, studied beyond first-order perturbation theory, and extended to a more general class of backgrounds. We also briefly explore connections with the stochastic inflation framework. Finally, we undertake a study of primordial non-Gaussianity of the local type, where we use our results to reconstruct and constrain the shape of the landscape potential with the help of Cosmic Microwave Background observations by the Planck telescope, and additionally, we explore prospects for observable quantities in the Large-Scale Structure of our universe towards constraining the primordial statistics of the universe. 

Should any of these signals be measured by upcoming cosmological surveys, we will have our understanding of early-universe physics greatly enhanced, as those observations may be readily connected to the dynamics of the inflationary perturbations.

}
\end{abstract2}

\cleardoublepage

\begin{dedicatoria} % opcional
To my parents, Nancy and Rodrigo, and to my sister, Irma.
\end{dedicatoria}

\cleardoublepage

\begin{thanks} % opcional

I wish to thank all the people who supported me throughout the development of this thesis. In particular, I wish to thank Gonzalo Palma, Spyros Sypsas, Xingang Chen, Domenico Sapone, Walter Riquelme, Rafael Bravo, Bastian Pradenas, Javier Silva, Bryan Sagredo, and Cristóbal Zenteno for useful discussions and comments throughout the development of this work. I would also like to thank Patricio Cordero, Luis Foà, Fernando Lund, and Nelson Zamorano for helping me further develop my career in other directions within physics.

Finally, I wish to acknowledge support from the Fondecyt Regular project number 1171811 (CONICYT), and from a CONICYT grant, number CONICYT-PFCHA/Mag\'{i}sterNacional/2018-22181513.

\end{thanks}
\cleardoublepage

\tableofcontents
%\listoftables % opcional

\cleardoublepage

\listoffigures % opcional

\cleardoublepage

\mainmatter

\begin{intro}

Throughout the history of humanity, our understanding of the universe has been continuously growing, if at times somewhat slowly from our current perspective. In our memories, the perhaps most important scientific revolutions have happened within the last three thousand years, if not the last four hundred. From ancient Babylon, Egypt, or Greece to modern times, the amount of knowledge and information we have acquired has been constantly growing, and so have our ideas and conceptualizations about the phenomena that provide us that information.

One of the earliest ideas about the nature of our universe, the environment wherein we live, was that we lived inside a rectangular box, where the ground on which humans stood was the floor and the side walls served as a support on which a river flowed, carrying the Sun and Moon in their barques\footnote{See, for instance, ``The Sleepwalkers'', by A. Koestler.}. The stars in the night sky were lamps suspended from the ceiling of the box, or carried by gods. While these ideas may seem bizarre from today's perspective, at the time they served their purpose as an explanatory narrative of why the Earth seems flat if one walks around a few kilometers at a time, and why the Sun, Moon, and stars moved the way they did.

In time, other individuals and cultures performed more detailed observations of each feature, often concluding that the actual explanation of the phenomenon was different than that suggested by other civilizations (or individuals). However, in the vast majority of cases the acquired knowledge was not discarded: the description of the universe surrounding us had to take into account all of the available information for it to be consistent. Contrary to some popular belief, the sphericity of the Earth has been an established fact for over two thousand years since Pythagoras and Parmenides introduced the idea in the 6th century BC and was supported with empirical evidence by Aristotle around 330 BC. About a hundred years later, Eratosthenes made a first calculation of the circumference of the Earth (see Figure~\ref{fig:circ-earth} for an illustration). Moreover, although the notion that the Earth orbits around the Sun instead of the other way around is typically attributed to Copernicus, the idea was also considered by Aristarchus of Samos in the 3rd century BC. 

\begin{figure}[t!]
\begin{center}
\includegraphics[scale=0.35]{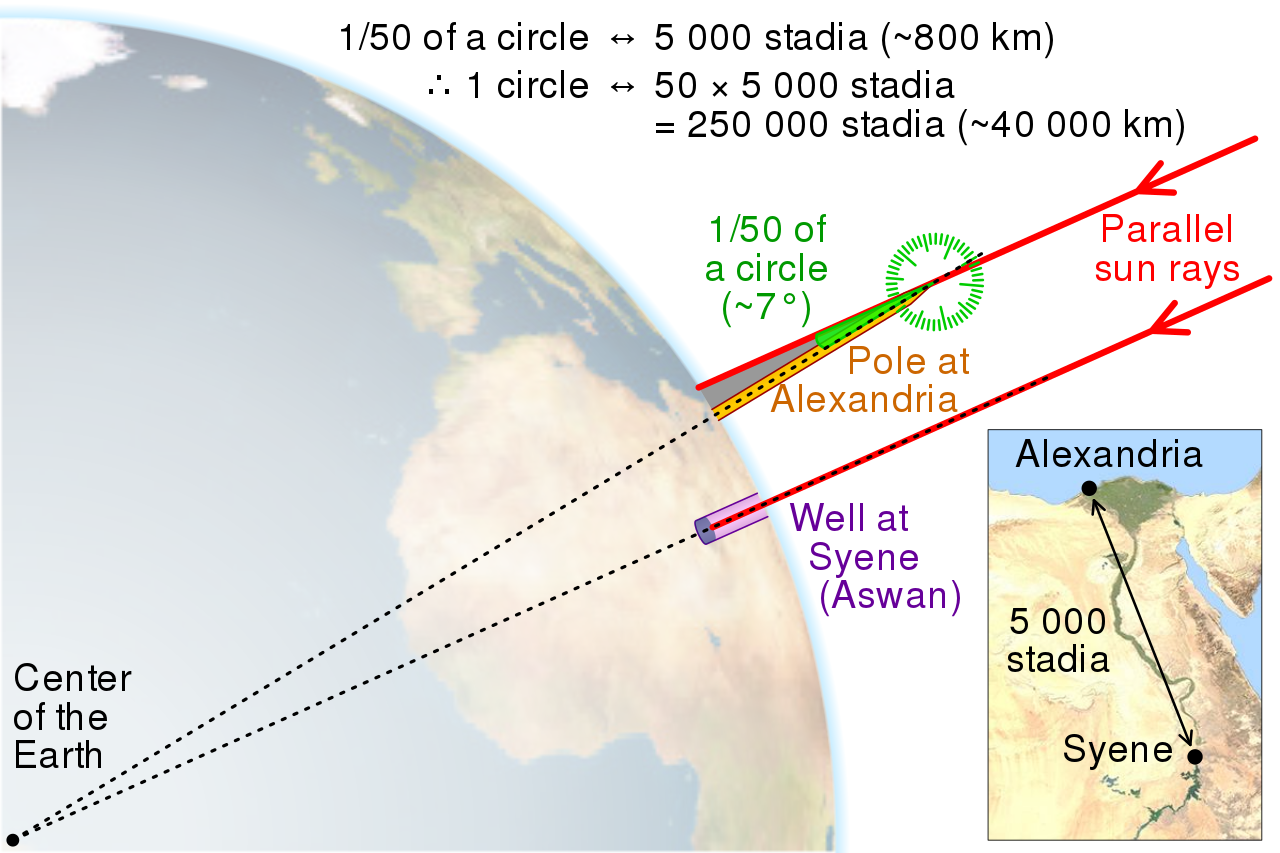}
\caption{Illustration showing a portion of the globe showing a part of the African continent. The sunbeams shown as two rays hitting the ground at Syene and Alexandria. Angle of sunbeam and the gnomons (vertical pole) is shown at Alexandria, which allowed Eratosthenes' estimates of radius and circumference of Earth.  \footnotesize{ Extracted from \href{https://commons.wikimedia.org/wiki/File:Eratosthenes\_measure\_of\_Earth\_circumference.svg}{https://commons.wikimedia.org/wiki/File:Eratosthenes\_} \href{https://commons.wikimedia.org/wiki/File:Eratosthenes\_measure\_of\_Earth\_circumference.svg}{measure\_of\_Earth\_circumference.svg} unaltered, licensed under the \href{https://creativecommons.org/licenses/by-sa/4.0/deed.en}{Creative Commons Attribution-Share Alike 4.0 International} license. Authored by cmglee, David Monniaux, jimht at shaw dot ca.  }  }
\label{fig:circ-earth}
\end{center}
\end{figure}

This is a useful point to stop and think about the evolution of science. From nowadays' point of view, it may seem ridiculous that the concept of the Earth's orbiting around the Sun, which today is regarded as obvious, was seemingly lost through the centuries for it to be rediscovered eighteen centuries later. However, it is highly likely that such a notion was disregarded on similar grounds to how an idea might be discarded today: by lack of empirical evidence supporting that specific theory. For a new notion, theory, or idea of our universe to supersede the previous, it not only has to account for all of the features that the earlier theory did, but it also has to make new predictions that could not have previously been met and to successfully withstand a comparison with empirical evidence.

The same reasoning can be applied to today's quest for a fundamental theory of our universe. Currently, the standard model of particle physics and the elementary particles it contains: the leptons, %(i.e., electrons, muons, and taus plus their associated neutrinos); 
quarks, 
%(up, down, charm, strange, top, bottom); 
and force-carrying gauge bosons, 
%(photons, gluons, W bosons, and Z bosons) 
along with the Higgs boson, comprise all the knowledge we have on the basic constituents of matter (see Figure~\ref{fig:standard-model}). This theory (which is within the framework of quantum mechanics) along with general relativity, each in its own domain of validity, accurately account for most of the observed phenomena nowadays, and are sitting in place until a better theory (in terms of predictive power and explanatory success) comes about. Among the challenges that remain to be solved, a consistent theory reconciling quantum mechanics and gravity within a single framework has eluded a satisfactory formulation. 

\begin{figure}[t!]
\begin{center}
\includegraphics[scale=0.20]{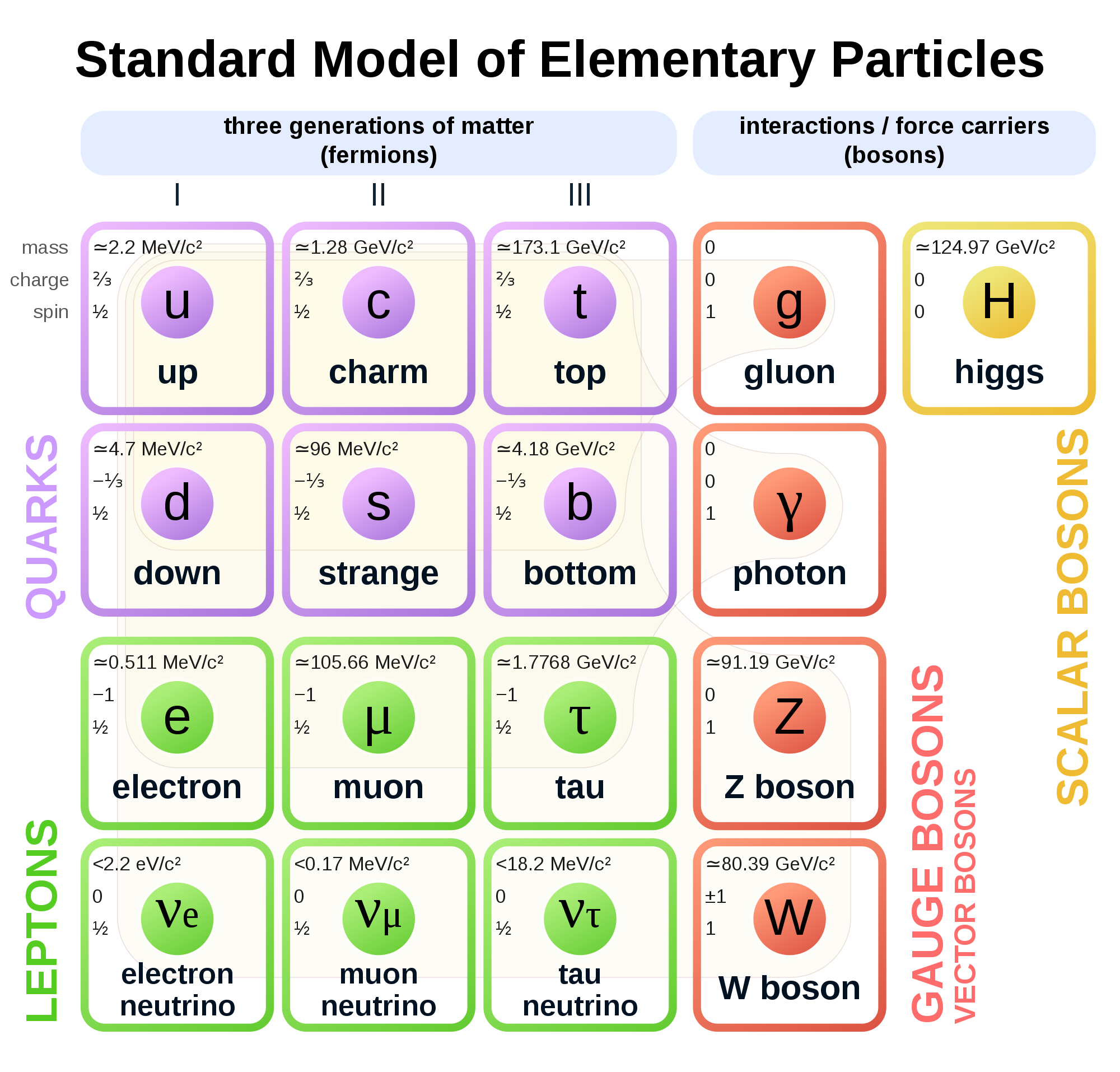}
\caption{The Standard model of elementary particles. The leptons (in green) along with the quarks (in purple) are the building blocks of matter, while the gauge bosons (in red) are the force-carrying particles that mediate the interactions between them. The Higgs Boson complements the model by giving mass to each particle.  \footnotesize{ Extracted from \href{https://commons.wikimedia.org/wiki/File:Standard\_Model\_of\_Elementary\_Particles.svg}{https://commons.wikimedia.org/wiki/File:Standard\_Model\_} \href{https://commons.wikimedia.org/wiki/File:Standard\_Model\_of\_Elementary\_Particles.svg}{of\_Elementary\_Particles.svg} unaltered, which has been released to the public domain by its author. }  }
\label{fig:standard-model}
\end{center}
\end{figure}

Probably, this is so in lack of empirical results pointing us in the right direction on where, and how, to formulate a theory of quantum gravity. Nonetheless, considerable efforts have been made in developing frameworks and mathematical tools to tackle the issue, but lacking experimental findings it has proven to be an arduous task. Thus, it is imperative to find and explore physical settings in which both gravity and quantum mechanics are equally important, and, at the same time, to derive concrete predictions from theories that pretend to account for the phenomena. This should, hopefully, allow physicists to discard theories discordant with observations, and therefore, to have a firmer grasp on the empirical requirements a theory of quantum gravity should satisfy.

Today, at least two areas of physics show promise regarding how to obtain and interpret such data. Over the last fifty years, a great deal of effort has been devoted to studying a setting wherein both gravitational and quantum mechanical effects should be relevant: the event horizon of a black hole. As such, black hole physics has become an ever-expanding field of research, with connections to thermodynamics, information theory, string theory, and many others besides the main subjects under discussion: quantum mechanics and general relativity. Consequently, it has attracted much attention from a significant portion of the physics community. 

The other area where a theory of quantum gravity may be probed is early universe Cosmology. Indeed, the inflationary paradigm~\cite{Guth:1980zm, Linde:1981mu, Albrecht:1982wi, Starobinsky:1980te, Mukhanov:1981xt} has attracted a lot of attention because having an initial phase of accelerated expansion implies that, at some point in the very early universe, the length scales that are now macroscopic were once small enough that quantum-mechanical effects were relevant\footnote{To be concrete, it is a fluctuation associated to some dynamical variable that which feels the stretching of lengths: in the absence of interactions, a variation of a quantity over a microscopic length scale will later become a fluctuation over a macroscopic length, because the physical distance between the endpoints of the variation will have increased.}. Moreover, inflation is required (by observations) to have happened over such an extended period of time that it is hoped that cosmological experiments could be sensible to physical length scales that approach the purported domain of String Theory, that is, the Planck length $\ell_P \approx 1.6 \times 10^{-35} \, {\rm m}$. This prospect to explore such high energy scales has led physicists to write down a myriad of distinct models where inflation can happen consistently with observational constraints and also give testable predictions for experiments to resolve in the years to come.

%One may wonder, however, whether quantum mechanics will intertwine at all with gravity.  . 

It is an already established idea that the cosmological fluctuations in the distribution of matter and energy were seeded by primordial quantum fluctuations: as much as we don't notice it in our daily lives, quantum mechanics was responsible for the universe as we see it today. Remarkably, this is not what inflation was engineered to solve~\cite{Guth:1980zm, Linde:1981mu}, but nonetheless predicts it. Moreover, the simplest models of inflation predict that the probability distribution associated to these fluctuations is very close to Gaussian. This means, and is equivalent to saying, that the statistics of the fluctuations can be completely characterized by their two-point correlation function. This assertion has been resoundingly confirmed by Cosmic Microwave Background (CMB) surveys up to the available precision.

However, most models of inflation do not predict purely Gaussian statistics. Indeed, departures from Gaussian statistics in the primordial quantum fluctuations of the gravitational potential (hereafter primordial non-Gaussianity) are among the main observables wherein to look for novel signatures of the physics of the primordial universe. With these motivations, the search for primordial non-Gaussianity (NG) has been guided by physicists' ability to make predictions within the inflationary paradigm, that is, our understandings and limitations of how to perform calculations.

In this context, Effective Field Theories (EFTs) have proved to be a valuable asset when studying the low-energy predictions of high-energy physics. Although the origin of the primordial fluctuations is associated to high-energy physics, the fact that their observable outcomes at the end of inflation are described in terms of distortions to the primordial gravitational potential, which comparatively corresponds to very low-energy physics, allows physicists to write down EFTs of inflation~\cite{Noumi:2012vr,Cheung:2007st,Senatore:2010wk} that account for the physics we observe.

As low-energy effective theories, the simplest models of inflation for the cosmological perturbations predict that the main departures from Gaussianity should be found in the form of small, but distinct from zero, three-point correlation functions of the primordial gravitational potential~\cite{Gangui:1993tt, Komatsu:2001rj, Acquaviva:2002ud, Maldacena:2002vr}, also dubbed as ``bispectrum''. This has led to most of the data-oriented literature on non-Gaussianities to focus on ways to constrain the 3-point correlation, and on the theoretical side, to produce detectable or distinguishable signals through the bispectrum. The 4-point function, dubbed as ``trispectrum'', has also been subject to some interest, as it complements an indicator of an asymmetric deviation from Gaussianity (the 3-point function), which treats perturbations of opposite sign differently, with an indicator of a symmetric departure from Gaussian statistics as the 4-point function.

However, this type of analysis precludes the possibility of more complicated kinds of non-Gaussianity. Indeed, to assume that the 3-point or the 4-point functions are a thorough indicator of the presence of NG is precipitate, as all of the higher order $n$-point functions might contain extra, independent information that is not captured at all by the 3- and 4-point correlations. This presents important questions: are there inflationary models that can naturally give rise to non-Gaussianity manifestly not well suited to be described by 3- or 4-point functions? And if there are, how can we search for such signals within the data available from this and next decade's cosmological surveys? Finally, if the answer to both is satisfactory, we may ask: what information will we be able to acquire about the primordial universe? and optimistically, how does this help us in the quest of describing the quantum mechanics of gravity?

The results developed in this work pretend to answer the first three questions in a concrete manner. We will present an explicit example of an inflationary model that satisfies the requirement of naturally delivering nontrivial\footnote{In the sense that the statistical outcomes cannot be reproduced by Gaussian statistics.} higher $n$-point correlations, which give rise to a characteristic signature in terms of the primordial gravitational potential's probability distribution that can be straightforwardly sought for in the data. Furthermore, in absence of other competing signals, these signatures can be directly related to properties of the inflationary landscape (a concept we will get to in Chapter~\ref{chap:n-point}). While this does not give a direct probe of quantum gravity, it might give observables that can be related to the fundamental quantum theory whence gravity emerges, e.g., String Theory. This, of course, relies on the existence of such physical mechanisms within that theory that account for both gravity and the extra degree(s) of freedom that give rise to inflation. 

Concretely, what we will pursue is a perturbative computation involving interactions between the fields present during inflation that are dynamically relevant, which we assume comes from a UV complete theory\footnote{A UV complete theory is one that is consistently defined from very low energies (say, 1 eV) to arbitrarily high energies beyond the Planck mass $\sim 10^{19}$ GeV.} of which our starting point is only a low-energy effective description. Generically, in this context low-energy means that the characteristic mass/energy scales of the effective theory are much smaller than the Planck mass $M_{\rm pl}$, which, one assumes, is a characteristic mass scale of the fundamental theory. However, as an EFT, our results apply to any high-energy theory of which its low-energy description for the nonlinear interactions of the field perturbations can be cast in terms of a scalar potential. The quadratic part of the theory, which generates linear equations of motion, is fixed from the symmetries of the inflationary background~\cite{Achucarro:2016fby,Langlois:2008mn}. Therefore, even though the results herein may seem model-dependent, we are not binding ourselves to any particular UV-completion, i.e., to any particular model of the higher-energy physics, and thus they should constitute a valuable, quasi model-independent tool to study the outcomes of perturbations during inflation. 

Let us stress that even though the developments we will here pursue are not presently simpler than other existing theories that \textit{are} statistically consistent with every cosmological observation so far, we should not let our lack of understanding sway us from pursuing seemingly more complicated calculations or more intricate abstractions, as they may ultimately prove to be part of a much more satisfying picture describing the universe we live in. If signals of the type suggested here are ever found and confirmed, we will have reduced the class of theories to explore to only those that have a low-energy effective description of the form proposed here.

This thesis is organized as follows: In Chapter~\ref{chap:search}, we give a brief overview on the history of our universe, starting from the early universe with inflation until its later stages such as the formation of structure, for instance, in how galaxies cluster. Afterwards we shortly discuss the current constraints and characterization of the CMB features in terms of the $\Lambda$CDM model, the nowadays ``standard model'' of cosmology.  A similar exposition of LSS features, plus some prospects regarding the detectability of signals in the near future, completes Chapter~\ref{chap:search}.%, attempting to give a broad picture of what will or might be observed and constrained soon. 

Chapter~\ref{chap:inflation} commences discussing the canonical setup for inflation in terms of a single scalar degree of freedom, to then explain how this construction accounts for the features we currently observe in our universe. Some necessary technical considerations are given at this point regarding which gauge we will be choosing to work with throughout the rest of this thesis. Then we proceed to introduce the more general framework of Multi-Field inflation, highlighting the motivations to pursue such a construction and in what way it is more satisfactory than single-field inflation. Inflationary isocurvature fluctuations are introduced at this point, and an exposition regarding previous work on how these perturbations may affect the primordial gravitational potential is developed. This lays the groundwork for the remainder of the thesis and also allows one to discuss the detectability prospects of non-Gaussian signals, which can be enhanced by the presence of extra degrees of freedom.

Chapter~\ref{chap:n-point} constitutes novel results, here dubbed ``Tomographic non-Gaussianity'', developed throughout the same time span of this thesis and first presented in~\cite{Chen:2018uul} and~\cite{Chen:2018brw}, where the influence and effects of weak nonlinearities in the dynamics of the isocurvature fields, in the form of a sinusoidal scalar potential, is communicated to the primordial field, the curvature perturbation, that seeded the initial conditions for the CMB and LSS to occur. Several technical details are included in the main discussion, such as the quantum-mechanical formalism of the interaction picture, or the explicit reconstruction of the 1-point PDF for the curvature perturbation. The notationally heavier computations are left to Appendix~\ref{chap:details}. Finally, a more general picture of the generation of this type of non-Gaussianity is given, presenting the explicit form of the curvature perturbation 1-point PDF for arbitrary potentials in the normal (orthogonal) direction to the inflationary trajectory.

Once the main results have been established, Chapter~\ref{chap:field} provides a different approach to computing the observables, be them $n$-point functions or probability densities, in the form of the explicit evolution of the quantum field to first order in perturbation theory, which is the starting point of Chapter~\ref{chap:CMB-LSS}. %A comparison with the canonical results in the search for non-Gaussianity is given, and the main qualitative differences with the possible outcomes of single-field inflation are listed. 
We also study the dynamics of the quantum field beyond first-order perturbation theory, for a certain range of scales during inflation. Stochastic Inflation is also considered as a point of comparison, and it is shown that, up to a certain level, the descriptions seem to be closely connected. Finally, we extend the results of Chapter~\ref{chap:n-point} to a more general class of backgrounds by exploiting the interaction picture of quantum mechanics, to first order in perturbation theory, allowing us to also incorporate an evolution of the background different than the typical de Sitter stage. The corrections at higher orders in perturbation theory are expounded in Appendix~\ref{chap:ininPDF}.

As the closing chapter of this thesis, Chapter~\ref{chap:CMB-LSS} discusses the search for the type of non-Gaussianity, studied herein in terms of observable quantities in the CMB or LSS. An analysis of the 1-point temperature PDF of the CMB is performed, and constraints are obtained on the primordial nonlinearities. It is also outlined how 2-point PDFs may improve the constraints already found. The same ideas are applied to the LSS matter density contrast to establish the effects of this non-Gaussianity. Specific probes of non-Gaussianity are examined in the case of LSS. For instance, the appearance of a particular scale dependence in the halo bias, i.e., in how the local density of matter/galaxy halos is modified by the presence of long modes in the primordial curvature fluctuations. 

Lastly, we outline the conclusions of this work and possible continuations of it in the near future. Among the main takeaway messages, this work establishes that there are sensible theoretical reasons to search for non-Gaussianity beyond the usually explored 3- and 4-point correlation functions, and, moreover, that this type of non-Gaussian fluctuations may be instrumental in characterizing the (more) fundamental field theory that gave rise to the early universe.

\end{intro}

\newpage
\thispagestyle{empty}
\mbox{}
\newpage

%\cleardoublepage

\chapter{Searching for the Initial Conditions of our Universe} \label{chap:search}

One of the most interesting questions to tackle in the realm of human knowledge, if perhaps the most  interesting, is to determine where we come from and how the phenomena we observe today started, if they ever did such a thing. Concretely, one would like to explain our observable universe from a set of initial conditions at the ``beginning'' of our universe, be them at a specific temporal distance from us or infinitely far away in the past.

There is much knowledge we have gained and that we can still continue gathering by means of experiments in laboratories on Earth about the local physics of our universe, going all the range from particle accelerators studying the elementary constituents of our universe to studying the nature of gravity at planetary scales by following orbital motions and setting satellites in place to probe the local gravitational field. However, the information we have to pursue a reconstruction of our universe's history is does not come from such a wide range of experiments. Indeed, such information is almost entirely stored in the light that arrives from every direction in the sky towards us, complemented by what we know about the nature of light and matter from ground-based experiments.

After Sir Arthur Eddington's expedition in 1919 confirmed the deflection of starlight by the Sun during the solar eclipse of May 29, continuing confirmation that Albert Einstein's theory of General Relativity accurately describes the dynamics of the gravitational field led to the concrete possibility that our universe was not static but expanding or contracting, and that possibility now had a concrete framework wherein to be studied. Surprisingly, within the next decades, astronomical observations began to signal that, on average, the other galaxies were moving away from the Milky Way, contrary to the then favoured notion of a static universe. Astronomers were able to determine this motion because the photons received from the respective galaxies appeared red-shifted\footnote{The term ``red-shift'' comes from the fact that light of low frequency in the visible spectrum is perceived to be of the color red; thus, if the frequency of a given light wave is diminished, we say it is red-shifted.} with respect to what it should be, with the only possible explanation being that a Doppler shift in the frequency of light had taken place, diminishing the frequency we observe in the incoming photons as they arrive. 

How could physicists and astronomers be sure that the light was red-shifted? As matter in other galaxies is composed by the same fundamental constituents as in our galaxy\footnote{Assuming that we do not live in a special place of the universe.}, the emission lines of Hydrogen (as well as that of every other element in the periodic table) are also the same. These are produced by electrons' lowering their energy level inside an atom or molecule, emitting photons in the process, with definite frequencies given by the difference between the energy levels of the atom or molecule. The high abundance of light elements like Hydrogen or Helium within galaxies allowed astronomers to obtain the corresponding Doppler shift by comparing the observed spectra of light received from other galaxies with the known emission/absorption lines of atoms, and determining the frequency shift necessary to convert one of them into other.

The average velocity with which galaxies move away from us was found to be well described by a simple expression, the Hubble-Lema\^{i}tre law
\be
v_{\rm recessional} = H_0 D,
\ee
where $D$ is the (physical) distance between the observed Galaxy and the Milky Way. In this equation, $H_0$ was introduced as the observed ratio between speed and distance of the observed galaxy with respect to the observer. Initially, $H_0$ was known as \textit{Hubble's constant}, but nowadays it is more commonly presented as \textit{Hubble's parameter}, because the physical quantity relating $v_{\rm recessional}$ and $D$ has been found to have evolved throughout the history of our universe. Although the values originally obtained by Edwin Hubble and Georges Lema\^{i}tre for $H_0$ were an order of magnitude greater than the currently determined value by experiments of this decade~\cite{Ade:2015xua,Riess:2016jrr,Aghanim:2018eyx}\footnote{Currently, there is statistical tension between the value of $H_0$ obtained from measurements in the local universe and that inferred from the CMB. There is ongoing debate on whether the cause of the discrepancy is a physical phenomena or systematic errors in the data acquisition.}, the observation that our universe was expanding withstood the test of time. Thus, studying a time-dependent metric for our universe became of paramount relevance. 

If we assume that no particular place in the universe is intrinsically special, then on the largest scales, where all the relevant quantities describing the evolution of the universe are averages, our universe must be spatially isotropic and homogeneous. In the framework of General Relativity, this means we should be able to write the metric of spacetime as
\be \label{FLRW}
ds^2 = g_{\mu \nu} dx^{\mu} dx^{\nu} = - c^2 dt^2 + a^2(t) \gamma_{ij} dx^i dx^j
\ee
where $t$ is a coordinate representing the physical time a local observer would measure with his clock as they travel through the universe at a fixed spatial coordinate. It is conventional to call the coordinates $x^i$ in~\eqref{FLRW} ``co-moving coordinates''. Conversely, given a metric one can compute distances at a fixed time by integrating the spatio-temporal interval $ds$ along a curve connecting two points. For instance, if we have a curve $\mathcal{C}$ parametrized by a parameter $\lambda \in [0,1]$ through $x^i(\lambda)$, its length is given by
\be
L[\mathcal{C}] = \int_{\mathcal{C}} ds = a(t) \int_0^1 \sqrt{\gamma_{ij} \frac{dx^i}{d\lambda} \frac{dx^j}{d\lambda} } d\lambda,
\ee
where we have used the metric of~\eqref{FLRW}.

On the other hand, $\gamma_{ij}$ defines the spatial part of the metric, which characterizes how ``curved'' space is. Homogeneity and isotropy impose that the underlying manifold\footnote{The mathematical nomenclature given to topological spaces that admit continuously differentiable mappings from and to $\mathbb{R}^n$. For a physically motivated overview of these spaces, see the classic General Relativity textbooks~\cite{Misner:1974qy,Wald:1984rg}.} be characterized by a single constant number: its curvature. This defines three possible metrics
\be
\gamma_{ij} = \delta_{ij} + \frac{k }{1 - k (x_i x^i)} x_i x_j
\ee
depending on the sign of $k$. If $k$ is positive one says the space is \textit{spherical}, while if $k$ is negative one says the space is \textit{hyperbolic}\footnote{The names ``spherical'' and ``hyperbolic'' are given because the corresponding manifolds can be realized by embedding a sphere and a hyperboloid in a higher-dimensional space respectively, and studying the induced metric on those hypersurfaces.}. If $k=0$, one says the space is \textit{Euclidean} or \textit{flat}. This metric is usually written in polar (spherical) coordinates $(r,\theta,\phi)$, as it becomes diagonal and one can write
\be
ds^2 = - c^2 dt^2 + a^2(t) \gamma_{ij} dx^i dx^j = -c^2 dt^2 + a^2(t) \left[ \frac{dr^2}{1 - kr^2} + r^2 d\Omega^2 \right],
\ee
where $d\Omega^2 = d\theta^2 + \sin^2 \theta d\phi^2$ is the angular differential of spherical coordinates. As advertised, this metric is manifestly isotropic. As we have chosen a set of coordinates centered at $r=0$, the metric is not manifestly homogeneous. However, at \textit{any} spacetime point we may choose polar (spherical) coordinates centered at that point and the metric will have the exact same form, because the underlying manifold \textit{is} exactly homogeneous.

This is all we can say without specifying the dynamics that the spacetime metric~\eqref{FLRW} undergoes as the universe evolves in time. However, with what we have discussed so far, we have enough tools to enter the description of its evolution. As we are about to enter a purely relativistic setting, which we will not leave for the remainder of this thesis, we set units such that $c = 1$; i.e., we will be measuring time in terms of lengths or vice-versa.

\section{The dynamics and content of our Universe}

According to Einstein's theory of General Relativity, what one perceives daily as the gravitational field is nothing more than a side-effect of us standing ``still'' on a curved spacetime\footnote{This is stressed by the \textit{Equivalence principle}: a freely falling observer feels (locally) no gravitational effects, and the physics they describe corresponds to that of a ``classical'' inertial observer.}. Thus, the dynamics of the gravitational field is actually that of the spacetime curvature, or equivalently, of the spacetime metric $g_{\mu \nu}$\footnote{One calls ``metric'' to $ds^2 = g_{\mu \nu} dx^\mu dx^\nu$ and $g_{\mu \nu}$ interchangeably.}. The equation of motion for the metric is usually written in terms of the \textit{Ricci} tensor
\be
R_{\mu \nu} \equiv \partial_\lambda \Gamma^{\lambda}_{\mu \nu} - \partial_\nu \Gamma^{\lambda}_{\mu \lambda} + \Gamma^\lambda_{\lambda \rho} \Gamma^{\rho}_{\mu \nu} - \Gamma^\rho_{\mu \lambda} \Gamma^{\lambda}_{ \nu \rho}
\ee
where $\Gamma^{\mu}_{\alpha \beta}$ are the \textit{Christoffel symbols}
\be
\Gamma^{\mu}_{\alpha \beta} \equiv \frac{1}{2} g^{\mu \lambda} ( \partial_\alpha g_{\beta \lambda} + \partial_\beta g_{\alpha \lambda} - \partial_\lambda g_{\alpha \beta} ).
\ee
It is also helpful to define $R \equiv R^\mu_\mu$.

Then, the Einstein equation may be written as
\be \label{Einstein}
R_{\mu \nu} - \frac{1}{2} R g_{\mu \nu} = 8 \pi G T_{\mu \nu},
\ee
where $G$ is Isaac Newton's gravitational constant, and $T_{\mu \nu}$ is the stress-energy tensor associated to the objects that inhabit spacetime. All forms of energy and matter have a definite stress-energy tensor, but we will be particularly interested in that of perfect fluids, with which we will be able to describe the matter content of our universe. Namely, a perfect fluid has
\be
T_{\mu \nu} = (\rho + p) u_\mu u_\nu - p g_{\mu \nu}
\ee
where $\rho$ is its density, $p$ its pressure, and $u^\mu$ its 4-velocity.

What is important about~\eqref{Einstein} is that this equation not only applies for local patches of spacetime filled with some energy and matter, but it should\footnote{This is an assumption, as gravity has not been thoroughly tested at large (super-galactic) scales. However, the fact that the final result will be consistent with CMB and LSS observations is nothing short of yet another success of Einstein's General Relativity.} also apply for the evolution of the entire universe. Thus, as at sufficiently large scales we may describe our universe's spacetime through~\eqref{FLRW}, we can insert it in Einstein's equation~\eqref{Einstein} and derive the equations of motion for $a(t)$, the scale factor of our universe, assuming it is inhabited by perfect fluids. One obtains
\bea
\left( \frac{\dot a}{a} \right)^2 &=& \frac{8 \pi G}{3} \sum_{i} \rho_i - \frac{k}{a^2}, \label{Friedmann-1} \\
\frac{\ddot a}{a} &=& -\frac{4\pi G}{3} \sum_i ( \rho_i + 3p_i), \label{Friedmann-2}
\eea
where the sum over $i$ is over the different fluids that inhabit spacetime. These equations, known as the \textit{Friedmann} Equations, completely describe the evolution of the scale factor $a(t)$. Assuming the different species $i$ do not interact, and that they are homogeneously and isotropically distributed in space, we can also write a continuity equation
\be \label{eq-cont}
\dot \rho_i + 3 \frac{\dot a}{a} (\rho_i + p_i) = 0,
\ee
which may be used alongside~\eqref{Friedmann-1} to make~\eqref{Friedmann-2} redundant.

Furthermore,~\eqref{eq-cont} completely determines how the density and pressure of the fluid evolve over time given an equation of state. There are three equations of state that are typically considered to describe the content of our universe:
\begin{enumerate}
\item Pressureless Matter: $p = 0$

This is what one usually refers to as ``matter'' when studying cosmology. It consists of all particles with pressure much smaller than their density $|p| \ll \rho$, which is the case for any gas of non-relativistic particles. Both Dark matter, that does not interact with photons (light), and Baryonic matter\footnote{In cosmology it is usual to refer to all visible matter by ``Baryonic'' matter, including other hadrons and leptons.}, that interacts with the electromagnetic field, are thought of as pressureless in this context.

According to the continuity equation~\eqref{eq-cont}, this type of fluid evolves as $\rho(t) a^3(t) = \rho(t_0) a^3(t_0) \iff \rho \propto a^{-3}$.

\item Radiation: $p = \frac{1}{3} \rho$

A gas of relativistic particles has an equation of state given by $p = \frac{1}{3} \rho$, and we will call any matter/energy species that obeys this equation of state ``radiation''. Photons, the (nearly massless) Neutrinos, and the (hypothetical) Gravitons constitute radiation.

According to the continuity equation~\eqref{eq-cont}, this type of fluid evolves as $\rho(t) a^4(t) = \rho(t_0) a^4(t_0) \iff \rho \propto a^{-4}$.

\item Dark Energy: $p = - \rho$

Finally, there is a mysterious form of energy, known as Dark Energy, that is currently favoured by data analysis to explain the observation that our universe is undergoing a phase of accelerated expansion. The requirement on this type of energy, for it to explain the acceleration $\ddot a > 0$, is that its energy density should be constant over time $\rho \propto a^0$, which as per the continuity equation~\eqref{eq-cont} is equivalent to $\rho = - p$, i.e., negative pressure.

Thus far, there is only one natural candidate for this type of energy: the energy density associated to the quantum fluctuations of the vacuum of empty space; that is to say, the energy associated to the ground state of the corresponding quantum system, which by definition contains no particles. This energy is nonzero due to Heisenberg's uncertainty principle. However, the natural energy scale predicted by Quantum Field Theory is 120 orders of magnitude greater than the value implied by cosmological observations. 

The simplest solution to this problem is to include an extra term in Einstein's Equations, dubbed a ``Cosmological Constant'', that compensates for the extremely high energy density of vacuum fluctuations. However, this means that the aforementioned cosmological constant would have to be fine-tuned to 120 orders of magnitude, which is not a satisfactory explanation. As a consequence of this fact, the nature of our universe's Dark Energy has been a subject of intense debate that has yet to be settled.

\end{enumerate}

Applying the continuity equation to each fluid, one arrives at the most common formulation of the Friedmann equation~\eqref{Friedmann-1}:
\be \label{Friedmann}
\frac{H^2}{H_0^2} = \left[ \Omega_{r,0} \left(\frac{a_0}{a}\right)^4 +  \Omega_{m,0} \left(\frac{a_0}{a}\right)^3 +  \Omega_{k,0} \left(\frac{a_0}{a}\right)^2 +  \Omega_{\Lambda,0} \right],
\ee
where the $_0$ subscripts indicate that the quantities are evaluated at a given time $t_0$ (usually set to be the present time). Here we have defined:
\begin{enumerate}
\item the Hubble parameter:
\be
H \equiv \frac{\dot a}{a},
\ee

\item the fractional density parameters $\Omega_{a}$ ($a \in \{r,m,k,\Lambda\}$):
\be
\Omega_{a} \equiv \frac{8\pi G \rho_a}{3 H^2},
\ee

\item and the curvature ``energy density'':
\be
\rho_k \equiv \frac{- 3k}{8\pi G a^2}.
\ee

\end{enumerate}

Now that we have introduced the basic constituents of our universe and established the dynamics the scale factor $a$ follows, we have the basic ingredients to describe how our universe came to be the way we nowadays see it, and how we can look into the past to search for the primordial seeds that set the initial conditions for galaxies and structure to emerge.

\section{A short history of our universe}

Nowadays, the leading model for the description of our universe, i.e., the one that best fits the data and explains the physics behind them, is known as $\Lambda$CDM. The name is an abbreviation of Dark Energy (symbolized by the letter $\Lambda$) plus cold Dark Matter (CDM). Thus far, every cosmological observation has been consistent with curvature $k = 0$, and therefore, the curvature ``energy density'' $\Omega_{k,0}$ is usually neglected and set by hand to zero.

Then, starting from today, we can use equation~\eqref{Friedmann} to go backwards in time and trace the evolution of the universe to understand how it became what we see today. This is because~\eqref{Friedmann} is completely determined by values of quantities at this time in the history of our universe.

\begin{figure}[t!]
\begin{center}
\includegraphics[scale=1]{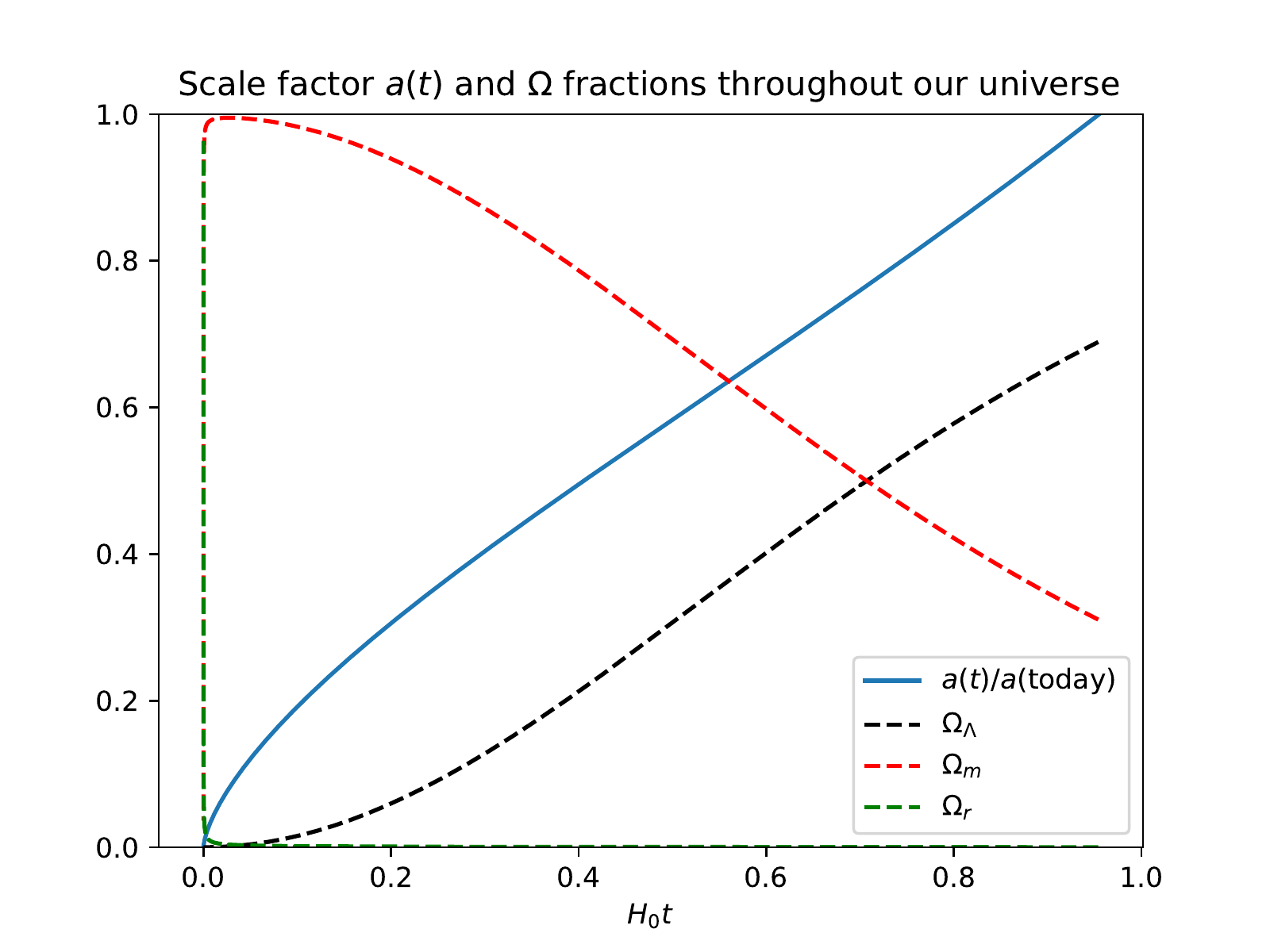}
\caption{Evolution of the scale factor $a$ and $\Omega$ over the history of our universe. As shown by the green dashed line, radiation was only relevant for the expansion of the universe at very early times.}
\label{fig:sclf}
\end{center}
\end{figure}

Using the best-fit cosmological parameters reported by the Planck Collaboration in 2018~\cite{Aghanim:2018eyx,Akrami:2018vks} to the $\Lambda$CDM model, including Baryon Acoustic Oscillations (BAO) data, we have $H_0 = 67.7  \,\rm{km } \, \rm{s}^{-1} \, \rm{ Mpc }^{-1}$, $\Omega_{m,0} = 0.31$, and $\Omega_{\Lambda,0} = 0.69$. $\Omega_{r,0}$ can be determined from the CMB average temperature and the Hubble parameter $H_0$ through the formula~\cite{Ade:2013zuv}:
\be
\Omega_{r,0} = \frac{\rho_\gamma + \rho_\nu}{\rho_{c,0}} = \frac{8 \pi^3 k_B^4 G}{45 \hbar^3 c^5 H_0^2} \left(1 + 3.046 \frac{7}{8} \left(\frac{4}{11} \right)^{4/3} \right) T_0^4 \approx 9 \times 10^{-5},
\ee
where we have used that $\rho_\gamma$ is the energy density of radiation from a blackbody at temperature $T_0$. Here $k_B$ is Boltzmann's constant, $\hbar$ is the reduced Planck's constant, and $T_0$ is the average CMB temperature today. With these quantities in hand, the scale factor evolves as shown in Fig.~\ref{fig:sclf} in standard Big-Bang cosmology: starting from an extremely small value of the scale factor $a(t)$, often taken to zero, the universe expands rapidly, initially pushed by radiation, but soon after becomes matter-dominated, slowing down the rate of expansion, but expanding all the same. This rate would have kept diminishing if Dark Energy were not present, which became dominant some 3 billion years ago and is now driving an accelerated expansion of our universe.

\begin{figure}[t!]
\begin{center}
\includegraphics[scale=0.15]{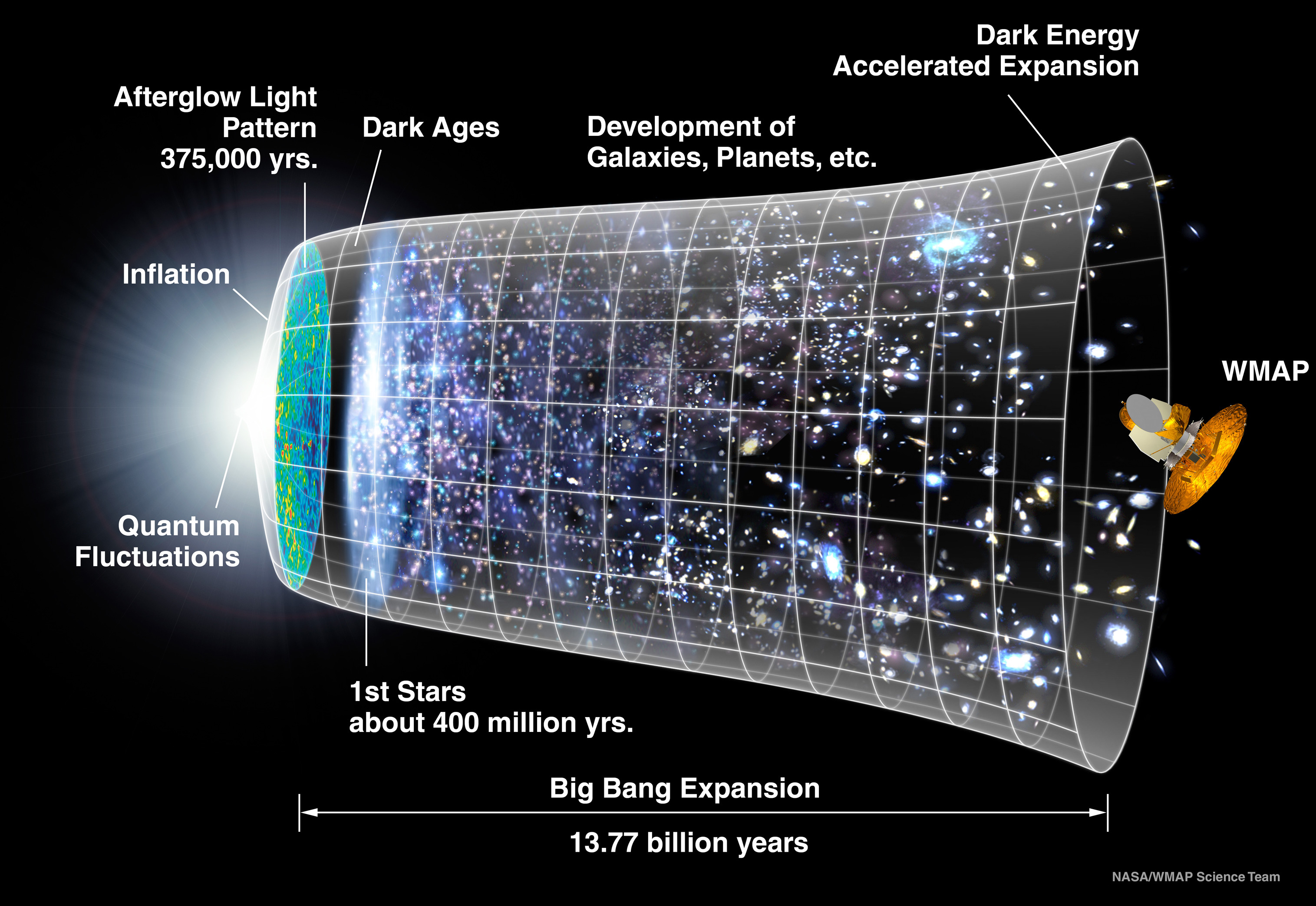}
\caption{Illustrative depiction of the history of the universe, starting from the primordial quantum fluctuations, going through the hot Big-Bang, and finally evolving to the formation of structure and the universe we presently observe. \footnotesize{Credits: National Aeronautics and Space Administration (NASA)/} \footnotesize{Wilkinson Microwave Anisotropy Probe (WMAP) Science Team} }
\label{fig:WMAP}
\end{center}
\end{figure}

Most formation of structure occurred in the matter-dominated phase, as represented graphically in Figure~\ref{fig:WMAP}. This is of paramount importance because an overly fast expansion could have been capable of preventing the occurrence of the matter distribution as we now see it. This domination would have been much shorter without Dark Matter, which accounts for $\sim 85$\%~\cite{Aghanim:2018eyx,Akrami:2018vks} of all pressureless matter.

Starting from a hot ``Big-Bang'', the epochs in the history of our universe can be summarised as follows, as the temperature $T \propto a^{-1}$ decreases: 
\begin{itemize}
\item Baryogenesis: The knowledge we have on fundamental physics, i.e., the Standard Model and Quantum Field Theory, implies that anti-particles exist, and gives no reason for there to be any less antimatter than matter. Thus, unless the initial conditions of the universe favored matter over antimatter, a physical mechanism is required to obtain a universe with the currently observed baryon-to-photon ratio. 

\item Electroweak phase transition: Once the thermal energy scale $k_B T$ drops below the mass of the Higgs boson $\sim 100$ GeV, particles become massive due to the Higgs mechanism and thus the $W$ and $Z$ bosons become massive, singling out a transition that distinguishes the photon, carrying the electromagnetic force, from the weak force carriers. 

\item QCD phase transition: Presumably, quarks were once weakly interacting particles as the thermal energy scale was above 150 MeV. Below this energy, quarks and gluons form bound states, constituting baryons and mesons.

\item Dark Matter and Neutrino decouplings: Neutrinos are only coupled to the rest of the Standard Model through the weak force carriers, the $Z$ and $W$ bosons. This implies that below $k_B T \sim 0.8 MeV$ neutrinos decouple from thermal equilibrium. Similarly, one expects dark matter to have decoupled from the other forms of energy by this point, as it must interact very weakly with ordinary matter.

\item Electron-positron annihilation: After the neutrinos decouple, the remainder of antimatter is annihilated in electron-positron processes, transferring their energy to photons. This happens at about 6 seconds into the history of our universe.

\item Nucleosynthesis: Light elements are formed around 3 minutes after the Big Bang. This happens at about $k_B T \sim 100 \, \rm{keV}$.

\item Matter-Radiation equality: At about 60000 years after the Big Bang, the energy density of pressureless matter starts to dominate over radiation. Although the universe is expanding, the rate of the expansion is decelerating $\ddot a < 0$, which makes the present-day acceleration an obvious signal that Dark Energy is present.

\item Recombination and Photon decoupling: After the universe cools down enough so that the disintegration of a Hydrogen atom by an incident photon becomes highly unlikely, \textit{Recombination} happens, where electrons and protons recombine to form stable Hydrogen atoms. Then, since photons are no longer able to interact with free electrons, they decouple and stream freely throughout the universe. It is this radiation that we now observe as the Cosmic Microwave Background.

\end{itemize}

Nonetheless, as it may be evident from the first stage (Baryogenesis), the universe requires initial conditions to begin with. Apart from hand-picking the initial conditions, the only other option, besides adding an extra previous stage to the universe's history, is to give it random initial conditions, in an appropriate sense, and try to find a patch within it where our observed universe developed. However, as it turns out, without a prior stage of accelerated expansion in the evolution of the universe, or another physical mechanism that accomplishes the same effects, it is extremely unlikely that our universe could have happened as it has.

The main issues that stand in the way were historically dubbed as the ``flatness problem'' and the ``horizon problem.'' The first is based on the observation that we nowadays see a negligible energy density associated to curvature. However, if we allow for a small contribution to it, which is technically natural, as there is no reason for it not to be there, we will now have
\be \label{Omega}
\Omega - 1 \equiv \Omega_r + \Omega_m + \Omega_\Lambda - 1 = \frac{k}{H^2 a^2} \neq 0.
\ee
Both during radiation- and matter-domination epochs $H^2$ decays with $a$ equally or faster than $a^{-3}$, which in turn implies that $|\Omega - 1|$ must have grown in time throughout most of the history of our universe. When numbers are plugged in, this would require an extremely small initial curvature energy density to maintain curvature below the observational constraints today. While nonetheless one could choose the initial value of that energy density so that it satisfies this requirement, it is a `fine-tuning' problem, as it seems highly unlikely that those numbers occurred naturally.

On the other hand, the horizon problem is one of causality: if we follow the path each CMB photon travelled from the surface of last scattering until today, and take a look at the dynamics that preceded their emission, we will find that the regions of the universe from which they were emitted could not have possibly communicated between them to achieve thermal equilibrium and thus source a uniform photon background radiation of the same average temperature, as we observe it today. That is, in the same way curvature is too small today, the universe is also too homogeneous for it to have started from the hot ``Big-Bang'' we described earlier. To solve these problems, an earlier phase in the history of our universe was postulated: Inflation. We will be specific about how Inflation starts and takes place in Chapter~\ref{chap:inflation}; for now, we only describe its consequences on our universe and how it solves the aforementioned problems.

Inflation posits that our universe began with a period of rapidly accelerating expansion, where the scale factor grew so fast that points in space that were once able to communicate with one another are pushed away in a manner that they are no longer able to ``talk'' to each other. This solves the horizon problem, as all the regions of space we now see in the CMB would have once been very close together, thus allowing them to have reached thermodynamic equilibrium, for then Inflation to push them away from each other and consequently appear as distant regions in our sky, but with the same thermodynamical properties.

The flatness problem is solved even more straightforwardly: in an accelerated expansion phase $\ddot a > 0$, $(aH)^{-1}$ is a decreasing function of time, and therefore the relative curvature energy density~\eqref{Omega} can be decreased. Thus, if Inflation lasts long enough, this energy density can be diminished to arbitrarily low values, and thus setting an adequate initial condition for our late-time universe. 

As we will see later, Inflation also accounts for the primordial fluctuations that generated the anisotropies in the CMB and seeded the initial conditions for the formation of structure in our universe, along with determining all of the cosmological parameters we measure today.

\begin{figure}[t!]
\begin{center}
\includegraphics[scale=0.35]{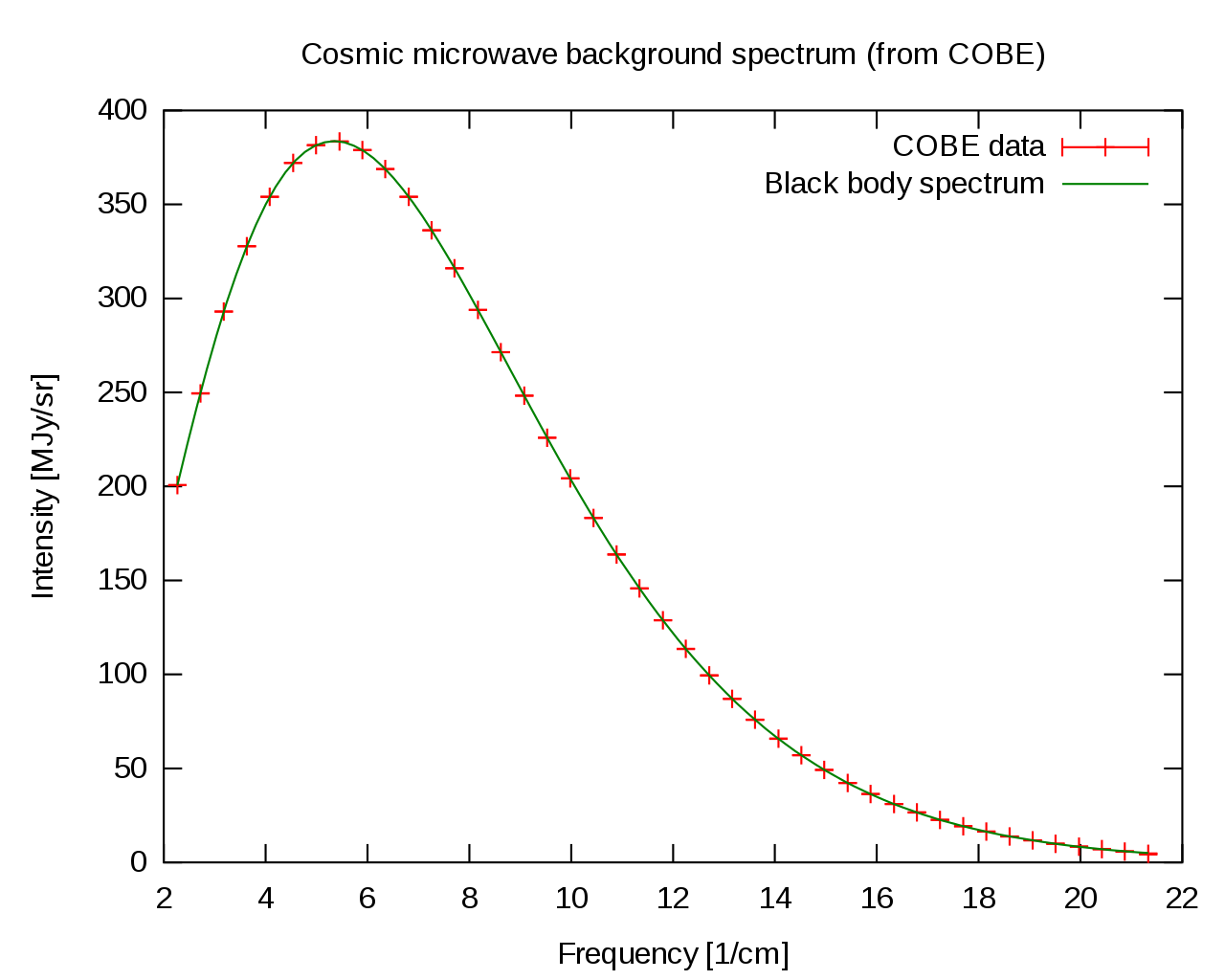}
\caption{Measurement of the Cosmic Microwave Background Spectrum by the COBE FIRAS Instrument. \footnotesize{Credits: National Aeronautics and Space Administration (NASA)/} \footnotesize{Cosmic Background Explorer (COBE), Far Infrared Absolute Spectrophotometer (FIRAS) team.} }
\label{fig:CMB-black}
\end{center}
\end{figure}

But how are those parameters obtained in the first place? The most reliable source of information we have about the evolution of the universe is the Cosmic Microwave Background (CMB). As we will see in the next section, the statistics of the CMB temperature fluctuations contain enough information to allow for a reconstruction of our universe's history. Large-Scale Structure (LSS) surveys, i.e., maps of the distribution of matter at super-galactic scales of our universe, may also be employed towards this end, as long as the wavenumber of the (typically averaged) fluctuations stays below $k_{NL} \sim 0.3 h {\rm Mpc}^{-1}$, so that non-linear effects in the evolution of the matter density profile may be treated perturbatively in the framework of an effective field theory~\cite{Carrasco:2012cv}.\footnote{$h$ is defined as $H_0/(100 \, \rm{km} \, \rm{s}^{-1} \, \rm{Mpc}^{-1} )$.}

\section{The Cosmic Microwave Background}

Arguably, the CMB has been the predominant source of information on the evolution of our universe for the past three decades. The steady stream of photons following the most precise blackbody spectrum ever measured, currently at a characteristic temperature of $T_0 = 2.725 \, 48 \pm 0.000 \, 57 \, {\rm K}$~\cite{Fixsen:2009} (see FIG~\ref{fig:CMB-black}), has provided a reliable and thorough mechanism to reconstruct the history of our universe. How is this so, if such a spectrum only provides a single number, the background temperature? It turns out, that upon a closer look, the Cosmic Microwave Background contains fluctuations about the average temperature that are four orders of magnitude smaller than their central value. Moreover, current observations of these fluctuations show that they are Gaussianly distributed, and that their correlations are scale-dependent, in the sense that different angular separations on the sky have different correlations.

\begin{figure}[t!]
\begin{center}
\includegraphics[scale=0.55]{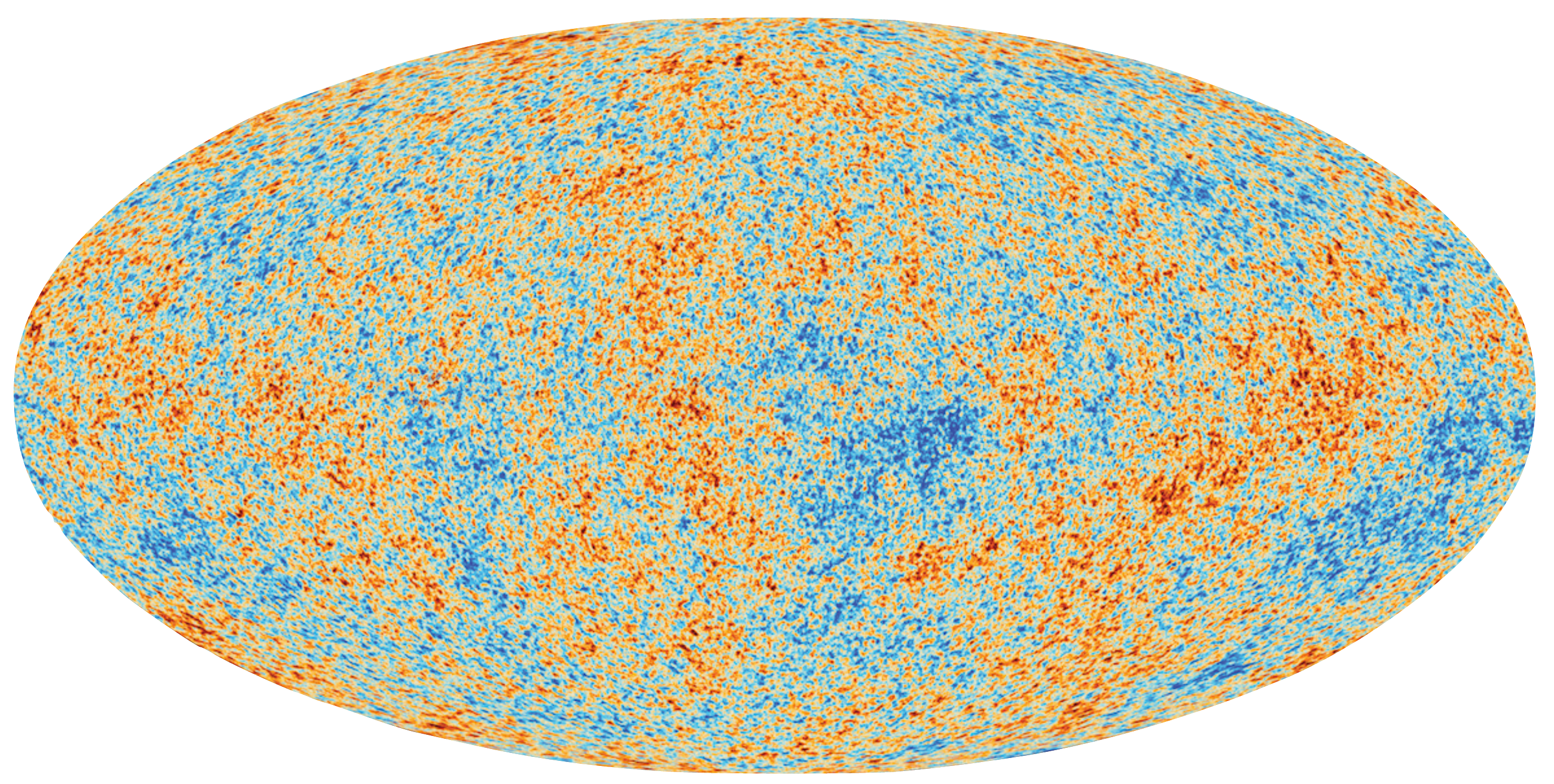}
\caption{Temperature fluctuations map from the 2018 Planck Telescope data release. \footnotesize{Credits: European Space Agency (ESA)/Planck Collaboration} }
\label{fig:CMB}
\end{center}
\end{figure}

Figure~\ref{fig:CMB} shows the full-sky map of temperature fluctuations presented by the Planck Collaboration on 2018~\cite{Akrami:2018vks} after processing data from the 53 months that the satellite recorded. The interpretation of such a map is simple: from each direction on the sky, photon radiation is received with a spectrum alike to that of a blackbody of temperature $T$, and the map is the result of subtracting $T_0$, the average CMB temperature, to the temperature inferred from each cell in the sky.

The standard analysis of the CMB is to take the temperature fluctuations as a function on the unitary sphere, and to assign a value of the fluctuation to each direction. That is to say, the function under study is $\Delta T = \Delta T(\n)$. Then the signal is decomposed into spherical harmonics:
\be
\frac{\Delta T(\n)}{T_0} = \sum_{\ell=0}^{\ell_{\rm max}} \sum_{m=-\ell}^\ell a_{\ell m} Y_{\ell m}(\n),
\ee
where the $a_{\ell m}$ coefficients are to be determined by observations and $Y_{\ell m}$ are spherical harmonics. In this notation, the $\ell$'s denote the different multipoles of the expansion and $m$ is a label for the $(2\ell +1)$ independent modes of fluctuation at that multipole. $\ell_{\rm max}$ is usually set by the angular resolution of the experiment, because no fluctuations at smaller scales (larger $\ell$) will be detected as they will be averaged out. 

Because of statistical isotropy, the correlations over an ensemble average are given by
\be
\langle a_{\ell m} a^*_{\ell' m'} \rangle = \delta_{\ell \ell'} \delta_{m m'} C_\ell.
\ee
This is equivalent to writing down the 2-point correlation function in terms of the temperature fluctuations
\be \label{2-point-temp}
\left\langle \frac{\Delta T(\n)}{T_0}  \frac{\Delta T(\n')}{T_0}  \right\rangle = \frac{1}{4\pi} \sum_\ell (2\ell + 1) C_\ell P_\ell(\mu = \n \cdot \n').
\ee
Under the assumption of Gaussianity, the statistics of the CMB are completely described by the $C_\ell$ coefficients, or equivalently, by the 2-point function~\eqref{2-point-temp}. As no deviation from Gaussian statistics has been confirmed so far, this has been the principal subject of study to try and acquire information about the evolution of the universe. Polarization maps are also of interest, and provide extra information with which to place tighter constraints on the parameters of the cosmological model at hand.

\begin{figure}[t!]
\begin{center}
\includegraphics[scale=0.85]{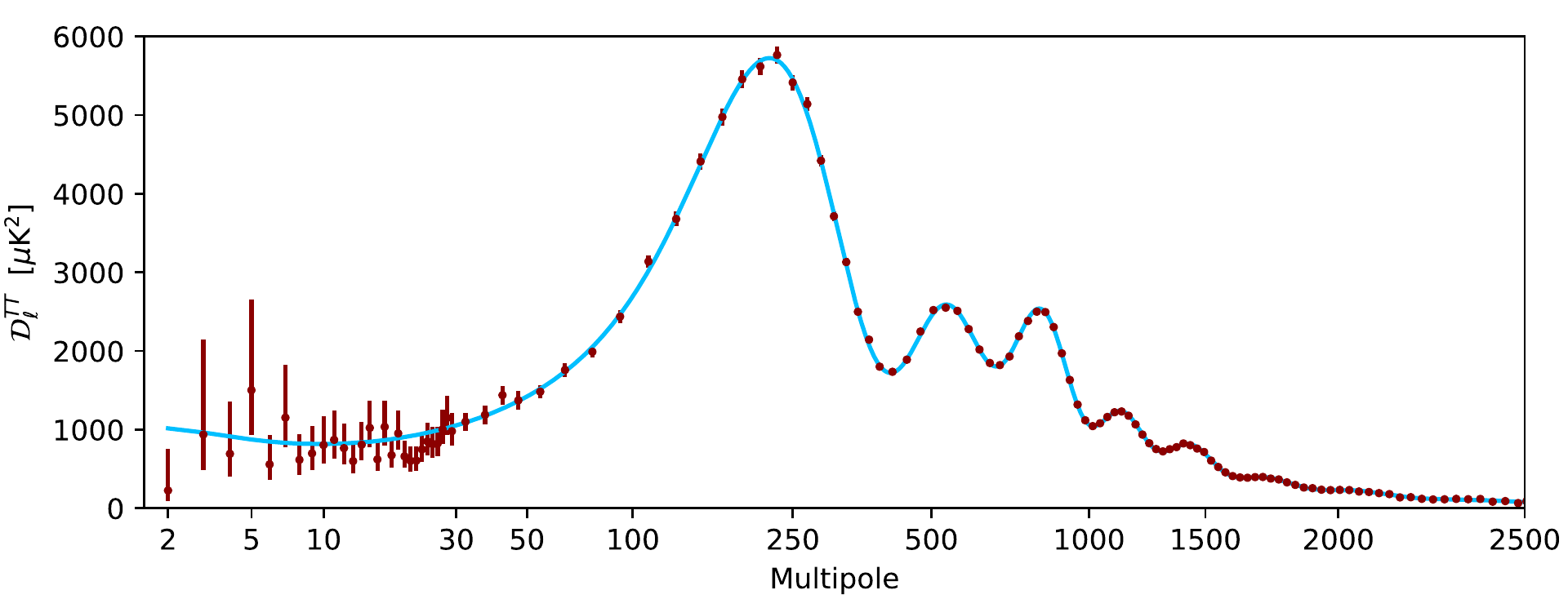}
\caption{Temperature-Temperature correlations at different angular scales on the sky, i.e., as a function of the multipoles $\ell$. From these results alone it is possible to fit and constrain the cosmological parameters of the $\Lambda$CDM model. \footnotesize{Credits: European Space Agency (ESA)/Planck Collaboration}~\cite{Aghanim:2018eyx}. }
\label{fig:CMB-spectra}
\end{center}
\end{figure}

Accordingly, measurements of the CMB temperature and polarization correlations as a function of angular scale have been of great interest and become increasingly precise over the last decades. On 2018, the Planck Collaboration~\cite{Aghanim:2018eyx} reported Figure~\ref{fig:CMB-spectra} as the result for the temperature angular correlations on the sky, in terms of $\mathcal{D}_\ell = \frac{\ell (\ell + 1)}{2\pi} C_\ell$, which allowed them to determine the $\Lambda$CDM model parameters with great accuracy.

However, we have yet to describe the physics of how the $\Lambda$CDM model is able to account for all the features within the 2-point correlation function in Figure~\ref{fig:CMB-spectra}. As this is not our main topic of interest, we will only give a brief overview of how this takes place, following~\cite{Weinberg:2008zzc}. The starting point for our universe, within the realm of standard cosmology, is to assume that at the initial time slice there are inhomogeneities on the gravitational potential field. We will later justify the presence of these fluctuations from an inflationary perspective; for now, we will only assume their existence as a source for the inhomogeneities of our universe.

To derive the effects of these primordial fluctuations, the usual procedure is to linearize the equations of motion about the homogeneous background solution, assuming the inhomogeneities are small. For our present purposes, this is precisely the case. If we let $\mathcal{R}_{\k}$ stand for the Fourier transform of the primordial gravitational potential, linear equations of motion on a homogeneous and isotropic background must yield
\be
\frac{\Delta T(\n)}{T_0} = \frac{1}{(2\pi)^3} \int d^3 k \mathcal{R}_{\k} e^{i\k \cdot \n r_L } F(k,\n \cdot \hat{k}),
\ee
where $F(k,\n \cdot \hat{k})$ relates the amplitude of the primordial fluctuations $\mathcal{R}$ to the temperature fluctuations. Here, $r_L$ is the co-moving radius of the surface of last scattering, as it is at this distance from us that photons will be emitted, and hence whence they will acquire their differences in energy. If one wants to obtain the ``transfer'' function $F(k,\n \cdot \hat{k})$ exactly, one has little choice but to numerically solve the Boltzmann transport equations that describe the evolution of the photon energy density.

However, after making appropriate considerations and approximations, it is possible to arrive at
\be
\frac{\Delta T(\n)}{T_0} = \frac{1}{(2\pi)^3} \int d^3 k \mathcal{R}_{\k} e^{i\k \cdot \n r_L } \left( F(k) + \n \cdot \hat{k} \, G(k) \right).
\ee
This is derived in~\cite{Weinberg:2008zzc} within the hydrodynamic approximation, in which the photons, along with the rest of matter evolve together as superimposed interacting fluids. %However, as a leading order approximation, one can assume that the gravitational field perturbations at last scattering are dominated by dark matter, neglecting the ratio of visible (baryonic) matter over dark matter.
From this approximation, and interpolating between the results in the short- and long- wavelength regime, a qualitatively correct result can be derived, which can be fitted to the available experimental data:
\bea
F(k) &=& \frac{e^{-\tau_{\rm r}}}{5} \left[3 R_L \mathcal{T} \! \left( \frac{k d_T}{a_L} \right)  - \mathcal{S} \! \left( \frac{k d_T}{a_L} \right) \frac{e^{-k^2 d_D^2/a_L^2}}{(1+R_L)^{1/4}} \cos \left( \frac{k d_H}{a_L} + \Delta \! \left( \frac{k d_T}{a_L} \right) \right) \right] \! , \label{F-def} \\
G(k) &=& - e^{-\tau_{\rm r}} \frac{\sqrt{3} e^{-k^2 d_D^2/a_L^2}}{5(1+R_L)^{3/4}}  \mathcal{S} \! \left( \frac{k d_T}{a_L} \right) \sin \left( \frac{k d_H}{a_L} + \Delta \! \left( \frac{k d_T}{a_L} \right) \right) \! . \label{G-def}
\eea
These functions involve a number of important quantities, which we now describe:
\begin{itemize}
\item Transfer functions: $\mathcal{T}$ is the matter transfer function, which accounts for the amplitude of the fluctuation in the matter density at each scale, while $\mathcal{S}$ and $\Delta$ are the functions that account for the transfer to the fluctuations in radiation energy density and its scalar velocity potential. Analytic fits to these functions do exist, but in general it is necessary to determine them numerically.
\item $\tau_{\rm r}$ is the optical depth of the reionized plasma at redshift $z_{\rm reion} = 7.64\pm 0.74$~\cite{Aghanim:2018eyx}, when ultraviolet light emitted by the first stars reionizes neutral hydrogen and releases free electrons. Accordingly, $e^{-\tau_{\rm r}}$ is the probability of CMB photons not being scattered by electrons in their way to us, which is exactly the fraction of photons that we do observe.
\item $a_L$ is the scale factor at last scattering. It is introduced so that all wavenumbers may be compared with lengths at the last scattering surface.
\item $d_T$ is a characteristic length scale of the transfer functions, given by $\frac{\sqrt{2} a_L}{a_{\rm EQ} H_{\rm EQ}}$. That is to say, it is the inverse of the co-moving Hubble radius $aH$ at matter-radiation equality, mapped to a physical length at the surface of last scattering.
\item $R_L$ is the ratio of baryonic matter energy density to photon energy density $R \equiv 3{\bar \rho}_B/4 \bar{\rho}_\gamma  $ at the last scattering surface. In equations~\eqref{F-def} and~\eqref{G-def}, the transfer functions $\mathcal{T}$ and $\mathcal{S}$ are corrected by functions of $R_L$, increasing the matter contribution if the relative density $R_L$ is larger, and conversely, enhancing the radiation contributions if $R_L$ is small.
\item $d_D$ is a damping length scale that radiation experiences before and at the surface of last scattering. Two contributions enter this damping: the first is due to scattering of photons through the baryonic plasma, and the second is because fluctuations are averaged at last scattering because of the finite duration of the process. Both suppress short-wavelength contributions to the temperature fluctuations.
\item Finally, $d_H$ is the acoustic horizon distance at last scattering. This term is present as a result of density waves in the baryonic plasma with a sound speed of $c/\sqrt{3(1+R)}$, thus generating correlations between features at a distance of the size of the horizon, because by definition the acoustic horizon is the distance between two points receiving, from opposite directions, a signal emitted at a given point in the universe's initial time slice. Therefore, this enhances the amplitude of fluctuations in the temperature map of wavenumber corresponding to $d_H$, as points separated by this distance will receive sound waves originating from the same primordial spacetime event and will be therefore correlated.

\end{itemize}

As the observed statistics of the CMB are Gaussian, $\mathcal{R}$ is usually assumed to be a realization of a Gaussian random field, with a power spectrum $P_{\mathcal{R}}(k)$ given by
\be \label{power-R}
\langle \mathcal{R}_\k \mathcal{R}_{\k'} \rangle = (2\pi)^3 \delta^3 (\k + \k') P_{\mathcal{R}}(k) = (2\pi)^3 \delta^3 (\k + \k') \frac{2\pi^2 A_s}{k^3} \left( \frac{k}{k_*} \right)^{n_s - 1},
\ee
and thus the 2-point correlation~\eqref{2-point-temp} is equal to
\be
\left\langle \frac{\Delta T(\n)}{T_0}  \frac{\Delta T(\n')}{T_0}  \right\rangle = \frac{A_s}{4\pi} \int \frac{d^3 k}{k^3} \left( \frac{k}{k_*} \right)^{n_s - 1} \!\!\!\!\!\! e^{i \k \cdot (\n - \n') r_L} \left( F(k) + \n \cdot \hat{k} \, G(k) \right) \left( F(k) - \n' \cdot \hat{k} \, G(k) \right)
\ee
from where the $C_\ell$'s can be computed and compared with experimental results. By fitting the observed power spectrum to this result one can infer all of the parameters in the $\Lambda$CDM model. Among the most famous results, Planck obtains~\cite{Aghanim:2018eyx} (see Table 1 therein):
\begin{align}
 \Omega_{m,0} &= 0.3147 \pm 0.0074,  &  H_0^2 &= 67.37 \pm 0.54,  & 100 \theta_{*} &= 1.04108 \pm 0.00031, \nn \\
  \tau_{\rm r} &= 0.0540 \pm 0.0074,  &  \ln(10^{10} A_s) &= 3.043 \pm 0.014, &  n_s &= 0.9562 \pm 0.0042,
\end{align}
where $\theta_*$ is the angular scale on the sky at which the acoustic oscillations are measured.

It is important to note that the quantity $\mathcal{R}$, which we previously defined as the primordial gravitational potential, may be identified as such depending on the choice of coordinates employed to describe spacetime, which in a general relativistic context is a \textit{gauge} freedom. $\mathcal{R}$, of course, is gauge invariant, because the temperature fluctuations $\Delta T$ and matter density fluctuations are observables which cannot depend on the gauge chosen to perform the computation. We will later specify how to compute $\mathcal{R}$ out of the inflationary background in Chaper~\ref{chap:inflation}. As it is this quantity that ultimately sources all\footnote{Primordial gravitational waves, which have yet to be detected, are sourced by tensor mode fluctuations that can be independent of $\mathcal{R}$.} of the observed inhomogeneities of our universe, it constitutes the initial condition we seek to determine.

We now proceed to briefly overview the other main observable that allows to gain information on $\mathcal{R}$ and constrain cosmological parameters.

%A final note in this section is in order regarding the assumption of Gaussianity, which is what we will tackle later on.

\section{Large-Scale Structure}

Although the CMB has been the most important source of information in the past decades, perhaps the most obvious evidence of the presence of primordial inhomogeneities is the fact that galaxies exist, and occupy specific positions in space. Therefore, it is natural to study the matter distribution of our universe to obtain information about the primordial fluctuations.

After recombination, baryons decouple from photons and effectively behave like Dark Matter in their dynamical evolution, and also play the same role in the formation of structure. Thus, one treats the fluctuations of both baryons and Dark Matter in a single quantity, the matter density contrast:
\be
\delta \equiv \frac{ \delta \rho_{DM} + \delta \rho_{B}}{{\bar \rho}_{DM}  + {\bar \rho}_{B} }
\ee
where ${\bar \rho}_{DM}$ and ${\bar \rho}_{B}$ are the average energy densities associated to Dark Matter and baryons respectively, while $\delta \rho_{DM}$ and $\delta \rho_{B}$ are their fluctuations about their average values. As long as these fluctuations are small, which will amount to studying long-wavelength fluctuations where the dynamics are well approximated by linear equations of motion, the matter density contrast will be linearly related to $\mathcal{R}$
\be
\delta(\k,t) = \frac{3}{5} \alpha(k) D(t) \mathcal{R}_\k,
\ee
where $D(t)$ is a linear growth factor derived from the fact that perturbations are on top of an expanding background, and $\alpha(k)$ is given by
\be
\alpha(k) = \frac{2 k^2 \mathcal{T}\left(\frac{k d_T}{a_L}\right)}{3 H_L^2 a_L^2 }.
\ee
In particular, this means that if the primordial fluctuations $\mathcal{R}$ are Gaussian, the matter density contrast $\delta$ is also Gaussianly distributed, with a power spectrum given by
\be
P_\delta(k) = \left( \frac{2 D(t_0)}{5 H_L^2 a_L^2} \right)^2 k^4 \left| \mathcal{T} \! \left( \frac{k d_T}{a_L}\right) \right|^2 P_{\mathcal{R}}(k),
\ee
at least well below momentum scales where the physics of the matter distribution becomes nonlinear $k \ll k_{\rm NL}$.

However, most of the structure we observe has undergone highly nonlinear processes: galaxies and the halos containing them are formed through the action of gravity, and as they are relatively compact objects, they involve density fluctuations of a high wavenumber where linear perturbation theory ceases to apply. Thus, $\delta$ cannot give the final result: it is only an initial condition for the growth of structure at smaller scales. To give a full treatment of these nonlinearities it is necessary to start sufficiently early in the matter-dominated era so that the perturbations are still small and the initial conditions be related to $\mathcal{R}$ linearly. Afterwards, numerical computations are usually required.

That does not amount to saying analytic progress has been overlooked. To account for nonlinear gravitational collapse, a typical approach is to model the process using insights from Newtonian gravity. For instance, it is common to estimate whether a distribution of mass will collapse or not depending on how much mass is inside a given radius. It has also proved useful to idealize the collapse dynamics as spherical, where the exact computation can be carried out to determine the final overdensity field starting from an initial over-density sourced by $\mathcal{R}$.

Nonetheless, at large scales (low wavenumber $k \ll k_{\rm NL}$) $\delta$ has observable effects in the nonlinear, short-wavelength dynamics. Consider, for instance, the halo number density $n[{\bar \rho}, \delta_S]$, which in principle depends on only two quantities: the average background density ${\bar \rho}_M$ and the local initial fluctuations $\delta_S$ set by short-scale fluctuations in $\mathcal{R}$. While it is true that whether a given matter distribution collapses or not into a halo depends mainly on the short-wavelength fluctuations that can give rise to regions with high mass density, long-wavelength perturbations $\delta_L$ also induce an extra contribution that adds to the average background density:
\be
n = n[{\bar \rho} + \delta_L, \delta_S] \sim n[{\bar \rho}, \delta_S] \times \left(1 + \frac{\partial \ln n}{\partial \delta_L} \delta_L \right) = n[{\bar \rho}, \delta_S] (1 + b_L \delta_L),
\ee
known as the linear bias $b_L$. Thus, given a model for halo formation $n[{\bar \rho}, \delta_S]$, one can study the correlations of the halo number density field with the long-wavelength fluctuations as a way of acquiring information of the primordial statistics.

With the advent of new LSS surveys, such as {\sc Lsst}~\cite{Lsst-web}, {\sc Euclid}~\cite{Euclid-web}, and {\sc Ska}~\cite{Bull:2018lat}, a new era of precision cosmology is beginning, with new opportunities for the study of fundamental physics. In particular, it presents the chance to constrain more tightly the statistics of $\mathcal{R}$, the initial condition of the inhomogeneities, and thus of all structure of our universe. How is this related to fundamental physics? Because $\mathcal{R}$ is explained as the direct result of the quantum-mechanical fluctuations of the field that is responsible for Inflation happening. 

An interesting and also relevant observation is that inflationary models typically predict small, but eventually observable, departures from what is currently accepted within the $\Lambda$CDM cosmology. In particular, we will later be focused on deviations from Gaussianity in the statistics of $\mathcal{R}$, which may be employed to constrain the parameters of each inflationary model. To describe how the quantum fluctuations take place, and how non-Gaussianities (NG) arise, we first need to take a look at the inflationary setup.

%\lipsum[1-3]
%\begin{defn}[ver \cite{KAR00}] Definición definitiva $$\frac{d}{dx}\int_a^xf(y)dy=f(x).$$\end{defn}

\cleardoublepage

\chapter{Models of Inflation and Primordial non-Gaussianity}  \label{chap:inflation}

As we discussed earlier, Inflation is the mechanism by which our universe got to be as homogeneous as we now observe it, but it is also responsible for the small fluctuations on top of that background. In this Chapter, we will discuss the canonical setup of inflation, describe how it takes place and what the requirements on the dynamical quantities are, to then set out the evolution of the primordial quantum fluctuations.

 After that, we will generalize the setup to a multi-field context, where there are many degrees of freedom that may play a role in the inflationary stage. Finally, we review previous work~\cite{Achucarro:2016fby} on the cumulative effects that can emerge for the perturbations in a multi-field setting, and how this impacts on the detectability of primordial non-Gaussianity. From this Chapter onwards, we will set the reduced Planck constant to unity, $\hbar = 1$.

\section{Single-Field Inflation}

The simplest model of inflation is the one obtained by considering the dynamics of a scalar field $\phi$ in the presence of a potential and letting its dynamics source the energy density required to generate an accelerated expansion. As it is apparent from equation~\eqref{Friedmann-2}, to have an accelerated expansion it is necessary that the dominant fluid inhabiting the universe have an equation of state such that $\rho + 3p < 0$. It turns out that, under certain conditions, a scalar field satisfies this condition.

The starting point to set up Inflation is to specify the action principle of the relevant degrees of freedom,
\be \label{Infl-action}
S = \int d^4 x \sqrt{-g} \L = \int d^4 x \sqrt{- g} \left(M_{\rm pl}^2 \frac{R}{2} - \frac{1}{2} \partial_\mu \phi \partial^\mu \phi - V(\phi) \right),
\ee
which considers the scalar field in a general-relativistic setting, where the spacetime metric $g_{\mu \nu}$ is also dynamical, and has the Einstein equations~\eqref{Einstein} as equations of motion implied by the presence of the Ricci scalar $R$ in the Lagrangian. $M_{\rm pl} = \sqrt{\frac{1}{8\pi G}}$ is the reduced Planck mass.
Also, from this Lagrangian one can derive the equation of motion for $\phi$
\be \label{eom-phi}
\ddot \phi + 3 H \dot \phi - \frac{1}{a^2} \nabla^2 \phi + V'(\phi) = 0,
\ee
where we have assumed that the metric $g$ is of the form of~\eqref{FLRW}, i.e., with a single function of time $a(t)$ (the scale factor) characterizing it. In the same manner as in classical Newtonian physics, $V(\phi)$ can be regarded as a potential generating a force $-V'(\phi)$ that pushes the value of the field towards the configuration of minimal potential energy. 

One can also write down the energy-momentum tensor of the scalar field
\be
T_{\mu \nu} = \partial_\mu \phi \partial_\nu \phi + g_{\mu \nu} \mathcal{L},
\ee
and identify its pressure and energy density by comparing with the expression for a perfect fluid
\bea
\rho_\phi &=& T_{00} = \frac{1}{2} \dot \phi^2 + V(\phi) + \frac{1}{2 a^2} (\nabla \phi)^2 \\
p_\phi &=& \frac{T^i_i}{3} = \frac{1}{2} \dot \phi^2 - V(\phi) - \frac{1}{6 a^2} (\nabla \phi)^2.
\eea
Now we assume, in agreement with current observations, that the primordial fluctuations that generated inhomogeneities in our universe were small, and that we may therefore split the field $\phi$ into two pieces
\be
\phi = \phi_0(t) + \delta \phi(\x,t),
\ee
where $\phi_0(t)$ is the ``classical'' homogeneous solution, which we assume to be completely determined by classical equations of motion, and $\delta \phi(\x,t)$ represents the fluctuations around $\phi_0(t)$, which will be computed within the framework of quantum mechanics. That is, we will show that all the observed inhomogeneities of our universe can be accounted for through primordial quantum fluctuations. We will come back to these perturbations later on; for now, we will focus on the evolution of the background metric.

\subsection{How inflation happens}

If we keep ourselves to studying the classical evolution of the field $\phi$, the energy-momentum tensor becomes
\bea
\rho_{\phi_0} &=& T_{00} = \frac{1}{2} \dot \phi_0^2 + V(\phi_0) \\
p_{\phi_0} &=& \frac{T^i_i}{3} = \frac{1}{2} \dot \phi_0^2 - V(\phi_0),
\eea
which can trivially satisfy $\rho_{\phi_0} + 3p_{\phi_0} < 0$ if the kinetic energy density $\dot \phi_0^2/2$ is much smaller than the potential energy density $V(\phi_0)$\footnote{As we are in a General Relativity context, the magnitude of the energy density is relevant.}, because it then follows that $\rho_{\phi_0} \approx - p_{\phi_0}$. Note that $\rho_{\phi_0}$ must be positive so that the Friedmann equation~\eqref{Friedmann-1} gives a positive value for $H^2$. That is to say, a scalar field with low kinetic energy implies an energy density that behaves as vacuum energy, and in the same way we nowadays infer the presence of Dark Energy, that energy density drives an accelerated expansion of the universe.

Now let us take a closer look at what are the requirements for this to happen. Since the equation of motion for $\phi_0$ is
\be \label{phi0-eom}
\ddot \phi_0 + 3 H \dot \phi_0 + V'(\phi_0) = 0,
\ee
we expect that a large value of $V'$ should drive an ``acceleration'' $\ddot \phi_0$ on the field value, which typically would lead to a growth in the kinetic energy that could, possibly, make it comparable to the potential energy. Therefore, if we want $\dot \phi_0^2/2 \ll V(\phi)$, we need a ``flat'' potential, where $\dot \phi_0$ does not grow significantly. Accordingly, we may approximate the equation of motion~\eqref{phi0-eom} by
\be \label{phi0-red-eom}
3 H \dot \phi_0 = - V'(\phi),
\ee
provided, of course, that $|\ddot \phi_0| \ll |3 H \dot \phi_0| \approx |V'(\phi)|$. Substituting the requirement $\dot \phi_0^2/2 \ll V(\phi)$ into~\eqref{phi0-red-eom} requires that
\be
\frac{V'^2}{V} \ll H^2,
\ee
and if we use the Friedmann equation $H^2 \approx 8\pi G V(\phi_0)/3$, this translates into $\frac{3}{8\pi G} \left(\frac{V'}{V} \right)^2 \ll 1$. To keep track of this inequality and enforce it, it is useful to define the first \textit{slow-roll} parameter $\epsilon$:
\be
\epsilon \equiv - \frac{ \dot H}{H^2} = - \frac{H \dot \phi_0 V'}{2 V H^2 } =   \frac{V'^2 }{6 V H^2 } = \frac{1}{16 \pi G} \left( \frac{V'}{V} \right)^2,
\ee
which therefore has to satisfy $\epsilon \ll 1$.

Finally, to take into account the consistency requirement $|\frac{\ddot \phi_0}{3 H \dot \phi_0} | \ll 1$, we derive the equation of motion~\eqref{phi0-red-eom} to obtain
\be
3 \dot H \dot \phi_0 + 3 H \ddot \phi_0 = - V''(\phi_0) \dot \phi_0 \implies \ddot \phi_0 = - \frac{\dot H}{H} \dot \phi_0 - \frac{V''(\phi_0) \dot \phi_0}{3 H},
\ee
and thus we require
\be
\left| \epsilon - \frac{V''(\phi_0)}{3 H^2}  \right| \ll 1,
\ee
which is satisfied if and only if
\be
\eta_\phi \equiv - \frac{\ddot \phi_0}{H \dot \phi_0} = - \epsilon + \frac{V''}{3H^2} 
\ee
fulfils $|\eta_\phi| \ll 1$. Equivalently, we can demand
\be
\eta \equiv \frac{\dot \epsilon}{H \epsilon} = -2 \eta_\phi + 2\epsilon
\ee
to satisfy $|\eta| \ll 1$. In any case, these requirements translate into the second derivative of the potential being small.
Therefore, we have established that inflation with $\rho_{\phi_0} \approx p_{\phi_0}$ is possible provided the potential is sufficiently flat and stays that way for a sufficiently long field range. Moreover, during this period we have
\be
\dot \epsilon, \, \dot \eta_\phi = \mathcal{O}(\epsilon^2, \eta_\phi^2),
\ee
and we may treat $\epsilon$ and $\eta_\phi$ as constants.

How does this solve the problems we described on Chapter~\ref{chap:search}? To first order in the slow-roll parameters (i.e., approximating $\dot \epsilon = 0$), we can use the definition of $\epsilon$ to solve for $H$ and the scale factor
\begin{align}
H(t) = \frac{H_I}{1+ \epsilon H_I t} & & a(t) = a_I (1 + \epsilon H_I t)^{1/\epsilon},
\end{align}
which in the limit $\epsilon \to 0$ converges to
\begin{align}
H(t) = H_I & & a(t) = a_I \exp (H_I t),
\end{align}
that is, an exponential expansion for the scale factor. Here we have denoted $H_I$ and $a_I$ as the values of $H$ and $a$, respectively, at the beginning of inflation. 

To solve the horizon problem, inflation needs to allow for causal contact to have happened between two points in opposite directions on our sky, i.e., at a distance of roughly $H_0$. During inflation, two points in space can communicate provided that the physical distance separating them is less than the physical Hubble radius,
\be
H_I^{-1} = \frac{a}{\dot a},
\ee
because this is the characteristic time of expansion, determining the maximum distance a particle can travel $d = cH_I^{-1}$ (with $c=1$) as the scale factor doubles (roughly), and thus a signal emitted from a point has a chance of reaching the other point only if their initial separation was less than $H_I^{-1}$.
Therefore, we require
\be \label{efold-cond}
H_0^{-1} \frac{a_F}{a_0} \frac{a_I}{a_F} \leq H_I^{-1},
\ee
i.e., that the co-moving coordinates of two points now separated at a distance $H_0^{-1}$ have been separated by a distance lesser than $H_I^{-1}$ at the beginning of inflation. $a_F$, the scale factor at the end of inflation, can be manipulated as $a_F/a_0 = T_0/T_F$ in terms of the temperature at the end of inflation $T_F$, and if we write $a_F = a_I e^N$, the requirement~\eqref{efold-cond} translates into
\be
\frac{T_0}{H_0} \frac{H_I}{T_F} e^{-N}  \leq 1\implies N \geq \ln \left( \frac{k_B T_0}{\hbar H_0} \right) - \ln \left( \frac{k_B T_F}{\hbar H_I} \right) \approx 67 + \ln \left( \frac{k_B T_F}{\hbar H_I} \right),
\ee
which, apart from the logarithmic dependence on $\frac{k_B T_F}{\hbar H_I}$ (which should be of order 1 given that the temperature during inflation should be proportional to $H_I$), gives, conservatively, $N \geq 60$.

Then, inflation can solve the horizon problem if at least 60 e-folds of expansion took place (i.e., an increase in the scale factor of $e^{60}$). Similarly, the flatness problem is resolved by decreasing the curvature energy density from a value of order $\mathcal{O}(1)$ at the beginning of inflation to a value of order $e^{-2N}$ at the end of inflation because during this period $a^2 H^2 = a_I^2 H_I^2 e^{2H_I t}$.

%The particle horizon, i.e., the distance a signal can travel starting from the initial time $t_I$, is given by
%\be
%r_H(t) = a(t) \int_{t_I}^t \frac{dt'}{a(t')},
%\ee
%which in the case of inflation can be approximated by $a(t)/H_I$ provided sufficient time has passed.

A final note is in order to explain how inflation ends. This period is called \textit{Reheating}, where the inflaton field $\phi$ decays to a minimum of the potential, about which it oscillates and radiates away energy, hypothetically, in the form of standard model particles. Most of this energy must be liberated in the form of radiation, so that the universe after inflation starts in a phase of radiation domination. In this way, the universe is ``reheated'' with radiation, and standard cosmology in the form of $\Lambda$CDM takes over.

\subsection{Fluctuation fields and horizon exit} \label{sec:exit}

Before proceeding any further, it is useful to stop and think about what will be the effect of primordial fluctuations on late-time cosmological observables, such as CMB or LSS, so as to compute meaningful quantities from perturbation theory on the inflationary background. In particular, a feature of paramount importance in both CMB and LSS is the characteristic wavelengths involved in their respective physics. To be explicit, fluctuations in the Cosmic Microwave Background have wavenumbers $k$ in between
\be \label{Range}
10^{-4} \, {\rm Mpc}^{-1} \leq k \leq  0.3  \, {\rm Mpc}^{-1},
\ee
whilst LSS modes span a range which is at least as large as that, but going beyond~\eqref{Range} requires to have more and more computational control on the nonlinear effects that start affecting the dynamics~\cite{Carrasco:2012cv}.

The lower limit on the range of $k$ corresponds, roughly to the inverse of the Hubble parameter today. That, is it is the size of today's Hubble radius. It turns out that the Hubble radius is precisely the quantity we need to characterize the evolution of fluctuation modes throughout the history of the universe. To illustrate this point, consider a massless field on an FLRW background, i.e.,
\be
S = \int d^4 x a^3(t) \left[ \frac{1}{2} \dot \varphi^2 - \frac{1}{2a^2} (\nabla \varphi)^2  \right]
\ee
with equation of motion (in its Fourier decomposition) given by
\be
\ddot \varphi_\k + 3 H \dot \varphi_\k + \frac{k^2}{a^2} \varphi_\k = 0.
\ee
There are two time scales in this equation: $H^{-1}$ and $a/k$. Qualitatively speaking, if $a H \ll k$, i.e., if the wavelength associated to the mode $k$ is well within the Hubble radius $(aH)^{-1}$, then we may drop the ``damping'' term $3H \dot \varphi_\k$ and obtain a parametric harmonic oscillator, which will describe (roughly) plane wave solutions. However, if $a H \gg k$, i.e., if the wavelength associated to the mode $k$ is outside the Hubble radius $(aH)^{-1}$, we may drop the last term and obtain a solution to the resulting equation for $\varphi_\k$. This gives a constant contribution plus an exponentially decaying piece. However, for the complete solution to assume an exponential decay, it must have had an initial condition with the exact pair of values $(\varphi_\k, \dot \varphi_\k)$. In particular, if the rate of change of $\varphi_\k$ was small enough at the initial time slice for a mode $k \ll aH$ well outside the Hubble radius, that mode will conserve its value until it re-enters the Hubble radius.

Physically, what happens is that modes of a given wavelength exit the Hubble radius during inflation, as $aH \propto H e^{Ht}$, acquire a ``frozen out'' value because of the existence of a constant solution to their equations of motion, and then re-enter our causal horizon during the late-time evolution of our universe, at which point we can perceive their consequences on the CMB and LSS. If we track the evolution of the modes~\eqref{Range} of a scalar $\varphi$ well into the past, we see that they exited the Hubble radius and froze to a given value nearly 60 e-folds before inflation ended. That is to say, the universe expanded by a factor of $e^{60}$ after the fluctuations we nowadays see in the CMB acquired the value that seeded initial conditions. This will later become relevant, as we will later be able to study the dynamics of the perturbations during inflation in the super-horizon limit.

\subsection{From primordial perturbations to the initial conditions of our universe}

Now we wish to study the fluctuations above this inflationary background, i.e., the dynamics of the perturbation $\delta \phi(\x,t)$ to the classical value of the inflaton $\phi_0(t)$. There is more to this than merely perturbing~\eqref{eom-phi}, as a fluctuation in the field $\delta \phi$ implies a perturbation to the stress-energy tensor, which in turn implies a fluctuation in the Einstein equations and therefore in the metric tensor $g_{\mu \nu} \to g_{\mu \nu} + \delta g_{\mu \nu}$, implying a correction to the equation of motion.

Consequently, one has to study the perturbed system for both quantities $\delta \phi$ and $\delta g_{\mu \nu}$ jointly. To do that, we introduce an explicit form for the metric, involving only scalar degrees of freedom as fluctuations\footnote{Tensor fluctuations are also present in principle, but for our present purposes, which is to derive the emergence of $\mathcal{R}$ from quantum fluctuations.}
\be
ds^2 = - a^2\left[(1 + 2 \Phi) d\tau^2 + 2\partial_i B d\tau dx^i + \left( (1 - 2\Psi) \delta_{ij} + \left(\partial_i \partial_j - \frac{1}{3} \delta_{ij} \nabla^2 \right) E  \right) dx^i dx^j \right]
\ee
where we have introduced the conformal time coordinate $\tau$, defined through $a d\tau = dt$. Thus, the fluctuating (scalar) degrees of freedom are presented in terms of four functions: $\Phi, \Psi, B$, and $E$. These quantities are dependent on our choice of coordinates, which in General Relativity, is actually a gauge freedom. When faced with this issue, i.e., that of describing equivalent physics with different sets of quantities (coordinates), one can either
\begin{enumerate}
\item Fix the gauge (choose the coordinate system) and perform all computations there, or
\item Identify gauge-invariant quantities, i.e., those that take the same value independently of the choice of coordinates, and use them as a guiding principle in the computation.
\end{enumerate}

As it turns out, there is a gauge invariant quantity $\mathcal{R}$ which corresponds to the gravitational potential $\Psi$ on co-moving hyper-surfaces\footnote{The hyper-surfaces with $\delta \phi = 0$ are called co-moving surfaces.}, given by
\be
\mathcal{R} = \Psi + H \frac{\delta \phi}{\dot \phi}.
\ee
$\mathcal{R}$ is therefore called the \textit{co-moving curvature perturbation}. It is also called the ``adiabatic'' perturbation, because of reasons that will become evident shortly.

A crucial property of single-field inflation is that the field perturbations $\delta \phi$ go over the same trajectory in field space as the background solution, i.e., $\phi_0(t) + \delta \phi(t,\x) $ is a value attained at some time by the homogeneous solution $\phi_0(t)$. This is equivalent to saying that the perturbations $\delta \phi$ are along the same direction of the inflationary trajectory. Moreover, once the gauge has been fixed (for concreteness, to any gauge with $\delta \phi \neq 0$), the dynamical equations that govern the evolution of $\delta \phi$ along with the rest of the perturbations are of second order on time derivatives on $\delta \phi$, which upon quantization at very early times give the only relevant degrees of freedom for the field equations, with the other quantities typically being realized as functionals of the inflaton fluctuation $\delta \phi$. For instance, in Newtonian Gauge, where $B = E = 0$, the Einstein equations imply $\Psi = \Phi$ and the field equations read~\cite{Weinberg:2008zzc}:
\bea
\dot \Psi_\k + H \Psi_\k &=& 4\pi G \dot \phi_0 \delta \phi_\k, \\
\delta \ddot \phi_\k + 3 H \delta \dot \phi_\k + V''(\phi_0) \delta \phi_\k + \frac{k^2}{a^2} \delta \phi_\k &=& - 2 \Psi_\k V'(\phi_0) + 4 \dot \Psi_\k \dot \phi_0, \\
\left( \dot H + \frac{k^2}{a^2} \right) \Psi_\k &=& 4\pi G \left( - \dot \phi_0 \delta \dot \phi_\k + \ddot \phi_0 \delta \phi_\k \right).
\eea
Of course, one could rearrange the equations to get $\delta \phi$ as a functional of $\Psi$, with the latter having equations of motion of second-order in temporal derivatives. The crucial point is that there is only one scalar degree of freedom that with independent dynamics, giving a single degree of freedom to be quantized\footnote{To do this, one has to choose a vacuum, which we will get to in time.}. For concreteness, we will take $\delta \phi$ as this degree of freedom the discussion that follows.

Therefore, as all perturbations may be written as a linear functional of $\delta \phi$, this implies that any variation in a scalar quantity $X$ can be expressed as
\be \label{adiabatic}
\delta X = \dot X \frac{1}{\frac{\delta \phi}{\delta t}} \delta \phi \implies \frac{\delta X}{\dot X} =  \frac{\delta \phi}{\dot \phi},
\ee
independently of what $X$ is. In particular, this means that all scalar perturbations, whether matter or radiation density fluctuations, are treated on equal footing. If one then goes back to the hydrodynamic approach of~\cite{Weinberg:2008zzc} and imposes this as a requirement for the initial condition at the start of the radiation-dominated era of our universe, one finds that the mentioned initial condition is completely determined by the scalar $\mathcal{R}$. Thus, all we need in order to make predictions about the late-time universe, in terms of the CMB and LSS, is the result for $\mathcal{R}$ out of inflation.

Mukhanov et. al.~\cite{Mukhanov:1990me} showed that this variable satisfies
\be
{\ddot{\mathcal{R}}}_\k + H {\dot{\mathcal{R}}}_\k + \frac{2}{z} {\dot z} {\dot{\mathcal{R}}}_\k + \frac{k^2}{a^2} \mathcal{R}_\k = 0,
\ee
where $z \equiv a^2 \dot \phi/ \dot a$. This is known as the \textit{Mukhanov-Sasaki equation}. In the limit where the slow-roll parameters are small, we may treat $\dot \phi$ as constant and obtain
\be
\ddot{\mathcal{R}}_\k + 3 H \dot{\mathcal{R}}_\k + \frac{k^2}{a^2} \mathcal{R}_\k = 0,
\ee
which is the equation of motion for a massless scalar field on an FLRW background. That is to say, these fluctuations will evolve within the inflationary epoch, and as discussed on Section~\ref{sec:exit}, will exit the horizon and become ``frozen'' to a fixed value, waiting until the late-time expansion of the universe imprints them on the CMB and LSS.

These equations of motion can be derived from the action principle
\be
S = \int d^4x a^3 \epsilon \left[  \dot{\mathcal{R}}^2 - \frac{1}{a^2} (\nabla \mathcal{R})^2 \right],
\ee
which is the action, to quadratic order in the fluctuations, describing the dynamics of the (adiabatic) curvature perturbation in an FLRW spacetime in co-moving gauge. In the particular case of a quasi-de Sitter inflationary phase (where $ 0 \neq \epsilon \ll 1$), we may approximate the evolution of the background $a(t)$ with an exact de Sitter space\footnote{The normalization factor is never meaningful; only fractional comparisons between scale factors at different times give relevant physical information.} $a = e^{Ht}$ keeping in mind that this amounts to neglecting corrections of order $\epsilon$.

The theory may then be quantized by introducing the canonically normalized field $u \equiv \sqrt{2\epsilon} a \mathcal{R}$ and moving to conformal time $d\tau = dt/a$, where $\tau \in (-\infty,0)$, $a(\tau) = - \frac{1}{H\tau}$, and the action integral is given by
\be
S = \frac{1}{2} \int d^3x d\tau \left[ u'^2 - (\nabla u)^2 + \frac{2}{\tau^2} u^2 \right],
\ee
wherein the prime denotes a derivative with respect to conformal time. One can then enforce canonical commutation relations $[u(\tau,\x), \Pi(\tau,\y)] = i \delta^{(3)}(\x - \y) $, where we have taken natural units in which $\hbar = 1$, to obtain that the solutions to the equations of motion are given in terms of ladder (creation/annihilation) operators
\be \label{mode-sol-R}
\hat{u}_\k(\tau) = u_k(\tau) a(\k) + u_k^*(\tau) a^{\dagger}(-\k),
\ee
where $[a(\k), a^\dagger(\k')] = (2\pi)^3 \delta^{(3)}(\k - \k')$ with the other commutators vanishing, and
\be
u_k(\tau) = \frac{1}{\sqrt{2k}} \left(1 - \frac{i}{k\tau} \right) e^{-ik\tau}
\ee
is the mode function corresponding to the Bunch-Davies vacuum. The only assumption here, i.e., in choosing this vacuum, is that boundary conditions at $\tau = -\infty$ have been imposed, enforcing that the mode functions for the canonical field $u$ are as if the background space were a flat Minkowski background.

What are the predictions of this theory? As it is a free quantum field theory, all the observable information is encoded within the power spectrum of the theory $P_{\mathcal{R}}$, defined as in~\eqref{power-R}:
\be
\langle R_\k R_{\k'} \rangle = (2\pi)^3 \delta^3 (\k + \k') P_{\mathcal{R}}(k).
\ee
If we compute this directly from the solution~\eqref{mode-sol-R}, considering a quantum expectation value rather than a classical ensemble average, we obtain
\be
P_{\mathcal{R}} = \frac{H^2}{4 \epsilon k^3}
\ee
up to corrections of higher order in $\epsilon$. In particular, deviations from a perfectly de Sitter inflationary phase induce a \textit{tilt} (also called \textit{spectral index}) in the power spectrum
\be
n_s - 1 \equiv \frac{d \ln \left( k^3 P_{\mathcal{R} }(k) \right)}{d \ln k},
\ee
which, when comparing with observations, is parametrized as
\be
\langle R_\k R_{\k'} \rangle = (2\pi)^3 \delta^3 (\k + \k') P_{\mathcal{R}}(k) = (2\pi)^3 \delta^3 (\k + \k') \frac{2\pi^2 A_s}{k^3} \left(\frac{k}{k_*} \right)^{n_s - 1},
\ee
where $k_*$ is a pivot scale (which is degenerate with $A_s$, and is thus set by hand to a meaningful value). To first order in the slow-roll parameters, single-field inflation predicts
\be
n_s - 1 = -2\epsilon - \eta_\phi,
\ee
which therefore measures deviations from the perfect de Sitter limit. It can also be related to the derivatives of the potential, and thus the spectral index is a valuable probe of what the early-universe physics can be. The latest 2018 article on cosmological parameters by Planck~\cite{Aghanim:2018eyx} reported
\be
n_s = 0.9649 \pm 0.0042,
\ee
that is, a nearly scale-invariant power spectrum\footnote{A power spectrum that is proportional to $k^{-3}$ is considered scale-invariant, because upon a linear rescaling of wavenumbers the measure of configuration space changes by precisely the factor needed to cancel the rescaling in the power spectrum. That is to say, at any given scale $k$ the higher- and lower-energy physics look the same.} with a definite tilt towards the infrared side of the spectrum.

This typical amplitude of fluctuations, the power spectrum, is all we need to account for all observations of the CMB and LSS thus far. However, the theory we have described up to this point is, by construction, linear on the perturbations, which in turn mathematically implies that the fluctuations will be completely characterized by their power spectrum. Is there any chance of measuring any departure from this?

In principle, there is. In the same manner as it is possible to write linear equations of motion for the perturbations during inflation, one can also write equations of motion in a power series of the field, and compute the next-to-leading order corrections as a perturbative expansion in the nonlinearity coefficients. However, within single-field slow-roll inflation, these coefficients are suppressed by the slow-roll parameters, making any departure from Gaussian statistics, from now on non-Gaussianities (NG), extremely small compared to the observational precision available nowadays. As an example, consider $f_{\rm NL}$~\cite{Komatsu:2001rj}, the first local nonlinearity parameter in a power series expansion about Gaussian statistics
\be
\mathcal{R} = \mathcal{R}_G + \frac{3}{5} f_{\rm NL} \mathcal{R}_G^2,
\ee
which implies a non-vanishing three-point function, also dubbed \textit{bispectrum}
\be
\langle \mathcal{R}_{\k_1} \mathcal{R}_{\k_2} \mathcal{R}_{\k_3} \rangle = (2\pi)^3 \delta^{(3)}(\k_1+\k_2+\k_3) \frac{6}{5} f_{\rm NL} \left[ P_{\mathcal{R}}(k_1) P_{\mathcal{R}}(k_2) + P_{\mathcal{R}}(k_2) P_{\mathcal{R}}(k_3) + P_{\mathcal{R}}(k_3) P_{\mathcal{R}}(k_1)  \right].
\ee
Maldacena's consistency relation~\cite{Maldacena:2002vr} gives that $f_{\rm NL}$, to first order in the slow-roll parameters, is related to the spectral index $n_s$ 
\be
f_{\rm NL} = - \frac{5}{12} (n_s - 1),
\ee
thus giving an apparently testable prediction for the years to come to verify. However, to give a quantitative criteria of how precise the measurements would need to be in order to measure a deviation from Gaussianity of this size, one must consider that $\epsilon$ should be at most of order $10^{-2}$, which, consequently, is the same estimate for the order of magnitude of $f_{\rm NL}$. Sadly, this is just too low to be measured within the foreseeable future. Current constraints provided by Planck~\cite{Akrami:2019izv} indicate $f_{\rm NL} = -0.9 \pm 5.1$, and the cosmological surveys of the next decade~\cite{Lsst-web,Euclid-web,Bull:2018lat,Abazajian:2016yjj} will only reduce $\sigma(f_{\rm NL})$ down to 1.

A similar point can be made for primordial non-Gaussianity of higher order in the power series expansion: it is likely to be present, but also too small to be measured if it comes from single-field slow-roll inflation. Thus, if a non-Gaussian primordial signal is found within the next decade, it would constitute a clear departure from the minimal models of inflation, and force cosmologists to reconsider the dynamics of the inflationary setup. As it turns out, interactions involving the inflaton and other fields could enhance the amplitude of the three-point or higher point correlation functions (see \cite{Bartolo:2004if,Liguori:2010hx,Chen:2010xka,Wang:2013eqj} for reviews). These interactions could be self-interactions of the inflaton or interactions of the inflaton with other degrees of freedom. It is at this point in which studying inflation with extra fields/degrees of freedom becomes of interest and of actual relevance. This will be the topic of the next section: Multi-Field Inflation.

Following Maldacena~\cite{Maldacena:2002vr}, from this point forward we will denote the co-moving curvature perturbation by the greek letter $\zeta$.

\section{Multi-Field Inflation}

As we mentioned in the previous section, upcoming cosmological surveys may require theories able to account for a non-vanishing bispectrum, and thus it is of interest to study and characterize them from an effective point of view. However, it is also an opportunity to test fundamental theories of nature. Indeed, the typical situation within String Theory (see, for instance, reference~\cite{Baumann:2014nda}) is to have many scalar fields emanating from the compactification\footnote{Compactification refers to the procedure by which extra spatial dimensions are made small enough that the physics of our macroscopic universe may be described by a temporal and three spatial dimensions. This is typically achieved through the extra dimensions being periodic in some sense, for instance, as a torus.} of extra spacetime dimensions, which prove to be necessary in order to have a consistent String Theory. This makes the study of multi-field inflationary scenarios all the more relevant, as a purportedly fundamental theory of quantum gravity could be constrained by studying cosmological observables.

\subsection{The classical inflaton trajectory}

Just as single-field inflation could be derived from the Lagrangian dynamics of a scalar field on a general relativistic background~\eqref{Infl-action}, multi-field inflation can be written in terms of an action principle
\be \label{action-multi}
S = \int d^4 x \sqrt{-g} \L = \int d^4 x \sqrt{- g} \left(\frac{R}{2} - \frac{1}{2} \gamma_{ab} \partial_\mu \phi^a \partial^\mu \phi^b - V(\phi) \right),
\ee
where we have introduced indices $a,b \in \{1, ..., N \}$, where $N$ is the number of scalar degrees of freedom active during inflation. Furthermore, there is a new object in this expression, the field-space metric\footnote{Depending on the reference/textbook at hand, it may also be referred to as ``target-space metric.''} $\gamma_{ab}$, which is a function of the field coordinates $\phi^a$, and characterizes the properties of the dynamical evolution of the scalars in their own target space. The equations of motion are easily obtained by varying the action integral and demanding $\delta S = 0$ according to Hamilton's principle, yielding
\be
\frac{1}{\sqrt{-g}}\partial_\mu \left( \sqrt{-g} \partial^\mu \phi^a \right) + \Gamma^a_{bc} \partial_\mu \phi^b \partial^\mu \phi^c + \gamma^{ab} \frac{\partial V}{\partial \phi^b} = 0,
\ee
where $\Gamma^a_{bc}$ are the Christoffel symbols in field space
\be
\Gamma^a_{bc} \equiv \frac{\gamma^{ad}}{2} \left( \frac{\partial \gamma_{dc}}{\partial \phi^b} + \frac{\partial \gamma_{db}}{\partial \phi^c} - \frac{\partial \gamma_{bc}}{\partial \phi^d} \right).
\ee

Analogously to what was done for the single-field case, we may first study the background solutions that account for inflation happening in the first place, and then study the equations for the cosmological perturbations. The energy-momentum tensor now reads
\be
T_{\mu \nu} = \frac{1}{2} \gamma_{a b} \partial_\mu \phi^a \partial_\nu \phi^b - g_{\mu \nu} V(\phi),
\ee
which, if we only look for homogeneous solutions $\phi^a = \phi_0^a(t)$ (independent of the space coordinate $\x$), the Friedmann equation~\eqref{Friedmann-1} reads
\be \label{Hubble-multi}
H^2 = \frac{8 \pi G}{3} \left( \frac{1}{2} \gamma_{a b} \dot \phi^a_0 \dot \phi^b_0 + V(\phi) \right),
\ee
and the equation of motion for the field, in the presence of such FLRW background~\eqref{FLRW} is given by
\be \label{multi-field-evol}
\frac{D \dot \phi_0^a}{dt} + 3 H \dot \phi^a + \gamma^{a b} \frac{\partial V}{\partial \phi^b} = 0,
\ee
where we have introduced the covariant derivative along the inflationary trajectory of a coordinate ``vector'' in field space
\be
\frac{D X^a}{dt} \equiv \dot X + \Gamma^a_{bc} \dot X^b \dot \phi_0^c.
\ee
At this point, if a unitary vector 
\be
T^a \equiv \dot \phi_0^a/\dot \phi_0
\ee
is defined, in which $\dot \phi_0^2 \equiv \gamma_{a b} \dot \phi^a_0 \dot \phi^b_0$, then $T^a$ points along the direction of the trajectory the fields $\phi^a$ follow in field space. It is thus the natural direction along which to identify the inflaton, in the sense that it defines the inflationary trajectory. Moreover, $T^a$ obeys
\be
\frac{D T^a}{dt} =  - \frac{\ddot \phi_0}{\dot \phi_0} T^a - \frac{1}{\dot \phi_0} \left(3 H \dot \phi_0^a + \gamma^{a b} \frac{\partial V}{\partial \phi^b} \right),
\ee
which can be decomposed into parallel and orthogonal directions\footnote{By construction, $T^a$ is orthogonal to its temporal derivative because $0 = \frac{D (T_a T^a) }{dt}  = 2 T_a \frac{D T^a}{dt} $.}:
\begin{align}
\ddot \phi_0 + 3 H \dot \phi_0 + T^b \frac{\partial V}{\partial \phi^b} = 0  & & \frac{D T^a}{dt} = - \frac{N^b}{\dot \phi_0} \frac{\partial V}{\partial \phi^b} N^a,
\end{align}
where $N^a$ is a unitary vector that pointing along the direction of $\frac{D T^a}{dt}$:
\be
N^a \equiv - \frac{D T^a}{dt} \bigg/ \sqrt{ \gamma_{ab} \frac{D T^a}{dt} \frac{D T^b}{dt} }.
\ee
In particular, there will be a ``turning'' of the trajectory induced by the potential $V$ if the derivative along the normal direction to the trajectory $N^a$ is non-vanishing. One may define a turning rate through
\be
\frac{D T^a}{dt} = - \Omega N^a \implies \Omega = \frac{N^b}{\dot \phi_0} \frac{\partial V}{\partial \phi^b},
\ee
which will be a central quantity later on. 

\subsection{Slow-roll conditions and perturbations}

Now let us turn to the slow-roll conditions. To start with, we need inflation to last long enough, which is equivalent to saying that the Hubble rate should remain nearly constant for a while during inflation. Using the definition of the first slow-roll parameter $\epsilon \equiv - \dot H / H^2$, combined with the Friedmann equations~\eqref{Friedmann-1} and~\eqref{Friedmann-2}, one can show
\be
\epsilon = \frac{1}{8\pi G} \frac{\dot \phi_0^2}{2H^2},
\ee
which in the light of~\eqref{Hubble-multi}, makes $\epsilon$ proportional to the quotient between the kinetic energy density of the scalar over is total energy density (including the potential). Thus, $\epsilon \ll 1$ ensures that the vacuum energy equation of state $p = -\rho$ is fulfilled. One also has to guarantee that this stays this way for a sufficient period of time. To that end, we can also use the second slow-roll parameter $\eta \equiv \dot \epsilon / H \epsilon$ to make this quantification. It turns out that here
\be
\eta_\phi \equiv - \frac{\ddot \phi_0}{H \dot \phi_0} \equiv - \frac{\gamma_{a b} \ddot \phi_0^a \dot \phi_0^b }{H \dot \phi_0^2} = - \frac{\gamma_{a b} \ddot \phi^a T^b }{H \dot \phi_0}
\ee
is also a good quantifier of this, because it also holds that $\eta = 2( \epsilon - \eta_\phi)$. Concretely, $\epsilon, \eta_\phi \ll 1 \implies \epsilon, \eta \ll 1$. However, the ``natural'' field vector whence $\eta_\phi$ comes from,
\be
\eta^a \equiv -\frac{1}{H \dot \phi_0} \frac{D \dot \phi_0^a}{dt} = \eta_\phi T^a + \frac{\Omega}{H} N^a,
\ee
which is decomposed into parallel and orthogonal pieces to the inflationary trajectory, need not be small. In particular, slow-roll conditions do not impose a constraint on the magnitude of $\Omega$, except for the requirement that the kinetic energy density $\dot \phi_0^2/2$ must stay small compared to the background potential. %This will become particularly useful later on.

It turns out that this new parameter $\Omega$ provides another coupling that can mediate the generation of non-Gaussianities. As the path of the inflaton turns, the perturbations along the inflationary trajectory, which are adiabatic in the same sense as in~\eqref{adiabatic}, mix with the perturbations orthogonal to the inflationary trajectory. The modes associated to the latter perturbations are called \textit{isocurvature} modes, because they perturb the solution away from the background solution, and therefore are not adiabatic in the sense of~\eqref{adiabatic}. This is a crucial observation, because currently the initial conditions of our universe, as studied by the Planck Collaboration on 2018~\cite{Ade:2015lrj,Akrami:2018odb} show no conclusive evidence for the presence of isocurvature modes. Nonetheless, they may have observable consequences if they couple to the adiabatic mode $\zeta$. A realization of this setup that has received some interest is quasi-single field inflation~\cite{Chen:2009we, Chen:2009zp}, where the isocurvature fields have masses of the same order of magnitude as the Hubble parameter during inflation $H$, and large non-Gaussianities may be generated in comparison with the canonical single-field scenario.

In general, the turning of the inflationary trajectory is reflected already at the quadratic level in the action integral. In a two-field model, the quadratic action obtained by expanding the action~\eqref{action-multi} about the (background) inflationary trajectory $\phi^a_0(t)$, in co-moving gauge where $\delta \phi^a T_a = 0$ is given by~\cite{Achucarro:2016fby}:
\be \label{2-field}
S = \int d^4 x a^3 \left[ \epsilon \dot \zeta^2 - 2 \epsilon \alpha \dot \zeta \psi - \frac{\epsilon}{a^2} (\nabla \zeta)^2 + \frac{1}{2} \left( \dot{\psi}^2 - \frac{1}{a^2} (\nabla \psi)^2 \right) - \frac{1}{2} m_\psi^2 \psi^2 \right],
\ee
where the parameter $\alpha$ is linearly related to $\Omega$ through
\be
\alpha \equiv - \frac{2 \Omega}{\sqrt{2\epsilon}}.
\ee
In this expression $\psi \equiv \delta \phi^a N_a$ as there is only one orthogonal degree of freedom to the inflationary trajectory. Its mass is derived from properties of the potential, the shape of the trajectory, and the Ricci scalar of the two-field manifold~\cite{Achucarro:2016fby}.

It is important to note that all of the contributions to~\eqref{2-field} will be present in general multi-field actions for the perturbations~\cite{Gordon:2000hv,GrootNibbelink:2000vx,GrootNibbelink:2001qt,Hetz:2016tes}, coupling the adiabatic mode to the perturbations orthogonal to the trajectory in the direction defined by $N^{a}$. Conversely, from an EFT standpoint it is possible to show that a coupling of the form $\dot \zeta \psi$ is the only interaction compatible with diffeomorphism invariance and the symmetries of the background~\cite{Achucarro:2016fby,Langlois:2008mn}, provided that $\zeta$ is the adiabatic mode.

%The system of linearized perturbations for a two-field model was investigated in~\cite{Achucarro:2016fby}, and can be derived from an action principle:

Furthermore, this Lagrangian density can be rearranged in a manner that is more revealing of the dynamics of the field perturbations, into an effective kinetic term $\propto (\dot{\zeta} - \alpha \psi)^2$ and an effective mass for the isocurvature perturbation
\be \label{quad-action}
S = \int d^4 x \, a^3 \left[ \epsilon (\dot{\zeta} - \alpha \psi)^2 - \frac{ \epsilon }{a^2} (\nabla \zeta)^2 + \frac{1}{2} \left( \dot{\psi}^2 - \frac{1}{a^2} (\nabla \psi)^2 \right) - \frac{1}{2} \mu^2 \psi^2 \right],
\ee
where $\mu \equiv m_\psi^2 + 4\Omega^2$ corresponds to the so-called ``entropy'' mass of $\psi$. This particular rearrangement will be of use in section~\ref{sec:linearth}; in fact, the action~\eqref{quad-action} constitutes the starting point for our subsequent computations.

\subsection{Cumulative Effects}  \label{sec:linearth}

As we have commented earlier, in a general multi-field setting the turning of a trajectory $\Omega$, defined implicitly through
\be
\frac{D}{dt} T^a = -\Omega N^a,
\ee
will generate a coupling between the quantum field fluctuations along the tangent direction $T^a$ with those along the normal direction $N^a$. Similarly, further couplings may appear because of the turning of the normal direction, generating an interaction term between the perturbations along the normal direction with those along the binormal direction, and the turning of its turning until all the dimensions of the field target space have been spanned. Therefore, in order to study the effects that the modes of isocurvature fluctuations will induce in the curvature fluctuations $\zeta$, it suffices to consider a two-field model, because
\begin{enumerate}
\item it is already sufficient to generate a nontrivial modification to the kinetic term of $\zeta$, and,
\item we can always regard the theory of the extra fields as having been integrated out to generate effective self-interactions of the isocurvaton $\psi$.
\end{enumerate}

Therefore, to study the dynamics of the perturbations in multi-field inflation, at the linear level it is sufficient to consider the action~\eqref{quad-action} as a generator of the field dynamics. In this type of models, the turning rate $\Omega$ is typically related linearly to the coupling that modifies the kinetic term $\alpha$. For instance, in the above exposition, presented with more detail in~\cite{Achucarro:2016fby,Chen:2018uul} $ \sqrt{2 \epsilon} \alpha = \Omega$. However, from an effective field theory point of view, this modification to the kinetic term is the only quadratic interaction allowed by the symmetries of the theory, and therefore the Lagrangian~\eqref{quad-action} need not come from the construction multi-field inflation we have outlined here, and $\alpha,\mu$ may be regarded as free parameters. In any case, the perturbation $\psi$ will constitute an isocurvature perturbation, as it is a degree of freedom that has no requirement on its adiabaticity. From now on, we will sometimes call it ``isocurvaton''.

%As in~\eqref{quad-action}, we will denote the isocurvature fluctuation (isocurvaton) by $\psi$.

%a first approximation we can consider

% Equation~\eqref{quad-action}

%- Restriction to bilinear coupling

%- Include lagrangians, hamiltonians

%Before computing the $n$-point correlation functions using the full nonlinear theory, let us have a look into the linear theory obtained in the limit $\Lambda^4 \to 0$. 

In the absence of nonlinear interactions, i.e., as in equation~\eqref{quad-action}, the resulting statistics for the curvature perturbation are Gaussian, and the only meaningful quantity to compute is the power spectrum for $\zeta$. This system was thoroughly investigated in Ref.~\cite{Achucarro:2016fby}, and here we show the main steps allowing one to deduce the value of the power spectrum. This result will be important later on when we include nonlinearities. To simplify any computation involving $\zeta$ and $\psi$ we will assume a purely de Sitter background, with $a(t)=e^{H t}$.

\subsubsection{Classical results and symmetry considerations} %\label{sec:cl-symm-2f}

In the long wavelength limit, $\psi$ satisfies the following equation of motion (obtained after integrating the equation of motion for $\zeta$ once)
\be
\ddot \psi + 3 H \dot \psi + \mu^2 \psi = 0 , \label{superhorizon-eom-psi}
\ee
from where it is possible to read that $\mu$ determines the mass of $\psi$ on superhorizon scales.
Notice that $\zeta$ and $\psi$ interact through the coupling $\alpha$ appearing in the special combination
\be
D_t \zeta \equiv \dot \zeta - \alpha \psi,
\ee
determining the kinetic term of $\zeta$. In general the coupling $\alpha$ depends on time, and, as we discussed earlier, its appearance may be understood as the consequence of bends of the inflationary trajectory in the multifield target space (or more precisely, nongeodesic motions in target space)~\cite{Chen:2009we,Chen:2009zp,GrootNibbelink:2000vx, GrootNibbelink:2001qt, Achucarro:2010jv, Achucarro:2010da}.

If the entropy mass vanishes ($\mu = 0$), the field $\psi$ becomes ``ultralight", and the system gains a symmetry given by
\bea
&&\psi \to \psi' = \psi + C , \label{symm-1} \\ \label{symm-2}
&&\zeta \to \zeta' = \zeta + C \int^t \!\!  dt \, \alpha ,
\eea
where $C$ is an arbitrary constant. %The consequences of this symmetry were investigated in Ref.~\cite{Achucarro:2016fby}. 
We summarize the findings of~\cite{Achucarro:2016fby} as follows: First, the symmetry of the action integral (\ref{quad-action}) under the transformation (\ref{symm-1}) ensures the existence of a constant solution for $\psi$. This can be seen directly in Eq.~(\ref{superhorizon-eom-psi}). This solution, say $\psi_*$, spontaneously breaks the symmetry, and dominates on superhorizon scales. Second, the symmetry of the Lagrangian under the transformation (\ref{symm-2}) implies that the constant solution $\psi_*$ will source the evolution of $\zeta$ on superhorizon scales. Concretely, if for simplicity we assume that $\alpha$ is nearly constant, on superhorizon scales one finds:
\be
\zeta \simeq \frac{\alpha}{H} \psi_* \Delta N \label{zeta-psi-N-1}
\ee
for a sufficient number of e-folds $ 2\epsilon \alpha^2 \Delta N^2/H^2 \gg 1$. If we conveniently identify $\psi_*$ as the value of the field at horizon crossing, then $\Delta N$ corresponds to the number of $e$-folds after that event. A given $n$-point correlation function is then given by:
\be
\langle \zeta^n \rangle \simeq \left( \frac{\alpha}{H} \Delta N \right)^n \langle \psi_*^n \rangle . \label{zeta-n-psi-n}
\ee
In the particular case of $n=2$, we obtain a relation between the power spectrum of $\zeta$ and the power spectrum of $\psi$:
\be
P_{\zeta} \simeq \frac{\alpha^2 \Delta N^2}{H^2} P_{\psi} . \label{Pzeta-Ppsi}
\ee
Moreover, it is possible to show that $\zeta$ has a negligible influence on the evolution of $\psi$, which behaves as a massless field before and after horizon crossing as long as $\mu \ll H$. This implies that the power spectrum for $\psi$ is given by
\be
P_{\psi} = \frac{H_*^2}{4 \pi^2} ,
\ee
where $H_*$ is the Hubble parameter evaluated at horizon crossing. Now the key issue to stress about Eqs.~(\ref{zeta-n-psi-n}) and~(\ref{Pzeta-Ppsi}) is that the statistics of the field $\zeta$ is completely determined by the statistics of $\psi$. In this case, given that we are only considering a quadratic Lagrangian without higher order self-interactions for $\psi$, the statistics of $\psi$ are found to be Gaussian. Then, the statistics inherited by $\zeta$ are also found to be Gaussian, with non-Gaussian deviations suppressed by slow-roll parameters as usual~\cite{Maldacena:2002vr} had we included the corresponding self-interactions of $\zeta$.

In order to give an explicit derivation of the above results, and to set the foundations for the consequent study of non-Gaussianities, we will now proceed to treat the perturbations quantum-mechanically.

\subsubsection{Quantization of the theory}

To start with, notice that this system consists of two canonical massless fields $u$ and $v$ coupled through a quadratic interaction Hamiltonian. To see this, define the canonical fields by
\be  \label{uv-def}
u  \equiv \sqrt{2\epsilon} a \zeta,  \,\,\,\,\,\,\, v \equiv  a \psi,
\ee
and switch coordinates from physical time to conformal time via $d\tau = dt / a = dt e^{-Ht}$. The action integral now reads
\be
S = \frac{1}{2} \int d^3 x d\tau  \Bigg[ \left( u'  + \frac{\lambda}{\tau}  v \right)^2 +\frac{2}{\tau^2} u^2    -  (\nabla u)^2   +  ( v' )^2 + \frac{2}{\tau^2} v^2    - (\nabla v)^2  + \lambda \frac{2}{\tau^2} u v \Bigg],  \label{action-quadratic-u-v}
\ee
where we have introduced the dimensionless coupling
\be \label{lambda}
\lambda\equiv\frac{\sqrt{2\epsilon}\alpha }{H},
\ee
which is taken to be non-vanishing in the de Sitter limit. From Eq.~(\ref{action-quadratic-u-v}), we infer that the canonical momenta associated with $u$ and $v$ are,  respectively, given by
\bea
\Pi_u &=& u'  + \frac{\lambda}{\tau}  v , \label{can-momenta-1} \\
\Pi_v &=& v' . \label{can-momenta-2}
\eea
These momenta satisfy the equal time commutation relations, given by
\bea
\left[u(\x, \tau), \Pi_u (\y, \tau)\right] &=& i \delta^{(3)} (\x - \y) , \label{commut-rel-1} \\
\left[v(\x, \tau), \Pi_v (\y, \tau)\right]  &=& i \delta^{(3)} (\x - \y)  ,\label{commut-rel-2}
\eea
with every other commutator vanishing. From (\ref{can-momenta-1}) and (\ref{can-momenta-2}) we see that the Hamiltonian of the system is given by
\be
\begin{split}
H \! = \! \frac{1}{2} \int_x \Bigg[ \Pi_u^2 + (\nabla u)^2 - \frac{2}{\tau^2} u^2 + \Pi_v^2 +  (\nabla v)^2 -\frac{2}{\tau^2} v^2 - \frac{2 \lambda}{\tau} v   \left(  \Pi_u + \frac{u}{\tau}  \right)  \Bigg] ,
\end{split}
\ee
where we have introduced the notation $\int_x = \int d^3x$. This naturally suggests a splitting $H = H_0 + H_\lambda$ where $H_0$ corresponds to the free Hamiltonian of the system, obtained in the limit $\lambda = 0$. As promised earlier, $H_0$ describes a system with two decoupled massless scalar perturbations
\be
H_0 =  \frac{1}{2} \int_x \left[ \Pi_u^2 + (\nabla u)^2 - \frac{2}{\tau^2} u^2 + \Pi_v^2 + (\nabla v)^2 - \frac{2}{\tau^2} v^2 \right], \label{H-0}
\ee
and on the other hand, $H_\lambda$ contains the interaction term proportional to $\lambda$,
\be
H_\lambda = - \int_x  \frac{\lambda}{\tau} v   \left(  \Pi_u + \frac{u}{\tau} \right), \label{H-mixing}
\ee
which is quadratic on the fields.

We may now quantize the system by adopting the interacting picture framework. That is, the quantum fields $u$ and $v$ are expressed as
\bea
u (\x , \tau) &=& U^\dag (\tau)  u_I (\x , \tau)  U (\tau),  \label{u-v-U-1} \\
v (\x , \tau) &=& U^\dag (\tau)  v_I (\x , \tau)  U (\tau), \label{u-v-U-2}
\eea
where $u_I (\x , \tau)$ and $v_I (\x , \tau)$ are the interaction picture fields, which evolve as quantum fields of the free theory. Explicitly, they are given by
\bea
u_I(\x , \tau) &=& \int_k \, \hat u_I (\k , \tau) \, e^{- i \k \cdot \x }, \\
v_I(\x , \tau) &=& \int_k \, \hat v_I (\k , \tau) \,  e^{- i \k \cdot \x },
\eea
with $\int_k = (2\pi)^{-3} \int d^3 k$, and where
\bea
\hat u_I (\k , \tau) &=& u_k(\tau) a_{-}(\k) + u_k^*(\tau) a_{-}^\dag(-\k)  , \label{Fourier-u-v-1} \\
\hat v_I (\k , \tau) &=& v_k(\tau) a_{+}(\k) + v_k^*(\tau) a_{+}^\dag(-\k) . \label{Fourier-u-v-2}
\eea
Here, the pairs $a_{\pm}(\k)$ and $a_{\pm}^\dag(\k)$ correspond to creation and annihilation operators satisfying the commutation relations:
\be
\left[ a_{b} (\k) , a_{c}^{\dag} (\k') \right] = (2 \pi)^3 \delta^{(3)} (\k - \k') \delta_{bc} , \label{cre-anh-def}
\ee
with \(b,c \in \{+, -\}\). The mode functions $u_k(\tau)$ and $v_k(\tau)$ are both given by
\be
u_k(\tau) = v_k(\tau) = \frac{1}{\sqrt{2 k}} \left( 1 - \frac{i}{k \tau} \right) e^{- i k \tau} , \label{k-mode-solutions}
\ee
which corresponds to the standard expression for a massless mode on a de Sitter spacetime with Bunch-Davies initial conditions. On the other hand, $U (\tau)$ is the propagator in the interaction picture, which is given by
\be
U (\tau) = \mathcal T \exp \left\{ - i \int_{-\infty_{+}}^{\tau} \!\!\!\!\!\! d \tau' H_I (\tau')  \right\} , \label{U-propagator}
\ee
where $\mathcal T$ stands for the time ordering symbol. In a given product of operators, $\mathcal T$ instructs us to put operators evaluated at later times on the left and operators evaluated at earlier times on the right. In addition, we have $\infty_{\pm} = \infty (1 \mp i \epsilon)$, where $\epsilon$ is a small positive number introduced to select the correct interaction picture vacuum. Finally, $H_I$ in Eq.~(\ref{U-propagator}) is given by
\be
H_I = H_I^{\lambda} = - \int_x  \frac{\lambda}{\tau} v_I   \left(  \Pi_u^{I} + \frac{u_I}{\tau}  \right). \label{int-alpha} 
\ee
Notice that in the previous expressions the canonical momenta $\Pi_u^{I}$ and $\Pi_v^{I}$ in the interaction picture are simply given by
\be
 \Pi_u^{I} = \frac{d}{d\tau} u_I , \qquad  \Pi_v^{I} = \frac{d}{d\tau} v_I .
 \ee

\subsubsection{Computing the observable: the $\zeta$ power spectrum}

\begin{figure}[t!]
\includegraphics[scale=0.38]{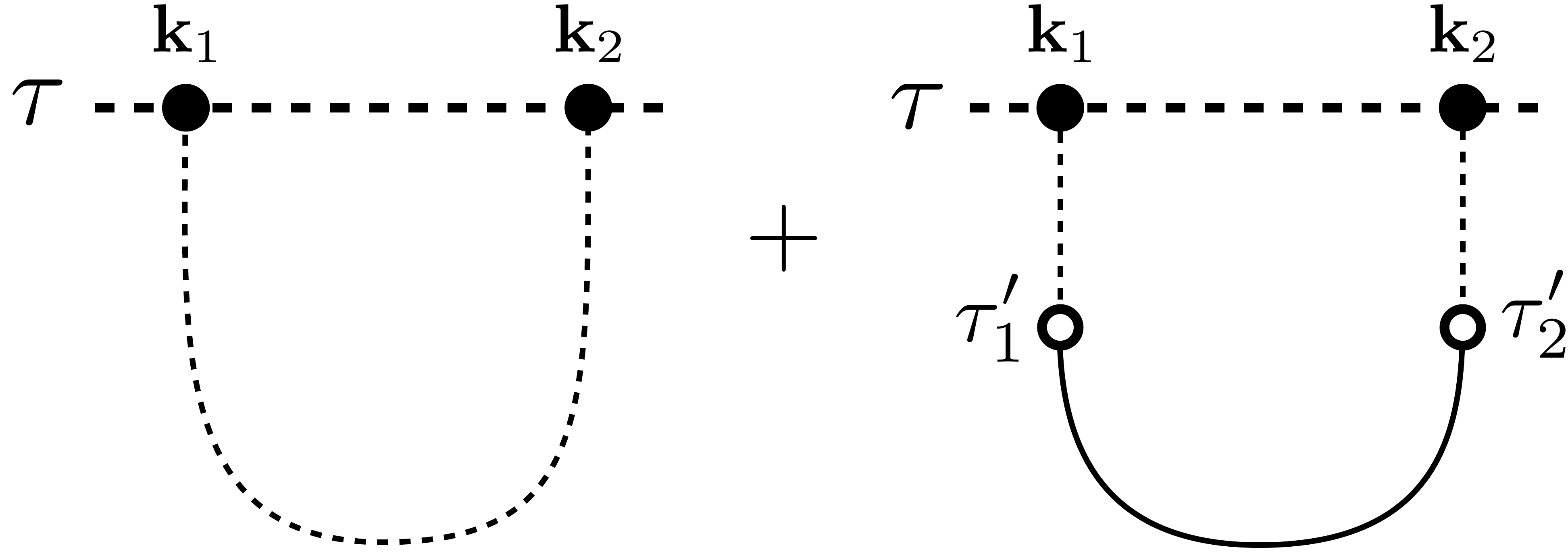}
\caption{The two diagrams contributing to the computation of the two-point function. The first diagram gives the standard power spectrum for $\zeta$, whereas the second diagram gives the correction due to the $\lambda$ derivative interaction.}
\label{fig:FIG_two-point-alpha}
\end{figure}

The power spectrum for $\zeta$ may be obtained by computing the two-point correlation function $\langle u (\x, \tau) u(\y, \tau) \rangle$. We will perform this computation up to order $\lambda^2$. Given a fluctuation $\varphi$, we define its power spectrum $P_\varphi (k)$ as
\be
\langle  \hat \varphi (\k_1) \hat \varphi (\k_2)  \rangle = (2 \pi)^3 \delta(\k_1 + \k_2) P_\varphi (k) .
\ee
To proceed with the computation of $P_{\zeta}$, we write $u (\x , \tau) = U^{\dag} (\tau) u_I (\x, \tau) U (\tau)$, where $U(\tau)$ is given by Eq.~(\ref{U-propagator}), with $H_I$ as in (\ref{int-alpha}). Up to second order in $\lambda$ this quantity is given by
\be
u (\x , \tau) = u_I (\x, \tau) +   i \int_{-\infty}^{\tau} \!\!\!\!\!\! d \tau' [ H_I^{\lambda} (\tau') , u_I (\x, \tau)]  
 -  \int_{-\infty}^{\tau} \!\!\!\!\!\! d \tau' \int_{-\infty}^{\tau'} \!\!\!\!\!\! d \tau'' [ H_I^{\lambda} (\tau'') , [ H_I^{\lambda} (\tau')  , u_I (\x, \tau)]] . \label{u-uI-lambda}
\ee
The two integrals may be solved in the superhorizon limit $k |\tau| \ll 1$. The first integral is found to be given by
\be
\begin{split}
 \int_{-\infty }^{\tau} \!\!\!\!\!\!\! d \tau' \left[ H_I^{\lambda} (\tau') , \hat u_I (\k , \tau) \right] = &  \lim_{k \bar \tau \to - \infty} \frac{ \lambda}{(2 k)^{3/2} \tau} \bigg( \gamma
 - 2 + \ln 2
 - \frac{i\pi}{2}  + 2 \ln (- k \tau) - \ln ( - k \bar \tau) \bigg)  a_{+}(\k)  \\ & \,\,\,\,\,\,\,\,\,\,\,\,\,\,\,\,\,\,\,\,\,\,\,\, + {\rm H.c.} (-\k), \label{comm-linear-1}
 \end{split}
\ee
whereas the second integral reads
\bea
 \! \int_{-\infty }^{\tau} \!\!\!\!\!\!\! d \tau' \int_{-\infty }^{\tau'} \!\!\!\!\!\!\! d \tau'' \left[ H_I^{\lambda} (\tau'') , \left[ H_I^{\lambda} (\tau') , \hat u_I (\k , \tau) \right]  \right]  &=& \lim_{k \bar \tau \to - \infty} \frac{i \lambda^2}{4 (2 k)^{3/2} \tau} \bigg( - 4 - \frac{\pi^2}{6} +
\Big[  \gamma - 2 - \frac{i \pi}{2}    \nn \\
&& +  \ln 2 + \ln (- k \bar \tau) \Big]^2   + 2 \Big[  \gamma - 2 - \frac{i \pi}{2} +  \ln 2  \nn \\ &&
+ \ln (- k \bar \tau) \Big] \Big[  \gamma - 2 - \frac{i \pi}{2} +  \ln 2 - \ln (- k \bar \tau) \nn
 \\ 
&& + 2 \ln (- k \tau) \Big] \bigg) a_{-}(\k) +  {\rm H.c.} (-\k) . 
\label{comm-linear-2}
\eea
These expressions, together with~(\ref{u-uI-lambda}), allow one to compute the two-point correlation function $\langle u (\x, \tau) u(\y, \tau) \rangle$. In momentum space, one obtains
\be \label{2pointcorrelation}
 \langle 0 | \hat u (\k_1 , \tau) \hat u (\k_2 , \tau) | 0 \rangle = (2 \pi)^3 \delta(\k_1 + \k_2) \frac{1}{2 k^3 \tau^2}   \bigg( 1 + \lambda^2 \Big[ A_1 - A_2 \ln (- k \tau) + \ln^2(- k \tau ) \Big]  \bigg) ,
\ee
where $A_1$ and $A_2$ are numbers given by
\bea
A_1 &=& - \frac{\pi^2}{6}  +  (3 - \ln 2) (1 - \ln 2) -  \gamma (4 - \gamma - 2 \ln 2)  , \qquad \\
A_2 &=&  4 -2 \gamma - 2 \ln 2 .
\eea
Their numerical values are $A_1 \simeq - 2.11$ and $A_2 \simeq 1.46$. Note that in putting together~(\ref{u-uI-lambda})-(\ref{comm-linear-2}) to compute the two-point correlation function, the divergent logarithms $\ln (- k \bar \tau)$ cancel out. The computation of the two-point correlation function of Eq.~(\ref{2pointcorrelation}) may be thought of as the result of the diagrammatic expansion of Fig.~\ref{fig:FIG_two-point-alpha}. The zeroth order contribution corresponds to the first diagram, whereas the contribution of order $\lambda^2$ corresponds to the second diagram, where the two external legs are mediated by a $v$ propagator.

Now, horizon crossing happens when $k |\tau| \simeq 1$. Thus, the number of $e$-folds after horizon crossing is given by
\be
\Delta N = - \ln (- k \tau) . \label{Delta-N-2-point}
\ee
It may be seen that after several $e$-folds the contribution to the power spectrum~(\ref{2pointcorrelation}) quadratic in $\Delta N$ dominates, and we obtain
\bea
&& \langle 0 | \hat u (\k_1 , \tau) \hat u (\k_2 , \tau) | 0 \rangle = (2 \pi)^3 \delta(\k_1 + \k_2) \frac{\lambda^2}{2 k^3 \tau^2} \Delta N^2  . \qquad \label{two-point-lambda-N}
\eea
For this to happen, we need $\lambda^2 \Delta N^2 \gtrsim 1$. The power spectrum for $v$ may be found through a similar computation, which gives
\bea
&& \langle 0 | \hat v (\k_1 , \tau) \hat v (\k_2 , \tau) | 0 \rangle = (2 \pi)^3 \delta(\k_1 + \k_2) \frac{1}{2 k^3 \tau^2}  ,
\eea
valid up to order $\mathcal O (\lambda^4)$. Combining these two results, we then derive the following relation between the two power spectra
\be
P_\zeta (k) =   \frac{\lambda^2}{2 \epsilon} \Delta N^2 P_\psi (k) , \label{power-zeta-psi}
\ee
where $P_\psi (k)$ is found to be given by
\be
P_\psi (k) = \frac{H^2}{2 k^3 } .
\ee
This result is consistent with the behavior shown in Eq.~(\ref{zeta-psi-N-1}) based on symmetry arguments. It shows that the power spectrum for $\zeta$ is proportional to the power spectrum of $\psi$, with a factor that grows with the number of $e$-folds. This does not spoil its near scale-invariance, as $\Delta N$ runs logarithmically on $k$. Concretely, we may write
\be
\Delta N(k) = - \ln \left( \frac{ e^{-60} k}{k_{\rm CMB}}  \right) = 60 \left( 1 - \frac{1}{60} \ln \left(\frac{ k}{k_{\rm CMB}} \right) \right) \sim 60 \times \left(\frac{k}{k_{\rm CMB}} \right)^{-1/60},
\ee
where we are implying that the longest mode on the CMB $k_{\rm CMB}$ spent 60 $e$-folds outside the horizon. Consequently, this yields a tilt $n_s - 1$ of $-1/30 = -0.033...$, which is close to (and consistent with) what is currently reported by the Planck collaboration~\cite{Aghanim:2018eyx} data $n_s = 0.9649 \pm 0.0042$.

Let us briefly comment on the validity of the result shown in Eq.~(\ref{two-point-lambda-N}). Our perturbative method consisted of separating the theory between a zeroth order Hamiltonian $H_0$ [given in Eq.~(\ref{H-0})] and an interaction Hamiltonian (given in Eq.~(\ref{int-alpha})), proportional to $\lambda$. On the one hand, we have argued that, in our final result for the power spectrum (\ref{2pointcorrelation}), we are allowed to retain as the dominant piece the term proportional to $\lambda^2$. On the other hand, notice that our perturbative method is valid as long as $\lambda^2 \ll 1$. These two statements are not in contradiction: the computation admits a cumulative effect that grows with the number of $e$-folds as $\lambda^2 \Delta N^2$, which may be larger than $1$ (after $\Delta N \simeq 60$). This effect was discussed in detail in Ref.~\cite{Achucarro:2016fby}, and it will play an important role in Sec.~\ref{sec:tom-NG}.

Also, the example of the derivative coupling we used here has a special property that, at superhorizon scales, the linear equation of motion for the isocurvaton field $\psi$ has no source term from the curvature mode $\zeta$. Therefore, $\psi$ does not grow once it exits the horizon, while $\zeta$ does. This means that, if we were to solve the coupled linear equation iteratively to all orders, the enhancement factor from $\Delta N$ would stay at the order $\Delta N^2$. Therefore, the requirement of the perturbation theory is only that $\lambda \ll 1$, and $\lambda \Delta N$ can be greater than 1.

Last but not least, even if the two conditions $\lambda^2 \Delta N^2 \gtrsim 1$ and $\lambda^2 \ll 1$ may seem to be fine-tuned, the condition $\lambda^2 \ll 1$ has only been adopted in order to be able to perform analytic computations. The requirement $\lambda^2 \Delta N^2 \gtrsim 1$ is valid independently of the perturbativity condition $\lambda^2 \ll 1$, and can already be inferred from the symmetry arguments around Eq.~(\ref{Pzeta-Ppsi}). In principle, Eq.~(\ref{Pzeta-Ppsi}) [or Eq.~(\ref{power-zeta-psi})] should be valid independently of the value of $\lambda$.

Thus, we have described how curvature perturbations in a generic multi-field setting are affected by isocurvature modes. We have verified that, at the linear level on the perturbations, the theory is Gaussian and the power spectrum is nearly scale invariant. Now we have a chance of exploring what are the consequences of the existence of isocurvature modes regarding the generation and detectability of primordial non-Gaussianity. 

\section{Detectability of Primordial non-Gaussianity}

As the statistics of the universe we observe are nearly Gaussian, deviations from Gaussianity, if there are any, are bound to be small. In this sense, it is natural to have a theory that predicts a negligibly small amount of non-Gaussianity. However, if it is too small to be measured in the foreseeable future, it is more rewarding to examine what theories may be ruled out (or confirmed) when the next batch of data arrives in the next decade.

In this sense, and as we have mentioned before, single-field slow-roll inflation predicts that the quantities controlling the non-linearities (or interactions) of the theory are the slow-roll parameters $\epsilon$ and $\eta$, which predicts a level of non-Gaussianity which is at least two orders of magnitude below the errorbar $\sigma(f_{\rm NL})$ that upcoming surveys~\cite{Lsst-web,Euclid-web,Bull:2018lat,Abazajian:2016yjj} will provide.

However, if we consider multi-field inflation, self-interactions of the isocurvature degree of freedom are not constrained by slow-roll parameters, and therefore if the isocurvaton $\psi$ interacts with the curvature perturbations $\zeta$ in the sense described above, it is conceivable that non-Gaussianities induced by the self-interactions of $\psi$ onto its own statistics are also transferred to the curvature perturbation $\zeta$. Conveniently, in this section we have described a mechanism with which this transfer can take place. In particular, equation~\eqref{zeta-psi-N-1} reveals that, provided a sufficient number of e-folds that the modes can spend outside the horizon so that the dominant contribution to $\zeta$ comes from $\psi_*$, the statistics of $\zeta$ should be completely induced by those of $\psi$. 

Throughout the next two chapters, we study the generation and phenomenology of this type of non-Gaussianity, dubbed here ``Tomographic non-Gaussianity''.

\cleardoublepage

\chapter{Tomographic non-Gaussianity: the $n$-point Function Approach} \label{chap:n-point}

Non-Gaussianity is one of the main probes in the search for signatures of new physics during the early stages of our universe. Any deviation from Gaussianity in the primordial fluctuations of the gravitational field would be a clear indicator that non-linear interactions were present at some point during the inflationary epoch.

However, as we commented earlier, within single-field slow-roll inflation the self-interactions of the adiabatic mode are suppressed by the slow-roll parameters, severely hindering measurability of their possibly observable effects. Nonetheless, multi-field inflation does provide mechanisms to generate larger amounts of non-Gaussianity, which could become measurable with the next generation of CMB~\cite{Abazajian:2016yjj} and LSS~\cite{Euclid-web,Lsst-web,Bull:2018lat} surveys. This, in turn, provides us with a window of opportunity to test and constrain theories that typically give rise to multi-field inflation, such as the various realizations of String Theory. Moreover, while current cosmic microwave background (CMB) constraints on the bispectrum are consistent with Gaussian statistics~\cite{Akrami:2019izv,Ade:2015ava}, it is possible that the method of three-point or higher-point correlation functions do not constitute the most efficient parametrization of primordial NG hidden in the data~\cite{Komatsu:2003fd, Buchert:2017uup}.

Thus, as we search for a new type of non-Gaussianity, besides providing a model in which such NG emerges, we need to use unconventional tools and observables to constrain such signals. In this section, we will do both: we will work out a model in which the departures from Gaussianity cannot be characterized by 3- or 4-point functions, and also provide the object that manages to do so. As we will later see, it is a probability density function. These developments were first reported in~\cite{Chen:2018uul,Chen:2018brw}.

\section{Correlation Functions and Deviations from Gaussianity}

It is useful to first overview the mathematics pertinent to the study of $n$-point functions, Gaussian statistics, and non-Gaussianity. That is to say, the statistics of fields. The main object of study throughout this Chapter will be $n$-point functions, denoted by
\be
\langle \zeta(\x_1) \cdots \zeta(\x_n)\rangle.
\ee
The bracket $\langle \cdot \rangle$ denotes an expectation value, i.e., an average taken with a certain prescription. For instance, Quantum Mechanics defines such a prescription given a state $\Psi$ upon which a field of operators (denoted by the same letter $\zeta$) act, 
\be
\langle \zeta(\x_1) \cdots \zeta(\x_n)\rangle = \bra{\Psi} \zeta(\x_1) \cdots \zeta(\x_n) \ket{\Psi},
\ee
and the expectation value is equal to the inner product of the states $\zeta(\x_1) \cdots \zeta(\x_n) \ket{\Psi}$ and $\ket{\Psi}$. On the other hand, the usual probabilistic way to obtain $n$-point functions is through a distribution $P[\zeta]$ that generates them by taking moments:
\be
\langle \zeta(\x_1) \cdots \zeta(\x_n)\rangle = \int D[\zeta] P [\zeta] \zeta(\x_1) \cdots \zeta(\x_n),
\ee
where functional integration is necessary as we are dealing with an observable $\zeta$ that takes values on a continuous domain: in order to compute an average, one needs to sum (integrate) over all possible outcomes (field configurations $\zeta(\x)$), which is equivalent to saying that it is necessary to integrate over all possible functions with an appropriate measure. The distribution $P[\zeta]$, which serves as a measure for the field configurations $\zeta(\x)$, is called the Probability Density Functional (PDF).

The archetypical example of such distribution is the Gaussian measure
\be \label{Gaussian}
P_G[\zeta] = \mathcal{N} \exp \left( -\frac{1}{2} \int_\x \int_\y \zeta(\x) \Sigma^{-1}(\x,\y) \zeta(\y) \right) 
\ee
where $\Sigma$ is the covariance matrix associated to the field distribution\footnote{The inverse is here meant with respect to the product defined by integrating along one of the arguments of $\Sigma$, and the unit element is the Dirac delta. That is to say, $\int_\z \Sigma(\x,\z) \Sigma^{-1}(\z,\y) = \int_\z \Sigma^{-1}(\x,\z) \Sigma(\z,\y) = \delta^{3} (\x - \y)$.  }. In this case, all $n$-point functions are completely determined by the 2-point function
\be
\langle \zeta(\x_1) \zeta(\x_2)\rangle = \Sigma(\x_1, \x_2),
\ee
because all the odd $n$-point functions vanish as the change $\zeta \to - \zeta$ leaves the measure invariant, while the even correlations are given by
\be
\langle \zeta(\x_1) \cdots \zeta(\x_{2n})\rangle = \sum_{\substack{ {\rm sets \, of \, pairings}  \\ \{(i_{2\alpha - 1}, i_{2\alpha}) \}_{\alpha = 1}^{n} }} \Sigma(\x_{i_1}, \x_{i_2} ) \cdots \Sigma(\x_{i_{2n - 1 }}, \x_{i_{2n}} ).
\ee
This is also known as \textit{Wick's theorem}.

When isotropy and homogeneity are incorporated, which is the typical physical situation, the Gaussian distribution can be further rewritten as
\be
P_G[\zeta] =  \mathcal{N} \exp \left( -\frac{1}{2} \int_\k \frac{\zeta(\k) \zeta(-\k)}{P_\zeta(k) }  \right),
\ee
where $P_\zeta(k)$ is called the \textit{Power Spectrum} of $\zeta$, and it is related to the (inverse) of the covariance matrix $\Sigma$ by
\be
\Sigma^{-1}(\x,\y) \equiv \int_\k \frac{e^{i \k \cdot (\x-\y) } }{P_{\zeta}(k)},
\ee
at which point we have recovered the setup that gives the leading-order predictions for the perturbations out of inflation
\be
\langle \zeta(\k_1) \zeta(\k_2)\rangle = (2\pi)^3 \delta^{3}(\k_1 + \k_2) P_\zeta(k_1).
\ee

How do we incorporate non-Gaussianity in this framework? Strictly speaking, any PDF that is not equivalent to~\eqref{Gaussian} defines non-Gaussian statistics. However, since we know that the primordial statistics for the curvature perturbation are nearly Gaussian, the appropriate approach is to look for small deformations of a Gaussian measure. We will tackle this problem directly in Chapter~\ref{chap:CMB-LSS}. For the moment, we will probe for departures from Gaussianity by computing $n$-point functions starting from the interaction picture of Quantum Mechanics 
\be
\langle \zeta(\x_1) \cdots \zeta(\x_n)\rangle = \bra{\Psi} U^\dagger \zeta(\x_1) \cdots \zeta(\x_n) U \ket{\Psi},
\ee
where $U$ is the corresponding temporal evolution operator. It will turn out that the type of NG that we will study can be fully characterized by a reduced PDF that only takes into account the single-point statistics of the field. That is to say, the object we will reconstruct in this chapter is a distribution $\rho \left( \zeta(\x) \right)$ such that
\be
\langle \zeta(\x)^n \rangle = \int_{-\infty}^{\infty} d \left(\zeta(\x) \right) \rho \left( \zeta(\x) \right) \left( \zeta(\x) \right)^n.
\ee
At the end of this Chapter we will also discuss how this result may be generalized to other (inflationary or non-inflationary) setups, along with other ways to probe for this type of primordial non-Gaussianity.

\section{Description of the mechanism sourcing the Perturbations}

In general situations we expect a self-interaction affecting the dynamics of the isocurvature field $\psi$ to be present. Thus, instead of the Lagrangian (\ref{quad-action}), we may consider the following extension:
\be
\mathcal L (\zeta , \psi) = a^3 \Big[\epsilon  ( \dot \zeta - \alpha \psi ) ^2 -   \frac{\epsilon}{a^2} (\nabla \zeta)^2 
+  \frac{1}{2} \dot \psi^2 - \frac{1}{2a^2} (\nabla \psi)^2  - \Delta V(\psi) \Big]. \label{Lagrantian-full-v}
\ee
Notice that the potential $\Delta V(\psi)$ is replacing the initial mass term of Eq.~(\ref{quad-action}). Without any concrete knowledge of $\Delta V(\psi)$ we would expect that it could be expanded in a power series of the form:
\be
\Delta V(\psi) \simeq \frac{1}{2} \mu^2 \psi^2 + \frac{1}{6} g \, \psi^3 + \cdots . \label{hierarchical-exp-V}
\ee
However, such an expansion assumes that the coefficients $\mu^2$, $g$, etc... are such that, for amplitudes of $\psi$ characteristic of horizon crossing, the higher order terms of the expansion remain suppressed. In this work we want to explore those situations where the fluctuations $\psi$ are such that we cannot disregard the structure of $\Delta V(\psi)$ by assuming the hierarchical expansion of (\ref{hierarchical-exp-V}). In this sense, $\Delta V(\psi)$ is the \textit{Landscape} of the perturbations.

Among the theories that give rise to multi-field inflation, it has been of interest to study setups involving particles known as ``axions'', because axion-like particles are a natural outcome of string theory~\cite{Marsh:2015xka,Cicoli:2013ana,Halverson:2018xge}, and among other things, they can accommodate the observed cosmological constant\footnote{See for instance~\cite{Bachlechner:2017zpb,Bachlechner:2017hsj,Bachlechner:2018gew,Bachlechner:2019vcb} for a realization of inflation with nothing but axionic directions in field space.}. In this context, an axion-like particle is a scalar field with a periodic potential. The potential that is typically considered is
\be
V(\psi) = \Lambda^4 \left[ 1 - \cos (\psi/f) \right],
\ee
where $\Lambda$ is the characteristic energy scale of the interaction, and $f$ would be the axion decay constant. Therefore, it is natural to consider an axion-like potential
\be
\Delta V(\psi) =  \Lambda^4 \left[ 1 - \cos \left( \frac{\psi}{f} \right) \right]. \label{potential-psi}
\ee

To continue, notice that the potential (\ref{potential-psi}) breaks the shift symmetry (for $\mu = 0$) of the Lagrangian (\ref{quad-action}) down to a discrete symmetry:
\bea
&& \psi \to \psi' = \psi +  2 \pi n f , \\
&& \zeta \to \zeta' = \zeta +  2 \pi n f  \int^t \!\!  dt \, \alpha .
\eea
This time, $\psi$ may acquire constant solutions that minimize the sinusoidal potential. On superhorizon scales, any of these solutions will dominate the behavior of $\psi$. Just as before, $\zeta$ will be sourced by $\psi$, but this time the enhancement will happen for those values of $\psi$ that minimize the potential. This result suggests that the statistics transferred from $\psi$ to $\zeta$ will continue to be operative in this new context, but in a manner that it will be enhanced at those values in which $\psi$ coincides with a minimum of the potential, and suppressed for those values in which $\psi$ coincides with a maximum. Therefore, the structure of the potential will be necessarily inherited by the PDF of the curvature perturbation $\zeta$, which becomes non-Gaussian.

Presumably, the Lagrangian~(\ref{Lagrantian-full-v}) is the result of perturbing a more fundamental multifield theory, with a scalar field potential of the form $V = V_0 + \Delta V$. We will be interested in the regime $\Lambda^4 /3 H^2 M_{\rm Pl}^2 \ll 1$, so the potential $\Delta V$ has little to say about the background dynamics of the full system, and inflation is driven by a piece $V_0$. Then, the background equations of motion require $V_0 \sim 3 H^2 M_{\rm Pl}^2$. In addition, if $\Lambda^4 / 3M_{\rm Pl}^2H^2 \ll 1$ then the symmetry breaking is mild and, for all practical purposes, before and during horizon crossing the field $\psi$ will behave as an ultralight field. This means that at horizon crossing $\psi$ will freeze, and Eq.~(\ref{zeta-psi-N-1}) will describe how $\psi$ transfers its statistics to $\zeta$. As times passes, the nonlinearities due to $\Delta V(\psi)$ will start to become accentuated, and one expects a nonlinear contribution to Eq.~(\ref{zeta-psi-N-1}) coming from the nonlinear evolution of $\psi$ that does not freeze. To leading order in $\alpha$ we expect that any level of nonlinearity will be communicated to $\zeta$ through a non-Gaussian contribution to the $n$-point correlation functions of the form
\be
\langle \zeta^n \rangle_{\rm NG} \propto \left( \frac{\alpha}{H} \Delta N \right)^n \langle \psi^n \rangle_{\rm NG} . \label{NG-transfer}
\ee
The reason behind this guess is the following: First, in the absence of interactions between the two fields ($\alpha = 0$), the fluctuation $\psi$ will acquire a non-Gaussian distribution due to its potential $\Delta V(\psi) = \Lambda^4 [1 - \cos (\psi / f)]$. The non-Gaussian contributions to the $n$-point correlation functions were computed in Ref.~\cite{Palma:2017lww} using the in-in formalism, and are found to be given by
\be
\langle \psi (\k_1, \tau)  \cdots \psi(\k_n , \tau) \rangle_c  =  (-1)^{n/2}  (2\pi)^3 \delta^{(3)}
\Big(\sum_j \k_j \Big) \frac{2}{3} \frac{\Lambda^4}{H^4 }  e^{ - \frac{\sigma_0^2}{2 f^2} }  \left( \frac{H^2}{2 f } \right)^{n}  \frac{  k_1^3 + \cdots + k_n^3 }{ k_1^3 \cdots k_n^3 } \Delta N, \label{n-point-psi}
\ee
where $\sigma_0^2$ is the variance of $\psi$ appearing from loop resummations. In the previous expression the subscript $c$ indicates that we are only taking into account the diagrammatic contributions due to the potential $\Delta V(\psi)$ that are fully connected (which is why there is a single overall Dirac-delta function on the right-hand side of (\ref{n-point-psi}) enforcing momentum conservation). This set of $n$-point correlation functions are generated during horizon crossing.

Second, going back to our current setup, let us turn on the coupling $\alpha \neq 0$. Then, because we are assuming that $\Delta V(\psi) / 3 H^3 M_{\rm Pl}^2 \ll 1$ the field $\psi$ is essentially massless and the linear relation (\ref{zeta-psi-N-1}) will remain valid on superhorizon scales, independent of the nonlinear dynamics. This implies that any non-Gaussianity gained by $\psi$ during horizon crossing will be transferred to $\zeta$ via Eq.~(\ref{NG-transfer}) after horizon crossing.  As we shall see, a detailed computation leads to
\be
\langle \zeta^n \rangle_{\rm NG}  \simeq \frac{1}{2} \left( \frac{\alpha}{H} \Delta N \right)^n \langle \psi^n \rangle_{\rm NG} , \label{n-points-zeta-psi}
\ee
where the factor $1/2$ comes from the interaction structure implied by certain nested integrals appearing in the computation of the $n$-point correlation functions using the in-in formalism.

The importance of Eqs.~(\ref{n-point-psi}) and~(\ref{n-points-zeta-psi}) is that they allow us to infer a probability distribution function for $\zeta$. This PDF is characterized by a class of non-Gaussianity that cannot be fully captured with three- or four-point correlation functions, as opposed to the case springing from the ansatz~(\ref{hierarchical-exp-V}). This PDF is found to be given by (see Sec.~\ref{subsec:PDF-derivation-from-n-points-full} for the derivation)
\be
\rho (\zeta) = \frac{e^{- \frac{\zeta^2}{2 \sigma_{\zeta}^2}}}{\sqrt{2\pi} \sigma_{\zeta}}  \left[ 1 + A^2 \int_0^{\infty} \!\!\frac{dx}{x} \, \mathcal{K}(x)  \left(  \frac{  \sigma_{\zeta}^2 }{ f_\zeta(x)^2 }   \cos\left( \frac{\zeta}{f_\zeta(x)} \right) -  \frac{  \zeta }{ f_\zeta(x) }   \sin\left( \frac{\zeta}{f_\zeta(x)}  \right) \right)  \right] , \label{distribution-0}
\ee
where $\sigma_{\zeta}$ is the variance of $\zeta$ parametrizing the Gaussian part of the distribution. In the previous expression $\mathcal K(x)$, $f_{\zeta}(x)$ and $A$ are given functions and quantities determined by parameters related to $\Delta V(\psi)$ that will be deduced in the next sections.

The main characteristic of the distribution function $\rho (\zeta)$ is that, in spite of the $x$ integral, it inherits the structure of the potential $\Delta V(\psi)$. That is, the probability of measuring $\zeta$ increases (decreases) if the field $\psi$ sourcing its amplitude is at a local minimum (maximum) of $\Delta V(\psi)$. The mechanism described here is certainly not exclusive to the potential $\Delta V(\psi)$ given in Eq.~(\ref{potential-psi}). It should be safe to suspect that any potential $\Delta V(\psi)$ with a rich structure (i.e., characterized by field distances $\Delta \psi$ smaller than $H$) will imply the existence of some level of non-Gaussianity for $\zeta$ revealing the structure of $\Delta V(\psi)$. In fact, we will prove this statement in Section~\ref{sec:gen-NG-Tom}. Thus, we see that the type of non-Gaussian departures discussed here in principle gives us nontrivial information about the landscape, offering us tomographic information about the shape of the multi-field potential.

Before finishing this section, let us mention that the field $\psi$ considered here is not expected to be a true axion as realized in QCD or string theory~\cite{Peccei:1977hh, Svrcek:2006yi} for the range of parameters that we are interested in, but only axion-like. The reason is that large fluctuations of $\psi$ traversing many minima of the potential would destabilize the radial field fixing the value of the axion decay constant $f$~\cite{Lyth:1992tx}. For this reason, we take the potential of Eq.~(\ref{potential-psi}) to be representative of systems with potentials with a rich structure, as expected in the string landscape. See Ref.~\cite{Linde:1993xx} for a previous work that has studied the system (\ref{Lagrantian-full-v}) with an axionlike potential (for the decoupled case $\alpha = 0$) analyzing issues related to the landscape.

\subsection{Diagrammatics of the Tomography} \label{sec:stage}

In order to probe for primordial NG arising from an axion-like potential, we need to compute $n$-point correlation functions of $\zeta$ resulting from a sinusoidal potential~\eqref{potential-psi} 
\be
\Delta V(\psi) = \Lambda^4 \left[ 1 - \cos (\psi/f) \right].
\ee 
To proceed, we follow the same procedure as in section~\ref{sec:linearth}, writing the theory~\eqref{Lagrantian-full-v} in terms of canonical fields $u$ and $v$, %introduce canonical fields $u$ and $v$ defined as:
%\be  \label{uv-def}
%u  \equiv \sqrt{2\epsilon} a \zeta,  \,\,\,\,\,\,\, v \equiv  a \psi .
%\ee
%To simplify any computation involving $u$ and $v$ we will assume a purely de Sitter background, with $a(t)=e^{H t}$. This means that our computations will have corrections of order $\epsilon$ not accounted for. It is also convenient to introduce conformal time $\tau$ (via $d\tau = dt / a = dt e^{-Ht}$) and write
%\be
%a(\tau)= - 1 / H\tau , \label{a-tau}
%\ee
%with $- \infty <\tau < 0$. Then, putting together Eqs.~(\ref{u-def})-(\ref{a-tau}) back into the action~(\ref{Lagrantian-quadratic}), we obtain
\be
\begin{split}
S = \frac{1}{2} \int d^3 x d\tau  \Bigg[ \left( u'  + \frac{\lambda}{\tau}  v \right)^2 +\frac{2}{\tau^2} u^2    -  (\nabla u)^2   +  ( v' )^2 + & \frac{2}{\tau^2} v^2    - (\nabla v)^2  + \lambda \frac{2}{\tau^2} u v  \\
 & - 2 a^4 \Lambda^4 \bigg( 1 - \cos \Big(\frac{v}{af} \Big) \bigg) \Bigg] ,  \label{action-not-quadratic-u-v}
\end{split}
\ee
which we now have to quantize.
%where we have introduced the dimensionless coupling
%\be \label{lambda}
%\lambda\equiv\frac{\sqrt{2\epsilon}\alpha }{H},
%\ee
%which is taken to be nonvanishing in the de Sitter limit. From Eq.~(\ref{action-quadratic-u-v}), we infer that the canonical momenta associated with $u$ and $v$ are,  respectively, given by
%\bea
%\Pi_u &=& u'  + \frac{\lambda}{\tau}  v , \label{can-momenta-1} \\
%\Pi_v &=& v' . \label{can-momenta-2}
%\eea
%These momenta satisfy the equal time commutation relations, given by
%\bea
%\left[u(\x, \tau), \Pi_u (\y, \tau)\right] &=& i \delta^{(3)} (\x - \y) , \label{commut-rel-1} \\
%\left[v(\x, \tau), \Pi_v (\y, \tau)\right]  &=& i \delta^{(3)} (\x - \y)  ,\label{commut-rel-2}
%\eea
%with every other commutator vanishing. From (\ref{can-momenta-1}) and (\ref{can-momenta-2}) we see that the Hamiltonian of the system is given by
%\be
%\begin{split}
%H \! = \! \frac{1}{2} \int_x \Bigg[ \Pi_u^2 + (\nabla u)^2 - \frac{2}{\tau^2} u^2 + \Pi_v^2 +  (\nabla v)^2 - &\frac{2}{\tau^2} v^2 - \frac{2 \lambda}{\tau} v   \left(  \Pi_u + \frac{u}{\tau}  \right) \\ & + 2 a^4 \Lambda^4 \bigg( 1 - \cos \Big(\frac{v}{af} \Big) \bigg) \Bigg] ,
%\end{split}
%\ee
%where we have introduced the notation $\int_x = \int d^3x$.

\subsubsection{Splitting the theory}

We may now split the Hamiltonian into three contributions as $H = H_0 + H_\lambda + H_{\Lambda}$, where $H_0$ corresponds to the free Hamiltonian of the system (obtained in the limit $\lambda = \Lambda = 0$). Notice that $H_0$ describes a system with two decoupled massless scalar perturbations
\be
H_0 =  \frac{1}{2} \int_x \left[ \Pi_u^2 + (\nabla u)^2 - \frac{2}{\tau^2} u^2 + \Pi_v^2 + (\nabla v)^2 - \frac{2}{\tau^2} v^2 \right] .
\ee
On the other hand, $H_\lambda$ contains the interaction term proportional to $\lambda$,
\be
H_\lambda = - \int_x  \frac{\lambda}{\tau} v   \left(  \Pi_u + \frac{u}{\tau} \right)  , 
\ee
and $H_\Lambda$ contains the self-interactions for $v$
\be
H_\Lambda =  \int_x a^4 \Lambda^4 \bigg( 1 - \cos \Big(\frac{v}{af} \Big) \bigg) .  \label{H-Lambda}
\ee
As before, we may quantize the system by adopting the interacting picture framework using the definitions introduced in~\ref{sec:linearth}, from~\eqref{u-v-U-1} until~\eqref{k-mode-solutions}. 
%That is, the quantum fields $u$ and $v$ are expressed as
%\bea
%u (\x , \tau) &=& U^\dag (\tau)  u_I (\x , \tau)  U (\tau),  \label{u-v-U-1} \\
%v (\x , \tau) &=& U^\dag (\tau)  v_I (\x , \tau)  U (\tau), \label{u-v-U-2}
%\eea
%where $u_I (\x , \tau)$ and $v_I (\x , \tau)$ are the interaction picture fields, which evolve as quantum fields of the free theory. Explicitly, they are given by
%\bea
%u_I(\x , \tau) &=& \int_k \, \hat u_I (\k , \tau) \, e^{- i \k \cdot \x }, \\
%v_I(\x , \tau) &=& \int_k \, \hat v_I (\k , \tau) \,  e^{- i \k \cdot \x },
%\eea
%with $\int_k = (2\pi)^{-3} \int d^3 k$, and where
%\bea
%\hat u_I (\k , \tau) &=& u_k(\tau) a_{-}(\k) + u_k^*(\tau) a_{-}^\dag(-\k)  , \label{Fourier-u-v-1} \\
%\hat v_I (\k , \tau) &=& v_k(\tau) a_{+}(\k) + v_k^*(\tau) a_{+}^\dag(-\k) . \label{Fourier-u-v-2}
%\eea
%Here, the pairs $a_{\pm}(\k)$ and $a_{\pm}^\dag(\k)$ correspond to creation and annihilation operators satisfying the commutation relations:
%\be
%\left[ a_{b} (\k) , a_{c}^{\dag} (\k') \right] = (2 \pi)^3 \delta^{(3)} (\k - \k') \delta_{bc} , \label{cre-anh-def}
%\ee
%with \(b,c \in \{+, -\}\). The mode functions $u_k(\tau)$ and $v_k(\tau)$ are both given by
%\be
%u_k(\tau) = v_k(\tau) = \frac{1}{\sqrt{2 k}} \left( 1 - \frac{i}{k \tau} \right) e^{- i k \tau} , \label{k-mode-solutions}
%\ee
%which corresponds to the standard expression for a massless mode on a de Sitter spacetime with Bunch-Davies initial conditions. 
Then $U (\tau)$, the propagator in the interaction picture, which is given by
\be
U (\tau) = \mathcal T \exp \left\{ - i \int_{-\infty_{+}}^{\tau} \!\!\!\!\!\! d \tau' H_I (\tau')  \right\} , \label{U-propagator-2}
\ee
where $\mathcal T$ stands for the time ordering symbol and $\infty_{\pm} = \infty (1 \mp i \epsilon)$, with $\epsilon$ is a small positive number introduced to select the correct interaction picture vacuum. Finally, $H_I$ in Eq.~(\ref{U-propagator-2}) is given by
\be
H_I = H_I^{\lambda}  + H_I^{\Lambda}   ,
\ee
where
\bea
H_I^{\lambda} &=& - \int_x  \frac{\lambda}{\tau} v_I   \left(  \Pi_u^{I} + \frac{u_I}{\tau}  \right) , \label{int-alpha-2} \\
H_I^{\Lambda} &=& \int_x a^4 \Lambda^4 \bigg( 1 - \cos \Big(\frac{v_I}{af} \Big) \bigg) . \label{int-Lambda}
\eea
%Notice that in the previous expressions the canonical momenta $\Pi_u^{I}$ and $\Pi_v^{I}$ in the interaction picture are simply given by
%\be
 %\Pi_u^{I} = \frac{d}{d\tau} u_I , \qquad  \Pi_v^{I} = \frac{d}{d\tau} v_I .
 %\ee
 In order to deal with $H_I^{\Lambda}$, we will consider the Taylor expansion of the cosine function as:
\be
H_I^{\Lambda}(\tau) = - \frac{ \Lambda^4 }{H^4 \tau^4} \sum_{m=1}^{\infty} \int_x   \frac{(-1)^{m}}{(2m)!} \Big(\frac{H \tau }{f} v_I(\x,\tau) \Big)^{2m} .  \label{cosine-expansion}
\ee
This expansion gives us an infinite number of vertices to deal with. As we shall see in Sec.~\ref{sec:npoints}, it will be possible to resum this expansion back into an exponential contribution, leading to nonperturbative results in terms of the ratio $H/f$.

\begin{figure}[t!]
\begin{center}
\includegraphics[scale=0.45]{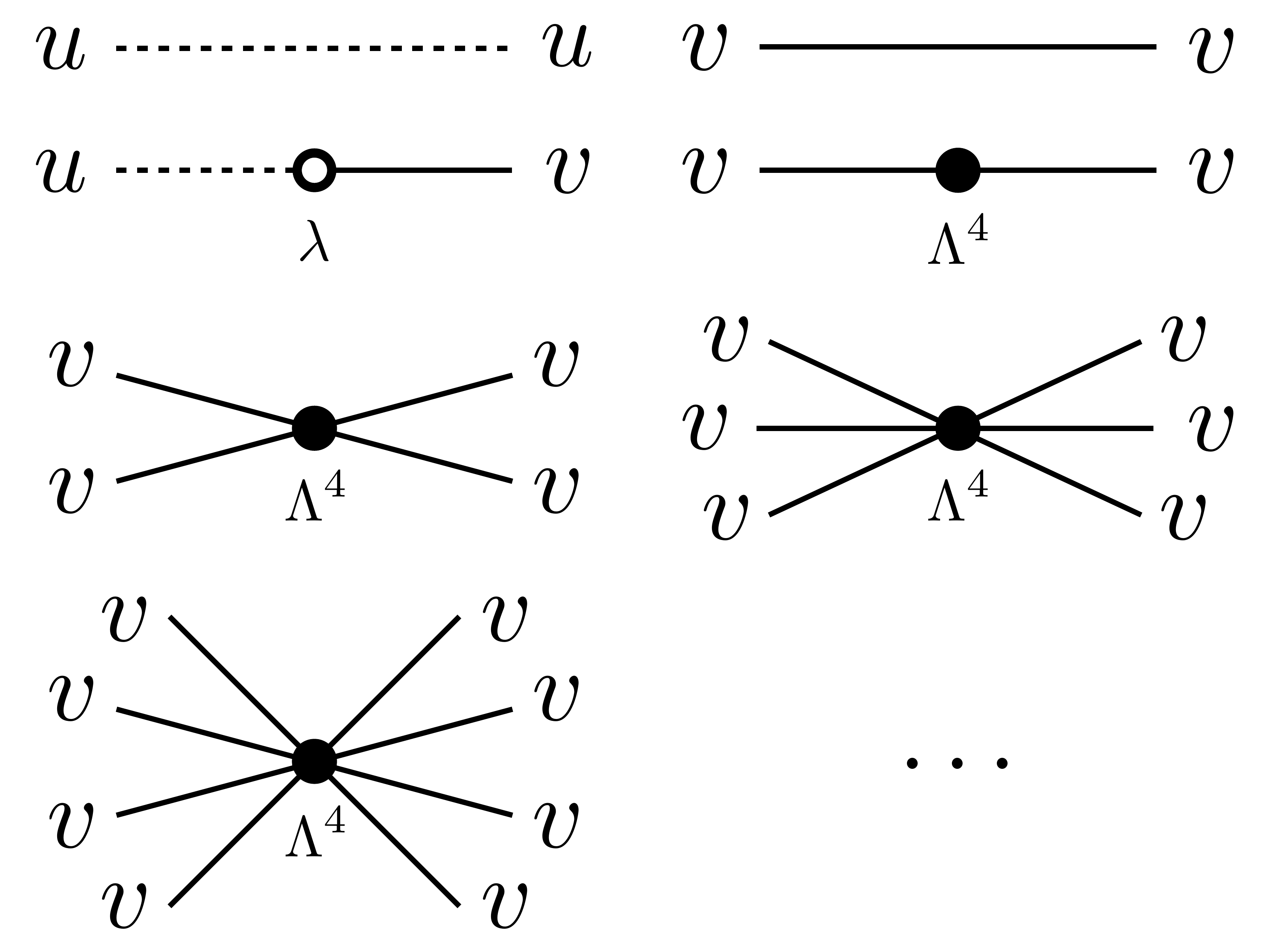}
\caption{The various diagrams of the theory. The empty circle denotes the two leg vertex offered by the Hamiltonian of Eq.~(\ref{int-alpha-2}). The diagrams with solid circles denote the vertices coming from the expansion (\ref{cosine-expansion}). }
\label{fig:FIG_diagrams-theory}
\end{center}
\end{figure}

The computation of $n$-point correlation functions of the following sections may be organized diagrammatically. These computations involve contractions of the Hamiltonians $H_I^{\lambda}$ and $H_I^{\Lambda}$ and the fields $u_I$ and $v_I$. In the present context, a contraction is the result of a commutation between creation and annihilation operators introduced in Eqs.~(\ref{Fourier-u-v-1}) and~(\ref{Fourier-u-v-2}) that results from their normal ordering.  A commutation involving the pair $a_{-}^\dag$ and $a_{-}$ is represented by a dashed line joining vertices (or external legs) labeled by the $u$ fields that contain these operators. Similarly, a commutation involving the pair $a_{+}^\dag$ and $a_{+}$ is represented by a solid line labeled by the $v$ fields that contain these operators. If the field participating in the commutation comes from $H_I^{\lambda}$, then the line joins a vertex with an empty circle. Otherwise, if the field participating in the commutation comes from $H_I^{\Lambda}$, then the line joins a vertex with a solid circle. Figure~\ref{fig:FIG_diagrams-theory} shows the various classes of diagrams appearing in the computation of $n$-point correlation functions.

\subsubsection{Perturbativity conditions} \label{pert-conds}

In the next sections, we compute the $n$-point correlation functions perturbatively. Given that we will resum the expansion in $H/f$ appearing in~(\ref{cosine-expansion}), we will not impose any condition on the size of $f$. On the other hand, it is worth counting with some criteria on how large $\lambda$ and $\Lambda$ can be in order to have a well-behaved perturbation theory. A naive estimation (that does not take into account renormalization) may be obtained by rewriting the Lagrangian~\eqref{Lagrantian-full-v} in a dimensionless form, weighting spacetime variables and fields by their characteristic values. In de Sitter, a characteristic length scale is given by $H^{-1}$. Moreover, the amplitude of massless scalar fields around horizon crossing is of order $H$. Thus, redefining spacetime and field variables as
\bea
t &\to& \bar t = t H ,  \\
\x &\to& \bar \x = \x H , \\
\psi &\to& \bar \psi = \frac{1}{H}\psi  , \\
\zeta &\to& \bar \zeta = \frac{\sqrt{2 \epsilon}}{H} \zeta ,
\eea
and after writing $ \bar {\mathcal L} = \mathcal L / H^4$, the Lagrangian~\eqref{Lagrantian-full-v} becomes
\be
 \bar {\mathcal L} (\bar \zeta , \bar \psi) = a^3\Bigg[\frac{1}{2} \left(\partial_{\bar t} \bar \zeta - \lambda \bar \psi \right) ^2 - \frac{1}{2} \frac{(\bar{\nabla}  \bar \zeta)^2}{a^2}  \nn \\
 +   \frac{1}{2}  ( \partial_{\bar t} \bar\psi)^2 -  \frac{1}{2}  \frac{(\bar \nabla \bar \psi)^2}{a^2}  -  \frac{1}{H^4} \Delta V (g \bar \psi) \Bigg], \nn
\ee
where $\lambda= \sqrt{2 \epsilon} \alpha / H$ is the dimensionless coupling already defined in Eq.~\eqref{lambda}. Here, derivatives are with respect to the dimensionless variables, and we have further defined the ratio $g\equiv H/f$. By asking that the dimensionless couplings remain small, we obtain the following perturbativity conditions
\be \label{perturbative-cond}
\frac{\Lambda^4}{H^4} \ll 1 \qquad \text{and} \qquad \lambda \ll 1,
\ee
for the potential~\eqref{potential-psi}. Note that the first condition is stronger than $\Lambda^4 / 3M_{\rm Pl}^2H^2\ll 1$, which is required for the background evolution not to be affected by the dynamics of $\psi$. Also, recall that we are not restricting the value of $g=H/f$, as the results of the next sections are nonperturbative with respect to this parameter.

As we shall see in Sec.~\ref{sec:npoints}, the loop corrections due to the resummation of the sinusoidal potential \eqref{cosine-expansion} will renormalize the bare coupling $\Lambda$. Moreover, derivative operations on the potential will be made, giving extra factors of $f$ to the final result. The consequence of this is that the correct perturbative parameter will turn out to be
\be
\frac{\Lambda^4_{\rm ren}}{H^2 f^2} = \frac{\Lambda^4}{H^2 f^2} e^{- \frac{\sigma_{S}^2}{2 f^2}} , \label{Lambda-ren}
\ee
where $\sigma_{S}^2$ is a short wavelength contribution to the variance of the field $\psi$ (we will compute this quantity in Sec.~\ref{subsec:n-point}).  In summary, our results will be perturbative in the couplings $\Lambda_{\rm ren}^4/(f^2 H^2)$ and $\lambda$, but nonperturbative in the parameter $H/f$ because \eqref{cosine-expansion} will be eventually re-summed.

\subsection{Computation of correlation functions}  \label{sec:npoints}

\begin{figure*}[t!]
\includegraphics[scale=0.45]{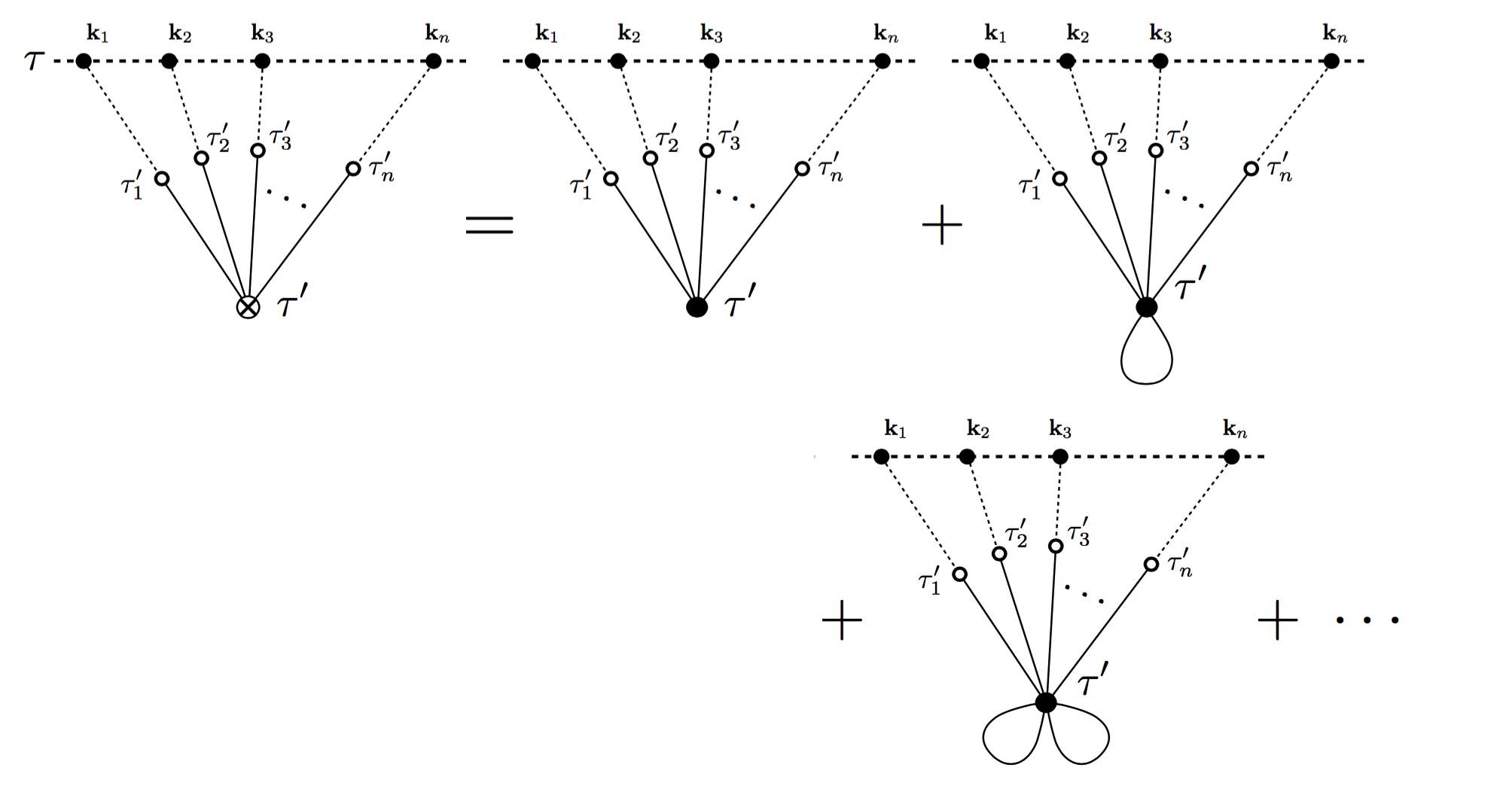}
\caption{All the connected diagrams contributing to $\tilde G^{(n)}$ at order $\Lambda^4$. The $\otimes$ vertex represents the resummation of all the loop contributions coming from the expansion of the cosine function shown in Eq.~(\ref{cosine-expansion}). In other words, for a given number of legs, the $\otimes$ vertex contains all the relevant effects due to the cosine. Because of the combinatorial factors of each diagram, and given that there are no external momenta running through the loops, the resummation reduces to a constant factor given by $\exp({- \sigma^2 / 2 f^2})$.}
\label{fig:FIG_diagram_res}
\end{figure*}

In this section, we describe how to compute the $n$-point correlation functions of $\zeta$ at the end of inflation (details of this computation are shown in Appendix~\ref{chap:details}). The quantity of interest corresponds to
\be
\tilde G^{(n)}_{\zeta} (\tau, \k_1 , \cdots , \k_n)  \equiv \langle \zeta (\k_1, \tau)  \cdots \zeta(\k_n , \tau) \rangle,
\ee
which, in terms of the canonically normalized fields introduced in Sec.~\ref{sec:stage}, is given by
\be
\tilde G^{(n)}_{\zeta}  = \left(  \frac{H |\tau| }{\sqrt{2 \epsilon}} \right)^n \langle \hat u (\k_1, \tau)  \cdots \hat  u(\k_n , \tau) \rangle . \label{n-point-u-1}
\ee
Our main goal is to obtain an expression for this function up to order $\Lambda^4$. Given that the interaction Hamiltonian~(\ref{int-Lambda}) determined by the potential $\Delta V(\psi)$ does not depend on $u_I$, the $n$-point correlation functions will acquire a dependence on $\Lambda^4$ only through the mixing Hamiltonian~(\ref{int-alpha-2}) involving the coupling $\lambda$. This means that the fully connected contribution to (\ref{n-point-u-1}) will necessarily involve at least one factor $\lambda$ per field $u(\x_n , \tau)$. In other words, the lowest order contribution to (\ref{n-point-u-1}) represented by fully connected diagrams, will be of order $\Lambda^4 \lambda^n$. The diagrammatic representation of this computation in momentum space is shown in Fig.~\ref{fig:FIG_diagram_res}. The $\otimes$ vertex denotes the exact vertex, up to order $\Lambda^4$, connecting the $n$ external legs participating in (\ref{n-point-u-1}). Because the expansion~(\ref{cosine-expansion}) contains an infinite number of vertices (with an even number of legs), the exact vertex consists of the sum of an infinite number of diagrams involving loops that start and finish on the same vertices. Due to overall momentum conservation, these loops do not carry external momenta.

To proceed with the computation of $\tilde G_\zeta^{(n)}$, we start by recalling that a given $\hat u (\k_j , \tau)$ entering the $n$-point function $\langle \hat u (\k_1, \tau)  \cdots \hat  u(\k_n , \tau) \rangle$ of Eq.~(\ref{n-point-u-1}) has the form  $ u (\k_j , \tau) = U^{\dag} (\tau) u_I (\k_j, \tau) U (\tau) $.  Let us for a moment disregard the $\epsilon$ prescription determining the integration limits $\infty_{\pm}$. Then, by expanding the propagator $U (\tau)$ in this expression, we obtain
\be
\begin{split}
u (\k_j , \tau) =& \,\, u_I (\k_j , \tau) +   i \int_{-\infty}^{\tau} \!\!\!\!\!\! d \tau' [ H_I (\tau') , u_I (\k_j , \tau)] 
\\ & -  \int_{-\infty}^{\tau} \!\!\!\!\!\! d \tau' \int_{-\infty}^{\tau'} \!\!\!\!\!\! d \tau'' [ H_I (\tau'') , [ H_I (\tau' ) , u_I (\k_j , \tau)]]  + \cdots .
\end{split}
\ee
We only need to keep terms up to order $\Lambda^4$. Then, because $H_I^{\Lambda}$ commutes with $u_I$, but not with $v_I$, the previous equation may be further reduced to
\be
\begin{split}
u (\k_j , \tau)  =& \,\, u_I (\k_j , \tau) +   i \int_{-\infty}^{\tau} \!\!\!\!\!\! d  \tau_j   [ H_I^{\lambda} (\tau_j) , u_I (\k_j , \tau) ]  \\
 & -  \int_{-\infty}^{\tau} \!\!\!\!\!\! d \tau_j \int_{-\infty}^{\tau_j} \!\!\!\!\!\! d \tau' [ H_I^{\Lambda} (\tau') , [ H_I^{\lambda} (\tau_j)  , u_I (\k_j , \tau) ]]  \\
 & + \sum_{p=1}^{n-1} i^{p+2} \int_{-\infty}^{\tau} \!\!\!\!\!\! d \tau_j \int_{-\infty}^{\tau_j} \!\!\!\!\!\! d \tau' \int_{-\infty}^{\tau'} \!\!\!\!\!\! d \tau'_1 \cdots   \int_{-\infty}^{\tau'_{p-1}} \!\!\!\!\!\! d \tau'_p  \\ & \,\,\,\,\,\, \times
  \left[
  H^{\lambda}_I (\tau'_{p}) , \cdots \left[
  H^{\lambda}_I (\tau'_1) , \left[ H^{\Lambda}_I (\tau') , \left[ H^{\lambda}_I  (\tau_j)  , u_I (\k_j , \tau) \right] \right]
   \right] \cdots
    \right]
  \\
 & + \cdots,  \label{u-u_I}
\end{split}
\ee
where the ellipses of the last line denote terms that will not contribute to the piece that we want to compute. For instance, by inserting another \(H^{\lambda}_I\) Hamiltonian (through a commutator) between \(\tau'\) and \(\tau_j\) we would be computing a correction to the \(\zeta\) propagator and not to the fully connected part of order \(\Lambda^4 \lambda^n\).

Now, Eq.~(\ref{u-u_I}) tells us that the structure of $u (\k_j , \tau)$ in terms of creation and annihilation operators $a_{\pm}$ is of the following form:
\be
\begin{split}
u \sim a_{-} + \lambda a_{+} & + \Lambda^4 \lambda \sum_{m} a_{+}^{2m-1}  + \Lambda^4 \lambda^2  a_{-} \sum_{m} a_{+}^{2m-2}   \\
& + \cdots + \Lambda^4 \lambda^n  a_{-}^{n-1} \sum_{m} a_{+}^{2m-n} + \cdots  .  \label{structure-u}
\end{split}
\ee
The computation of $n$-point correlation functions requires us to plug this form of $u$ back into (\ref{n-point-u-1}) and perform every possible contraction between creation and annihilation operators of the various terms appearing in (\ref{structure-u}). The final result that we are pursuing is an expression containing only terms of order $\Lambda^4 \lambda^n$, and thus many of the contractions correspond to loops involving pairs of $a_{+}$ operators. The diagrammatic expansion of this computation is shown in Fig.~\ref{fig:FIG_diagram_res}. Since we are computing the fully connected contribution to (\ref{structure-u}), in every contraction involving $a_{+}$ operators, at least one of them must come from a term of order $\Lambda^4$ in (\ref{structure-u}).

The details of this computation are shown in Appendix~\ref{chap:details}. The final result is found to be given by
\be
 \tilde G_{\zeta}^{(n)} (\tau, \k_1 , \cdots , \k_n) =  (-1)^{n/2} (2 \pi)^3 \delta^{(3)} \Big(\sum_j \k_j \Big) \frac{\Lambda^4  e^{- \frac{\sigma_0^2}{2 f^2}} }{3H^4}  \left( \frac{\lambda H^2 \Delta N}{2 f \sqrt{2 \epsilon}} \right)^n  \frac{k_1^3 + \cdots + k_n^3}{k_1^3 \cdots k_n^3} \Delta N , \label{n-point-final-res-zeta}
\ee
for even $n$, as it vanishes for odd $n$ because the potential is even under $\psi \to -\psi$. Here $\sigma_0$ is the variance of $\psi$, defined through the relation
\be
\sigma_0^2 \equiv \langle \psi^2 \rangle = H^2 \tau^2 \int_k u_{k} (\tau) u_{k}^* (\tau)  . \label{sigma_0-def}
\ee
The factor $\exp(- \sigma_0^2 / 2 f^2)$ of Eq.~(\ref{n-point-final-res-zeta}) appears as the consequence of the resummation shown in Fig.~\ref{fig:FIG_diagram_res}. This result may be compared to that shown in Eq.~(\ref{n-point-psi}) and obtained in Ref.~\cite{Palma:2017lww}. This comparison proves the result quoted in Eq.~(\ref{n-points-zeta-psi}).

In Eq.~(\ref{n-point-final-res-zeta}), $\Delta N$ corresponds to the number of $e$-folds elapsed since a reference time $\tau_0$ around which the set of modes $k_l$ crossed the horizon ($k_l \tau_0 \sim 1$)
\be
\Delta N = \ln \left( \frac{\tau_0}{\tau} \right) .
\ee
This definition coincides with that of Eq.~(\ref{Delta-N-2-point}) in the case of the two-point function.
Notice that the shape in momentum space of the non-Gaussianity parametrized by these $n$-point correlation functions is of the local type. In obtaining this result we have assumed that the condition $ \Delta N \gg 1$ holds. If instead one has a relatively small $\Delta N $, then other terms that were neglected might need to be included back. Given that our perturbativity conditions demand $\lambda \ll 1$, this is in agreement with the power spectrum shown in (\ref{power-zeta-psi}), valid for $\lambda^2 \Delta N^2 \gtrsim 1$.

\subsubsection{About horizon exit} \label{sec:npointshorizon}

Here let us make a remark on the value of $\Delta N$. Since the $\psi$ field in our model example is exactly massless, $\Delta N$ should be evaluated from when modes exit the horizon until the end of inflation. Because longer modes exit the horizon earlier, $\Delta N$ is not exactly a constant and has a logarithmic dependence on the magnitude of relevant momenta. Here we ignore this weak momentum dependence and approximate $\Delta N \sim 60$ as a constant. On the other hand, for our purpose we do not have to regard the mass of the scalar field $\psi$ to be exactly zero (but still require $\mu \ll H$, so our model conditions are satisfied). Such a field would have decayed before it stays at the superhorizon for the entire 60 $e$-folds, as it happens in the lower mass range case of the quasi-single-field inflation models \cite{Chen:2009we,Chen:2009zp}. In this case, all modes of $\psi$ will stay for the same amount of $e$-folds, $\Delta N$, after the horizon exit and before the decay, and $1<\Delta N<60$ is now exactly a constant.

\subsubsection{Regularization: IR and UV cutoffs}

Before we finish this section, let us briefly come back to Eq.~(\ref{sigma_0-def}). If we replace the mode solutions $u_k(\tau)$ of Eq.~(\ref{k-mode-solutions}) back into~(\ref{sigma_0-def}), and define the dimensionless integration variable $q = k |\tau|$, we obtain
\be
\sigma_0^2 = \frac{H^2}{4 \pi^2} \int \!  dq  \left( q + \frac{1}{q} \right) . \label{variance-0}
\ee
Observe that $\sigma_0$ is independent of time $\tau$. However, it contains divergences coming from the integration limits $q\to0$ and $q \to +\infty$.
We may therefore introduce infrared and ultraviolet cutoff scales $q_{\rm IR}$ and  $q_{\rm UV}$, respectively, and obtain
\be
\sigma_0^2 = \frac{H^2}{4 \pi^2}   \left( \frac{1}{2} (q_{\rm UV}^2 - q_{\rm IR}^2) + \ln (q_{\rm UV}/q_{\rm IR}) \right).
\ee
Notice that the variable $q = k |\tau| = k / a H$ is the physical momenta per unit of the Hubble scale.
The UV cutoff refers to a scale that is deep inside the horizon and is the scale of new physics and the limit of low energy effective theory. This cutoff contributes to the renormalization of the coupling $\Lambda^4$ as in what happens in flat spacetime. The logarithmic IR divergence is due to the random walk of the massless field in the dS space. Actual observations do not have access to all the scales, and so the IR cutoff should be set by the size of the observable Universe. We will come back to this issue in Sec.~\ref{subsec:n-point}, where we consider the need of defining the variance of modes available to cosmological observers.

\subsection{Tomographic non-Gaussianity} \label{sec:tom-NG}

The expression for the $n$-point correlation functions given by (\ref{n-point-final-res-zeta}) may be Fourier transformed back into coordinate space as
\be
G_{\zeta}^{(n)} (\tau, \x_1 , \cdots ) = \int_{k_1} \!\!\!  \cdots \int_{k_n} \!\!\! e^{- i \sum_j \k_j \cdot \x_j}  \tilde G_{\zeta}^{(n)} (\tau, \k_1 , \cdots ) . \label{G-xi}
\ee
This expression may be used to deduce the probability distribution function of measuring an amplitude $\zeta({\bf x})$ at a given position ${\bf x}$. To this end, we need to compute the moments $\langle \zeta^n \rangle$ that are given by evaluating all the coordinates in (\ref{G-xi}) at a common value $\x$
\be
 \langle \zeta^n \rangle_c \equiv G_{\zeta}^{(n)} (\tau, \x , \cdots , \x) , \label{zeta-n-G}
\ee
where the subscript $c$ denotes that $\langle \zeta^n \rangle_c$ comes from fully connected diagrams, hence it is proportional to a Dirac-delta that conserves momentum. Due to this momentum conservation, $ \langle \zeta^n \rangle_c$ defined in Eq.~(\ref{zeta-n-G}) is independent of $\x$. In the following subsections, we first obtain a concrete expression for the $n$-point functions of Eq.~(\ref{zeta-n-G}) valid for long wavelength modes, and then we proceed to derive the PDF from where these $n$-point functions are computed.

\subsubsection{$n$-point functions for long wavelength modes} \label{subsec:n-point}

The quantity $\tilde G_{\zeta}^{(n)} (\tau, \k_1 , \cdots , \k_n)$ of Eq.~(\ref{n-point-final-res-zeta}) only shows the leading IR contribution to the full $n$-point correlation function. For the same reason, in (\ref{G-xi}) we cannot integrate along the entire momentum space, and we are forced to introduce a cutoff momentum $k_L$. This is not a technical limitation, but all the contrary. We are interested in making predictions of inflation valid for superhorizon perturbations (that will later on reenter the horizon after inflation), and so we want to compute correlation functions of long wavelength $\zeta$ modes. This is normally done by introducing window functions selecting the relevant scales for the computation of correlation functions in coordinates space. For simplicity, here we consider a window function with a hard cutoff $k_L$. With this purpose in mind, we introduce the cutoff in terms of physical momentum $q_{\rm phys} \equiv k |\tau|$ (per unit of $H$) instead of comoving momentum $k$. That is, we choose a hard cutoff momentum $q_L$ and split the curvature perturbation as
\be
\zeta = \zeta_S + \zeta_L ,  \label{scale-division}
\ee
where $\zeta_L$ only includes modes of wavelengths larger than some fixed value $2 \pi / q_L$. Horizon crossing happens at $q_{\rm phys} = 1$, and so we must impose 
\be
q_L \leq 1 .
\ee
In other words $\zeta_L$ contains superhorizon contributions (at the end of inflation) between the physical cutoff scales $q_L$ and $q_{\rm IR}$. Explicitly, $\zeta_L$ is given by
\be
\zeta_L (\x, \tau) = \int_{k<k_L} \zeta(\k,\tau) e^{- i\k \cdot \x} ,
\ee
where $k_L = q_L / |\tau|$. Thus, we will compute a more restricted version of (\ref{zeta-n-G}) given by
\be
 \langle \zeta_L^n \rangle_c = G_{\zeta,L}^{(n)} (\tau, \x , \cdots , \x) , \label{zeta-n-G-L}
\ee
where $G_{\zeta,L}^{(n)} (\tau, \x_1 , \cdots , \x_n)$ reads as in (\ref{G-xi}), but now with the momenta integrated up to $k_L$. Explicitly, we have
\be %\label{n-point-corr-space-cut}
\begin{split}
 G_{\zeta,L}^{(n)} (\tau, \x_1 , \cdots , \x_n) =  (-1)^{n/2} (2 \pi)^3 \frac{\Lambda^4}{3H^4}  \int_{k_1 < k_L}  \cdots \int_{k_n < k_L} \delta^{(3)} \Big(\sum_j \k_j \Big) e^{- i \sum_j \k_j \cdot \x_j} \\
 \times e^{- \frac{\sigma_0^2}{2 f^2}} \left( \frac{\lambda H^2}{2 f \sqrt{2 \epsilon}} \Delta N \right)^n  \frac{k_1^3 + \cdots + k_n^3}{k_1^3 \cdots k_n^3} \Delta N . \label{n-point-final-res-coord}
\end{split}
\ee
The division of scales (\ref{scale-division}) forces us to split $\sigma_0^2 = \langle \psi^2 \rangle$, introduced in Eq.~(\ref{variance-0}), into short and long wavelength contributions as $\sigma_0^2 = \sigma_S^2 + \sigma_L^2$, in such a way that $\sigma_S^2$ and  $\sigma_L^2$ receive contributions larger and smaller than $k_L$ respectively. From Eq.~(\ref{variance-0}) we see that $\sigma_S^2$ and $\sigma_L^2$ are given by
\bea
\sigma_S^2 &=&\frac{H^2}{4 \pi^2}   \left( \frac{1}{2} (q_{\rm UV}^2 - q_{L}^2) + \ln (q_{\rm UV}/q_{L}) \right) \simeq \frac{H^2}{8 \pi^2}  q_{\rm UV}^2  , \label{sigma-S-cut-off} \\
\sigma_L^2 &=&  \frac{H^2}{4 \pi^2}   \left( \frac{1}{2} (q_{L}^2 - q_{\rm IR}^2) + \ln (q_{L}/q_{\rm IR}) \right) \simeq \frac{H^2}{4 \pi^2} \ln \xi  , \label{sigma-cut-off}
\eea
where we have introduced the ratio
\be
\xi = \frac{q_L}{q_{\rm IR}} = \frac{k_L} {k_{\rm IR}}, \label{beta-def}
\ee
which is a measure of the range of scales spanned by the long mode contributions $\zeta_L$.  The logarithmic dependence of (\ref{sigma-cut-off}) suggests that $\sigma_L^2 \simeq H^2$. For instance, if we take $q_L = 1$, then $\xi$ corresponds to the ratio between the largest wavelength and the Hubble radius at the end of inflation. Then $\ln \xi \simeq 60$, and one obtains $\sigma_L^2 \simeq H^2$. Notice, however, that, in general, $\xi$ parametrizes the window function selecting the scales, hence its value should be determined by the range of momenta available to cosmological observations. In the particular case of the CMB, this ratio is approximately given by $\ln \xi \sim 8$.

Next, to obtain an expression for $\langle \zeta^n_L \rangle_c $, we evaluate the arguments of (\ref{n-point-final-res-coord}) at a single coordinate value $\x$. Because of momentum conservation, the argument of the exponential vanishes, and we are left with the following expression
\be
 \langle \zeta^n_L \rangle_c =  (-1)^{n/2}  g_n A^2  e^{- \frac{\sigma_L^2}{2 f^2}} \left[ \frac{\lambda \sigma_L^2}{f \sqrt{ 2 \epsilon}  } \Delta N   \right]^n  , \label{n-point-final-res-coord-sol1}
\ee
where
\be
A^2 \equiv \frac{\Delta N}{6 \sigma_L^2 } \frac{\Lambda^4_{\rm ren}}{H^2} . \label{A-def}
\ee
Here, $\Lambda^4_{\rm ren} = e^{- \sigma_{S}^2 / 2 f^2} \Lambda^4$ is the renormalized coupling  introduced in Eq.~(\ref{Lambda-ren}) resulting from the loop resummation introduced in Sec.~\ref{sec:npoints}. Because this resummation is always induced by $\Lambda$, the combination $\Lambda_{\rm ren}^4 $ will be present at all orders in perturbation theory (disregarding higher order loop corrections carrying external momenta that start appearing at order $\Lambda^8$). In the work presented in Ref.~\cite{Chen:2018brw}, which is detailed in Section~\ref{sec:gen-NG-Tom}, we discuss the renormalization of $\Delta V(\psi)$ more generally, paying special attention to the running of the parameters defining $\Delta V(\psi)$ in order to make observables independent of the cutoff scales. To continue, the coefficient $g_n$ in (\ref{n-point-final-res-coord-sol1}) is defined as
\be
g_n \equiv \frac{  (2 \pi)^3 }{( 2 \sigma_L^2 / H^2 )^{n-1}} I_n, \label{gn-def}
\ee
for even $n$ and zero otherwise because $\tilde{G}_{\zeta}^{(n)}$ vanishes if $n$ is odd. Here $I_n$ corresponds to the following integral:
\be
I_n \equiv   \int_{k_1 < k_L}  \!\!\!\!\!\!\!\!   \cdots \int_{k_n < k_L} \!\!\!\!\!\!\! \delta^{(3)} \Big(\sum_j \k_j \Big) \frac{k_1^3 + \cdots + k_n^3}{k_1^3 \cdots k_{n}^3} .  \label{full-I-n}
\ee
Equation (\ref{n-point-final-res-coord-sol1}) is written in terms of the variance $\sigma_L^2$ associated with the probability distribution function of $\psi$. It will be more useful to write $\langle \zeta^n_L \rangle_c$ in terms of the variance $\sigma_{\zeta}^2$ instead of $\sigma_L^2$. Recall that in Sec.~\ref{sec:linearth} we derived the power spectrum of $\zeta$ in terms of the power spectrum of $\psi$, given in Eq.~(\ref{power-zeta-psi}). When $\lambda^2 \Delta N^2 \gtrsim 1$, this result implies
\be
\sigma_\zeta^2 = \sigma_L^2 \frac{\lambda^2}{2 \epsilon} \Delta N^2 . \label{sigma-zeta}
\ee
Then, by defining $f_\zeta$ as
\be
f_{\zeta} \equiv f \frac{\sigma_\zeta}{\sigma_L} = f  \frac{\lambda}{\sqrt{2 \epsilon}} \Delta N  ,  \label{f-sigma-zeta}
\ee
it is direct to find
\bea
&& \langle \zeta^n_L \rangle_c =  (-1)^{n/2} g_n A^2 e^{- \frac{\sigma_\zeta^2}{2 f_\zeta^2}} \left[ \frac{\sigma_\zeta^2}{f_\zeta}   \right]^n  .  \label{zeta-n-1}
\eea
This is the general form of the $n$-point correlation function that we need in order to reconstruct the tomographic PDF for $\zeta$. It is important to emphasize here that we can also consider the regime $\lambda^2 \Delta N^2 < 1$, where one finds $\sigma_\zeta^2 = \sigma_L^2 / 2 \epsilon$. This case would give us a slightly different expression for (\ref{zeta-n-1}) but would not change the form of the reconstructed PDF (except for the way in which some parameters appear). For simplicity, we stick to the regime $\lambda^2 \Delta N^2 \gtrsim 1$.

\subsubsection{Dependence of the $n$-point functions on the cutoff scales}

The integral $I_n$ determining the form of the factor $g_n$ [through Eq.~(\ref{gn-def})] is a function of the order $n$ and the ratio $\xi$ introduced in Eq.~(\ref{beta-def}). Indeed, in Appendix~\ref{app_about_I_n}, we show that $I_n$ can be written in terms of a single integration variable as
\be
I_n (\xi) = \frac{n }{(2 \pi^2)^{n+1}} \int_0^{\infty} \!\! \frac{dx}{x} G(\xi,x) \left[ F(\xi , x) \right]^{n-1}   ,  \quad \label{full-I_n} 
\ee
where $G(\xi , x)$ and $F(\xi , x)$ are given by
\bea
G(\xi , x) &=&[ \sin (x) - x \cos (x) - \sin (x / \xi)  + (x / \xi) \cos (x / \xi) ] , \qquad \\
F(\xi , x) &=& {\rm Ci} (x) - \frac{\sin (x)}{x}  - {\rm Ci} ( x / \xi ) + \frac{\sin ( x / \xi )}{ x / \xi} . \qquad
\eea
Here $ {\rm Ci} (x)$ is the cosine integral function. In Appendix~\ref{app_about_I_n}, we also show that in the formal limit $\xi \to \infty$, the integral asymptotes to a simple function $I_n^{0}  (\xi)$ given by
\be
 I_n^{0}  (\xi) \equiv \frac{n \pi}{2 (2 \pi^2)^{n+1}} ( \ln \xi )^{n-1} . \label{asympt-I_n}
\ee
Then, given that $\ln \xi = 4 \pi^2 \sigma^2_L / H^2$ [which can be seen from Eq.~(\ref{sigma-cut-off})], one obtains $g_n = n$, implying a very simple general expression for the $n$-point functions $\langle \zeta^n_L \rangle_c$. However, given that $k_L$ is at most the horizon exit scale, and that $k_{\rm IR}$ is the largest scale available to present observers, we have the bound
\be
\ln \xi \leq 60,
\ee 
which implies that $\ln \xi$ is too small to allow us to take $I_n$ as $I_n^{0}$. The reason for this is that, with $\ln \xi \simeq 60$, the correction $\Delta I_n = I_n - I_n^{0}$ is already one-tenth of $I_n^0$ for $n \sim 35$. This in turn implies that the PDF derived with $ I_n^{0}$ starts to deviate significantly from the one derived with $I_n$ when $f$ is smaller than $\sigma_L \sim H$, which is precisely the interesting region of parameters that we wish to explore. A proof of this statement is given in Appendix~\ref{app_about_I_n}.

%%%%%%%%%%%%%%%%%%%%%%%%%%%%%%%%%%%%%%
%%%%%%%%%%%%%%%%%%%%%%%%%%%%%%%%%%%%%%
%%%%%%%%%%%%%%%%%%%%%%%%%%%%%%%%%%%%%%

\section{Reconstructing the Landscape}

In what follows, we devote ourselves to reconstruct the PDF out of the $n$-point function $\langle \zeta^n_L \rangle_c$ given in Eq.~(\ref{n-point-final-res-coord-sol1}). We will first do this in Sec.~\ref{subsec:PDF-derivation-from-n-points} for the case in which $I_n$ is taken to be as $I_n^{0}  (\xi)$ shown in Eq.~(\ref{asympt-I_n}). Then, in Sec.~\ref{subsec:PDF-derivation-from-n-points-full},  we will show how to obtain the PDF for the full expression for $I_n (\xi)$ shown in Eq.~(\ref{full-I_n}). Before deriving these two PDFs we describe the general idea behind its reconstruction.

\subsection{PDF reconstruction: general idea}

Recall that we have focused our interest on the computation of the fully connected contributions to the $n$-point correlation functions. Had we focused instead on the full $n$-point correlation functions, including disconnected diagrams, we would have arrived at the more general expression
\be
\langle \zeta_L^n \rangle = \sum_{m=0}^{n/2} \frac{n!}{m! (n-2m)! 2^m} \sigma_{\zeta}^{2m} \langle \zeta^{n-2m}_L \rangle_c . \label{full-n-point}
\ee
Here, the factors $\sigma_{\zeta}^{2m}$ come from propagators connecting pairs of external lines. The combinatorial factor ${n!} / {m! (n-2m)! 2^m}$ consists of the total number of ways to connect the $n$ external legs in such a way that $2m$ of them are connected by propagators, and the rest  are connected to the $\Lambda^4$ vertex.

The probability distribution function $\rho(\zeta)$ that we are searching for must be such that
\be
\langle \zeta_L^n \rangle = \int \! d\zeta \, \rho(\zeta) \, \zeta^n .
\ee
To find $\rho(\zeta)$, it is useful to notice that the term $m=n/2$ in Eq.~(\ref{full-n-point}) is given by
\be
\langle \zeta_L^n \rangle_{G} = \frac{n!}{(n/2)! 2^{n/2}} \sigma_{\zeta}^{n} , \label{Gaussian-n-points}
\ee
which corresponds to the $n$-point correlation function of a Gaussian distribution. This means that $\rho (\zeta) $ is given by a leading Gaussian distribution with a non-Gaussian correction proportional to $\Lambda^4$. Thus, to find $\rho (\zeta)$ we may try the following ansatz
\be
\rho (\zeta) = \rho_G (\zeta) + \Delta \rho (\zeta) , 
\ee
where 
\be
\rho_G (\zeta) = \frac{e^{- \frac{\zeta^2}{2 \sigma_{\zeta}^2}}}{\sqrt{2\pi} \sigma_{\zeta}} ,
\ee
is the Gaussian part giving rise to the subset of $n$-point functions given in Eq.~(\ref{Gaussian-n-points}). The piece $\Delta \rho (\zeta) $ corresponds to the correction resulting from the nonlinear interactions proportional to $\Lambda^4$ (or equivalently, $A^2$). 

In what follows, we determine the form of  $\Delta \rho (\zeta) $ due to $\langle \zeta^n_L \rangle_c$ shown in Eq.~(\ref{n-point-final-res-coord-sol1}). The procedure crucially depends on knowing how $\langle \zeta^n_L \rangle_c$ depends on $n$, which requires us to deal with $I_n(\xi)$ of Eq.~(\ref{full-I_n}). To proceed, we find it useful to first show how to deduce $\Delta \rho (\zeta) $ in the case where $I_n(\xi)$ is given by its asymptotic form $I_n^0(\xi)$. This will then allow us to deal easily with the more general situation in which $I_n(\xi)$ is given by the full expression given in~(\ref{full-I_n}).

\subsection{Asymptotic reconstruction} \label{subsec:PDF-derivation-from-n-points}

If we take Eq.~(\ref{n-point-final-res-coord-sol1}) with $I_n (\xi)$ replaced by its asymptotic form $I_n^0(\xi)$ given in Eq.(\ref{asympt-I_n}), then $g_n = n$ when $n$ is even, and in this case we simply have
\bea
&& \langle \zeta^n_L \rangle_c =  (-1)^{n/2} n A^2 e^{- \frac{\sigma_\zeta^2}{2 f_\zeta^2}} \left[ \frac{\sigma_\zeta^2}{f_\zeta}   \right]^n  .  \label{zeta-n-2}
\eea
(and zero if $n$ is odd). This equation may be used to derive the PDF $\rho(\zeta)$ determining the probability of measuring a certain value of the curvature perturbation at an arbitrary position. To find $\Delta \rho (\zeta)$ we may try the following ansatz
\be
\Delta \rho (\zeta) = \frac{e^{- \frac{\zeta^2}{2 \sigma_{\zeta}^2}}}{\sqrt{2\pi} \sigma_{\zeta}}  \Bigg[ \sum_{m=0} B_m \zeta^{2m} \cos\left( \frac{\zeta}{f_\zeta} \right) + \sum_{m=0} C_m \zeta^{2 m +1}  \sin\left( \frac{\zeta}{f_\zeta}  \right) \Bigg]  , \label{PDF-ansatz}
\ee
where we have used the fact that $\rho(\zeta)$ must be even under the change $\zeta \to - \zeta$, for only the even moments are nonvanishing. It is direct to find that $B_0$ and $C_0$ are the only nonvanishing coefficients and, therefore, that the full PDF is given by~\cite{footnote-PDF-comp}
\be
\rho (\zeta) = \frac{e^{- \frac{\zeta^2}{2 \sigma_{\zeta}^2}}}{\sqrt{2\pi} \sigma_{\zeta}}  \left[ 1 + A^2   \frac{  \sigma_{\zeta}^2 }{ f_\zeta^2 }   \cos\left( \frac{\zeta}{f_\zeta} \right) - A^2  \frac{  \zeta }{ f_\zeta }   \sin\left( \frac{\zeta}{f_\zeta}  \right)  \right] . \label{PDF-1}
\ee
This PDF satisfies Eq.~(\ref{full-n-point}), and given that it corresponds to a small, absolutely continuous deformation of a Gaussian distribution, it is unique [that is, it is the only possible reconstruction from the moments of Eq.~(\ref{zeta-n-2})].

\begin{figure}[t!]
\begin{center}
\includegraphics[scale=0.6]{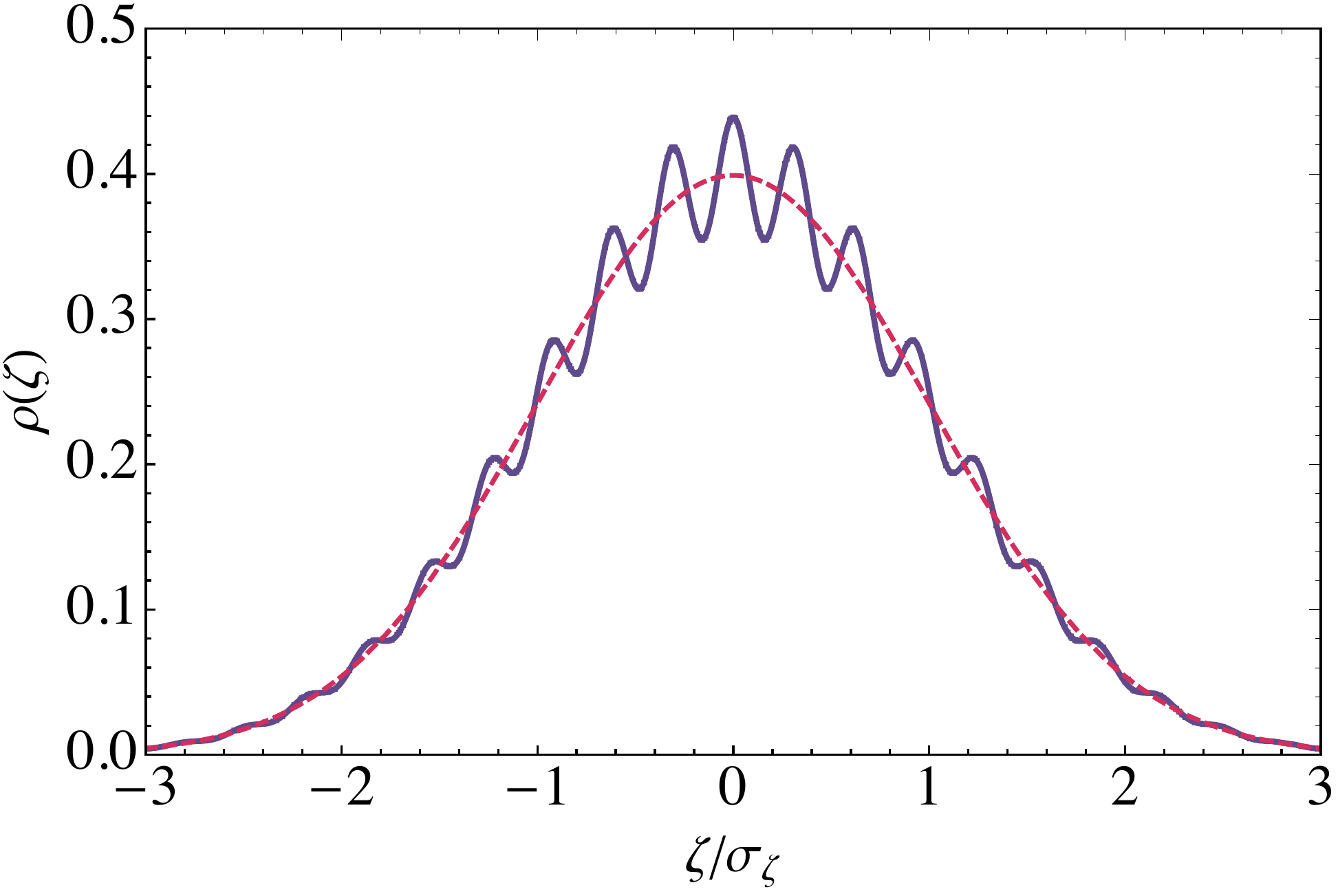}
\caption{An example of the PDF of Eq.~(\ref{PDF-1}) for the choice of parameters $f_\zeta / \sigma_\zeta = 5\times10^{-2}$ and $A^2 = 2.5 \times 10^{-4}$ (solid curve). For comparison, we have plotted a Gaussian distribution of variance $\sigma_\zeta$ (dashed curve).}
\label{fig:FIG-PDF}
\end{center}
\end{figure}

The probability distribution function~(\ref{PDF-1}) is valid in the formal limit $\xi \to \infty$. If $\sigma_L \gg H$ we could trust this result for the case $f \lesssim \sigma_L$, in which case the PDF shows nontrivial structures in the form of superimposed oscillations. However, given that $\sigma_L \sim H$ (because $\ln \xi \leq 60$), we cannot trust the regime  $f \lesssim \sigma_L$ (see Appendix~\ref{app_about_I_n}), and we are forced to consider the more general case in which $I_n$ is given by its full form shown in~(\ref{full-I_n}). In spite of this limitation, Eq.~(\ref{PDF-1}) constitutes one of our main results.  It gives a simple non-Gaussian probability distribution function for $\zeta$ in terms of various parameters related to the landscape shape. It may be verified that the PDF is already normalized as $\int \rho(\zeta) d\zeta = 1$. This probability distribution function is plotted in Fig.~\ref{fig:FIG-PDF} for specific values of $f_\zeta/ \sigma_\zeta$ and $A^2$. Notice that the second term inside the squared parenthesis in Eq.~(\ref{PDF-1}) accounts for the increase in probability of finding values of $\zeta$ that are sourced by those values of $\psi$ which minimize the cosine potential of Eq.~(\ref{potential-psi}). On the other hand, the third term, linear in $\zeta$ may be interpreted as a contribution accounting for the diffusion of $\zeta$ (one could in fact absorb the third term into the second term by slightly shifting $f_{\zeta}$).

\subsection{Full reconstruction} \label{subsec:PDF-derivation-from-n-points-full}

We now consider the task of deriving the full PDF, valid for any value of $\ln \xi > 0$. To proceed, it is helpful to realize that the most important aspect about the reconstruction performed in the previous section was the dependence of $\langle \zeta^n_L \rangle_c$ on $n$ as shown in Eq.~(\ref{zeta-n-2}). In the general case, if we consider the $x$ integral of Eq.~(\ref{full-I_n}) explicitly, we see that the dependence of $\langle \zeta^n_L \rangle_c$ on $n$ is exactly the same, except that this time it happens for each value of $x$. Then, a simple comparison with~\eqref{n-point-final-res-coord-sol1} shows that this time the reconstruction amounts to identifying an $x$ dependent decay constant 
\be
f_{\zeta}(x)\equiv f_{\zeta} \dfrac{\ln \xi }{F(\xi, x)} \geq f_{\zeta} , \label{f-x-def}
\ee
that satisfies $f_{\zeta}(0) = f_{\zeta}$. Hence, keeping track of all the numerical factors, we find
\be
\rho (\zeta) = \frac{e^{- \frac{\zeta^2}{2 \sigma_{\zeta}^2}}}{\sqrt{2\pi} \sigma_{\zeta}}  \left[ 1 + A^2 \int_0^{\infty} \!\!\frac{dx}{x} \, \mathcal{K}_\xi(x)  \left(  \frac{  \sigma_{\zeta}^2 }{ f_\zeta(x)^2 }   \cos\left( \frac{\zeta}{f_\zeta(x)} \right) -  \frac{  \zeta }{ f_\zeta(x) }   \sin\left( \frac{\zeta}{f_\zeta(x)}  \right) \right)  \right] ,  \label{PDF-full}
\ee
where the kernel $\mathcal{K}_\xi(x)$ is given by
\be
\mathcal{K}_\xi(x) \equiv \frac{ 2 G(\xi , x) \ln \xi }{\pi F(\xi, x)} \exp \left( -\frac{\sigma_{\zeta}^2 \left(f_{\zeta}^2(x) - f_{\zeta}^2\right)}{2 f_{\zeta}^2 f_{\zeta}^2(x)} \right). \label{def-kernel}
\ee
The result shown in Eq.~(\ref{PDF-full}) is our main result. It gives us the PDF for any value of the ratio $\xi =  k_{L}/k_{\rm IR} $. In particular, we can trust this result well inside the regime $f_\zeta < \sigma_\zeta$ for values $\ln \xi \simeq 8$, which corresponds to the range of scales available to CMB observations [as opposed to the case of the PDF of Eq.~(\ref{PDF-1})]. 

\begin{figure}[t!]
\begin{center}
\includegraphics[scale=0.6]{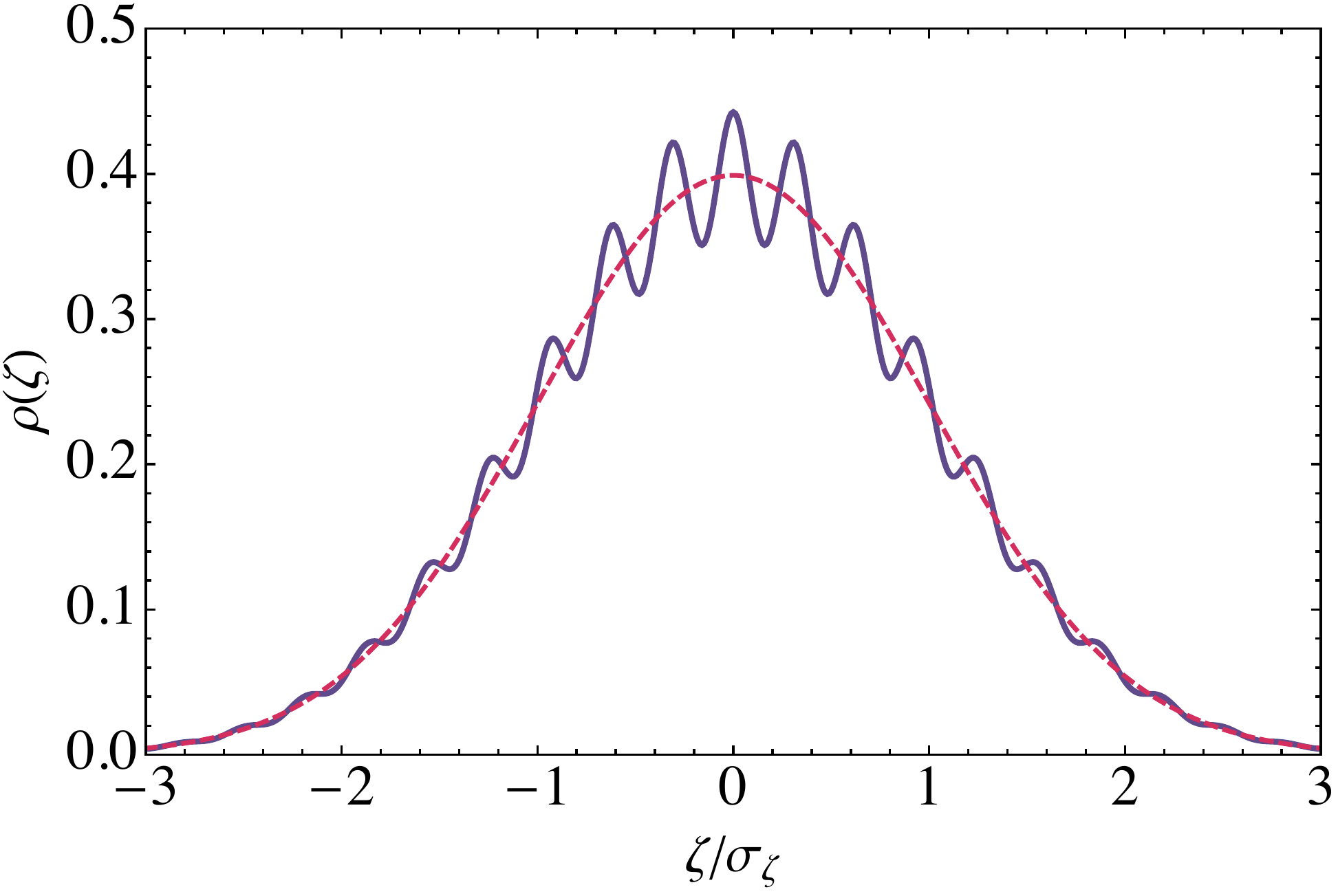}
\caption{An example of the PDF of Eq.~(\ref{PDF-full}) for the choice of parameters $f_\zeta / \sigma_\zeta = 5 \times10^{-2}$, $A^2 =  2.5 \times 10^{-2}$ and $\ln \xi = 8$ (solid curve). For comparison, we have plotted a Gaussian distribution of variance $\sigma_\zeta$ (dashed curve). Notice that $A^2$ is 100 times larger than the value used to plot Fig.~\ref{fig:FIG-PDF}.}
\label{fig:FIG-PDF-1}
\end{center}
\end{figure}

\begin{figure}[t!]
\begin{center}
\includegraphics[scale=0.6]{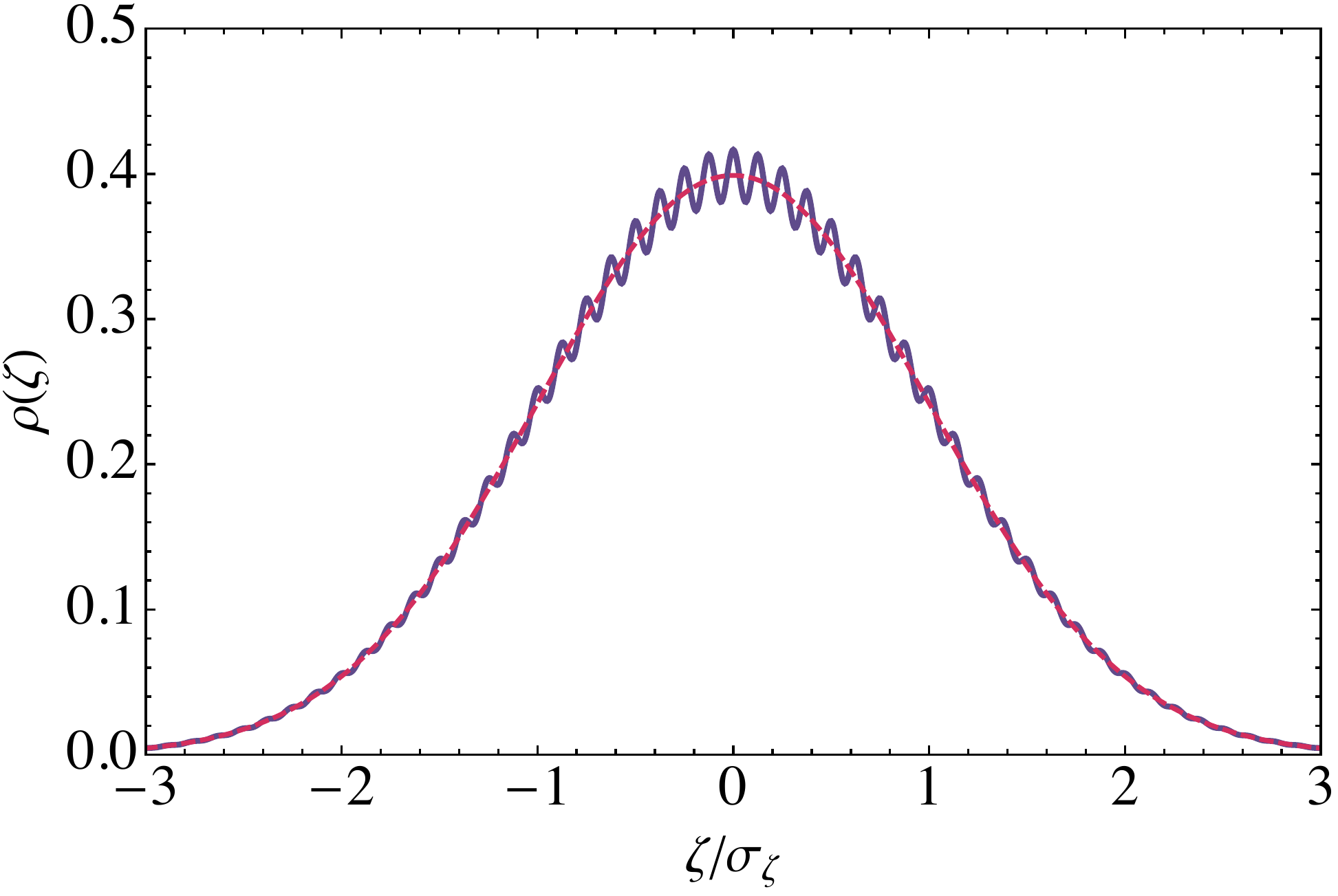}
\caption{An example of the PDF of Eq.~(\ref{PDF-full}) for the choice of parameters $f_\zeta / \sigma_\zeta = 2 \times10^{-2}$, $A^2 =  2.5 \times 10^{-2}$ and $\ln \xi = 8$ (solid curve). For comparison, we have plotted a Gaussian distribution of variance $\sigma_\zeta$ (dashed curve). Here $A^2$ has the same value as that of Fig.~\ref{fig:FIG-PDF-1}, but $f_\zeta / \sigma_\zeta$ is smaller.}
\label{fig:FIG-PDF-2}
\end{center}
\end{figure}

An outstanding property of~(\ref{PDF-full}) is that it preserves the oscillatory structure of the potential in a strikingly similar manner as~\eqref{PDF-1}. The main difference, is that now there is a filtering function that accounts for the effects that arise when one considers only the bounded region of $k$ space which we are able to probe. The consequences of this filtering can be appreciated by looking at Fig.~\ref{fig:FIG-PDF-1} (plotted for $\ln \xi = 8$). There we see by comparison to Fig.~\ref{fig:FIG-PDF} that the amplitude of the oscillations in the full PDF is suppressed, since the value chosen for $A^2$ in this last plot is $100$ times larger than the previous one. Moreover, as illustrated in Fig.~\ref{fig:FIG-PDF-2} (also plotted for $\ln \xi = 8$), decreasing the value of $f_{\zeta}$ from this point does not enhance the amplitude as we might have thought while looking at Eq.~(\ref{PDF-1}) but the opposite: the amplitude actually gets smaller, mostly because of the exponential factor in the kernel $\mathcal{K}_\xi(x)$.

Note that in the formal limit $\ln  \xi \to \infty$ the probability distribution function~(\ref{PDF-full}) becomes independent of $\xi$, and we recover~\eqref{PDF-1}. This corresponds to the ideal situation whereby the entire range of momenta is available to observers. In that case, one can directly infer the parameters $\Lambda^4_{\rm ren}$ and $f$ of the landscape potential by performing statistics with observations of primordial curvature perturbations. Otherwise, as long as observers can only have access to a limited amount of modes, as parametrized by $\ln \xi$, the filtering function appearing in the kernel $\mathcal K_\xi(x)$ will wash out the structure of the potential. This is simply because $\ln \xi$ restricts the number of modes in momentum space that can add up to increase the effect of nonlinearities due to $\Delta V$ on the PDF in coordinate space: momentum conservation through the vertex implies that  while certain modes are probing large (small) momenta limited by $q_L$ ($q_{\rm IR}$), other momenta will probe more restricted regions in momentum space. As a result, to extract information about $\Delta V$ one has to take into account the role of $\ln \xi$. %We will come back to this issue in Ref.~\cite{Chen:2018brw}. 

Let us make a final note on the perturbativity conditions required for the first-order computation to hold. As is evident from~\eqref{PDF-full}, the amplitude of the NG contribution to the PDF is proportional to $A^2 \sigma_L^2/f^2$, which implies that one must demand
\be
\frac{\Lambda^4_{\rm ren} \Delta N}{H^2 f^2} \ll 1
\ee
as the true perturbativity condition.

%%%%%%%%%%%%%%%%%%%%%%%%%%%%%%%%%%%%%%
%%%%%%%%%%%%%%%%%%%%%%%%%%%%%%%%%%%%%%
%%%%%%%%%%%%%%%%%%%%%%%%%%%%%%%%%%%%%%

\section{Aspects of the Tomography}

We have examined a regime of multifield inflation where the shape of the landscape potential in the isocurvature direction can be probed using non-Gaussianity of primordial density perturbations. At the level of $n$-point correlation functions (or polyspectra), these non-Gaussianities take the local form, as in all multifield models \cite{ Achucarro:2016fby,Gordon:2000hv, GrootNibbelink:2000vx, GrootNibbelink:2001qt, Enqvist:2004bk, Lyth:2005fi, Seery:2005gb, Rigopoulos:2005ae, Alabidi:2005qi, Battefeld:2007en, Byrnes:2008wi, Byrnes:2009qy, Battefeld:2009ym, Elliston:2011et, Mulryne:2011ni, McAllister:2012am, Byrnes:2012sc, Bjorkmo:2017nzd, Langlois:1999dw,  Amendola:2001ni, Bartolo:2001rt, Wands:2002bn, Tsujikawa:2002qx, Byrnes:2006fr, Choi:2007su, Cremonini:2010sv, Cremonini:2010ua} and quasi-single-field models \cite{Noumi:2012vr, Chen:2009we,Chen:2009zp, Baumann:2011nk, Chen:2012ge, Sefusatti:2012ye, Norena:2012yi,  Gong:2013sma, Emami:2013lma, Kehagias:2015jha, Arkani-Hamed:2015bza,  Dimastrogiovanni:2015pla,Chen:2015lza, Chen:2016cbe,Lee:2016vti,Chen:2016qce, Meerburg:2016zdz,Chen:2016uwp,Chen:2016hrz, An:2017hlx,Iyer:2017qzw,An:2017rwo, Kumar:2017ecc, Franciolini:2017ktv,Tong:2018tqf, MoradinezhadDizgah:2018ssw, Saito:2018xge,Franciolini:2018eno,Chen:2018sce} with sufficiently light isocurvatons. However, a novel point of this work is that the information about the shape of the potential is not manifest in the individual $n$-point correlation functions, but rather in the re-summed probability distribution function given in Eq.~(\ref{PDF-full}). This shows that local non-Gaussianity may have a rich structure inherited by the self-interactions of isocurvature fields, together with a derivative coupling common to multifield models.

Although the mechanism of statistical transfer examined herein is based on the derivative coupling $\mathcal L_{\rm int} \propto \dot \zeta \psi$, our results are likely to be more general. We therefore expect that other classes of interactions between the curvature perturbation and other scalar fields scanning the landscape lead to similar conclusions.

Also, as previously mentioned, the particular form of the potential $\Delta V(\psi)$ should not be so crucial. While it is true that the cosine function used in this work comes with the right properties making the loop resummations possible, more general potentials are in fact not intractable. We will show this in detail in section~\ref{sec:gen-NG-Tom}. 
In what follows, we discuss various relevant aspects related to our main result so far.

\subsection{Relation to previous works}

Our analysis has some similarities (but also important differences) with previous works studying the implications of isocurvature fields on the production of primordial non-Gaussianity. In quasi-single-field inflation models, the isocurvature field is assumed to be massive and to have some interactions, such as the cubic self-interaction as in Eq.~(\ref{hierarchical-exp-V})
\be
\Delta V(\psi) = \frac{1}{2} \mu^2 \psi^2 + \frac{1}{6} g \psi^3 .  \label{QSF-pot}
\ee
In these models, the mass of $\psi$ in principle can be a free parameter and plays an important role: it controls the extent to which the fluctuations of $\psi$ can survive at superhorizon scales and interact with $\zeta$. This is because the amplitude of $\psi$ decays after horizon crossing as
\begin{equation}
\psi \sim e^{- \frac{3}{2} (1 - R) \Delta N},
\end{equation}
where $R$ is the real part of $ \sqrt{1 - 4 \mu^2 /9 H^2}$.

Although many results apply for general $\mu$, the most interesting cases in the quasi-single-field literature \cite{Noumi:2012vr, Chen:2009we,Chen:2009zp,Baumann:2011nk,Chen:2012ge, Sefusatti:2012ye, Norena:2012yi,  Gong:2013sma, Emami:2013lma, Kehagias:2015jha, Arkani-Hamed:2015bza,  Dimastrogiovanni:2015pla,Chen:2015lza, Chen:2016cbe,Lee:2016vti,Chen:2016qce, Meerburg:2016zdz,Chen:2016uwp,Chen:2016hrz,An:2017hlx,Iyer:2017qzw,An:2017rwo, Kumar:2017ecc, Franciolini:2017ktv,Tong:2018tqf, MoradinezhadDizgah:2018ssw, Saito:2018xge,Franciolini:2018eno,Chen:2018sce} are those with $\mu \sim H$.
Since the potential \eqref{QSF-pot} is assumed to hold within a field range much larger than the amplitude of $\psi$, ${\cal O}(H)$,
the isocurvaton field $\psi$ is confined within this potential and does not fluctuate outside to explore the fuller structure of the landscape. In fact, the PDF of the density perturbation of the quasi-single-field inflation model may be worked out in a similar fashion and it should encode the shape of the potential, although it has much less rich structure.
The main predictions for non-Gaussianities coming from~(\ref{QSF-pot}) are some nontrivial polyspectra, such as bispectra and trispectra.

In the case studied in this work, the field $\psi$ does not decay. Note that, classically, $\psi$ has a mass coming from the cosine potential of Eq.~(\ref{potential-psi}),
\be
\mu^2 = \frac{\Lambda^4}{f^2} ,
\ee
and this quantity may be even larger than $H^2$. But given that the potential barriers are small ($\Delta V^{1/4} \sim \Lambda \ll H$ at tree level) the $\psi$ fluctuations will still be able to traverse vigorously the barriers of the potential and explore the potential landscape.  In other words, the $\psi$ field is effectively massless at the leading order.
In this case, the classical mass term is only part of the rich structure in the small perturbation $\Delta V$ and appears as the first term in the series expansion of $\Delta V$. Therefore, this series expansion needs to be re-summed in the final result.
In the case of the cosine potential studied here, these aspects are summarized in the fact that all the vertices depend on just two parameters ($\Lambda^4$ and $1/f$), and so every vertex contributes decisively in the computation of the $n$-point correlation functions.

\subsection{After inflation}

So far we have established how a scalar field $\psi$ with a non-Gaussian distribution function can transfer its statistics to the curvature perturbation $\zeta$ during inflation. The mechanism by which the statistics is transferred is cumulative: $\psi$ transfers its statistics (both Gaussian and non-Gaussian) to $\zeta$ as long as $\lambda \neq 0$, and the non-Gaussianity of $\zeta$ becomes more accentuated as time passes. After a long enough period, the statistics of $\zeta$ (which in the absence of the $\lambda$ coupling would be nearly Gaussian) becomes completely dominated by that of the curvature perturbation $\psi$.

Three main things could happen after such a period that bring this mechanism to an end:
(1) As mentioned, if $\psi$ is not exactly massless, after some $e$-folds at superhorizon it will naturally decay, as in quasi-single-field inflation models.
(2) The potential $\Delta V(\psi)$ changes drastically. In more realistic scenarios we would expect that the potential $\Delta V(\psi)$ depends explicitly on time due to its dependence on the background. Before the end of inflation $\Delta V(\psi)$ could introduce a new scale that makes $\psi$ a very massive field within the relevant amplitude range $\sigma_L \simeq H$. In that case, the amplitude of $\psi$ would quickly decay (due to the kinematics of a massive field in an expanding Universe) and $\psi$ would not be able to source $\zeta$ any more. (3) The third possibility is that $\lambda$ effectively vanishes before the end of inflation [before even $\Delta V(\psi)$ changes]. Here, even though the amplitude of $\psi$ has not decayed, the sourcing offered by $\psi$ ends.

In all of the previous cases, the non-Gaussian statistics of $\zeta$ persist, simply because on superhorizon scales $\zeta$ remains constant (after $\psi$ has done its job of sourcing its statistics). In other words, the statistics transferred by $\psi$ while $\lambda \neq 0$ and $\psi \neq 0$ serves as the initial condition for a posterior phase where $\lambda = 0$ and/or $\psi = 0$. Thus, because the statistical transfer is cumulative, the new phase with $\lambda = 0$ and/or $\psi = 0$ would not imply that the non-Gaussian statistics of $\zeta$ is erased. All the contrary, if $\lambda =0$ or $\psi$ becomes massive, then $\zeta$ would kinematically decouple from $\psi$ and would continue to evolve independently, with a frozen amplitude, preserving its non-Gaussian statistics. Of course, the statistics of $\zeta$ would then survive reheating until horizon entry, fixing the initial conditions for perturbations in the hot Big-Bang era.

\subsection{Bispectrum constraints on the PDF}

We can constrain the level of non-Gaussianity in the probability distribution function~(\ref{PDF-full}) by looking into current bounds on the trispectrum~\cite{Alabidi:2005qi, Okamoto:2002ik, Boubekeur:2005fj, Kogo:2006kh, Seery:2006vu, Seery:2006js} set by Planck, as this model has an identically vanishing bispectrum [see Eq.~\eqref{n-point-final-res-zeta}], and consequently, we cannot use it to constrain the resulting PDF.

Specifically, Planck is able to constrain the parameter $g_{\rm NL}^{\rm local}$ that appears in the following relation involving the four-point function for $\zeta$, and its power spectrum:
\be
\langle \zeta_{\k_1}\zeta_{\k_2}\zeta_{\k_3}\zeta_{\k_4} \rangle \equiv (2\pi)^3\delta^{(3)}\left(\sum \k_{i} \right)\frac{54}{25}g_{\rm NL}^{\rm local} \left[P_{\zeta}(k_1)P_{\zeta}(k_2)P_{\zeta}(k_3)+3 \, \text{perm.} \right].
\ee
This expression may be compared with our general expression~(\ref{n-point-final-res-zeta}) for the specific case $n=4$, which is given by
\be
 \langle \zeta_{\k_1}\zeta_{\k_2}\zeta_{\k_3}\zeta_{\k_4} \rangle =   (2 \pi)^3 \delta^{(3)} \Big(\sum_j \k_j \Big) \frac{\Lambda^4}{3H^4} e^{- \frac{\sigma_0^2}{2 f^2}} \left( \frac{\lambda H^2 \Delta N}{2 f \sqrt{2 \epsilon}} \right)^4 \left[ \frac{1}{k_1^3 k_2^3 k_3^3}  +3 \, \text{perm.}  \right]\Delta N .
\ee
To compare both expressions, it is necessary to recall, from the discussion around Eq.~(\ref{Pzeta-Ppsi}), that the power spectrum for $\zeta$ is given by
\be
P_\zeta (k) =  \frac{\lambda^2 H^2 \Delta N^2}{4  \epsilon k^3 } . \label{power-spectrum-zeta}
\ee
Then, it follows that $g_{\rm NL}^{\rm local}$ is given by
\bea
g_{\rm NL}= \frac{25}{54} \frac{A^2 (2\epsilon)}{\lambda^2\Delta N^2}  \frac{\sigma_{L}^2}{f^4}  e^{- \frac{\sigma_L^2}{2 f^2}}.
\eea
We can turn this expression into a more useful result by recalling, from Eqs.~(\ref{sigma-zeta}) and (\ref{f-sigma-zeta}), that $\sigma_\zeta^2  = \sigma_L^2 \lambda^2 \Delta N^2 / 2 \epsilon$ and $f_{\zeta} / \sigma_{\zeta} = f / \sigma_L$. We obtain
\bea
g_{\rm NL}= \frac{25}{54} A^2 \frac{ \sigma_{\zeta}^2}{f_{\zeta}^4}  e^{- \frac{\sigma_\zeta^2}{2 f_{\zeta}^2}}.
\eea

\begin{figure}[t!]
\begin{center}
\includegraphics[scale=0.53]{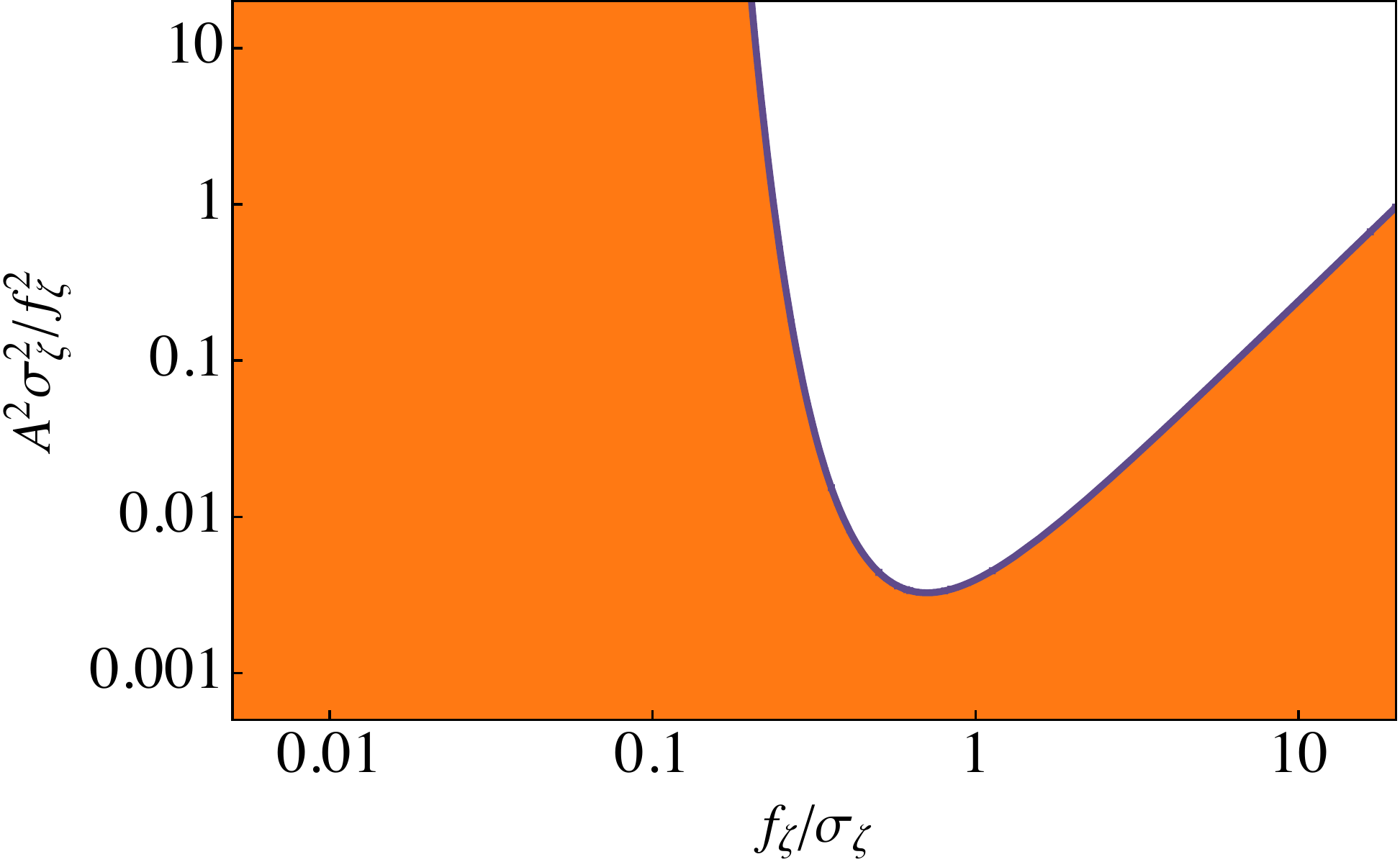}
\caption{The allowed values for the parameters $A^2$ and $f_\zeta / \sigma_\zeta$ deduced from current constraints by Planck on the trispectrum (at 95\% C.L.). The allowed region (in orange) is located bellow the solid curve. Notice that the combination $A^2 \sigma_\zeta^2 / f_\zeta^2$ corresponds to the coefficient in front of the cosine function in Eq.~(\ref{PDF-full}) at $x=0$.}
\label{fig:FIG-TRI}
\end{center}
\end{figure}

With the help of Eq.~(\ref{sigma-cut-off}) we see that $\sigma_{\zeta}^2$ is related to the power spectrum through
\be
\sigma_{\zeta}^2 =   \frac{P_\zeta (k) k^3}{2 \pi^2} \ln \xi ,
\ee
(recall that $\xi = q_L / q_{\rm IR}$). Planck observations on 2015~\cite{Ade:2015lrj} constrained the amplitude of the power spectrum as $P_\zeta (k) k^3 / 2 \pi^2 = (2.196 \pm 0.079) \times 10^{-9}$. Then, by setting $\ln \xi = 8$, the range of scales corresponding to the CMB, we may write $\sigma_{\zeta}^2 \simeq 1.3 \times 10^{-7}$. Then $g_{\rm NL}$ becomes
\bea
g_{\rm NL} \simeq 3.5 \times 10^6 \, A^2 \frac{ \sigma_{\zeta}^4}{f_{\zeta}^4}  e^{- \frac{\sigma_\zeta^2}{2 f_{\zeta}^2}}.
\eea
Furthermore, current constraints on the primordial trispectrum by Planck on 2015~\cite{Ade:2015ava} imply $g_{\rm NL} = ( - 9.0 \pm 15.4) \times 10^4$ at $95$\% C.L. It then follows that $A^2$ and the ratio $ f_{\zeta}^2 / \sigma_\zeta^2$ must satisfy the following restriction:
\be
A^2 \frac{ \sigma_{\zeta}^4}{f_{\zeta}^4}  e^{- \frac{\sigma_\zeta^2}{2 f_{\zeta}^2}} < 2.1 \times 10^{-3} .
\ee
Figure~\ref{fig:FIG-TRI} shows the allowed values for the parameter space spanned by $A^2$ and $f_\zeta / \sigma_\zeta$. It may be seen that $A^2$ becomes less constrained for both, large and small values of $f_\zeta / \sigma_\zeta$. It is interesting to note that the values used to plot both Fig.~\ref{fig:FIG-PDF-1} ($f_\zeta / \sigma_\zeta = 5 \times 10^{-2}$ and $A^2 \sigma_\zeta^2 / f_\zeta^2 = 10$) and Fig.~\ref{fig:FIG-PDF-2} ($f_\zeta / \sigma_\zeta = 2 \times 10^{-2}$ and $A^2 \sigma_\zeta^2 / f_\zeta^2 = 62.5$) are well within the allowed region. However, it is hard to conceive that peaks in the non-Gaussian PDF larger than those shown in Fig.~\ref{fig:FIG-PDF-1} are not excluded. This means that even strong constraints on the four-point function would not compete with other methods aiming to constrain the shape of the probability distribution function, such as the ones we will develop in Chapter~\ref{chap:CMB-LSS} with the help of the next section's general results.

%%%%%%%%%%%%%%%%%%%%%%%%%%%%%%%%%%
%%%%%%%%%%%%%%%%%%%%%%%%%%%%%%%%%%

\section{The general stage for Tomographic non-Gaussianity} \label{sec:gen-NG-Tom}

The purpose of this section is to extend the previous derivation to an arbitrary analytic potential $\Delta V$, setting the foundations to show how it is possible to reconstruct its shape with current and/or future cosmological data. %Our main claim is that, if the primordial landscape had a rich structure, then its shape (around the inflationary trajectory) could be stored in the statistics of $\zeta$ through a type of NG (tomographic NG) that cannot be fully parametrized with the bispectrum alone.
Starting from the same Lagrangian as before, describing $\zeta$ and $\psi$ ($M_{\rm Pl} = 1$):
\be
\mathcal L = a^3 \Big[\epsilon  ( \dot \zeta - \alpha \psi ) ^2 -   \frac{\epsilon}{a^2} (\nabla \zeta)^2  +  \frac{1}{2} \dot \psi^2  - \frac{1}{2a^2} (\nabla \psi)^2   -  \Delta V (\psi) \Big], \label{Lagrantian-full-v-2}
\ee
it is possible to compute the 
$n$-point correlation functions induced by a generic $\Delta V$, written in terms of its Taylor expansion as
\be \label{potential-taylor}
\Delta V \left( \psi \right) = \sum_{m} \frac{c_m}{m!} \psi^m.
\ee
This expansion gives us an infinite number of $m$ legged vertices, each one of order $c_m$.  Using the in-in formalism, the computation of $\langle \zeta_{{\bf k}_1 ... {\bf k}_n}^n \rangle_c$ requires us to consider the sum of each Feynman diagram proportional to $c_{n + 2m}$ with $m \geq 0$. In any such diagram, $n$ legs become $\zeta$ external legs (due to the $\alpha$ coupling), whereas $2m$ legs become $m$ loops. Finally, $\langle \zeta_{{\bf k}_1 ... {\bf k}_n}^n \rangle_c$ is the result of summing all of these diagrams after taking into account the appropriate combinatorial factors.

%which 
% for an arbitrary potential we first consider the following Taylor expansion

Following the computation that led to the $n$-point functions stemming from the sinusoidal potential, one sees that the only modification is in the coefficients of the Taylor expansion of $H_I$, which are time-independent, therefore only modifying the number emerging from~\eqref{re-summation}. Thus, the correlations are also of the local type and given by
\be \label{n-point-local}
\langle \zeta_{{\bf k}_1 ... {\bf k}_n}^n \rangle_c =  (2\pi)^3 \, h_n \, \delta^{(3)} \Big( \sum_{i=1}^n {\bf k}_i \Big) \frac{k_1^3 + \cdots + k_n^3}{k_1^3 \cdots k_n^3} ,
\ee
where the subscript $c$ informs us that we are only keeping fully connected contributions (in the language of perturbation theory). The set of amplitudes $\{h_n\}_n$ is given by
\be
h_n =-\left( \frac{\alpha H \Delta N}{2 } \right)^n \frac{\Delta N}{3H^{4}} \sum_{m=0}^{\infty} \frac{c_{n+2m}}{m!}  \left(\frac{\sigma_0^2}{2}  \right)^{m} , \label{h-n-V}
\ee
where we have now made explicit the sum that generalizes~\eqref{re-summation}.
%$\sigma_0^2  \equiv  (2\pi)^{-3} \! \int d^3 k \, \psi_k^*(t) \psi_k(t)$, appearing because of the loops, is the variance of the field $\psi$.
Let us remind the reader that the variance $\sigma_0^2$ is time independent~\cite{Palma:2017lww}, allowing to separate the sum and factor it out of the temporal integrals, which are consequently the same as in the sinusoidal potential scenario.

To deduce expressions that allow to reconstruct $\Delta V$, it is useful to perform the sum in Eq.~(\ref{h-n-V}), obtaining
\be
\sum_{m=0}^{\infty} \frac{c_{n+2m}}{m!}  \left(\frac{\sigma_0^2}{2}  \right)^{m} = e^{\frac{\sigma_0^2}{2} \partial_\psi^2} \frac{\partial^n}{\partial \psi^n} \Delta V \bigg|_{\psi = 0} . \label{general-resummation}
\ee
As before, notice that $\sigma_0^2$ is formally infinite, and hence, the infrared (IR) and ultraviolet (UV) physical momentum cutoffs are necessary. We remind the reader that the UV cutoff $q_{\rm UV}$ corresponds to a wavelength well inside the horizon ($q_{\rm UV} \gg H$), whereas the IR cutoff $q_{\rm IR}$ corresponds to the wavelength of the largest observable mode. In addition to these scales, it is convenient to introduce an arbitrary intermediate momentum $q_L$ that splits $\sigma_0^2$ into two contributions: $\sigma_0^2 = \sigma_S^2 + \sigma_L^2$, from short and long modes, respectively. This splitting allows us to define a renormalized potential 
\be
\Delta V_{\text{ren}} (\psi) \equiv \exp \left( \frac{\sigma_S^2}{2} \frac{\partial^2}{\partial \psi^2} \right) \Delta V(\psi),
\ee 
which coincides with our previous notion of renormalization in the case of the sinusoidal potential:
\be
\exp \left( \frac{\sigma_S^2}{2} \frac{\partial^2}{\partial \psi^2} \right) \Lambda^4 \cos \! \left(\psi/f \right) = \exp \left( - \frac{\sigma_S^2}{2 f^2} \right) \Lambda^4 \cos \! \left(\psi/f \right) = \Lambda^4_{\rm ren} \cos \! \left(\psi/f \right).
\ee 
In this way, observables can only depend on $\Delta V_{\text{ren}}$, which is independent of $q_{\rm UV}$.  

According to Eq.~(\ref{general-resummation}), this renormalization procedure simply corresponds to defining 
\be \label{potential-ren-resum}
\Delta V_{\rm ren} (\psi) = \sum_{m=0}^{\infty} \frac{1}{m!} c_m^{\rm ren} \psi^m,
\ee
where the coefficients $c_m^{\rm ren}$ are related to the bare couplings $c_m$ as
\be
\sum_{m=0}^{\infty} \frac{c_{n+2m}}{m!}  \left(\frac{\sigma_0^2}{2}  \right)^m = \sum_{m=0}^{\infty} \frac{c_{n+2m}^{\text{ren}}}{m!}  \left(\frac{\sigma_L^2}{2}  \right)^m. \label{coeff-ren}
\ee
This result allows us to identify $\Delta V_{\text{ren}}$ as the potential obtained by integrating out the high energy momenta beyond the scale $q_L$, just as in the Wilsonian approach of QFT. Now, it is crucial to notice that the $n$-point function of Eq.~(\ref{n-point-local}) is an observable, and so it cannot depend on $q_{L}$. This implies that $h_n$ is independent of $\sigma_L$. For this to be possible, the coefficients $c_{m}^{\rm ren}$ defining $\Delta V_{\rm ren}$ must run in such a way so that the entire expression (\ref{h-n-V}) remains independent of $\sigma_L$. Equation~(\ref{coeff-ren}) reveals how the coefficients $c_m^{\rm ren}$ run as more (or fewer) modes participate in $\sigma_L^2$ (again, in agreement with the Wilsonian picture).

To continue, using the Weierstrass transformation, the right hand side of Eq.~(\ref{general-resummation}) can be rewritten as
\be
\exp \left( \frac{\sigma_L^2}{2} \frac{\partial^2}{\partial \psi^2} \right) \frac{\partial^n}{\partial \psi^n} \Delta V_{\rm ren} \bigg|_{\psi = 0} = \int \! d\psi \frac{e^{- \frac{\psi^2}{2 \sigma_L^2}} }{\sqrt{2 \pi} \sigma_L}  \frac{\partial^n}{\partial \psi^n}  \Delta V_{\rm ren} .
\ee
Then, by performing several partial integrations, we finally obtain the following expression for $h_n$:
\be \label{h-n-V-2}
h_n =  \frac{1}{n}  \left( \frac{\alpha H \Delta N}{2 \sigma_L} \right)^n \frac{ \Delta N}{3H^{4}}  \int \!  d\psi \frac{e^{-\frac{\psi^2}{2\sigma_L^2}}}{\sqrt{2\pi} \sigma_L} {\rm He}_n\left( \psi /\sigma_L \right) 
 \left( \sigma_L^2 \frac{\partial^2}{\partial \psi^2} - \psi \frac{\partial}{\partial \psi} \right) \Delta V_{\text{ren}} (\psi) ,
\ee
where ${\rm He}_n(x) \equiv \exp( - \frac{1}{2} \frac{d^2}{dx^2}) x^n$ is the $n$th ``probabilist's'' Hermite polynomial. In the particular case where $\Delta V (\psi) = \Lambda^4 \left[ 1 - \cos (\psi / f) \right]$, Eq.~(\ref{h-n-V-2}) allows us to recover the expression for $\langle \zeta_{{\bf k}_1... {\bf k}_n}^n \rangle_c$ obtained for the sinusoidal potential.

We now compute the $n$-th moment $\langle \zeta^n \rangle$ for a particular position ${\bf x}$. Because of momentum conservation, the specific value of ${\bf x}$ is irrelevant. In practice, we only have observational access to a finite range of scales, implying that the computation of $\langle \zeta^n \rangle$ must consider a window function selecting that range. As before, we use a window function with a hard cutoff, and write
\be
\zeta_L =  \frac{1}{(2 \pi)^3}  \int_{ k < k_{L}} \!\!\!\!\!\!\!\!\! d^3k \,  \zeta_{\bf k}  \, e^{- i {\bf k} \cdot {\bf x}} .
\ee
Notice that we have chosen to cut the integral with the same cutoff $k_{L} = a \, q_L$ introduced to split $\sigma_0^2 = \sigma_S^2 + \sigma_L^2$. Up until now, $q_L$ was an arbitrary scale introduced to select the scales integrated out to obtain $\Delta V_{\rm ren}$. However, we can now choose $q_L$ to coincide with the physical cutoff momentum setting the range of modes contributing to the computation of $\langle \zeta^n_L \rangle$. Given that we are interested in a $q_L^{-1}$ larger than the horizon, we can write
\be
\sigma_L^2 = ( H^2 / 4 \pi^2) \ln \xi , \label{sigma_L-res}
\ee
where $\xi \equiv k_L / k_{\rm IR}$. Following our previous results, the $n$th moment of $\zeta_L$ is given by
\bea
\langle \zeta_L^n \rangle_c &=&  (2\pi)^3 \, h_n \, I_n (\xi) , \label{n-point-x-space}  \\
I_n (\xi) &=& \frac{n}{ (2 \pi^2)^{n+1}} \int_0^{\infty} \!\! \frac{dx}{x} G_\xi(x) \left[ F_\xi(x) \right]^{n-1}, \label{I_n-integral}
\eea
where, importantly, the function $F_\xi (x)$ satisfies $F_\xi(0) = \ln \xi$, and $F_\xi (x) \leq \ln \xi$. As before, we look for a PDF $\rho(\zeta)$ such that 
\be
\langle \zeta_L^n \rangle = \int \! d \zeta \, \rho (\zeta) \zeta^n ,
\ee
where $\langle \zeta_L^n \rangle$ is the full $n$th moment, including disconnected contributions. 
%where $G_\xi(x) =  \int^1_{\xi^{-1}}   d z z x^2 \sin (z x)$ and $F_\xi(x)=\int^1_{\xi^{-1}}\frac{d y}{y} \frac{\sin (y x)}{y x} $. 
%, related to $\langle \zeta_L^n \rangle_c$ by $\langle \zeta_L^n \rangle = \sum_{m=0}^{\lfloor n/2 \rfloor} \frac{n!}{m! (n-2m)! 2^m} \sigma_\zeta^{2m} \langle \zeta_L^{n-2m} \rangle_c$. 
%Recall that $\sigma_\zeta$ is given by $\sigma_{\zeta}^2 = \alpha^2 \Delta N^2 \sigma_{L}^2 / H^2 = \left[\alpha^2 \Delta N^2  / \left(4 \pi^2\right)\right] \ln \xi $. Planck~\cite{Ade:2015xua} fixes $\sigma_{\zeta}^2/ \ln \xi = P_{\zeta}(k) k^3 / \left(2 \pi^2\right) = (2.196 \pm 0.158 )\times 10^{-9}$.
To derive $\rho (\zeta)$ we just need to focus on the $n$ dependence of $\langle \zeta^n \rangle_c$. According to Eq.~(\ref{n-point-x-space}), this dependence has the form $X^n {\rm He}_n (Y)$, where $X$ and $Y$ are given quantities. The presence of the integrals does not alter this argument, as they can be factored out. 

This alone allows us to infer the PDF for $\zeta$, which is found to be given by
\bea
\rho (\zeta) &=&\frac{1}{\sqrt{2\pi}\sigma_\zeta } e^{-\frac{\zeta^2}{2\sigma_\zeta^2}}  \left[ 1 + \Delta(\zeta) \right],  \label{main-result-1} \\
 \Delta (\zeta) &\equiv& \int_0^{\infty} \!\! \frac{dx}{x}  \mathcal{K}(x) \! \int_{-\infty}^{\infty} \!\!\!\! d \bar \zeta \,\, \frac{\exp \Big[{-\frac{ \left(\bar \zeta - \zeta(x) \right)^2}{2\sigma_\zeta^2 (x)} } \Big] }{\sqrt{2\pi  } \sigma_\zeta (x)} \frac{\Delta N}{3 H^4}  \left( \sigma_\zeta^2 \frac{\partial^2}{\partial {\bar \zeta}^2} - {\bar \zeta} \frac{\partial}{\partial {\bar \zeta}} \right) \Delta V_{\text{ren}}  \! \left(\psi_{\bar \zeta}\right) . \label{main-result-2}
\eea
In the previous expression, $\Delta(\zeta)$ parametrizes the NG deviation. To write it, we defined the following quantities: 
\bea
\zeta(x) &\equiv&   [ F_\xi(x) /\ln \xi ]  \zeta, \\
\sigma_{\zeta}^2 (x) &\equiv&  \sigma_{\zeta}^2 (1 - [ F_\xi(x) /\ln \xi]^2 ), \\
\mathcal{K}(x) &\equiv& 4\pi G_\xi(x) / F_\xi(x), \\
\psi_{\zeta} &\equiv& (\alpha \Delta N/ H)^{-1} \zeta. 
\eea
These definitions satisfy $|\zeta (x)| \leq |\zeta |$ and $0 \leq \sigma_{\zeta}^2 (x) \leq \sigma_{\zeta}^2$. The latter ensures that the PDF is well-behaved and together with the former define the filtering effects that the observables emanating from the potential acquire.

\begin{figure}[t!]
\includegraphics[scale=1.2]{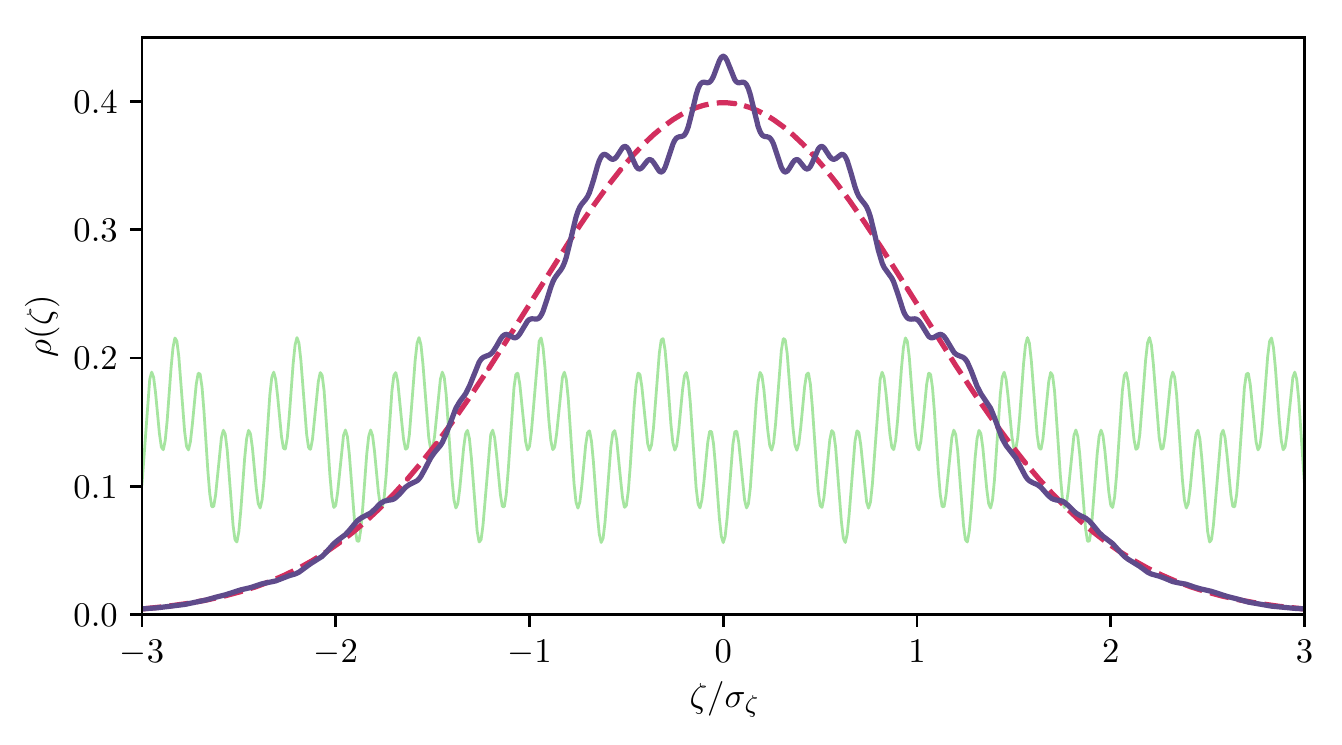}
\caption{The PDF (violet) resulting from a potential $\Delta V (\psi) \propto \left[ 2 - \cos (\psi/f_1) - \cos (\psi/f_2) \right]$, with $f_1 = 0.1 \sigma_L$ and $f_2 = 0.02 \sigma_L$ (light green). Both contributions have the same amplitude; however, $f_2$ contributes less than $f_1$. A Gaussian PDF is plotted for comparison (red, dashed). To plot the figure we used the relation $\psi_\zeta / \sigma_L = \zeta / \sigma_\zeta$.}
\label{fig:PDF_example}
\end{figure}

Equation~(\ref{main-result-1}) gives us the PDF of $\zeta$ at the end of inflation. By direct inspection, it is possible to verify that the perturbativity condition on the potential is $\Delta V_{\rm ren} / H^4 \ll 1$, and that the next to leading order term is of order $\mathcal O (\Delta^2)$ (see Ref.~\cite{Chen:2018uul}). The presence of the derivative operator acting on $\Delta V_{\rm ren}$ implies that the probability of measuring $\zeta$ at a given amplitude increases at those corresponding values $\psi_{\zeta}$ that minimize the potential. In addition, the $x$ dependence of $\zeta(x)$ and $\sigma_{\zeta}^2 (x)$ has the effect of filtering the structure; sharper structures contribute less to the PDF. To illustrate this, Figure~\ref{fig:PDF_example} shows the PDF obtained for a potential $\Delta V_{\rm ren} (\psi) \propto \left[ 2 - \cos (\psi/f_1) - \cos (\psi/f_2) \right]$. In this example there are two sinusoidal contributions with field scales $f_1 = 0.1 \sigma_L$ and $f_2 = 0.02 \sigma_L$. Both contributions have the same amplitude, however, the NG deformation implied by $f_2$ is smaller than that of $f_1$, indicating that sharper features get suppressed as the window of observation gets smaller.

With this, we have enough tools to constrain and attempt to reconstruct $\Delta V$ from observations. The (non-Gaussian) PDF~\eqref{main-result-1} is completely characterized by its fully connected moments~\eqref{n-point-x-space}, which aside from numerical factors, is the Hermite polynomial expansion of
\be
\left( \sigma_L^2 \frac{\partial^2}{\partial \psi^2} - \psi \frac{\partial}{\partial \psi} \right) \Delta V_{\text{ren}} (\psi).
\ee
If the potential does not grow too fast as $\psi \to \pm \infty$ (concretely, if $\Delta V$ is within the function space $L^2$ with a Gaussian kernel in the inner product), this expansion is unique and thus the coefficients $h_n$ completely determine the function $\Delta V$. This means that there is a one-to-one correspondence between the PDF~\eqref{main-result-1} and the underlying landscape potential, and therefore a reconstruction of the primordial curvature perturbations' PDF is also a reconstruction of the inflationary landscape.

We will later use this important result to constrain the primordial isocurvature potential $\Delta V$ using CMB observations in Chapter~\ref{chap:CMB-LSS}.

%This concludes our discussion on the observable statistics generated by the Hamiltonian evolution of quantum fields. Now we proceed to shift the focus to the field variable itself.

\cleardoublepage

\chapter{The Landscape of Tomographic non-Gaussianity} \label{chap:field}

With what we have done so far, we have fully characterized the departures from Gaussianity expected in multi-field inflationary models with turning trajectory, or any model of the primordial universe that can be expressed in terms of an EFT with an equivalent Lagrangian, to first order in the source for the nonlinearities of the isocurvature mode. However, although instructive, the method of computing $n$-point functions may be somewhat obscure at times. As it is possible to derive the same results in a conceptually clearer manner, we will now do so. We will accomplish that by computing the evolution of the quantum field operator directly, without making references to expectation values.

We will also explore links with the stochastic formulation of inflation that can be readily established from the relations we will obtain for the quantum field operator. Furthermore, we will explore higher-order corrections by pushing the approximations we have been making thus far a bit further. Finally, we will generalize the Landscape Tomography to other backgrounds and more general self-interactions, deriving general formulas that can be used to gain information on the past dynamics of the fluctuations from the final time slice $n$-point functions.

%\section{$\zeta$ and Quantum Mechanics}

\section{Evolving the Quantum Field}

The generation of tomographic non-Gaussianity may be traced back to the self interactions of an isocurvature field $\psi$, and these self-interactions are then transferred to the curvature perturbations thanks to a bilinear interaction coupling both fields [see Eq.~(\ref{Lagrantian-full-v})]. To appreciate how self-interactions give rise to this form of non-Gaussianity at the level of field operators, we may start by recalling that in the Interaction picture, the evolution of the field $\psi (\x,\tau)$ is given by
\be
\psi (\x,\tau) = U (t,t_0) \psi_I (\x,\tau) U^\dag (t,t_0).
\ee 
%From a quantum-mechanical perspective in the Heisenberg picture, one can write
%\be
%\psi(\x,\tau) = \psi(\x, \tau_0) + i \int^{\tau}_{\tau_0} d\tau' [H(\tau'), \psi(\x,\tau')].
%\ee
At first order in time-dependent perturbation theory, the evolution of the field is given by
\be \label{first-order-evolution}
\psi(\x,\tau) = \psi_I(\x, \tau) + i \int^{\tau} d\tau' [H_I(\tau'), \psi_I(\x,\tau)].
\ee
where the subscript $I$ informs us that the corresponding operator is of the interaction picture and evolves as a free field. If, for simplicity, we take de Sitter spacetime as a background, the interaction picture Hamiltonian reads 
\be
H_I(\tau) = \int_x a^4(\tau) \Delta V \left(\psi_I(\x,\tau) \right)
\ee
with $a(\tau) = -1/H\tau$. With this, we may work on our previous equation to obtain
\be
\psi(\x,\tau) = \psi_I(\x, \tau) + i \int^{\tau} d\tau' \int_x a^4(\tau')  [\psi_I(\x',\tau'),\psi_I(\x,\tau)] \frac{\partial \Delta V}{\partial \psi}(\psi_I(\x',\tau')).
\ee

With the help of canonical commutation relations for the appropriate field $v \equiv a \psi$~\cite{Achucarro:2016fby,Chen:2018uul}, which we introduced earlier, the commutator $[\psi_I(\x',\tau'),\psi_I(\x,\tau)]$ is just a number and we can carry out the integral over $\tau'$ explicitly provided that:
\begin{enumerate}
\item The quantum field $\psi_I(\x',\tau')$ in the argument of $\partial \Delta V/\partial \psi$ may be treated as a constant over the integration time $\tau'$.
\item The range of modes under consideration involves super-horizon modes only $|k\tau'| \lesssim 1$.
\end{enumerate}
In fact, if the range of modes satisfies $|k\tau'| \ll 1$ for all $k$, then the first condition is implied by the second. 

Let us give some comments about these conditions: The first point seems natural in the sense that the statistics of $\psi_I$ do not evolve over time: it may be seen as a Gaussian random field with a definite covariance. The second point, although appealing, is both physically and mathematically suspect, since in principle the interaction picture Hamiltonian involves every mode (i.e. every momentum scale). Nonetheless, from an EFT perspective this is perfectly acceptable, as long as the potential $\Delta V$ is responsible for describing the physics at those scales. Moreover, this is the appropriate course of action when studying CMB or LSS modes that spent a large number of e-folds outside the horizon, because they do satisfy $|k\tau'| \ll 1$ through most of their history (practically for every time after horizon crossing this condition is fulfilled).

Using these considerations, one obtains
\be
[\psi_I(\x',\tau'),\psi_I(\x,\tau)] = (H^2 \tau \tau') \int_k e^{i \k \cdot (\x' - \x)} D(\tau',\tau,k) \approx \frac{i H^2}{3} (\tau^3 - \tau'^3) \int_k e^{i \k \cdot (\x' - \x)},
\ee
where $D(\tau',\tau,k) \equiv v_k(\tau') v_k^*(\tau) - v_k^*(\tau') v_k(\tau) $, and by the means of assumption 1.,
\be \label{local-isocurvature}
\psi(\x,\tau) = \psi_I(\x, \tau) - \frac{\Delta N}{3H^2} \int_{\x'} \int_k e^{i \k \cdot (\x' - \x)} \frac{\partial \Delta V}{\partial \psi}(\psi_I(\x',\tau))
\ee
where
\be
\Delta N = - \int^\tau \frac{d\tau'}{\tau'} \approx \int^\tau d\tau' \frac{\tau^3 - \tau'^3}{\tau'^4} 
\ee
is the number of e-folds spent outside the horizon by the range of modes of consideration, which we take to satisfy $\Delta N \gg 1$. While the former may seem to be a heavy restriction, if we consider that the currently observable range of scales in the CMB satisfies $\ln (k_{S}/k_L) \simeq 8$~\cite{Aghanim:2018eyx} and $\Delta N \sim 60$, we see that approximating $\Delta N$ to a single value for all modes in the considered range is justified.

Note that the operator $\int_{\x'} \int_k e^{i \k \cdot (\x' - \x)} ( \cdot )$ projects the function $\Delta V$ onto the modes under consideration for the field $\psi(\x)$, thus solving the concern one might have had about a product of field operators generating contributions with larger wavenumbers than those allowed in the effective description. That is to say, the integral over wavenumbers $\k$ has implicit cutoffs
\be
\int_k (\cdot) = \frac{1}{(2\pi)^3}  \int d\Omega \int_{k_L^\psi}^{k_S^\psi} dk k^2 (\cdot),
\ee
that prevent us from directly replacing the projecting operator for a Dirac delta. Schematically, however, if we remember that the function $\frac{\partial \Delta V}{\partial \psi}$ only contributes to the range of modes prescribed by the theory, we may write
\be \label{local-isocurvature-schem}
\psi(\x,\tau) = \psi_I(\x, \tau) - \frac{\Delta N}{3H^2} \frac{\partial \widetilde{\Delta V} }{\partial \psi}(\psi_I(\x,\tau)),
\ee
where the tilde ``$\widetilde{\,\,\,\,\,\,\,\, } $'' is there to remind us of the presence of the projection operator. This is exactly what generates the filtering in the position-space $n$-point functions obtained in the previous chapter.

%Of course, as was earlier suggested, a rigorous proof of this requires to go through every $n$-point function, performing an adequate renormalization procedure to make the computations consistent at every bounded range of momenta. This was thoroughly dealt with in previous works~\cite{Chen:2018uul, Chen:2018brw}, although stopping just short of writing down equation~\eqref{local-isocurvature}. 

The curvature perturbation $\zeta(\x,\tau)$ may be obtained in a completely analogous manner, only that we now have to consider an extra commutator to account for the quadratic mixing term~\eqref{Lagrantian-full-v}:
\be \label{zeta-perturbated}
\zeta (\x,\tau) = \zeta_I(\x,\tau) + i\int^{\tau} d\tau'  [H_I^{\alpha}(\tau'), \zeta_I(\x,\tau)]  - \int^{\tau} d\tau' \int^{\tau'} d\tau'' [H_I^{V}(\tau''), [H_I^{\alpha}(\tau'), \zeta_I(\x,\tau)]] + \cdots,
\ee
where $H^{V}_I$ is the term of the interaction picture hamiltonian that contains the $\psi$ self-interactions and $H^{\alpha}_I$ contains terms associated to the quadratic mixing. The ellipsis $\cdots$ stand for higher order terms.

It is of crucial importance to obtain the correct result to notice that the commutators in the last term of~\eqref{zeta-perturbated} only give a nonzero result when the pieces of the interaction Hamiltonian are written in that order. This, alongside the time ordering, yields an additional $1/2$ factor for the statistical transfer of the nonlinear perturbation $\Delta V$. After a calculation analogous to the one that led us to~\eqref{local-isocurvature}, with the same working assumptions, one obtains
\be \label{zeta-perturbated-2}
\zeta (\x,\tau) = \zeta_I(\x,\tau) + \frac{\alpha}{H} \Delta N \left( \psi_I (\x,\tau) - \frac{1}{2} \frac{\Delta N}{3 H^2} \int_{\x'} \int_k e^{i \k \cdot (\x' - \x)}  \Delta V' \left( \psi_I (\x',\tau) \right) \right),
\ee
or schematically,
\be \label{zeta-perturbated-2-schem}
\zeta (\x,\tau) = \zeta_I(\x,\tau) + \frac{\alpha}{H} \Delta N \left( \psi_I (\x,\tau) - \frac{1}{2} \frac{\Delta N}{3 H^2}  \widetilde{\Delta V}' \left( \psi_I (\x',\tau) \right) \right),
\ee

\begin{figure}[t!]
\begin{center}
\includegraphics[scale=0.22]{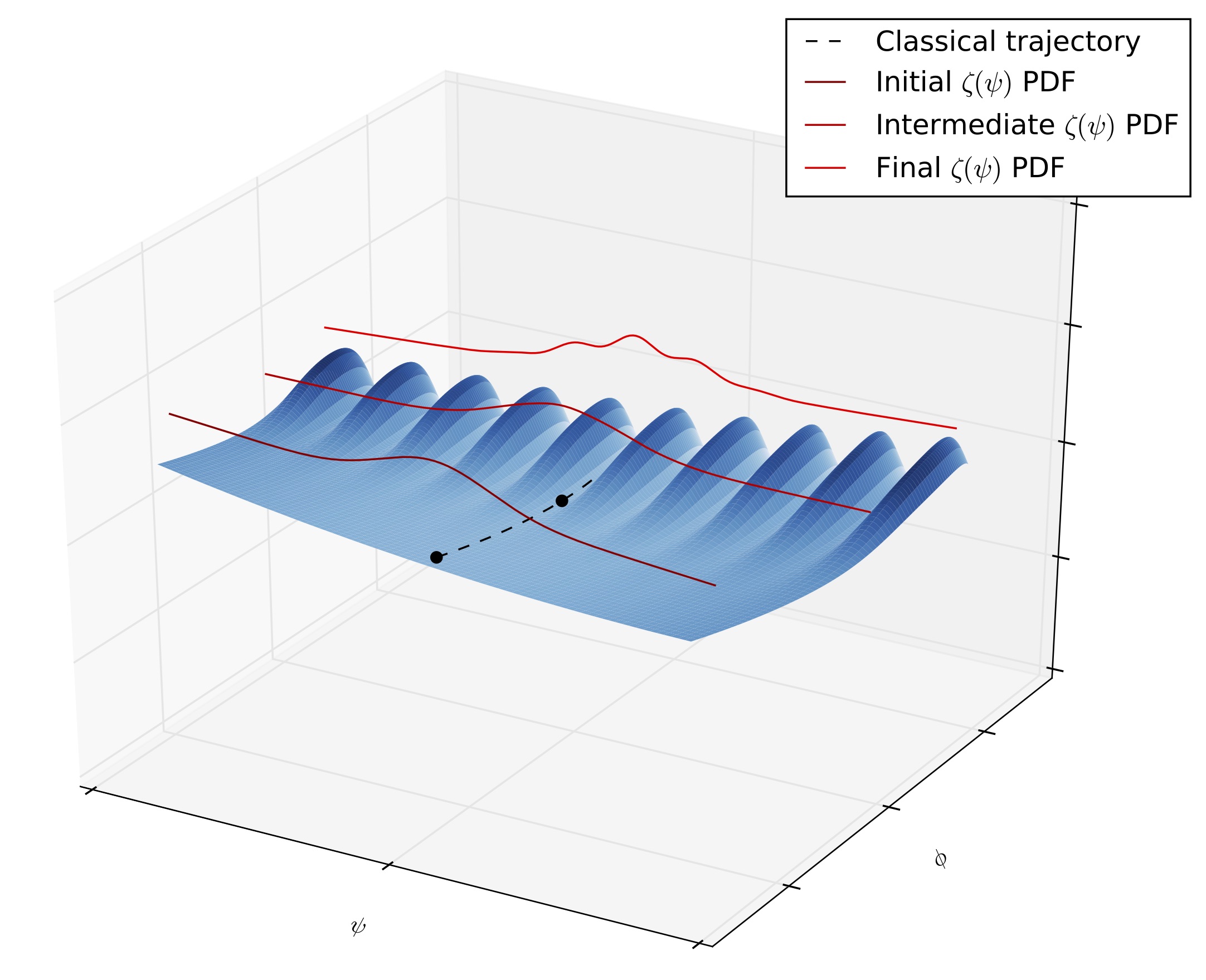}
\caption{The temporal evolution of the 1-point PDF of the curvature perturbation $\zeta$, sourced by $\psi$, here represented schematically along the isocurvature direction, orthogonal to the inflationary trajectory.}
\label{fig:FIG_evol-field}
\end{center}
\end{figure}

Equation~\eqref{zeta-perturbated-2} can accommodate a variety of regimes for the parameter $\alpha \Delta N/H$. However, we choose to work in a situation wherein  the linear transfer from $\psi$ to $\zeta$ dominates. Even though this last equation is a perturbative result, $\zeta = \frac{\alpha \Delta N}{H} \psi + \zeta_0 $ is an exact solution of the equations of motion on superhorizon scales~\cite{Achucarro:2016fby}. This allows us to neglect the first term, as the $\Delta N$ factor can grow enough that $\frac{\alpha \delta N}{H} \psi_I$ becomes large in comparison with $\zeta_I$, and therefore we consider
\be \label{zeta-perturbated-3-schem}
\zeta (\x,\tau) = \frac{\alpha}{H} \Delta N \left( \psi_I (\x,\tau) - \frac{1}{2} \frac{\Delta N}{3 H^2}  \widetilde{\Delta V}' \left( \psi_I (\x',\tau) \right) \right),
\ee
which we can write as
\be \label{zeta-perturbated-4-schem}
\zeta (\x,\tau) = \zeta_G(\x,\tau) - \frac{\Delta N}{6 H^2}  \widetilde{\Delta V}' \left( \frac{H \zeta_G (\x',\tau)}{\alpha \Delta N} \right),
\ee
where we have defined the effective induced field $\zeta_G = \frac{\alpha \Delta N}{H} \psi_I $, which has Gaussian statistics and we therefore denoted it by $\zeta_G$. In absence of the potential $\Delta V$, this reproduces the Power Spectrum derived in Chapter~\ref{chap:inflation}, and in its presence the $n$-point functions we obtained in Chapter~\ref{chap:n-point}. We have thus fulfilled our purpose of writing Tomographic NG in terms of equations for the field configurations.

%Furthermore, since conventionally $\zeta$ should satisfy $\langle \zeta \rangle = 0$, from this point forward we will consider
%\be \label{zeta-perturbated-3}
%\zeta (\x,\tau) \approx \frac{\alpha}{H} \Delta N \left( \psi_I (\x,\tau) - \frac{1}{2} \frac{\Delta N}{3 H^2}  [\Delta V' \left( \psi_I (\x,\tau) \right) - \langle \Delta V' \left( \psi_I (\x,\tau) \right) \rangle ] \right),
%\ee
%or equivalently, omitting the temporal coordinate
%\be \label{zeta-perturbated-4}
%\zeta (\x) = \zeta_G (\x) - \frac{\alpha^2 \Delta N^2}{2H^2} \frac{\Delta N}{3 H^2}  \left[ \frac{\partial}{\partial \zeta} \left( \Delta V \left( \frac{H \zeta_G (\x)}{\alpha \Delta N} \right) \right) - \left\langle \frac{\partial}{\partial \zeta} \left( \Delta V \left( \frac{H \zeta_G (\x)}{\alpha \Delta N} \right) \right) \right\rangle \right],
%\ee
%where we have written $\zeta_G$ instead of $\psi_I$ to stress the nature of our result: $\zeta$ is made up from a Gaussian contribution plus a local non-Gaussian term. 

\begin{figure}[t!]
\begin{center}
\includegraphics[scale=0.11]{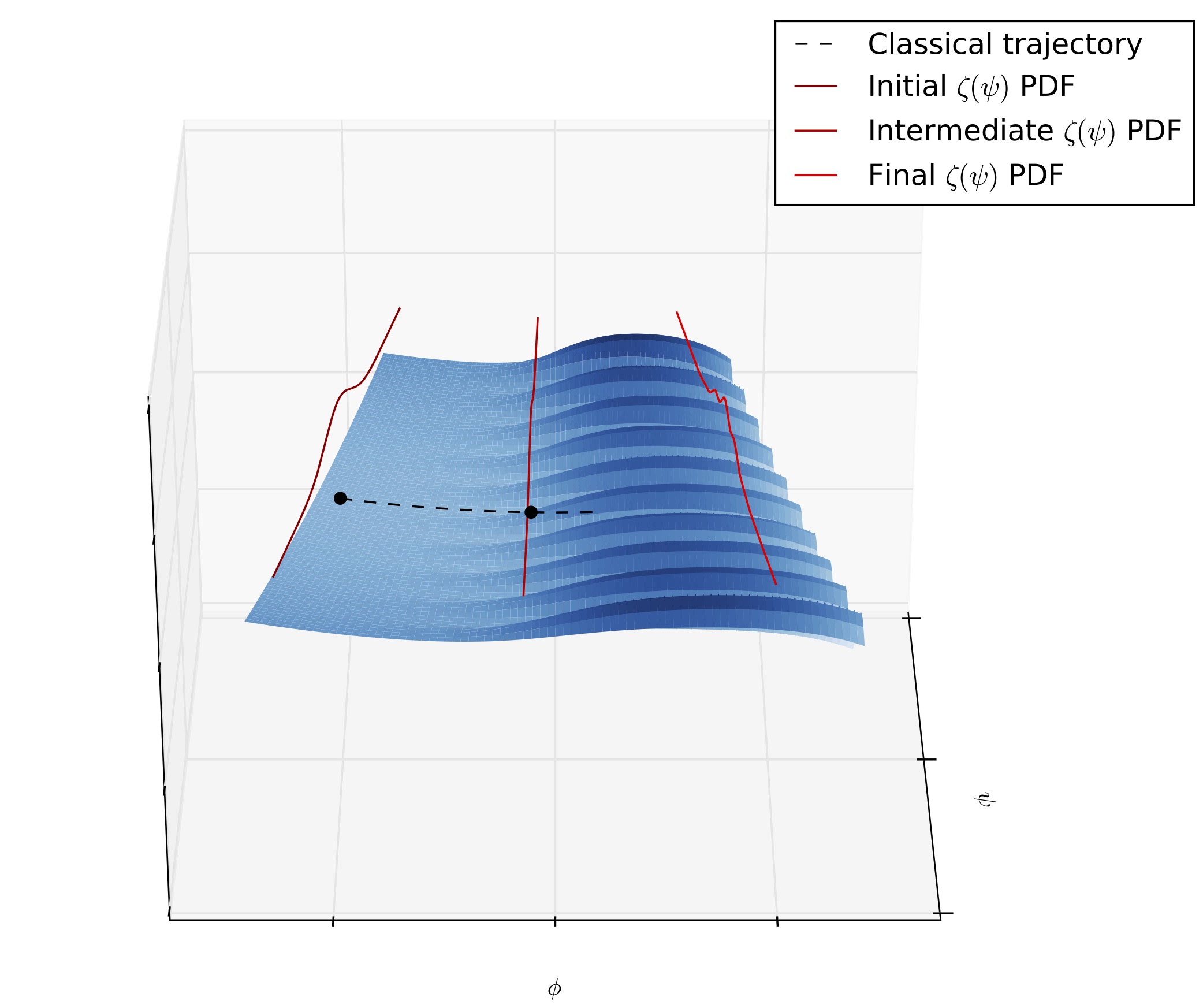}
\includegraphics[scale=0.11]{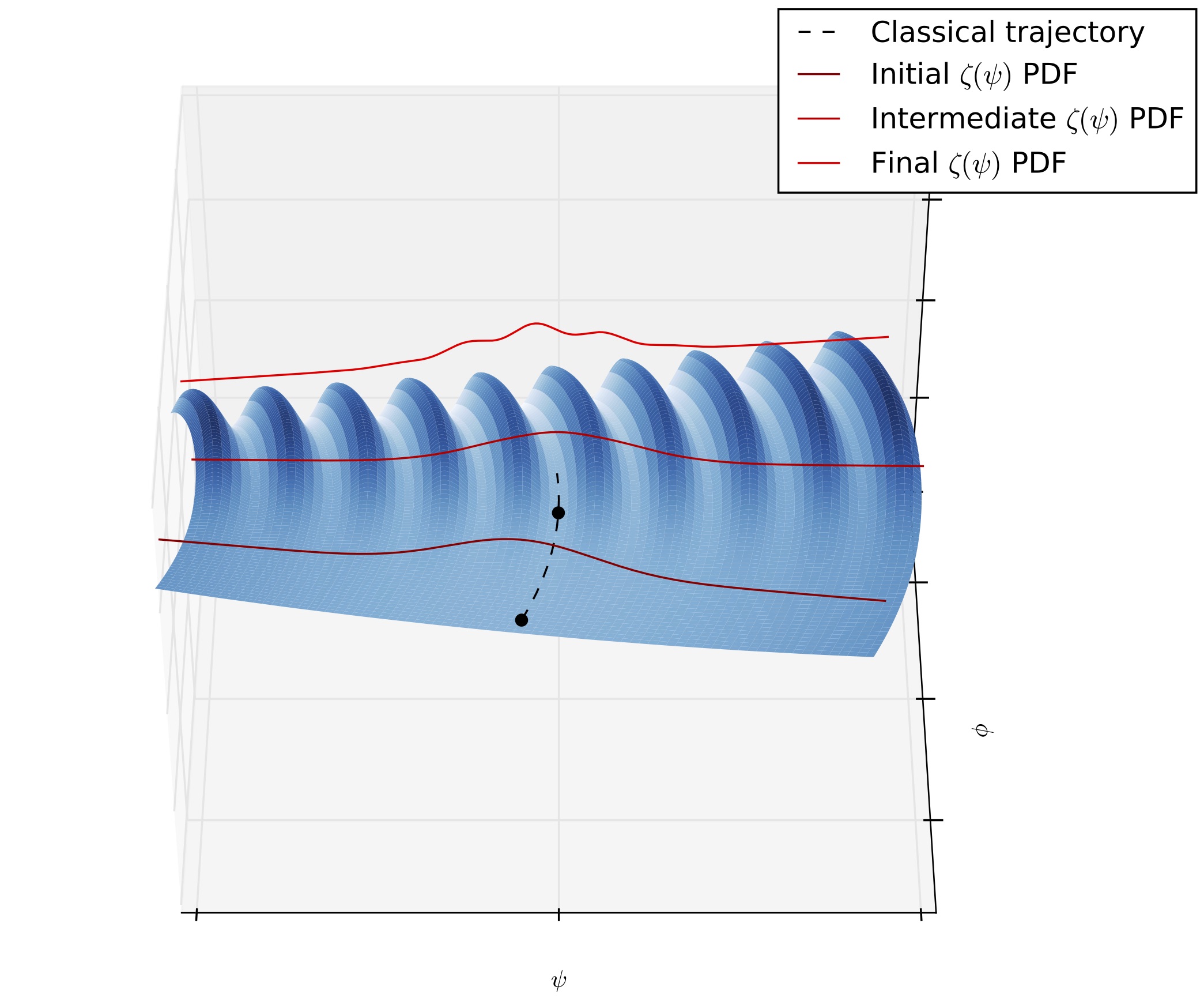}
\caption{Figure~\ref{fig:FIG_evol-field} as would be seen from different angles.}
\label{fig:FIG_evol-field2}
\end{center}
\end{figure}

Pictorially, we may describe the evolution of the quantum field $\zeta$ with Figure~\ref{fig:FIG_evol-field}. In the picture, the black dot describes the trajectory the inflaton $\phi_0^a$ follows in the multi-field target space. We have included the "turning" of the trajectory graphically to reflect the change in the tangent vector $T^a$ as felt by the inflaton while traveling on the curved target-space metric. Figure~\ref{fig:FIG_evol-field2} shows the same picture but from different angles. In these figures, we plot the qualitative evolution of the (single-point) PDF of the curvature perturbation $\zeta$, displayed along the axis of the isocurvature mode so as to simultaneously describe its temporal evolution and the influence the potential of the isocurvature mode has on it. We see that ``ripples'' start to form in the central region of the Gaussian distribution as the inflaton enters a narrow ``valley'' of the potential, increasing the probability at field values of $zeta$ when the corresponding value of $\psi = \psi_\zeta = \frac{H \zeta}{\alpha \Delta N}$ is at a minimum of the potential and decreasing it when $\psi_\zeta$ is at a maximum of the potential.

Recapitulating, the physical situation is the following: once a given set of modes (wavelengths) of the adiabatic mode $\zeta$ cross the Hubble radius and stop being able to interact with themselves directly, the only mechanism available to modify their amplitude is them being sourced by an extra field $\psi$, an isocurvature perturbation, which indeed does happen because of the bilinear coupling induced by the turning of the trajectory. Moreover, as the corresponding modes of the isocurvature field have already crossed the horizon, the only way in which their self-interactions can affect them is by modifying their local value, and to first order in perturbation theory, they do so by shifting the value of the field (see equations~\eqref{local-isocurvature},~\eqref{local-isocurvature-schem}) in the direction implied by the ``force'' in the equation of motion, $-\partial \Delta V/\partial \psi$, pushing the probability distribution of the free field $\psi_I$ towards the minima of the potential $\Delta V$. Therefore, as $\zeta$ is sourced by $\psi$, it also gets its probability distribution shifted towards the minima of the potential induced by the linear transfer $\zeta = \frac{\alpha \Delta N \psi}{H}$. This happens as long and for as long as the potential is active while the modes affected by it are at super-horizon scales.

In summary, with this study we have provided theoretical motivations for searching primordial non-Gaussianity in a general local setting:
\be \label{general-loc-NG-2}
\zeta(\x) = \zeta_G(\x) + \widetilde{F}(\zeta_G(\x)),
\ee
where $\widetilde{F}$ is defined by
\be
\widetilde{F} \left( \zeta_G(\x) \right) =  \int_{\x'} \int_k e^{i \k \cdot (\x' - \x)} F\left( \zeta_G(\x') \right).
\ee
We will come back to this in Chapter~\ref{chap:CMB-LSS}, where we will study the signatures that this kind of NG implies, with special focus on the prospects for upcoming cosmological surveys.

For the remainder of this chapter, we set our attention specifcally on the nonlinearities affecting the isocurvature mode, and we will keep the transfer mechanism to first order, citing the explicit expression for $\zeta$ only if the relationship with $\psi$ is different than~\eqref{zeta-perturbated-3-schem}.

\section{Links with Stochastic Inflation}

Although we have made our derivations quantum-mechanical perspective, similar results to~\eqref{local-isocurvature-schem} can be obtained within other frameworks. For instance, stochastic inflation~\cite{Starobinsky:1986fx,Pinol:2018euk,Tsamis:2005hd}, where the equation of motion for the coarse-grained (averaged over local patches of the universe) inflaton is subject to a source of stochastic noise. In a multi-field context, this is usually written as a Langevin equation
\be \label{stochastic}
\frac{d \phi^a}{dt} = - \frac{\gamma^{ab}}{3H} \frac{\partial V}{\partial \phi^b} + \Xi^a,
\ee
where $\Xi^a$ is a stochastic noise. The physical picture here is that the stochastic equation of motion attempts to capture the evolution of the long-wavelength modes, and the stochastic noise $\Xi^a$ appears because of short-wavelength modes exiting the Hubble radius and entering the effective description for long-wavelength modes, which feel a potential $V(\phi^I)$.

This can be connected to~\eqref{multi-field-evol} in the following manner: if we consider super-horizon modes only, then by inspecting the mode functions $u_k(\tau)$ for a massless scalar in a de Sitter space one finds that the temporal derivatives are suppressed by a factor of $|k\tau|$, thus allowing to neglect the derivative acting on $\dot \delta \phi^a$, i.e., for the fluctuating piece. On the same grounds, the laplacian $\frac{1}{a^2} \nabla^2 \phi^a$ can also be neglected. Finally, on the other end of the fluctuations, the contribution of the short modes of the fields is included with an extra stochastic term $\Xi^a$. As for the ``background'' inflationary trajectory (without spatial dependence), in the simplest slow-roll approximation it is typical to neglect the whole temporal derivative $D/dt$ acting on $\dot \phi^a$, implicitly assuming that the turning rate of the trajectory is also small\footnote{As we discussed earlier, for slow-roll inflation to happen it is only required that $\eta_\phi$ is small, and thus one should only neglect the contribution that is parallel to the inflationary trajectory.}. This gives the stochastic equation of motion~\eqref{stochastic}.

Let us now make a concrete connection to the setup we have been studying. In the two-field example presented in~\cite{Achucarro:2016fby,Chen:2018brw} for an inflationary turning trajectory, two fields $\phi^1 = \mathcal{X}$ and $\phi^2 = \mathcal{Y}$ sit on a 2-dimensional hyperbolic manifold, with target-space metric
\be
\gamma_{ab} = \begin{pmatrix}
e^{2\mathcal{Y}/R_0 } & 0 \\
0 & 1
\end{pmatrix}.
\ee
Here a solution to the background equations of motion has $\mathcal{X} = \mathcal{X}(t)$, while $\mathcal{Y} = \mathcal{Y}_0$ is preserved constant. Although this may seem like a straight line, the turning emerges because the variation of the tangent vector must be taken with the covariant derivative $D/dt$, which emphasizes and enforces that the turning must be regarded as a turn with respect to the metric of the field space. Making the identification $\psi = \mathcal{Y} - \mathcal{Y}_0$ as the fluctuating piece, the resulting equation of motion now reads\footnote{In this case, it is correct to replace this into~\eqref{stochastic} while still preserving a turning trajectory because the effects of $\zeta$ on $\psi$ through the bilinear coupling are suppressed by the first slow-roll parameter $\epsilon$. Thus, we may discard the turning effects on the equation of motion for $\psi$, and therefore use~\eqref{stochastic}, which assumes a weak turning.}
\be
\dot \psi = - \frac{1}{3H} \frac{\partial V}{\partial \mathcal{Y} }(\mathcal{Y}_0 + \psi) + \Xi^a,
\ee
and by means of the identification $V(\mathcal{Y}_0 + \psi) \to \Delta V(\psi)$, one obtains
\be \label{stoch-psi}
\dot \psi = - \frac{1}{3H} \frac{\partial  \Delta V}{\partial \psi } + \Xi^a,
\ee
where the potential should, for consistency, be taken with the projection prescription we introduced in the previous section. We omit it in this discussion to ease the notation.

Assuming an initially Gaussian super-horizon field at time $t_0$, i.e., that originally there were no self-interactions and that we will only be keeping track of it for the modes that exited the Hubble radius \textit{after} a given time $t_0$, we can integrate equation~\eqref{stoch-psi} to obtain
\be
\psi(\x,t) = \psi_G(\x,t_0) \int_{t_0}^t dt' \Xi^a - \int_{t_0}^t dt' \frac{1}{3H} \frac{\partial  \Delta V}{\partial \psi },
\ee
If the stochastic noise is nearly Gaussian for every mode that exits the horizon, as is expected from a typical QFT, their sum is also Gaussian. And even if the QFT does not give Gaussian predictions, the central limit theorem implies that, under the assumption that the modes that exit at different times are statistically independent, which they typically are as they have different wavenumbers, and if they satisfy certain properties, the sum of $\Xi^a$ over time will also converge to a Gaussian. This allows us to replace the first two terms by $\psi_G(\x,t)$, a Gaussian random field with statistics determined by sub-horizon physics, which is in turn determined by the inflationary background and the interactions of the field at those scales, plus the initial condition.

The second term can be evaluated in the following manner: go to conformal time, where $dt = a d\tau $, which in a de Sitter background $a(\tau) = -1/(H\tau)$ yields
\be
- \int_{\tau_0}^\tau \frac{(-1) d \tau' }{H \tau} \frac{1}{3H} \Delta V'(\psi(\x,\tau')),
\ee 
which can be further simplified if we assume $\psi \approx \psi_G$ in the argument of the potential. The equality holds if the potential were absent, so if we consider a weak potential (i.e., of small amplitude) this is a good approximation. If we now use our knowledge on the evolution of a massless field in de Sitter space, in the form of the mode functions $u_k(\tau)$, we may treat $\psi_G$ in the argument of $\Delta V$ as nearly constant over its super-horizon evolution and equal to its final configuration, allowing us to integrate over $\tau'$ directly and get
\be
\psi(\x,t) = \psi_G(\x,t) - \frac{ \Delta N }{3H^2} \Delta V'(\psi_G(\x,t)).
\ee
Thus, the $n$-point function statistics for $\psi$ is analogous to what we had previously obtained. However, when making the connection between this stochastic setup and the quantum-mechanical approach one must always keep in mind the assumptions that implicit in their respective formulations.

\section{Beyond First-Order Perturbation Theory}

Although we have here studied thoroughly the evolution of the quantum field to first order in time-dependent perturbation theory, one can wonder whether it is possible to extend some of our results to higher orders in perturbation theory. The full perturbation series for the isocurvature field $\psi$ can be written as
\be \label{all-order-evolution}
\psi(\x,\tau) = \psi_I(\x, \tau) + i \int^{\tau} \!\!\! d\tau_1 \, [H_I(\tau_1), \psi_I(\x,\tau)] + i^2 \int^{\tau} \!\!\! d\tau_1 \! \int^{\tau_1} \!\!\! d\tau_2 \, [H_I(\tau_2), [H_I(\tau_1), \psi_I(\x,\tau)] ] + \cdots,
\ee
where an extra (time-ordered) integral must be included at each order in perturbation theory with appropriate factors and commutators. In general, the $N$-th term of the perturbative series is equal to
\be \label{N-th-order}
i^N \int_{[\tau > \tau_1 > \tau_2 > \ldots > \tau_N > \tau_0 ]} \!\!\!\!\!\!\!\!\!\!\!\!\!\!\!\!\!\!\!\!\!\!\!\!\!\!\!\!\!\!\!\!\!\! d\tau_1 \ldots d\tau_N \,\,\,\,\,\,\,\, [ H_I(\tau_N), [ \ldots , [H_I(\tau_2), [H_I(\tau_1), \psi_I(\x,\tau)] ] \ldots ]].
\ee
Now we may proceed to evaluate the commutators one by one. The first commutator, in analogy with the first-order computation, is easily found
\be
[H_I(\tau_1), \psi_I(\x,\tau)] = \int_{\x_1} a^4(\tau_1) [\psi_I(\x_1,\tau_1),\psi_I(\x,\tau)] \frac{\partial \Delta V}{\partial \psi}(\psi_I(\x_1,\tau_1)),
\ee
and approximating the field in the argument of the potential by its late-time value, in accordance with $|k\tau_i| \ll 1$, obtain
\be
[H_I(\tau_1), \psi_I(\x,\tau)]  \underset{|k_1 \tau_1| \ll 1 }{\approx} \frac{i}{3 H^2} \int_{\x_1} \int_{k_1} e^{i \k_1 \cdot (\x_1 - \x)} \frac{\tau^3 - \tau_1^3}{\tau_1^4}  \frac{\partial \Delta V}{\partial \psi}(\psi_I(\x_1,\tau)).
\ee
The next commutators may be evaluated by using 
\be
[F(\psi_I(\x',\tau')), G(\psi_I(\x,\tau)) ] = [\psi_I(\x',\tau'),\psi_I(\x,\tau)] F'(\psi_I(\x',\tau')) G'(\psi_I(\x,\tau)), \ee 
where we have not taken issue with the ordering of operators at different times, because we will approximate all times in the argument of fields at which the potential is evaluated to the final time slice $\tau$, and therefore the functions of operators $F'$ and $G'$ commute. If we denote $\psi_m \equiv \psi_I(\x_m, \tau_m)$ and $\psi_I(\x,\tau) \equiv \psi_0$, this implies that the action of a commutator on the result of the previous commutator, $\mathcal{O}(\{\psi_l\}_{l=0}^m)$, is of the form
\be
[\Delta V( \psi_{m+1}), \mathcal{O}(\{\psi_i\}_{i=1}^m) ] = \Delta V'(\psi_{m+1}) \sum_{i=0}^m [\psi_{m+1}, \psi_i]  \frac{\partial}{\partial \psi_i}  \mathcal{O}(\{\psi_l\}_{l=0}^m),
\ee
which upon insertion of the spatial integral of the interaction Hamiltonian gives
\be
\begin{split}
[H_I(\tau_{m+1}), \mathcal{O}(\{\psi_i\}_{i=1}^m) ] \underset{|k_j\tau_j| \ll 1}{\approx} & \frac{i H^2 (\tau_i^3 - \tau_{m+1}^3 )}{3} \\ & \times \sum_{i=0}^m \int_{\x_{m+1}} \int_{\k_{m+1}} e^{i \k_{m+1} \cdot (\x_{m+1} - \x_i) } \Delta V'(\psi_{m+1}) \frac{\partial \mathcal{O}(\{\psi_l\}_{l=0}^m)}{\partial \psi_i}
\end{split}
\ee
As mentioned earlier, we are assuming that at every order in the perturbative series, the final result contains the self-interaction and derivatives of the self-interaction evaluated at the final time slice, and thus the ordering of operators is inconsequential at that level.

To make the rest of the computation tractable, we will make one further approximation: performing the temporal integrals, one obtains
\be
\int^{\tau_{m-1}}_{\tau_0} d\tau_{m} \left(\frac{\tau_i^3}{\tau_m} - 1 \right) \frac{1}{\tau_m} = \frac{\tau_i^3}{3} \left( \frac{1}{\tau_0^3} - \frac{1}{\tau_{m-1}^3} \right) - \ln \left( \frac{\tau_{n-1}}{\tau_0} \right) ,
\ee
where the absolute value of the first term in the parenthesis is always less than $1/3$, while the second term can be as large as $\ln (\tau_0/\tau ) $. Therefore, we will neglect the first term, and what we will get is the result in a leading logarithm approximation (similar approximations have been employed in an inflationary setup~\cite{Tsamis:2005hd,Woodard:2005cw}), tracking only the highest power of $\ln (\tau_0/\tau ) = \Delta N $ in the final result. Performing all of the temporal integrals, one obtains that~\eqref{N-th-order} is equal to
\be \label{result-Nth-order}
\begin{split}
\frac{1}{N!} \left( - \frac{\Delta N}{3 H^2} \right)^N  \int_{\x_{N}} \int_{\k_N} \cdots \int_{\x_{1}} \int_{\k_1} &\left( \sum_{i=0}^{N-1} e^{i \k_N \cdot (\x_N - \x_i)} \Delta V'(\psi_N) \frac{\partial}{\partial \psi_i} \right) \times \cdots \\ &
\cdots \times \left( \sum_{i=0}^{1-1}  e^{i \k_1 \cdot (\x_1 - \x_i)} \Delta V'(\psi_1) \frac{\partial}{\partial \psi_i} \right) \psi_0,
\end{split}
\ee
where the only remaining task to get a closed expression would be to carry out the sums over $i$, so that it becomes possible to perform the sum over $N$. However, even in the limit where the cutoffs disappear $k_L^\psi \to 0$, $k_S^\psi \to \infty$, violating the conditions with which we were able to make approximations, summing over $N$ gives an exponential of a differential operator acting on the ``free'' field
\be
\psi(\x,\tau) = \exp \left[-\frac{\Delta N}{3 H^2} \frac{\partial \Delta V}{\partial \psi} \frac{\partial}{\partial \psi} \right] \psi \bigg|_{\psi = \psi_I(\x,\tau)},
\ee
which is actually a solution to $\dot \psi = - \frac{1}{3H} \frac{\partial \Delta V}{\partial \psi} $ if we use the same approximations and assume they hold true. Thus summing over $N$ to get the formal solution in terms of a differential operator does not appear to give any useful information for practical computations so as to compare with observations.

Nonetheless, as we stressed before, the approximations we are employing require a restricted range of momenta, and moreover, the physics of interest for the CMB and LSS lies precisely in such a range of scales. In addition to that, since we know that the primordial statistics are nearly Gaussian, it should be sufficient to keep the leading order corrections. For instance, to third order in the potential, at the final time slice we get 
\be
\begin{split}
\psi(\x) = \psi_I(\x) & - \frac{\Delta N}{3 H^2} \reallywidetilde{\frac{\partial \Delta V}{\partial \psi}}(\x) + \frac{\Delta N^2}{18 H^4} \reallywidetilde{ \reallywidetilde{\frac{\partial \Delta V}{\partial \psi}} \frac{\partial^2 \Delta V}{\partial \psi^2} }(\x) \\ & - \frac{\Delta N^3}{162 H^6} \left( \reallywidetilde{ \reallywidetilde{ \reallywidetilde{\frac{\partial \Delta V}{\partial \psi}} \frac{\partial^2 \Delta V}{\partial \psi^2}}  \frac{\partial^2 \Delta V}{\partial \psi^2} }  (\x) + \reallywidetilde{ \reallywidetilde{\frac{\partial \Delta V}{\partial \psi}} \reallywidetilde{\frac{\partial \Delta V}{\partial \psi}} \frac{\partial^3 \Delta V}{\partial \psi^3} } (\x) \right) + \cdots
\end{split}
\ee
and higher order terms can be included directly from~\eqref{result-Nth-order} as needed. We have used the notation that was introduced earlier for the projection operator $\int_\y \int_\k e^{i \k \cdot (\y - \x)}$, concretely,
\be
\widetilde{F}(\x) = \int_\y \int_\k e^{i \k \cdot (\y - \x)} F(\y),
\ee
keeping in mind that the position dependence of each function, which are derivatives of the potential, is through the position argument of the field operator at which it is evaluated. 

Now we can write the same expression for $\zeta$: at each order in the perturbation expansion on $\Delta V$~\eqref{result-Nth-order}, the only difference that appears if we include an extra commutator containing an interaction Hamiltonian $H^\alpha_I$ to transfer the nonlinearity is that a factor of $\alpha \Delta N/H$ appears and a numeric factor changes. Concretely, the $N$-th order term on $\Delta V$ is given by
\be \label{result-Nth-order-zeta}
\begin{split}
\frac{\alpha \Delta N}{H} \frac{1}{(N+1)!} \left( - \frac{\Delta N}{3 H^2} \right)^N  \int_{\x_{N}} \int_{\k_N} \cdots \int_{\x_{1}} \int_{\k_1} &\left( \sum_{i=0}^{N-1} e^{i \k_N \cdot (\x_N - \x_i)} \Delta V'(\psi_N) \frac{\partial}{\partial \psi_i} \right) \times \cdots \\ &
\cdots \times \left( \sum_{i=0}^{1-1}  e^{i \k_1 \cdot (\x_1 - \x_i)} \Delta V'(\psi_1) \frac{\partial}{\partial \psi_i} \right) \psi_0,
\end{split}
\ee
and therefore the explicit result to third order is given by
\be
\begin{split}
\zeta(\x) = \frac{\alpha \Delta N}{H} \bigg( \psi_I(\x) & - \frac{\Delta N}{6 H^2} \reallywidetilde{\frac{\partial \Delta V}{\partial \psi}}(\x) + \frac{\Delta N^2}{54 H^4} \reallywidetilde{ \reallywidetilde{\frac{\partial \Delta V}{\partial \psi}} \frac{\partial^2 \Delta V}{\partial \psi^2} }(\x)  \\ &  - \frac{\Delta N^3}{648 H^6} \left( \reallywidetilde{ \reallywidetilde{ \reallywidetilde{\frac{\partial \Delta V}{\partial \psi}} \frac{\partial^2 \Delta V}{\partial \psi^2}}  \frac{\partial^2 \Delta V}{\partial \psi^2} }  (\x) + \reallywidetilde{ \reallywidetilde{\frac{\partial \Delta V}{\partial \psi}} \reallywidetilde{\frac{\partial \Delta V}{\partial \psi}} \frac{\partial^3 \Delta V}{\partial \psi^3} } (\x) \right) + \cdots \bigg).
\end{split}
\ee

It is straightforward to see that we have a very well-posed perturbation series if all the derivatives of the potential are small in the sense that
\be \label{perturbativity-strong}
\left| \frac{\Delta N}{3 H^3} \frac{\partial^{N} \Delta V}{\partial (\psi/H)^{N}} \right| \ll 1,
\ee
for all $N \geq 1$. This can always be achieved for a finite number of derivatives provided that the amplitude of the potential, say $\Lambda^4$, is sufficiently small. One might be worried, however, that in order to use a sinusoidal potential such as $\Delta V = \Lambda^4 \cos (\psi/f)$ with $f < H$, the perturbation series may not converge. However, it does, as~\eqref{perturbativity-strong} is a general criterion on convergence, and in order to be as general as it is, it is too strong to be able to rule out convergence. 

In the specific case of $\Delta V = \Lambda^4 \cos (\psi/f)$, studying the $N$-th order term reveals that as there are $2N - 1$ derivatives acting upon the $\Delta V$ functions, the term of order $N$ in $\Lambda^4$ is proportional to
\be
\propto f \left( \frac{\Delta N \Lambda^4}{3 H^2 f^2}  \right)^N,
\ee
thus identifying the perturbative parameter of this theory as
\be
\frac{\Delta N \Lambda^4}{3 H^2 f^2},
\ee
thus setting no restriction of the value of $f$ provided $\Lambda^4$ is small enough. Even if this is not the case, at all orders the series is guaranteed to converge as the $1/N!$ term will always catch up to the $N$-th power of any number. Thus, it is also possible to explore combinations of parameters that give a perturbative parameter larger than one, only that then one needs to take more terms into the expansion.

%\be
%\frac{\partial^{N} \Delta V}{\partial (\psi/H)^{N}} \propto \left( \frac{H}{f} \right)^N \overset{{\rm when} N \gg 1}{\gg} 1.
%\ee
%On the other hand, we will have the perturbative amplitude $\frac{\Delta N \Lambda^4}{3 H^3 f}$, that will be present at every order of the perturbative series, to the power of $N$ for the $N$-th term. 

This concludes our discussion on higher-order corrections. Now we will go back to first-order perturbation theory and extend our results to arbitrary potentials and a more general class of backgrounds.

\section{Generalizing the Tomography}

So far, we have discussed how the coupling of isocurvature modes affect the generation of non-Gaussianity in the adiabatic mode of the primordial curvature perturbations by transferring the nonlinearities in the isocurvature field dynamics. However, most of the tools used herein do not require to consider the specific setup in which we embedded the perturbations.\footnote{See~\cite{ScheihingHitschfeld:2019abs} for a more extensive derivation of the results presented in this section.}

Indeed, the interaction picture of Quantum Mechanics provides a way with which to study, perturbatively, the effects of a potential $V$ that generates non-linear dynamics in the $n$-point functions, or equivalently, in the PDF. For simplicity, we will consider a single real scalar field $\varphi$ in the presence of a self-interaction, which is the situation felt by the isocurvature mode $\psi$. Transferring the result to an extra field, say $\zeta$, assuming that the transfer mechanism is linear, can be done in a straightforward manner by modifying the covariance matrix introduced later~\eqref{cov}, and including appropriate numerical factors accompanying the potential $V$, corresponding to the result of performing the $\tau'$ integral in what would be $\int^{\tau} d\tau' \int^{\tau'} d\tau'' [H_I^{V}(\tau''), [H_I^{\alpha}(\tau'), \zeta_I(\x,\tau)]]$ of~\eqref{zeta-perturbated}. Equation~\eqref{zeta-perturbated-4-schem} reveals this in the simplest manner for the case we have discussed so far.

In absence of a potential $V$, $\varphi$ evolves in time following the Euler-Lagrange equations associated to a Lagrangian density $\L_0$, or equivalently to the Hamilton equations associated to a Hamiltonian density $\H_0$, with solutions that can be written as superposition of plane waves with appropriate coefficients
\bea
\varphi(\x , t) &=& \int_k \, \hat \varphi (\k , t) \, e^{- i \k \cdot \x } \\  \hat \varphi (\k , t) &=& \varphi_k(t) a(\k) + \varphi_k^*(t) a^\dag(-\k), 
\eea
and the operators $a(\k)$ and $a^\dag(\k)$ are creation and annihilation operators that satisfy commutation relations
\be
\left[ a (\k) , a^{\dag} (\k') \right] = (2 \pi)^3 \delta^{(3)} (\k - \k'),
\ee 
which in turn define the operator algebra of the quantum theory.

This is precisely the setup we had for the perturbation fields on a de Sitter space throughout Chapter~\ref{chap:n-point}, where the mode functions $\varphi_k(t)$ were given by $u_k(\tau)$. These functions contain all the information on the background's influence (the unperturbed metric around which one computes fluctuations) on the field dynamics. In writing the mode functions, we have implicitly assumed an isotropic and homogeneous background, as they only depend on the modulus of the wavenumber $k = |\k|$. The statistics for $\varphi$ can straightforwardly shown to be Gaussian, with a propagator $\Delta$ and covariance $\sigma$ given by
\bea
\Delta(t,t',p) &\equiv& \varphi^{}_p(t)\varphi^*_p(t') \\  \sigma^2(t,t',r) &\equiv& \frac{1}{(2\pi)^3} \int d^3p \; \Delta(t,t',p) e^{i\p \cdot \r} \nonumber \\ &=& \braket{0|\varphi_I(t,\x) \varphi_I(t',\y) |0},
\eea
where $r = |\x - \y|$.

In the presence of a potential $V$, which we take to be turned off in the infinite past (the furthest point in the past that the temporal coordinate can reach), one can use the interaction picture of Quantum Mechanics to compute a PDF for the scalar field $\varphi$ at a given time slice $t$. We take the self-interaction $V$ to be of the form
\be \label{taylorV-main}
V(\varphi(\r,t),\r,t) = \sum_{m=0}^{\infty} \frac{c_m(\r,t)}{m!} \varphi(\r,t)^m,
\ee
i.e., a local function of the field, including possible dependencies on the background. We present the full derivation of the PDF in Appendix~\ref{chap:ininPDF}, and here we only quote the result
\be \label{Path-integral-main}
\begin{split}
\rho_{\varphi} =& \int \! D\varphi_+ D\varphi_-  \exp \left\{ +i\int^{t}_{t_0 - i\epsilon |t_0|} \!\!\!\!\!\!\!\!\! d t' \int_{\r} V(\varphi_-(\r,t'),\r,t') \right\}  \! \\ & \times \frac{\exp \left( - \frac{1}{2} \varphi \cdot \left( {\Sigma}^{-1} \right) \cdot \varphi \right)}{\sqrt{|\text{det} (2\pi \Sigma) |}}   \exp \left\{ - i \int^{t}_{t_0 + i\epsilon |t_0|} \!\!\!\!\!\!\!\!\! d t' \int_{\r} V(\varphi_+(\r,t'),\r,t') \right\},
\end{split}
\ee
which is given in terms of functional integrals and the covariance of the free theory $\sigma$, here written in terms of a covariance matrix $\Sigma$ with definite prescriptions\footnote{See Appendix~\ref{chap:ininPDF} for details.}.

What is to be gained from here? To first order in perturbation theory, that is, taking the potential $V$ to be small, one can derive a generic results, analogous to what we obtained in~\eqref{main-result-1}, and use them to get information on the potential by measuring the resulting statistics. We now describe these results.

%to include extra factors of $\alpha \Delta N/H$ for each occurrence of the interaction picture isocurvature field $\psi_I$ and replace the nonlinear potential $\Delta V$ with $ \frac{\alpha \Delta N}{ 2 H} \Delta V$, which is what will be felt 

% - 1-point PDF results
%- k-space PDF, probe mode by mode

%In this section we list two generic quantities that can be derived from our previous results and may find a number of applications. The first considers the computation of a probability distribution function for a scalar field $\varphi$, starting from the free theory vacuum $\ket{0}$ as the \textit{in} state. The second provides a probability density for the mode amplitudes in momentum space, with the same initial conditions. They are complementary, as they quantify different aspects of the system's observable quantities, but ultimately encode the same information: the structure of the self-interaction.

%Both results can be regarded as Born approximations, in the sense that they are valid provided the potential is of small amplitude and that their results are linear on the self-interaction. This fact may prove useful for future studies of systems that have statistics that are very close to a Gaussian distribution and probe small departures from Gaussianity, simply because linear operations are easier to handle and, potentially, to be inverted. This opens the door to obtaining information on the underlying potential directly, thus shedding light on the fundamental structure of the theory that describes the phenomenon at hand.

\subsection{1-point and multi-point PDFs to first order}

Our matter of interest is to find the probability of a quantum field acquiring a certain amplitude at a given spacetime position, or the probability of a given field configuration. Let us here take a situation in which the system is originally in the vacuum state of the free theory, and an interacting term is turned on from $t=t_0$ onwards.

Since it is more natural to do so, we will start discussing how to obtain general $n$-point functions to first order, and then make the reduction to the 1-point case. Let us consider a situation in which we are interested in probing the values of the field at a given set of spacetime positions $\{\r_i\}_{i=1}^n$, and let us denote the corresponding values of the field by $\{\varphi_i\}_{i=1}^n$. The sensible question is then to ask: what is the probability of measuring a given configuration $(\varphi_1,...,\varphi_n)$ at time $t$ after the interaction is turned on?

If we take the term $N=1$ in equation~\eqref{PDF}, that is, the first order correction coming from the potential $V$, and consider it as a deviation $\Delta \rho = \rho - \rho_G$ from a Gaussian distribution, we find
\be \label{rpsifull}
 \Delta \rho(\varphi_1,...,\varphi_n;t) = \int_{\r} \int_{t_0}^t dt' \, 2\text{Im}  \left\{ \int d\phi  \frac{\exp\left(-\frac{1}{2} \pmb{\varphi}^T \cdot \pmb{\Sigma}^{-1} \cdot \pmb{\varphi} \right)}{\sqrt{(2\pi)^{n+1} |\text{det} \, {\pmb\Sigma|}}}  V\left(\varphi, \r, t' \right) \right\},
\ee
where \(\varphi_i\) represents the amplitude of the field at $(\r_i,t)$, which we do not write as \(\varphi(\r_i, t)\) in order to emphasize their being real numbers, i.e., they are real variables when one computes an expectation value from this distribution. Furthermore, we have denoted \(\pmb{\varphi}^T =  (\phi \,\,\, \varphi_1 \,\,\, ... \,\,\, \varphi_n  )\), and \(\pmb{\Sigma}\) is the corresponding covariance matrix
\be \label{cov}
  \pmb{\Sigma}=
  \left[ {\begin{array}{cccc}
   \sigma^2(t',t',0) & \sigma^2(t,t',r_{01}) & \cdots & \sigma^2(t,t',r_{0n})\\
   \sigma^2(t,t',r_{01}) & \sigma^2(t,t,0) & \cdots & \sigma^2(t,t,r_{1n})\\
   \sigma^2(t,t',r_{02}) & \sigma^2(t,t,r_{12}) & \ddots & \sigma^2(t,t,r_{2n})\\
   \vdots & \vdots & \cdots & \vdots\\
   \sigma^2(t,t',r_{0n}) & \sigma^2(t,t,r_{1n})  & \cdots & \sigma^2(t,t,0)\\
  \end{array} } \right],
\ee
with $r_{ij} = |\r_i - \r_j|$ and $\r = \r_0$. This answers our question and provides the desired distribution, which can be computed and compared with the observed statistics of a field.

In the case $\r_1 = ... = \r_n \equiv \x$, then the Gaussian measure effectively reduces itself to two field coordinates, one internal and another external. Consequently, the covariance matrix becomes a $2\times2$ matrix. If we further let $r = |\r|$, we arrive at
\be \label{rpsi2}
\begin{split}
\rho(\varphi,\x,t) = \frac{e^{-\frac{\varphi^2}{2\sigma^2(t,t,0)}}}{\sqrt{2\pi \sigma^2(t,t,0)}} \left[ 1 + \int_{\r} \int_{t_0}^t dt'  \int_{-\infty}^{\infty} \!\!\!\!\! d\phi \,\, 2 \, \text{Im}  \left\{  \frac{e^{-\frac{(\phi - R(t,t',r)\varphi)^2}{2\kappa^2(t,t',r)}}}{\sqrt{2\pi \kappa^2(t,t',r)}} \right\} V\left(\phi, \r + \x, t' \right) \right].
\end{split}
\ee
where we have defined
\bea
\kappa^2(t,t',r) &\equiv& \frac{\sigma^2(t,t,0) \sigma^2(t',t',0)-\sigma^4(t,t',r)}{\sigma^2(t,t,0)} \\ R(t,t',r) &\equiv& \frac{\sigma^2(t,t',r)}{\sigma^2(t,t,0)}.
\eea

Equation~\eqref{rpsi2} is the promised result: a 1-point PDF for the scalar field at $(\x,t)$. In general, this can be evaluated numerically in a straightforward manner, typically once the (co)variances have been already regularized by setting cutoffs in place. %However, if we had access to solving the integral over $t'$ at the $n$-point function level, the reconstruction of the PDF should even be more revealing of the underlying physics (see~\cite{Chen:2018uul, Chen:2018brw} for a realization of this type of analysis). 
In fact, if both the potential and the single-point variance of the field $\sigma^2(t,t,0)$ are independent of the spacetime coordinates, it is possible to retrieve information of the self-interaction directly from the connected $n$-point functions in terms of a Hermite polynomial expansion, and ultimately, to reconstruct the potential. This is a consequence of equation~\eqref{npointk} in the following section.

\subsection{A $\k$-space PDF for the amplitude of the fluctuations}

One can also wonder what type of PDF is able to keep track of the fluctuation amplitude at a fixed wavelength. While one might be worried that developing the statistics associated to the fluctuations at a fixed wavelength is futile, since one can only measure one mode of a given wavenumber in an observed field configuration, in the presence of isotropy and homogeneity their statistics only depend on the modulus of the wavenumber $k = |\k|$, and thus one can obtain a large dataset with which to constrain the parameters of the model that defines $V$.

In order to get a PDF in momentum space, we first need to determine the structure of the $n$-point functions in this representation. To that end, we first write down the fully connected $n$-point function explicitly at first order in $V$ and time $t$, which, to this order in the perturbation, is equal to the fully interacting contribution:
\be \label{ndpoint}
\begin{split}
\langle \varphi_{\r_1,...,\r_n}^n \rangle_c = \int_{t_0}^t dt' \int_{\r} \int_{-\infty}^{\infty} \!\!\!\! d\varphi & \frac{e^{-\frac{\varphi^2}{2\sigma^2(t',t',0)}}}{\sqrt{2\pi \sigma^2(t',t',0)}}
 \frac{\partial^n V}{\partial \varphi^n}\left(\varphi,\r,t' \right) \\ & \times  2 {\rm Im} \left\{ \sigma^2(t,t',|\r - \r_1|) \;...\; \sigma^2(t,t',|\r-\r_n|) \right\}.
 \end{split}
\ee
Integrating by parts over $\varphi$ and taking a Fourier transform to momentum space, we get
\be \label{npointk}
\begin{split}
\langle \varphi_{\k_1,...,\k_n}^n \rangle_c = \int_{t_0}^t dt' \int_{\r} \int_{-\infty}^{\infty} \!\!\!\! d\varphi &  \frac{e^{-\frac{\varphi^2}{2\sigma^2(t',t',0)}}}{\sqrt{2\pi \sigma^2(t',t',0)}} {\rm He}_n\left(\frac{\psi}{\sigma(t',t',0)} \right) \! V \! \left(\varphi,\r,t' \right) \\ & \times  2 {\rm Im} \left\{ \frac{\Delta(t,t',k_1) e^{i\k_1\cdot \r}}{\sigma(t',t',0)} \;...\;  \frac{\Delta(t,t',k_n) e^{i\k_n\cdot \r}}{\sigma(t',t',0)}  \right\}.
 \end{split}
\ee
On the other hand, when one thinks about measuring different modes of a given field, one has to take into account the experimental limitations. To that end, we define
\be
\bar{ \varphi}_{\k} \equiv \frac{3}{4\pi k_{\rm IR}^3} \int_{|\q - \k| < k_{\rm IR}} \!\!\!\!\!\!\!\!\!\!\!\!\!\!\!\!\!\!\! d^3 q \, \varphi_{\q}
\ee
where $k_{\rm IR}$ is an infrared cutoff, or a coarse-graining, that accounts for our not being able to measure arbitrarily large length scales (note that there is no $(2\pi)^{-3}$ factor beside the integral).

In the case of a quadratic theory, all of the relevant information is contained within the two-point function, which is also called the \textit{Power Spectrum}. Presently, it is given by
\be
\langle \bar{\varphi}_\k \bar{\varphi}_{-\k} \rangle_{\rm free} =  \frac{(2\pi)^3}{\left(\frac{4\pi}{3} k_{\rm IR}^3 \right)^2} \int_{|\q - \k| < k_{\rm IR}} \!\!\!\!\!\!\!\!\!\!\!\!\!\!\!\!\!\!\! d^3 q \, \Delta(t,t,q).
\ee
which, depending on the background metric, may be time-dependent. However, in the presence of an interaction term, the theory no longer has Gaussian statistics, and consequently the amplitude of the modes is no longer determined only through the two-point function.

For our present purposes, since the amplitude of a mode can be characterized by $|\varphi_\k|^2 = \varphi_\k \varphi_\k^{\dagger} = \varphi_\k \varphi_{-\k}$, the natural quantity to try and compute is
\be \label{fullnpointk}
\langle (\bar{\varphi}_\k \bar{\varphi}_{-\k})^n \rangle = \sum_{m=0}^{n} \frac{n!^2 \langle \bar{\varphi}_\k \bar{\varphi}_{-\k} \rangle_{\rm free}^{n-m}}{m!^2 (n-m)!}  \langle (\bar{\varphi}_\k \bar{\varphi}_{-\k})^m \rangle_{c},
\ee
where the combinatorial factor arises from the counting of all possible contractions to form diagrams with $n$ external momenta evaluated at $\k$ and an additional set of $n$ momenta evaluated at $-\k$. Notice that~\eqref{fullnpointk} requires $|\k| > k_{\rm IR}/2$ for consistency, so that the free theory contractions can only join $\k$ with $-\k$.

If we define
\be
\mathcal{F}_{n}[V;\r,t] \equiv \int_{-\infty}^{\infty} \!\!\!\! d\varphi \, \frac{e^{-\frac{\varphi^2}{2\sigma^2(t,t,0)}}}{\sqrt{2\pi \sigma^2(t,t,0)}}
{\rm He}_n\left(\frac{\varphi}{\sigma(t,t,0)} \right)  V\left(\varphi,\r,t \right),
\ee
we can readily write down the fully connected contributions
\be
\begin{split}
\langle (\bar{\varphi}_\k \bar{\varphi}_{-\k})^n \rangle_{c} = \int_{t_0}^t dt' & \!\! \int_\r \mathcal{F}_{2n}[V;\r,t'] \, \\ & \times 2 \text{Im} \left\{ \left( \frac{1}{(\frac{4\pi}{3} k_{\rm IR}^3)^2}  \int_{q,q' < k_{\rm IR}} \!\!\!\!\!\! \frac{\Delta(t,t',|\k-\q|) \Delta(t,t',|\k-\q'|)}{\sigma^2(t',t',0)} e^{i(\q - \q') \cdot \r}  \right)^n  \right\},
\end{split}
\ee
and thus what remains is a problem of finding the PDF that generates~\eqref{fullnpointk}. In essence, we want a distribution $\mathcal{K}$ such that $\int d(|\bar{\varphi}_\k|^2) \mathcal{K}_{\k}(|\bar{\varphi}_\k|^2) (|\bar{\varphi}_\k|^2)^n = \langle (\bar{\varphi}_\k \bar{\varphi}_{-\k})^n\rangle $. 

In the interests of notational simplicity, let us define
\bea
y &=& y(t,t',\r;\k) \equiv \frac{\int_{q,q' < k_{\rm IR}} \!\!\!\!\!\! \frac{\Delta(t,t',|\k-\q|) \Delta(t,t',|\k-\q'|)}{ \sigma^2(t',t',0)} e^{i(\q - \q') \cdot \r}}{(\frac{4\pi}{3} k_{\rm IR}^3)^2 \langle \bar{\varphi}_\k \bar{\varphi}_{-\k} \rangle_{\rm free}} , \\
x &\equiv& \frac{|\bar{\varphi}_\k|^2}{\langle \bar{\varphi}_\k \bar{\varphi}_{-\k} \rangle_{\rm free}}
\eea
where $\langle \bar{\varphi}_\k \bar{\varphi}_{-\k} \rangle_{\rm free}$ is the two-point function of the free theory.
Furthermore, we define a functional $T[V;\r,t](x)$ by
\be
\begin{split}
\int_0^{\infty} dx \, e^{-x} L_n\left(x \right) T[V;\r,t](x) = \frac{(-1)^n}{n!} \mathcal{F}_{2n}[V;\r,t].
\end{split}
\ee
where $L_n$ is the $n$-th Laguerre polynomial. This definition is always possible provided that $V(\varphi,\cdot,\cdot)$ be square integrable with respect to the Gaussian measure with variance $\sigma^2(t,t,0)$ because both sides of the expansion define the coefficients of square integrable functions in their respective Hilbert spaces: $(\{{\rm He}_n(x)\}_n, \frac{e^{-x^2/2}}{2\pi})$ and $(\{L_n(x)\}_n, e^{-x})$.

Then, it follows from the preceding definitions and some functional-algebraic manipulations that
\be
\mathcal{K}_{\k}(x) = e^{-x} \left[1 + \int^t_{t_0} dt' \int_\r \int_{0}^{\infty} \!\! dz \, \frac{e^{ -\frac{xy+z}{1-y} }}{1-y} I_0\left( \frac{2\sqrt{xyz}}{1-y} \right) T[V;\r,t'](z) \right]
\ee
where $I_0$ is a modified Bessel function of the first kind, and
\be
T[V;\r,t](z) = \int_{-\infty}^{\infty} \!\!\!\! d\varphi \, \frac{e^{-\frac{\varphi^2}{2\sigma^2(t,t,0)}}}{\sqrt{2\pi \sigma^2(t,t,0)}}
 V\left(\varphi,\r,t \right) \frac{1}{2\pi i} \int_{\mathcal{C}} \frac{dv}{v} \exp \left( \frac{v\varphi}{\sigma(t,t,0)} - \frac{v^2}{2} + \frac{z}{v^2+1} \right),
\ee
with $\mathcal{C}$ a counterclockwise integration contour encircling the three singularities of the integrand: $0,+i,-i$. Finally, we may write the PDF in terms of $| \bar{\varphi}_\k|$
\be \label{k-PDF}
\begin{split}
\mathcal{K}_{\k}(| \bar{\varphi}_\k|) = \frac{ 2 | \bar{\varphi}_\k| e^{-\frac{| \bar{\varphi}_\k|^2}{\langle \bar{\varphi}_\k \bar{\varphi}_{-\k} \rangle_{\rm free}} }}{ \langle \bar{\varphi}_\k \bar{\varphi}_{-\k} \rangle_{\rm free} } \left[1 + \int^t_{t_0} dt' \right. & \left. \int_\r \int_{0}^{\infty} \!\! dz  \frac{e^{ -\frac{y| \bar{\varphi}_\k|^2 +z \langle \bar{\varphi}_\k \bar{\varphi}_{-\k} \rangle_{\rm free}}{(1-y) \langle \bar{\varphi}_\k \bar{\varphi}_{-\k} \rangle_{\rm free}} }}{1-y} \right. \\
& \times \left. I_0\left( \frac{2 |\varphi_{\k}|}{(1-y) } \sqrt{\frac{yz}{\langle \bar{\varphi}_\k \bar{\varphi}_{-\k} \rangle_{\rm free}}} \right) T[V;\r,t'](z) \right],
\end{split}
\ee
as a distribution over $|\varphi_{\k}| \in (0,\infty)$.

This result may find applications, for instance, when generating initial conditions for the evolution of the universe after inflation, or even probing the landscape potential that generated those initial conditions through CMB or LSS statistics, in an analogous manner to what was done in~\cite{Chen:2018uul, Chen:2018brw} for the CMB. This result has both advantages and disadvantages over the approach implemented in those works. On the one hand, because all the modes in $k$-space are linearly independent, the result will be subject to far less intrinsic noise. But on the other hand, its analytical expression is more cumbersome and it will presumably require more data from smaller scales on the sky, as it only would be able to reconstruct a PDF (assuming isotropy) by counting occurrences of the fluctuations' amplitude over a sphere at fixed $|\k|$. 

As the final note of this section, it is worth mentioning that given a reconstruction of the $k$-space PDF $\mathcal{K}(x)$ from actual data, one can recover information about the even Hermite moments of the self-interaction $\mathcal{F}_{2n}$ through
\be
\begin{split}
\int_0^\infty & \!\! dx L_n(x) \mathcal{K}_{\k}(x) \\
& = \int_{t_0}^t dt' \int_{\r}  \mathcal{F}_{2n}[V;\r,t'] \, 2 {\rm Im} \left\{ y^n(t,t',\r;\k) \right\},
\end{split}
\ee
which, in the same spirit of last section's conclusions, is readily useful if $\mathcal{F}_{2n}[V;\r,t']$ does not depend on $\r$ nor $t'$. It must be noted that the PDF~\eqref{k-PDF} only contains information on the even part of the potential $V$, and therefore another complementary observable should be used to obtain information on the odd part of the potential.

%Copy last section of chap 3 here

%\section{Landscape Tomography v/s Canonical Single-Field Inflation}

%How the tomography circumvents some restrictions and has impacts in our observable universe.

%Maldacena's consistency relation; squeezed limit of observables

%\lipsum[50-60]

\cleardoublepage

\chapter{Searching for Tomographic non-Gaussianity in the CMB and LSS} \label{chap:CMB-LSS}

In the previous Chapters we have thoroughly studied a setup that generates a highly non-trivial class of non-Gaussianity, and characterized its consequences on the curvature perturbation. We now take one step further and study the impact of these departures from Gaussianity on the late-time universe, therefore providing the opportunity to constrain parameters of the fundamental theory that gave rise to the universe as we now know it. 

\section{Statistics of late-time observables}

Our observable universe is consistent with an extremely simple set of initial conditions. For all practical purposes, the observed cosmological inhomogeneities were seeded by a primordial curvature fluctuation $\zeta$ distributed according to a Gaussian profile~\cite{Akrami:2019izv,Komatsu:2003fd}, parametrised by an almost scale invariant power spectrum~\cite{Akrami:2018odb}. The confirmation of this state of affairs by future surveys would reinforce our confidence on the single-field slow-roll inflation paradigm, that is, the idea that $\zeta$ was the consequence of quantum perturbations of a single scalar fluid (the inflaton) that evolved adiabatically during inflation~\cite{Guth:1980zm, Starobinsky:1980te, Linde:1981mu, Albrecht:1982wi, Mukhanov:1981xt}. Tiny deviations from non-Gaussianity (NG), due to small non-linear self-interactions affecting $\zeta$, are known to emerge in single-field inflation but these are predicted to be too small to be observed in the near future. On the other hand, large non-Gaussianity (within current bounds) may emerge in models of inflation beyond the single-field paradigm, resulting from non-trivial self-interactions and/or interactions with other degrees of freedom. The observation of non-Gaussianity would therefore offer a unique opportunity to characterize the class of interactions that affected $\zeta$ during inflation, allowing us to pin down certain fundamental aspects about the period of inflation and consequently glimpse on the structure of the ultra-violet (UV) framework where it is realized.

However, although a well-established idea, inflation is not unique in explaining the observed inhomogeneities~\cite{Chen:2018cgg}. Other scenarios, such as the ekpyrotic~\cite{Lehners:2007ac,Khoury:2001wf} and bouncing universes~\cite{Wands:1998yp,Finelli:2001sr}, have been proposed as early universe models able to reproduce our universe's nearly Gaussian initial conditions.  Thus, by adopting an agnostic perspective about the details involved in the description of the pre-Big-Bang dynamics, we should agree that the main outcome from any model pretending to explain the initial conditions of our universe, consists of a relation giving us back the profile of $\zeta (\x)$ written in terms of a purely Gaussian random field $\zeta_G(\x)$. Such a relation must be of the form
\be \label{general-asatz}
\zeta(\x) = \zeta_G(\x) + {\mathcal F}\left( \zeta_G(\x), \nabla \right),
\ee
where $\mathcal F$ represents a non-linear function of the field $\zeta_G(\x)$ and spatial gradients $\nabla$ acting on it. In Fourier space, the previous relation may be reexpressed as the following expansion in powers of $\zeta_G$, starting at quadratic order
\be \label{general-NG-k}
\zeta_\k = \zeta_\k^G + (2 \pi)^3 \sum_{n=2} \frac{1}{n!}\int_{\p_1} ... \int_{\p_n} \delta^{(3)} \Big(\k - \sum_{i=1}^n \p_i \Big)  {\mathcal F}_n\left( \p_1 , ... , \p_n \right)\zeta^G_{\p_1} ... \zeta^G_{\p_n}, 
\ee
where $\int_\p \equiv (2 \pi)^{-3} \int d^3 p$, and where ${\mathcal F}_n\left( \p_1 , ... , \p_n \right)$ are functions of the momenta, symmetric under their permutations. This relation is sufficiently general to describe any form of primordial non-Gaussianity, and can be formally obtained in a generic manner from a quantum-mechanical framework. 

In inflation, the primordial curvature perturbation field $\zeta$ is sourced by quantum fluctuations of the inflaton field, or possibly other degrees of freedom such as isocurvaton fields. As a general statement, one can write down that the field operator $\zeta(\x)$ at the final time slice $t$ is given by 
\be
\zeta(\x,t) = U^\dagger(t,t_0) \zeta_I(\x,t) U(t,t_0),
\ee
where $U$ is the temporal evolution operator in the interaction picture of quantum mechanics, and $\zeta_I$ the interaction picture field, which follows the dynamics of the free theory. 

Naturally, the field $\zeta$ will generate a specific set of $n$-point functions when computing expectation values. As we are interested only in measuring the field $\zeta$, because in the late-time observables we study in this work it is the only degree of freedom with sensible consequences, one can construct a PDF $\rho[\zeta]$ that generates these statistics through functional integration
\be
\langle \zeta(\x_1) \cdots \zeta(\x_n) \rangle = \int D\zeta \rho[\zeta] \zeta(\x_1) \cdots \zeta(\x_n)
\ee
over the field configurations $\zeta(\x)$.

Quantum mechanics does provide the tools with which to determine $\rho[\zeta]$ directly, at least in principle. The operation $\zeta = U^\dagger \zeta_I U$ can be reframed in terms of a functional expression
\be
\zeta(\x) = \mathcal{O}[\zeta_I, \Pi_I, \{\psi^I_i, \Pi^I_i\}_i ](\x)
\ee
that depends on the whole spatiotemporal evolution of the interaction-picture fields, $\zeta_I$ and other degrees of freedom $\psi_i^I$, and that of their conjugate momenta, $\Pi_I$ and $\Pi_i^I$. In principle, one can compute correlations directly from this expression. However, if the dependence of $\mathcal{O}$ on the interaction-picture fields is known, one can determine the PDF of $\zeta$ by integrating over all possible configurations:
\be
\rho[\zeta] = \int D\zeta_I D\Pi_I \left( \prod_i D \psi_i^I D \Pi_i^I \right) P_G[\{\psi^I_i, \Pi^I_i\}_i] \left( \prod_\x \delta \left( \zeta(\x) - \mathcal{O}[\zeta_I, \Pi_I, \{\psi^I_i, \Pi^I_i\}_i ](\x) \right) \right),
\ee
where $P_G$ is a Gaussian measure, with appropriate prescriptions to take into account the ordering of operators in $\mathcal{O}$. The measure is guaranteed to be Gaussian because the free fields evolve linearly in time, and therefore the contraction of quantum field operators obeys Wick's theorem, which is equivalent to saying that the statistics are Gaussian. 

Once $\rho[\zeta]$ is obtained, the statistics of $\zeta$ are fully determined. Computationally, however, it is useful to have a probability distribution from which one knows how to obtain expectation values. On the other hand, one knows that the observed statistics for $\zeta$ are consistent with Gaussianity, and that deviations, if any, must be small. This motivates finding a functional map $\zeta_G(\x) = G[\zeta](\x)$, with inverse $\zeta(\x) = \zeta_G(\x) + \mathcal{F}[\zeta_G](\x)$, such that $\zeta_G$ has Gaussian statistics, i.e., such that
\be
\rho_G[\zeta_G] = \rho[\zeta_G + \F[\zeta_G]  \times {\rm det} \left( \frac{\delta \zeta}{\delta \zeta_G} \right).
\ee
The difficulty, of course, lies in finding such a mapping. Afterward, one can include another mapping, that makes the power spectrum of $\zeta_G$ to be consistent with current observations, if this is not the case already.

%Thus, we have justified writing $\zeta$ in terms of a Gaussian field $\zeta_G$,
%\be
%\zeta(\x) = \zeta_G(\x) + \mathcal{F}[\zeta_G](\x),
%\ee
%where the functional $\F$ is, in principle, arbitrary, and should be determined from the specifics of the inflationary model at hand.

Therefore, the $\mathcal F_n$ functions parametrize the non-Gaussian deviations generated by non-linear interactions to which $\zeta$ were subject, and may be deduced from a particular model of inflation by studying the evolution of $\zeta_\k$ from sub-horizon scales up until the end of inflation (\emph{e.g.} using the in-in formalism). In fact, any $n$-point correlation function for $\zeta$ may be computed out of~(\ref{general-NG-k}). For instance, at tree-level, the bispectrum parametrizing the amplitude of the three point function is found to be given by $B(\k_1 , \k_2 , \k_3) = \left[ P_{\zeta}(k_1) P_{\zeta}(k_2)   {\mathcal F}_2 (\k_1, \k_2)  + \textrm{perm} \right]$, where $P_{\zeta}$ is the power spectrum of the Gaussian field $\zeta^G_{\k}$. In the particular case of single field slow-roll inflation, up to first order in the slow-roll parameters, the bispectrum is recovered as long as ${\mathcal F}_2$ is given by 
\be \nn%\label{F_2-single-field-case}
{\mathcal F}_2 (\p_1, \p_2) = \frac{1}{2} (\eta - \epsilon) + \frac{\epsilon}{2} \frac{p_1 + p_2}{|\p_1 + \p_2|} + 2 \epsilon \frac{p_1^2 + p_2^2}{|\p_1 + \p_2| (p_1 + p_2 + |\p_1 + \p_2|)}, 
%+ \cdots,
\ee
where $\epsilon$ and $\eta$ are the usual slow-roll parameters describing the steady evolution of the FLRW background during inflation. %In the previous expression, the elipses stand for terms that cancel out in $B(\k_1 , \k_2 , \k_3)$ once momentum conservation is imposed.  
The effective field theory (EFT) of inflation approach~\cite{Cheung:2007st} to study models beyond the canonical single-field paradigm will also yield a specific form of ${\mathcal F}_2 (\p_1, \p_2)$, in which the sound speed of $\zeta$ plays an important role. In addition, the well known local ansatz~\cite{Gangui:1993tt, Komatsu:2001rj, Acquaviva:2002ud, Maldacena:2002vr, Bartolo:2004if, Liguori:2010hx, Chen:2010xka, Wang:2013eqj} %given by $\zeta = \zeta_G + \frac{3}{5} f_{\rm NL}^{\rm local} \zeta_G^2$, 
related to the presence of multifield dynamics, corresponds to another particular instance of this relation, where ${\mathcal F}_2 = \frac{6}{5} f_{\rm NL}^{\rm local}$.

One could take upon the challenge of directly reconstructing the form of $\mathcal F$ in (\ref{general-asatz}) ---or equivalently, the functions ${\mathcal F}_n$ appearing in (\ref{general-NG-k})--- out of cosmological data. This would constitute a bottom-up approach to determine the properties of the model that gave origin to the initial conditions.
To guide such a reconstruction, one could consider restricting the functions $\mathcal F_n$ according to certain rules dictated by the symmetries of the alleged bulk model that led to~(\ref{general-NG-k}) at the end of the pre-Big Bang period. 
For instance, scale invariance of the spectra is equivalent to the invariance of $\mathcal F_n$ under the simultaneous rescaling of its momenta: $\mathcal F_n (\lambda \p_1 , ... , \lambda \p_n) = \mathcal F_n ( \p_1 , ... ,  \p_n)$. Furthermore, the validity of soft theorems (under certain circumstances) would require some relations among $\mathcal F_n$ of different order in the limit where one of the momenta vanishes.
 
An objection to this proposal (the direct reconstruction of $\mathcal F$) is that perturbation theory applied to the study of the evolution of $\zeta$ implies that the expansion involved in the writing of Eq.~(\ref{general-NG-k}) is hierarchical. That is, given a coupling constant $g$ parametrizing the self-interactions experienced by $\zeta$ during inflation, the $\mathcal F_n$ functions are naturally expected to satisfy
\be \label{fn-intro}
\mathcal F_n \propto g^{n-1}.
\ee 
For instance, in the case of single-field slow-roll inflation $g$ happens to be of order $\epsilon$ and $\eta$ and so, non-Gaussianity is expected to be well parametrized by the bispectrum. In non-canonical single-field models described by the EFT of inflation, where the $\zeta$ fluctuations propagate with a reduced sound speed $c_s < 1$, the coupling $f$ is enhanced by a factor $c_s^{-2}$, but in order to trust perturbation theory, one still requires that $g$ stays sufficiently suppressed. Based on this argument, we could say that most efforts to characterise non-Gaussianity so far have focused on a truncated version of~(\ref{general-NG-k}), where only $\mathcal F_2 ( \p_1 , \p_2 )$ and $\mathcal F_3 ( \p_1 , \p_2 , \p_3 )$ (which at tree-level give the bispectrum and trisprectrum) are considered with the Planck data implying weak constraints on the form of $\mathcal F_2 ( \p_1 , \p_2 )$ with the help of the so called local, equilateral, orthogonal and folded templates.

In this chapter, we wish to argue in favor of reconstructing the full function $\mathcal F$ of Eq.~(\ref{general-asatz}) from cosmic microwave background (CMB) and large-scale structure (LSS) observations without necessarily assuming a hierarchical dependence of the functions $\mathcal F_n$ on a given coupling constant $g$. We posit that the search of non-Gaussianity focused on low $n$-point correlation functions may miss the existence of richer types of non-Gaussianity~\cite{Chen:2018uul,Chen:2018brw,Leblond:2010yq,Flauger:2016idt} that we may be unable to predict by following standard perturbation theory techniques. To make this discussion tractable, we will focus on the specific case where $\mathcal F$ in Eq.~(\ref{general-asatz}) is simply a filtered function of $\zeta_G$, but not a function of its gradients:
\bea \label{general-asatz-restricted}
\zeta(\x) &=& \zeta_G(\x) + \int_\y \int_{\k} e^{i\k \cdot (\x - \y) } F\left( \zeta_G(\x) \right) \\
&=& \zeta_G(\x) + {\mathcal F}\left[ \zeta_G \right](\x).
\eea
This is exactly the type of NG we derived in the previous chapter, and presented in Equation~\eqref{general-loc-NG-2}. In this more constrained case, the $\mathcal F_n$ functions appearing in (\ref{general-NG-k}) are directly given by $\mathcal F_n = \partial_{\zeta}^n \mathcal F |_{\zeta = 0}$ and the well known local ansatz is recovered by retaining the first term of the expansion. The reason why we keep the integrals $\int_\y \int_{\k}$ explicit is because observations only have access to a restricted range of scales, or equivalently, of wavenumbers $\k$ that we can probe. Therefore, in order to have a theory for the observable modes of $\zeta(\x)$, we write equation~\eqref{general-asatz-restricted} in this manner, as we will only derive predictions for this range of scales.

The main point of keeping every order of the Taylor expansion is that at scales relevant for the observable universe, non-Gaussianity may need to be parametrized by a function ${\mathcal F}\left[\zeta_G(\x) \right]$ with a rich structure that cannot be captured by the lowest order terms $\mathcal F_2$ and $\mathcal F_3$. Inflation can produce non-Gaussianity with such characteristics. In~\cite{Chen:2018uul,Chen:2018brw} it was shown that the interaction of $\zeta$ with an isocurvature field $\psi$ around horizon crossing can lead to a class of non-Gaussianity (tomographic non-Gaussianity) whereby the probability distribution functions may display a rich structure that cannot be parametrized by low $n$-moments.

%In Appendix~\ref{sec:gla} we show that in these classes of models the function $\mathcal F$ is found to be proportional to the gradient of the landscape potential $V$ controlling the dynamics of $\psi$. One then cannot discard the possibility of having potentials $V$ characterized by a rich structure, that is, consecutive minima separated by field distances smaller than the Hubble expansion rate during inflation. In these examples, the bispectrum would not constitute an efficient tool to parametrize non-Gaussianity~\cite{Chen:2018uul,Chen:2018brw}. We wish however to keep the discussion as general as possible and thus, in the main part of this work, we will not refer to any specific origin of this parametrization, focusing instead on its implications for cosmological observables.

Future LSS surveys such as {\sc Lsst}~\cite{Lsst-web}, {\sc Euclid}~\cite{Euclid-web} and {\sc Ska}~\cite{Bull:2018lat}, will revitalize the observational search of primordial non-Gaussianity. The proliferation of modes due to the three-dimensional volume probed by galaxy surveys is expected to yield constraints on primordial non-Gaussianity that might not only complement the current CMB ones but even surpass them. Among the most prominent effects of non-Gaussianity on the matter distribution is the celebrated observation that a nonzero skewness leads to an enhanced abundance of collapsed structures and a scale dependent correction in the halo bias~\cite{Dalal:2007cu}, a result which has brought LSS surveys in the front line of the search for non-Gaussianity. Furthermore, there has been an intense effort to study how UV physics can show up in the matter power spectrum and bispectrum. All these objects shaping the distribution of matter in the universe can be indeed derived from the ansatz Eq.~(\ref{general-asatz-restricted}).

%\subsection{The starting point: $\zeta$ statistics}

Our main object of study is the observable statistics that can be obtained by considering small, yet nontrivial, departures from Gaussianity of the local type~\eqref{general-asatz-restricted}. Since such a departure must be small, in accordance with current observational constraints from the CMB~\cite{Akrami:2018odb}, we will be able to perform perturbative computations around a Gaussian probability distribution functional and explore the consequences these deviations may have in both CMB and LSS and what are the prospects in this regard for upcoming surveys.

\subsection{Probability Distribution Functional: ``bare'' theory}

The first task we can perform, and the most readily available to us thus far, is to attempt to use Eq.~\eqref{general-asatz-restricted} to its full potential and derive directly the complete functional distribution that governs the $\zeta$ statistics: given that we know how a Gaussian random field is distributed, we may simply perform a change of variables (as in~\cite{Matarrese:2000iz}) to obtain the probability distribution functional $\rho$ for $\zeta(\x)$ such that,
\be
\langle \zeta(\x_1) \cdots \zeta(\x_n)\rangle = \int D[\zeta] \rho [\zeta] \zeta(\x_1) \cdots \zeta(\x_n) .
\ee
That is to say, the resulting distribution must account, to leading order in the perturbation $\F$, for every conceivable correlation function that may be constructed from the field $\zeta(\x)$ and for every expectation value of function(al)s of $\zeta(\x)$. To start with, the Gaussian random field $\zeta_G$ is drawn from the following functional distribution:
\be
\rho_G[\zeta_G] = \mathcal{N} \exp \left[-\frac{1}{2} \int_\x \int_\y \zeta_G(\x) \Sigma^{-1}(\x,\y) \zeta_G(\y) \right]= \mathcal{N} \exp \left[-\frac{1}{2} \int_\K \frac{\zeta_G(\K) \zeta_G(-\K)}{P_\zeta(k) }  \right] ,
\ee
where $\mathcal{N}$ is an overall normalisation constant, while $\Sigma^{-1}(\x,\y)$ and $P_{\zeta}(k)$ are the covariance matrix and the power spectrum respectively, related as
\be \label{power-cov}
\Sigma^{-1}(\x,\y) \equiv \int_\K \frac{e^{i \K \cdot (\x-\y) } }{P_{\zeta}(k)}.
\ee
In implementing the transformation $\zeta(\x) = \zeta_G(\x) + \F[\zeta_G](\x)$, there are two contributions that emerge: one is composed by the terms that come out of the Taylor expansion of the Gaussian distribution by regarding the perturbation $\F$ as small, and the other arises from the functional determinant of the transformation. The latter is given by
\bea
{\rm det}\left( \frac{\delta \zeta(\y) }{\delta \zeta_G(\x) } \right) = \exp \left[ {\rm tr} \left( \ln \left( \frac{\delta \zeta(\y) }{\delta \zeta_G(\x) } \right) \right) \right] &=&   \exp \left[ {\rm tr} \left( \int_{\k} e^{i \k \cdot (\x - \y)} \ln \left\{ 1 + \frac{d F}{d \zeta}(\zeta_G(\x)) \right\} \right) \right] \nonumber \\
&=& \exp \left[ \int_\x \int_{\k} \ln \left\{ 1 + \frac{d F}{d \zeta}(\zeta_G(\x)) \right\} \right].
\eea
As long as the determinant exists (concretely $d F/d\zeta > -1$) and is nonzero, we may, in principle, find an inverse to the relation $\zeta = \zeta_G + \F[\zeta_G]$ and denote it with $\zeta_G(\x) = G[\zeta](\x)$. Then we may change variables from the PDF associated to $\zeta_G$,
\be
\rho_G[\zeta_G] = \mathcal{N} \exp \left[-\frac{1}{2} \int_\x \int_\y \zeta_G(\x) \Sigma^{-1}(\x,\y) \zeta_G(\y) \right] = \mathcal{N} \exp \left[-\frac{1}{2} \int_\k \frac{\zeta_G(\k) \zeta_G(-\k)}{P_\zeta(k) }  \right],
\ee
to find
\be \label{exact-PDF}
\rho[\zeta] = \mathcal{N} \exp \left[ -\frac{1}{2} \int_\x \int_\y G[\zeta](\x) \Sigma^{-1}(\x,\y) G[\zeta](\y) -  \int_\x \int_{\k} \ln \left\{ 1 + \frac{d F}{d \zeta}(G[\zeta](\x)) \right\} \right].
\ee
This is an exact result.

However, if $F$ is small in comparison to the typical scales of the background theory on which the fluctuation field $\zeta$ lies, and so is $d F/d\zeta$, we can approximate the logarithm in the exponential with the first term in its power series expansion, and furthermore, we may approximate the inverse mapping by $G[\zeta] \approx \zeta - \F[\zeta]$. This yields, to first order in $\F$,
\be \label{exp-PDF-1}
\rho[\zeta] = \rho_G[\zeta] \times \exp \left[ \int_\x \int_\y \zeta(\x) \Sigma^{-1}(\x,\y) \F(\zeta(\y))  -  \int_\x \frac{\delta}{\delta \zeta(\x)} \int_{\k} \int_y e^{i \k \cdot (\x - \y)}  F(\zeta(\y)) \right].
\ee
Using the definition of $\F$ in terms of $F$, and using that $\int_\k \int_\y e^{i \k \cdot (\x - \y) } \F[\zeta](\y) = \F[\zeta](\x) $, we may write the PDF $\rho$, using the definition of the Power Spectrum~\eqref{power-cov}, as
\be \label{exp-PDF-2}
\rho[\zeta] = \rho_G[\zeta] \times \exp \left[ \int_\x \int_{\k} \int_\y e^{i \k \cdot (\x - \y) } \left(  \frac{\zeta(\x)}{P_\zeta(k)} - \frac{\delta}{\delta \zeta(\x)}  \right) \F[\zeta](\y) \right].
\ee

This PDF should serve as the guiding principle for all subsequent results, as even in a first-order approximation, the probability distribution is always positive. However, to make computations tractable, we find it convenient, because it is formally equivalent in a first-order approximation, to also make a power series expansion out of the non-Gaussian factor.

Then, to first order in $\F$, we find that 
\bea \label{PDF-F}
\rho_F [\zeta] &=& \rho_G[\zeta] \times  \left[ 1 -  \int_\x \int_\y \int_\k e^{ i \k \cdot (\x - \y)}  \left( \frac{\delta}{\delta \zeta(\x)}  - \frac{\zeta(\x)}{P_\zeta(k)}  \right) \F[\zeta](\y) \right] \nonumber \\
&=& \rho_G[\zeta] \times \left[ 1 -  \int_\y \int_\k e^{-i \k \cdot \y}  \left( \frac{\delta}{\delta \zeta(-\k)}  - \frac{\zeta(\k)}{P_\zeta(k)}  \right) \F[\zeta](\y) \right].
\eea

The functional distribution~\eqref{PDF-F} has support at all the scales where the underlying theory does, or at least, at the scales where the corresponding EFT is presumed to hold true. However, observable quantities do not typically involve all of the scales, and therefore it may be that the ``bare'' departure from Gaussianity $\F$ is not the most adequate quantity to describe them. We thus now turn to the discussion of using window functions and how to integrate scales out.

\subsection{Probability Distribution Functional: running and renormalizing} \label{sec:PDF-run-ren}

When making predictions, any EFT will force us to recognise certain scales at which the theory is no longer well-suited to describe the physical observables. This typically implies a high-energy scale, where the theory has to be cut off. Thus, we will set $k_{\rm UV}$ as the maximum possible wavenumber the mode expansion of the curvature perturbation can have. Similarly, while it is not always introduced, one can make the same assertion with the very-long wavelengths. As much as our theory may have predictions concerning phenomena happening at 10 times the present Hubble radius, they are currently unobservable. Therefore, in establishing predictions for currently observable quantities it seems natural to integrate out those scales, so that they are properly incorporated into the final, effective result. Because of this, we will take a conservative attitude and also define an infrared cutoff $k_{\rm IR}$, which can be thought of as the inverse of the current Hubble radius, thus bounding the domain of the theories we will be studying to $k \in (k_{\rm IR}, k_{\rm UV})$.

%Usually, it will be necessary to bound the range of momenta that contribute to the integral, and therefore to the mode expansion of the curvature perturbation. We formally set $k_{\rm IR} < k < k_{\rm UV}$, with $k_{\rm IR}$ and $k_{\rm UV}$ infrared and ultraviolet cutoffs respectively, setting the limiting scales at which we expect that the theory is well described by a known power spectrum $P_{\zeta}(k)$.

%For most practical purposes, such as CMB or LSS analyses, it is sufficient to set $k_{\rm IR}$ to the inverse of our current Hubble radius and $k_{\rm UV}$ as a scale mode that was well outside the horizon by the end of inflation. Then the probability distribution functional is given by

However, in realistic situations the experiment at hand may not allow us to access every value for the momentum scale $k$ evenly. In those cases we may wish to introduce a window function $W(k)$ to filter our results and give more weight to some scales. Accordingly, one would be interested in the statistics of the filtered field
\be \label{z-filtered}
\zeta_W (\x) \equiv  [W \star \zeta](\x) = \int_\k \int_\y e^{i{\k} \cdot (\x - \y)} W(k) \zeta(\y).
\ee
To derive the probability distribution functional of the field $\zeta_W (\x)$ it is enough to perform a change of variables $\zeta_W({\k} ) = W(k) \zeta({\k})$ in~(\ref{PDF-F}). This then yields
\be \label{PDF-windowed}
\begin{split}
\rho_{W}[\zeta_W] = \mathcal{N}_W & \exp \left[ -\frac{1}{2} \int_\k \frac{\zeta_W(\k) \zeta_W(-\k)}{P_W(k) } \right] \\ & \times  \left[ 1 -  \int_\y \int_\k  e^{-i \k \cdot \y} W(k) \left( \frac{\delta}{\delta \zeta_W(-\k)}  - \frac{\zeta_W(\k)}{P_W(k)}  \right) \F \left[ \int_q  \frac{e^{-i\q \cdot \y} \zeta_W(\q)}{W(q)} \right](\y) \right],
\end{split}
\ee
where $P_W(k) \equiv W^2(k) P_\zeta(k)$. Leaving aside the argument of the deviation from Gaussianity $\F$ for a moment, this PDF has the same structure as the unfiltered PDF of Eq.~(\ref{PDF-F}). 

This expression for $\rho_{W}$ poses an interesting question: what if the window function of choice is defined (as usual) with hard cutoffs, just as if we were redefining the limits of our EFT? That is to say, how does $\rho_{W}$ look if we have
\be \label{flat}
W(k) = \begin{cases}
1 & \text{if} \,\, k \in (k_L, k_S) \\
0 & \text{if} \,\, k \not\in (k_L, k_S),
\end{cases}
\ee
as the window function?

It turns out that, for the functional integral to be well-defined, we have to integrate out of the theory all the modes which will not take part in our observable quantities, in order to avoid dealing with $\zeta_W(\q)/W(\q)$, which from the perspective of the theory with the window function would be ill-defined for the scales where $W=0$, but is perfectly finite (and equal to $\zeta(\q)$) from the perspective of the original theory. Let us take the original functional distribution~\eqref{PDF-F} (with $k$-space variables) and integrate out the modes outside the support of the window function $W$. We may write this as 
\be \label{ren-PDF}
\rho_{W}[\zeta] = \int D\zeta_{k \notin (k_L, k_S)} \rho[\zeta].
\ee
Upon integration over the prescribed range of modes, the purely Gaussian term in~\eqref{PDF-F} gives a reduced Gaussian measure that considers only $k \in (k_L, k_S)$. To integrate the term containing $\F$ it is convenient to separate the integral over momentum space in~\eqref{PDF-F} into two contributions:
\be \label{in-out}
\begin{split}
\int_\k e^{-i \k \cdot \y}  \left( \frac{\delta}{\delta \zeta(-\k)} - \frac{\zeta(\k)}{P_\zeta(k)}  \right) \F[\zeta](\y) =& \int_{|\k| \in (k_L,k_S)} e^{-i \k \cdot \y}  \left( \frac{\delta}{\delta \zeta(-\k)}  - \frac{\zeta(\k)}{P_\zeta(k)}  \right) \F[\zeta](\y) \\ & + \int_{|\k| \not\in (k_L,k_S)} e^{-i \k \cdot \y}  \left( \frac{\delta}{\delta \zeta(-\k)}  - \frac{\zeta(\k)}{P_\zeta(k)}  \right) \F[\zeta](\y).
\end{split}
\ee
If we now perform the functional integration over $\zeta(\k)$, we see that the first term will only involve knowing how to deal with the quantity\footnote{Note that when considering the functional integration of the first line of Eq.~\eqref{in-out}, the differential operator $\left( \frac{\delta}{\delta \zeta(-\K)}  - \frac{\zeta(\K)}{P_\zeta(k)}  \right)$ involves scales $|\K| \in (k_L, k_S)$, while the functional integration goes over modes with $|\K| \notin (k_L, k_S)$. We may thus pull it out of the integral.}
\be \label{renorm}
 \left( \frac{\delta}{\delta \zeta(-\k)}  - \frac{\zeta(\k)}{P_\zeta(k)}  \right)  \int D\zeta_{|\k| \notin (k_L, k_S)} \exp \left[-\frac{1}{2} \int_\k \frac{\zeta(\k) \zeta(-\k)}{P_\zeta(k)}  \right] \int_\y \int_{|\k| \in (k_L, k_S)} \!\!\!\!\!\!\!\! e^{-i \k \cdot \y} \F[\zeta](\y),
\ee
whereas the second term vanishes after performing a functional partial integration with the functional derivative $\delta/\delta\zeta(-\k)$, because
\be
\frac{\delta}{\delta \zeta(-\k)} \exp \left[-\frac{1}{2} \int_\k \frac{\zeta(\k) \zeta(-\k)}{P_\zeta(k) }  \right] = - \frac{\zeta(\k)}{P_\zeta(k)}  \exp \left[-\frac{1}{2} \int_\k \frac{\zeta(\k) \zeta(-\k)}{P_\zeta(k)}  \right].
\ee
Thus, all what remains is knowing how to compute~\eqref{renorm} so as to see if, and how, the interaction is renormalized. It turns out that if we define
\be \label{def-bar-F}
{\bar F}(\zeta_W(\x)) \equiv \int D\zeta_{k \notin (k_L, k_S)} \exp \left[-\frac{1}{2} \int_{|\k| \notin (k_L,k_S)} \!\!\!\!\!\! \frac{\zeta(\k) \zeta(-\k)}{P_\zeta(k)}  \right] F(\zeta(\x)) = \int_{-\infty}^{\infty} d{\bar \zeta } \frac{e^{-\frac{(\zeta_W(\x) - {\bar \zeta})^2 }{2 \sigma_{\rm out}^2 }  }  }{\sqrt{2 \pi \sigma_{\rm out}^2 } } F(\bar \zeta),
\ee
then we may identify 
\be
\bar \F[\zeta](\x) = \int_{\y} \int_{|\k| \in (k_L, k_S)} \!\!\!\!\!\!\!\! e^{i \k \cdot (\x - \y) } \bar F(\zeta(\y) ) 
\ee
as the effective self-interaction, because we would have integrated out all the scales that are outside the range of interest and still leave the other scales within the measure $\rho$ of the PDF, while maintaining its analytic structure.

The last equality of~\eqref{def-bar-F}, the Weierstrass transform of $F(\zeta)$, can be obtained in numerous manners. If one were to follow standard diagrammatic perturbation theory, it arises from summing back every ``loop'' contraction performed by the Gaussian measure of $\F$ with itself. Since the momenta flowing through those loops is bounded by the range being integrated out, and there is no ``external'' momenta flowing through the diagrams, we have that their numerical value is the same for every loop and equal to the variance
\be
\sigma_{\rm out}^2 = \int_{k \not\in (k_L, k_S)} \!\!\!\!\!\! P_\zeta(k) 
\ee
that was subtracted from the Gaussian field statistics when the modes $k \not\in (k_L, k_S)$ were integrated out of the theory.

Therefore, by using these results in Eq.~\eqref{ren-PDF}, we obtain
\be
\rho_{\F,W}[\zeta] = \rho_{\bar \F}[\zeta_{k \in(k_L, k_S)}],
\ee
i.e., that $\rho_{W}[\zeta]$ for the restricted variable~\eqref{z-filtered} has the same functional form as the original PDF, with the only modification that now the departure from Gaussianity is given by a ``filtered'' interaction $\bar \F$ instead of the ``bare'' interaction $\F$.

In practice, there is more than one way of how to represent the running of $F$ depending on the scales one wants to include in the theory. Perhaps the most ethereal representation, but at the same time the most revealing of the theory's structure is through the differential expression of the Weierstrass transform, as in
\be \label{W-ren-out}
{\bar F}(\zeta) = \exp \left( \frac{\sigma_{\rm out}^2 }{2} \frac{\partial^2}{\partial \zeta^2 } \right) F(\zeta),
\ee
which makes it clear how the theory runs by removing more or less scales, as well as the fact that the transformation rule between $F$ at different scales follows an adequate composition property: integrating out ranges of momenta $A$ and $B$ is implemented via $\sigma_{\rm out,A}^2$ and $\sigma_{\rm out,B}^2$, and doing so yields the same result independently of the order in which one subtracts the modes from the theory. Furthermore, this shows that the functional form of $\F$ and $\bar \F$ is, in the sense that the quantities that determine its concrete expression are exactly the same: the only thing that the window does is to restrict the range of modes entering in the observables.

Conversely, just as the PDF may be recast in an analogous manner to that of the original theory, the field with modes between $k_L$ and $k_S$ may also be written down as a local departure from Gaussianity
\be \label{local-windowed}
\zeta_W(\x) = \zeta^G_W(\x) + {\bar \F}\left[ \zeta_W^G \right](\x),
\ee
merely because it follows statistics analogous to $\zeta$. Here we have to remind the reader that this is only so for $W(k)$ of the form~\eqref{flat}. Other window functions still give rise to an explicit PDF, namely equation~\eqref{PDF-windowed}, but the deviations from Gaussianity may no longer be written as concisely as in~\eqref{local-windowed}. The difference lies in that a general window function does not render irrelevant some degrees of freedom of the theory; it only gives them dissimilar weights in the final result. However, in order to obtain the function ${\bar \F}$ it is crucial that we reduce the number of independent variables in our theory, as all of them will leave their signature, if small, in any given correlation function.

We now have the basic tools to proceed to studying late-time observables: Thus far, we have established how we may write the probability distribution functional of our theory depending on the scales under consideration, and also how we may incorporate window functions into the distribution functional. These results set the foundations for studying the transfer of these perturbations to the observables in the sky: both CMB temperature fluctuations $\Delta T(\hat{n})/T_0$ and the matter density contrast $\delta(\x) = \delta \rho(\x)/\rho_0$ are sourced by the primordial statistics of $\zeta$.

Before passing to simpler statistical estimators stemming from this functional, we now discuss the partition function.

\subsection{Partition Function and $n$-point Correlators}

In practice, using the full probability distribution functional directly on cosmological data proves to be difficult, as there is only one realization of our universe to probe and conduct measurements in, so it is not possible to take a frequentist approximation to its statistics. While this suggests the use of Bayesian statistics to find the most probable $\F$ given the data by the means of the functional~\eqref{PDF-F}, it also reveals why one typically chooses to work with correlation functions to probe departures from Gaussianity: they can be computed from many Fourier modes on the sky, whose past history is presumably independent (at least if the nonlinearities are turned off), and therefore averages may be performed and compared with the theoretical predictions for the expectation values or correlations.

Fittingly, there is an object that encapsulates the information of all the correlation functions in a perhaps clearer way than the full probability density functional $\rho$. This is the partition function $Z[J]$, which would be the object that generates the $n$-point functions via functional differentiation
\be \label{Z-corr}
\langle \zeta({\k}_1 ) \cdots \zeta({\k}_n )\rangle = \frac{\delta^n Z[J]}{(i\delta J(-{\k}_1 )) \cdots (i \delta J(-{\k}_n)) }\bigg|_{J = 0},
\ee
or equivalently, as the functional Fourier transform of the PDF, which in the context of probability is called the characteristic function
\be \label{Z-Fourier}
Z[J] = \int D \zeta \, \rho[\zeta] \, e^{i\int_\k \zeta({\k}) J(-{\k})}.
\ee

Both expressions may be employed to obtain $Z[J]$: the first requires to know all of the $n$-point functions beforehand and reconstruct the object that has them as its functional derivatives, while the second requires to know an explicit expression for the probability distribution functional. Since we have the latter, we may carry out this computation explicitly\footnote{The details of this derivation are presented in Appendix~\ref{sec:Partition}.} to first order in $\F$, obtaining
\be \label{partition}
\begin{split}
Z[&J]  = \, \exp \left[-\frac{1}{2} \int_\k J(\k) J(-\k) P_\zeta(k) \right] \;\times\\  &  \left( 1 - \int_\x \frac{\int_\k e^{i\k \cdot \x} J(-\k)}{ \int_\k e^{i\k \cdot \x} J(-\k) P_\zeta(k) } \int_{\bar \zeta} \frac{\exp \left[ - \frac{\left({\bar \zeta} - i\int_\k e^{i\k \cdot \x} J(-\k) P_\zeta(k) \right)^2}{2 \sigma_\zeta^2} \right] }{\sqrt{2 \pi} \sigma_\zeta} \left(\sigma_\zeta^2 \frac{\partial }{\partial \bar \zeta} - \bar{\zeta}\right)  F(\bar \zeta) \right).
\end{split}
\ee
Here we have defined $\sigma_\zeta^2 \equiv \int_k P_\zeta(k)$ as the 1-point variance associated to the power spectrum for the relevant range of momenta. Window functions are easily incorporated by substituting $J(\k)$ with $J(\k) W(k)$, as this procedure will add a factor of $W(k)$ to every external leg in any given diagram.

Now that we have equation~\eqref{partition}, we may compute the $n$-point functions directly, without having to resort to functional integration as we would with~\eqref{PDF-F}. Moreover, the structure that will emerge in these correlations is more closely related to~\eqref{partition}, as is demonstrated by their explicit expressions in position space
\be \label{npt-position}
\langle \zeta_W({\x}_1 ) \cdots \zeta_W({\x}_n )\rangle_c = f_{n-1} \sum_{i=1}^n \int_{\x}  \frac{\int_{\k} W(k) e^{i\k \cdot (\x_i - \x) } }{\int_\k e^{i\k \cdot (\x_i - \x) } W(k) P_\zeta(k)  } \left( \prod_{i=1}^n \int_\k W(k) P_\zeta(k) e^{i \k \cdot (\x_i - \x) } \right),
\ee
where the subscript $c$ indicates the result only considers the fully connected piece. For completeness, we write down their counterparts in momentum space 
\be \label{n-pt-f}
\langle \zeta_W({\k}_1 ) \cdots \zeta_W({\k}_n )\rangle_c = f_{n-1} \, (2\pi)^3 \delta^{(3)} \left( \sum_{i=1}^n \k_i \right)  \left( \prod_{j=1}^n W(k_j) P_\zeta(k_j)  \right) \sum_{i=1}^n \frac{1}{P_\zeta(k_i)} , 
\ee
where the coefficients\footnote{In a more standard notation, the first few terms would correspond to $f_2=f_{\rm NL}$, $f_3=g_{\rm NL}$, etc.} $f_n$ are given by Hermite moments of $F$:
\be
f_n \equiv - \frac{1}{\sigma_\zeta^{n}} \int_{-\infty}^{\infty} d\zeta \frac{e^{-\frac{\zeta^2}{2\sigma_\zeta^2 } } }{\sqrt{2\pi  }\sigma_\zeta } {\rm He}_{n}\! \left( \frac{\zeta}{\sigma_\zeta} \right) F (\zeta).
\ee
The coefficients $f_n$ are quantities of mass dimension $1-n$, which are invariant under the renormalisation procedure~\ref{sec:PDF-run-ren} in a very fitting sense: because $\{f_n\}_{n=2}^{\infty}$ is also a set of coefficients of a Hermite polynomial expansion\footnote{We omit $n=0,1$ in the Hermite expansion because we assume $\langle \zeta \rangle = 0$ and $\langle \zeta \zeta \rangle $ to be set by the free theory and matching the Power Spectrum of observations. To put it differently, with this definition of the local ansatz, due to orthogonality properties of the Hermite polynomials, the power spectrum of $\zeta$ is not modified to first order in the nonlinearity parameters~\cite{Smith:2011ub}.}, we have
\be \label{F-herm}
F(\zeta;\sigma_\zeta^2) = - \sum_{n=2}^{\infty} \frac{f_n}{n!} \, \sigma_\zeta^{n} \, {\rm He}_n \! \left( \frac{\zeta}{\sigma_\zeta} \right) = - \sum_{n=2}^{\infty} \frac{f_n}{n!} \exp \! \left[ - \frac{\sigma_\zeta^2}{2} \frac{\partial^2 }{ \partial \zeta^2} \right] \zeta^n,
\ee
where we have introduced the variance $\sigma_\zeta^2$ as an argument of $F$ in order to emphasize that the associated field has the corresponding amplitude for its fluctuations. This means that $\bar F$ takes the following expansion:
\be
{\bar F}(\zeta) = \exp \! \left[ + \frac{\sigma_{\rm out}^2}{2} \frac{\partial^2 }{ \partial \zeta^2} \right] \F(\zeta;\sigma_\zeta^2) = F(\zeta;\sigma_\zeta^2 - \sigma_{\rm out}^2),
\ee
where the coefficients $\{f_n\}_{n}$ remain unchanged; only the variance gets reduced to its new value after integrating some modes out.

%$\F^{(m)}(0) \equiv \frac{\partial^{m}\F}{\partial\zeta^{m}}\Big|_{\zeta=0}$.

Nonetheless, it is important to point out that each individual $n$-point function does not encapsulate all of the non-Gaussian information contained in $F$. Indeed, each correlation function only yields one term of an infinite series expansion of $F$, all of which are independent, at least in principle. Therefore, it is natural to try and find objects that are able to keep all of this information, without having to compute an infinite number of quantities. For that reason, we now turn to exploring 1- and 2-point probability density functions, which we will later apply, in particular to the CMB.

\subsection{Fixed-point Probability Distribution Functions}

In the presence of a generic deviation from Gaussian statistics, involving both local and non-local terms, to assume that it is possible to capture all non-Gaussian information by looking at the single-point statistics of a field (i.e., correlations with all of the spatial coordinates at the same position) seems misguided, as the restriction to a single point is likely to mix local and non-local effects, making it difficult to disentangle them. However, if we restrict ourselves to local deviations from Gaussianity only, it is indeed possible to capture all such information.

In this subsection we write down explicit 1-point and 2-point PDFs for the curvature perturbation. Since the first was already derived~\cite{Chen:2018brw}, and is analogous to the derivation of the partition function, we will not give many details. On the other hand, the derivation of the 2-point PDF is quite lengthy, and therefore, in order to alleviate the discussion, we shall leave the details for Appendix~\ref{sec:2-point-det}.

\subsubsection{1-point Probability Distribution Function}

Now we set ourselves to derive the simplest distribution that can be obtained within this framework: a density function for the 1-point statistics. It is defined as the distribution $\rho(\zeta;\x)$ that satisfies
\be
\langle \zeta^n(\x) \rangle = \int_{-\infty}^{\infty} d \zeta \rho(\zeta; \x) \zeta^n.
\ee
Given that we assume a homogeneous universe, $\rho(\zeta;\x)$ cannot depend on $\x$. Thus, we write $\rho(\zeta;\x) = \rho(\zeta)$. From a functional perspective, it is given by
\be \label{1pt-condition}
\rho(\bar{\zeta};\x) = \int D\zeta \, \rho[\zeta] \, \delta(\zeta(\x) - \bar{\zeta}),
\ee
which may be evaluated in the same way as the partition function $Z[J]$ by writing down the Dirac delta as $\delta(\zeta(\x) - \bar{\zeta}) = \int_\gamma e^{i \gamma (\zeta(\x) - \bar{\zeta})}$ and noticing that what will be left over in the functional integral is exactly $Z[J(-\k) = \gamma e^{-i\k \cdot \x}] $. Then the remaining integral over $\gamma$ may be carried out by completing squares.

Thus, in the same spirit as the probability distribution functional we obtained earlier represents a first-order correction to Gaussian statistics, the 1-point probability density function may also be written as a slight departure from Gaussianity. Moreover, this density function resembles more closely the structure of $Z[J]$ than that of $\rho[\zeta]$ because marginalizing over all the other positions in presence of a finite range of wavelengths induces a filtering, which is manifest in $Z[J]$ but not so in $\rho[\zeta]$. Given the various applications it may find in LSS or in Primordial Black Holes (PBH) formation, it is of interest to write it for an arbitrary window function. Smoothing the field and its variance as in Eq.~\eqref{z-filtered}, the resulting expression is
\be \label{one-point}
\rho (\zeta_W) =\frac{1}{\sqrt{2\pi}\sigma_{W} } e^{-\frac{\zeta_W^2}{2\sigma_{W}^2}}  \left[ 1 + \Delta_W(\zeta_W) \right], 
\ee
where
\be \label{one-point-D}
 \Delta_W (\zeta) \equiv \int_0^{\infty} \!\! d x \;  \frac{4\pi x^2 W(x)}{{\rm s}^2(x)} \! \int_{-\infty}^{\infty} \!\!\!\! d \bar \zeta \,\, \frac{\exp \Big[{-\frac{ \left(\bar \zeta - {\rm b}(x) \zeta \right)^2}{2\sigma_{W}^2 (x)} } \Big] }{\sqrt{2\pi  } \sigma_{W} (x)}  \left( {\bar \zeta}  - \sigma_\zeta^2 \frac{\partial}{\partial {\bar \zeta}} \right) \! F \! \left({\bar \zeta}\right). %\label{main-result-2}
\ee
Here we have written $W(x) = \int_{\K} e^{i \K \cdot \x} W(k)$, the position-space representation of the window function $W$, with $x = |\x|$, and we have also defined the (co)variances
\be \label{var-1-defs-t}
\sigma_{W}^2 \equiv \int_k W^2(k) P_\zeta(k)\;, \quad {\rm s}^2(x) \equiv \int_k e^{i\k \cdot \x} W(k) P_\zeta(k) \;, \quad
\sigma_{W}^2(x) \equiv \sigma_\zeta^2 -  {\rm b}^2(x) \sigma_{W}^2,
\ee
with a ``bias'' factor\footnote{This parameter is not the usual bias. If we think of the window as setting the scale of a tracer, we can define a bias as the ratio of field-tracer to the tracer-tracer correlation functions. ${\rm b}$ here is the ratio of the field-tracer to the 1-point tracer-tracer correlation functions.} given by
\be \label{b-1-defs-t}
{\rm b}(x) \equiv \frac{{\rm s}^2(x)}{\sigma_W^2}.
\ee

In words, this expression means that, given a range of modes that defining the cutoffs of the theory, the perturbative correction to the PDF scans for the structure of the effective interaction at those scales through the action of $( {\bar \zeta} - \sigma_\zeta^2 \partial_{\bar \zeta} )$, and then filters it according to the difference between the variance and the correlation implied by incorporating window functions.
This observable can account for all the information contained within the local function $F$. This can be seen from the fact that all of the information concerning $F$ is stored within the $f_n$ coefficients, which can be retrieved by looking only at the 1-point statistics (c.f. Eq.~\eqref{npt-position}). Indeed, an analysis to constrain the departure from Gaussianity $F$ by the means of the 1-point CMB temperature distribution has already been performed in~\cite{Chen:2018brw} .

Before passing to the 2-point functional, and in order to make contact with current literature, let us comment on the relation of this PDF to the Edgeworth representation. If we use the Hermite polynomial expansion of $\F$, given by~\eqref{F-herm}, the effect of the Gaussian filtering of Eq.~\eqref{one-point-D} becomes transparent:
\be \label{W-effect}
\int_{-\infty}^{\infty} \!\!\!\! d \bar \zeta \,\, \frac{\exp \Big[{-\frac{ \left(\bar \zeta - {\rm b}(x) \zeta \right)^2}{2\sigma_{W}^2 (x)} } \Big] }{\sqrt{2\pi  } \sigma_{W} (x)} \! F(\bar\zeta;\sigma_\zeta) = F\left( {\rm b}\zeta;{\rm b}^2\sigma_W^2 \right) = \sum_{n=2}^\infty \frac{f_{n}}{n!} \,   {\rm He}_{n} \! \left( \frac{\zeta_W}{\sigma_W} \right) \frac{{\rm s}^{2n}(x)}{\sigma_W^{n}},
\ee
that is, it replaces the field variable with the biased one (we have suppresed the $x$ dependence of the bias). Using the fact that the Weierstrass trasform commutes with derivatives, the latter now being evaluated at the biased field, and that $(\delta-\sigma^2\partial_\delta){\rm He}_{n}=\sigma{\rm He}_{n+1}$, we can rewrite the NG deviation of the 1-point PDF as
\bea \label{Edgw}
\Delta_W (\zeta_W) &=&  \sum_{n=2}^\infty \frac{f_{n}}{n!} \,   {\rm He}_{n+1} \! \left( \frac{\zeta_W}{\sigma_W} \right) \frac1{\sigma_W^{n+1}} \int_0^{\infty}  \!\! d x \;   4\pi x^2 W(x) {\rm s}^{2n}(x) \nonumber \\
&=&  \sum_{n=2}^\infty \frac{{\rm He}_{n+1} \! \left( \frac{\zeta_W}{\sigma_W} \right)}{(n+1)!} \, \frac{\langle \zeta_W^{n+1} \rangle_c}{\sigma_W^{n+1}},   
\eea
where $\langle \zeta_W^{n} \rangle_c$ is given by Eq.~\eqref{n-pt-f} integrated over momenta.
This is exactly the Edgeworth expansion of a non-Gaussian PDF truncated to first order in the couplings $f$, since we have restricted our derivation of the PDF~\eqref{one-point-D} to first order\footnote{In principle, there is nothing stopping us from computing the Edgeworth expansion to any order in the couplings; indeed, we can expand the exact PDF of Eq.~\eqref{exact-PDF} to any order in $F$. We choose, however, for simplicity to truncate the series to first order.} in $F$. This is a physically motivated case where the truncation of the expansion to first order quantities should be a good approximation~\cite{Sellentin:2017aii}.

\subsubsection{2-point Probability Distribution Function}

Now we would like to write down an observable able to account for all of the non-Gaussianities that can emerge within this model (meaning we cannot look at any single $n$-point function) and does not integrate out the information about correlations in the sky. Therefore it cannot be a single-point PDF. Thus, we try to do the next least complicated thing: a 2-point PDF $ \rho (\zeta_1, \zeta_2; \x_1, \x_2)$. This function satisfies
\be \label{2-pr-corr}
\langle \zeta_W^n(\x_1) \zeta_W^m(\x_2) \rangle = \int d\zeta_1 d\zeta_2 \rho_W (\zeta_1, \zeta_2; |\x_1 - \x_2|) \zeta_1^n \zeta_2^m,
\ee
where we have written $ \rho (\zeta_1, \zeta_2; \x_1, \x_2) = \rho (\zeta_1, \zeta_2; |\x_1-\x_2|)$ because we are assuming our Universe to be statistically homogeneous. One way to obtain such an object is by conditioning in two points in a manner analogous to Eq.~\eqref{1pt-condition},
\be \label{2pt-condition}
\rho(\zeta_1, \zeta_2; |\x_1 - \x_2|) = \int D\zeta \, \rho[\zeta] \, \delta(\zeta(\x_1) - \zeta_1) \delta(\zeta(\x_2) - \zeta_2) ,
\ee
where again, the final result can only depend on the spatial coordinates through the distance between the two positions $\x_1$ and $\x_2$. With this in mind, let us define the distances 
\be
r\equiv|\x_1 - \x_2|,\quad r_1\equiv|\x - \x_1|,\quad r_2\equiv|\x - \x_2|.
\ee

A similar computation, though arguably trickier than that of the 1-point function, leads to a two-point distribution analogous to what was obtained in~\cite{Chen:2018brw}, but with two points defining the filtering instead of one:
\be \label{two-point}
\begin{split}
\rho_W (\zeta_1,\zeta_2,r) = & \; \rho_{G,W}(\zeta_1,\zeta_2,r)   \Bigg[ 1 - \int_\x \int_{-\infty}^{\infty} \!\!\!\! d \bar \zeta   \frac{\exp \Big[{-\frac{ \left(\bar \zeta - \zeta_W(r,r_1,r_2) \right)^2}{2\sigma_W^2 (r,r_1,r_2)} } \Big] }{\sqrt{2\pi  } \sigma_W (r,r_1,r_2)}  \\ &  \times \left\{  \frac{W(r_1)}{s^2(r_1)} \!  \left( G_{11} \frac{\partial}{\partial {\bar \zeta}} - G_{12} \right)      +  \frac{W(r_2)}{s^2(r_2)} \!  \left( G_{21} \frac{\partial}{\partial {\bar \zeta}} - G_{22} \right) \right\} F  \! \left({\bar \zeta}\right)  \Bigg],
\end{split}
\ee
where $\rho_{G,W}(\zeta_1,\zeta_2,r)$ is the bivariate Gaussian measure, with a covariance matrix given by the $2\times2$ bottom right block of $\pmb{\Sigma}$, defined below in~\eqref{cov-W}.
Let us go through this expression: the first thing to notice is the presence of two points, $\x_1$ and $\x_2$, defining a filtering through the same function as in the 1-point case. The second important aspect is that now the Gaussian that is convoluted with $F$ has a different mean and variance. However, they emerge in the same manner as $\sigma_W^2(x)$ and $\zeta_W(x)$ emerge in the 1-point case: $\sigma^2_W(r,r_1,r_2)$ and $\zeta_W(r,r_1,r_2)$ are the variance and mean of $\bar \zeta$ after conditioning on the values of $(\zeta_1, \zeta_2)$, starting from a joint Gaussian distribution for $({\bar \zeta}, \zeta_1, \zeta_2)$ with covariance matrix
\be \label{cov-W}
  \pmb{\Sigma}=
  \left[ {\begin{array}{ccc}
   \sigma_\zeta^2 & s^2(|\x - \x_1|) & s^2(|\x - \x_2|) \\
   s^2(|\x - \x_1|) & \sigma_W^2 & \sigma_{W,{\rm ext}}^2(|\x_1 - \x_2|)\\
   s^2(|\x - \x_2|) & \sigma_{W,{\rm ext}}^2(|\x_1 - \x_2|) & \sigma_W^2 \\
  \end{array} } \right],
\ee
where we have written the covariance between the two externally chosen points $\x_1$ and $\x_2$ as
\be
\sigma_{W,{\rm ext}}^2(|\x_1 - \x_2|) = \int_k e^{i\k \cdot (\x_1 - \x_2)} W^2(k) P_\zeta(k).
\ee
The functions $G_{ij}$ also appear in a similar way: $G_{i1}$ and $G_{i2}$ are ``rotated'' versions of $\sigma_\zeta^2$ and $\bar \zeta$, involving combinations of the free theory covariances that make the overall expression reduce to that of the 1-point PDF as $\x_1 \to \x_2$. Their precise definitions are listed in Appendix~\ref{sec:2-point-det}. In there, we delineate how to obtain the 2-point PDF: by starting from correlators of the type $\langle \zeta_W^n(\x_1) \zeta_W^m(\x_2) \rangle$, we deduce the function from which they emanate, corresponding to~\eqref{2-pr-corr}. Moreover, by exploiting the analyticity of the function $F$, it is possible to again write this as an Edgeworth expansion in bivariate Hermite polynomials~\cite{Sellentin:2017aii}, truncated to first order in the nonlinearity parameters.

This PDF contains all the information of the free theory, as having two points allows to scan over all the range of distances in the sky, thus probing, among others, the two-point correlation function completely, which is the defining object of a Gaussian theory. Even though Eq.~\eqref{two-point} has its non-Gaussian features encoded in a perhaps more complicated fashion than its 1-point counterpart~\eqref{one-point}, both contain the same information about the underlying function $\F$. Indeed, one can obtain Eq.~\eqref{one-point} by integrating over one of the field variables in Eq.~\eqref{two-point}. However, observationally, it might be more efficient to have information on the scale, since then we can, for example, disentangle different momentum shapes of correlation functions.

\section{An analysis of the Cosmic Microwave Background}

We are finally in position to undertake the task we promised to carry out at the end of Chapter~\ref{chap:n-point}. We start by exploring the generalities of how to map the curvature perturbation onto the celestial sphere. Then, we reconstruct the primordial isocurvature potential $\Delta V$, following the same notation as in Chapter~\ref{chap:n-point}, only that now we will introduce the transfer function, obtain a PDF, and from there constrain the potential based on the information provided by Planck in 2015~\cite{Ade:2015hxq}. Finally, we outline how a constraint from the 2-point PDF could be obtained, and what information could be gained in comparison with the 1-point approach.

\subsection{$\Theta$ Statistics} \label{sec:theta-stat}

It is also of interest to write down testable quantities that we can obtain by looking at the primordial information that can be stored in spherical shells on the sky, such as the CMB. For instance, the probability distribution functional $\rho[\zeta]$ and its associated partition function $Z[J]$ may be projected onto the celestial sphere to yield distributions of, say, the temperature fluctuations $T(\hat{n})$.

Consider a generic linear transfer function $T(\k,\n)$ from the primordial perturbations $\zeta$ to an observable defined on the sphere $\Theta(\n)$ such that
\be
\Theta(\n) = \int_\k T(\n,\k) \zeta(\k).
\ee
Then, if we take $\Sigma(\n,\n')$ to be the covariance matrix defining the observable's correlations between different directions in the sky $(\n,\n')$, i.e., 
\be
\langle \Theta(\n) \Theta(\n') \rangle = \Sigma(\n,\n') ,
\ee
and $\Sigma^{-1}(\n,\n')$ as its inverse matrix, we find that the probability distribution functional for $\Theta$ is given by
\be
\begin{split}
\rho[\Theta] = \mathcal{N}_{\Theta} e^{-\frac{1}{2} \int_{\n} \int_{\n'} \Theta(\n) \Sigma^{-1}(\n,\n') \Theta(\n') } & \left[ 1 - \int_\x  {\rm Ker_1}(\x) \frac{\partial F_\Theta}{\partial \zeta }(\zeta_{\Theta}(\x);\x) \right. \\ & \,\,\,\,\,\,\, \left. + \int_\x {\rm Ker_2}(\Theta; \x) F_\Theta(\zeta_\Theta(\x);\x) \right],
\end{split}
\ee
where now ${\rm Ker_1}(\x)$, ${\rm Ker_2}(\Theta;\x)$, $\zeta_{\Theta}(\x)$, and $F_\Theta(\zeta;\x)$ depend implicitly on the transfer function and, when noted, the variable of interest $\Theta(\n)$. All of these quantities may be intuitively understood as the result of projecting a field defined on three spatial dimensions over a two-dimensional spherical surface. For instance, $\zeta_{\Theta}(\x)$ is given by
\be
\zeta_{\Theta}(\x) \equiv \int_\k \int_{\n} \int_{\n'} T(\n,\k) e^{i\k \cdot \x} P_\zeta(k) \Sigma^{-1}(\n,\n') \Theta(\n')  ,
\ee
which basically amounts to saying: take the original statistics of your theory, i.e., $e^{i\k \cdot \x} P_\zeta(k)$, project them onto the sphere by applying $T(\n,\k)$ and then correlate it with the field of interest $\Theta(\n')$ by means of its inverse covariance matrix $\Sigma^{-1}(\n,\n')$. The integration kernels have similar definitions:
\bea
{\rm Ker_1}(\x) &\equiv & \int_\k \int_{\k'} \int_{\n} \int_{\n'} T(\n,\k) e^{i\k \cdot \x} P_\zeta(k) \Sigma^{-1}(\n,\n') T(\n',\k') e^{i\k' \cdot \x} , \\
{\rm Ker_2}(\Theta; \x) &\equiv& \int_\k \int_{\n} \int_{\n'} T(\n,\k) e^{i\k \cdot \x} \Sigma^{-1}(\n,\n') \Theta(\n').
\eea
However, the function $F_\Theta$ is a slightly different object than before. As a result of the projection, it acquires a spatial dependence, whose exact nature in terms of the primordial departure from Gaussianity $F$ is given by the Weierstrass transform
\be
F_\Theta(\zeta; \x) \equiv \exp \left[ \frac{\sigma_\zeta^2 - \sigma_{\zeta \Theta}^2(\x)}{2} \frac{\partial^2 }{\partial \zeta^2} \right] F(\zeta)  = \int_{\bar \zeta} \frac{\exp \Big[{-\frac{ \left(\bar \zeta - \zeta \right)^2}{2(\sigma_\zeta^2 - \sigma_{\zeta \Theta}^2(\x))} } \Big] }{\sqrt{2\pi (\sigma_\zeta^2 - \sigma_{\zeta \Theta}^2(\x)) } } F(\bar \zeta),
\ee
where $\sigma_{\zeta \Theta}^2(\x)$ may be understood as a position-dependent effective variance of $\zeta$, modified by projection effects:
\be
\sigma_{\zeta \Theta}^2(\x) \equiv \int_\k \int_{\k'} \int_{\n} \int_{\n'} T(\n,\k) e^{i\k \cdot \x} P_\zeta(k) \Sigma^{-1}(\n,\n') T(\n',\k') e^{i\k' \cdot \x} P_\zeta(k').
\ee
In all of the above, we have written $\Sigma(\n,\n')$ as a general function of the direction on the sphere. However, if we take into account that our universe is homogeneous, it must be possible to write it as a function of the angle between the two vectors, or equivalently, in terms of their scalar product $\n \cdot \n'$. We will not overemphasize this in what follows, as the notation we deem natural to treat $\n$ and $\n'$ is with them as separate directions, because they will have to be multiplied with another vector, the integration variable $\x$ that will appear in the NG kernel that modifies the PDF, which makes using $\n \cdot \n'$ notationally heavier than just using $(\n,\n')$.

Now we have to find ways of using this. One option would be to compare how likely is our present-day CMB given a certain primordial deviation from Gaussianity of the local type $F$ by using Bayesian statistics. However, in order for this to be useful at its maximum capacity, it is likely that one would first have to establish a definitive imprint of primordial NG and be forced to introduce extra parameters into the effective description because more often than not a model comparison will favour the one with less parameters. Therefore, we turn to the observable we have been largely focusing on thus far, which, to our knowledge, has been largely unexplored and may offer valuable constraints on the nature of primordial NG: 1- and 2-point PDFs.

\subsection{1-point PDF constraints}

Let us attempt to reconstruct $\Delta V_{\rm ren}$ of~\eqref{potential-ren-resum} out of the CMB data, using 
\be
F(\zeta) = - \frac{\Delta N}{6 H^2} \Delta V_{\rm ren}' \left(\frac{H \zeta}{\alpha \Delta N} \right),
\ee
i.e., the local departure from Gaussianity that emerges from that multi-field inflation setup. This requires us to deal with the observed temperature fluctuation $\Theta \equiv \Delta T / T$, instead of $\zeta$ at the end of inflation. Thus, we introduce a linear transfer function to write $\Theta ( {\bf k} , \hat n) \equiv T(k, \mu) \zeta_{\bf k}$, with $\mu = \hat n \cdot \hat k$, where $\hat n$ is the direction of sight of an observer standing at ${\bf x}$.  

In general, this transfer function would have to be computed by solving the Boltzmann transport equations that describe the evolution of the photon energy density throughout the history of the Universe after the Big Bang occurred and the primordial seeds left in place. However, as quoted in Chapter~\ref{chap:search}, Weinberg~\cite{Weinberg:2008zzc} has shown that in the hydrodynamic approximation the transfer function is well described by
\be
T({\bf k}, \hat n) = e^{i\k \cdot \n r_L } \left( F(k) + \n \cdot \hat{k} \, G(k) \right),
\ee
where the functions $F(k)$ and $G(k)$ are given as in Chapter~\ref{chap:search}:
\bea
F(k) &=& \frac{e^{-\tau_{\rm r}}}{5} \left[3 R_L \mathcal{T} \! \left( \frac{k d_T}{a_L} \right)  - \mathcal{S} \! \left( \frac{k d_T}{a_L} \right) \frac{e^{-k^2 d_D^2/a_L^2}}{(1+R_L)^{1/4}} \cos \left( \frac{k d_H}{a_L} + \Delta \! \left( \frac{k d_T}{a_L} \right) \right) \right] \! , \label{F-def-2} \\
G(k) &=& - e^{-\tau_{\rm r}} \frac{\sqrt{3} e^{-k^2 d_D^2/a_L^2}}{5(1+R_L)^{3/4}}  \mathcal{S} \! \left( \frac{k d_T}{a_L} \right) \sin \left( \frac{k d_H}{a_L} + \Delta \! \left( \frac{k d_T}{a_L} \right) \right) \! . \label{G-def-2}
\eea
According to Dicus and Weinberg~\cite{Weinberg:2008zzc}, the matter transfer function $\mathcal T(\kappa)$ is well fitted to 2\% by
\be
\mathcal T(\kappa) \approx \frac{\ln [1+  (0.124 \kappa)^2 ]}{(0.124 \kappa)^2 } \left[ \frac{1 + (1.257 \kappa )^2 + (0.4452 \kappa)^4 + (0.2197 \kappa)^6 }{1 + (1.606 \kappa )^2 + (0.8568 \kappa)^4 + (0.3927 \kappa)^6} \right]^{1/2}, \label{T-matter-transfer}
\ee
while the functions $\Delta$ and $\mathcal S$ have fitting formulas of similar accuracy given by
\be
\mathcal S (\kappa) \approx \left[ \frac{ 1 + (1.209\kappa)^2 + (0.5116\kappa)^4 + \sqrt{5}(0.1657\kappa)^6}{1 + (0.9459\kappa)^2 + (0.4249\kappa)^4 + (0.1657\kappa)^6}\right]^2 \label{S-transfer}
\ee
\be
\Delta (\kappa) \approx \left[ \frac{(1.1547\kappa)^2 + (0.5986\kappa)^4 + (0.2578\kappa)^6}{1 + (1.723\kappa)^2 + (0.8707\kappa)^4 + (0.4581\kappa)^6 + (0.2204\kappa)^8} \right]^{1/2} \label{Del-transfer}.
\ee
When evaluating the relevant cosmological quantities to evaluate these functions, we will use (again, following Weinberg)
\begin{align}
d_T &= 0.1331 \, {\rm Mpc}, & d_D &= 0.008130  \, {\rm Mpc}, & d_H &= 0.1351 \, {\rm Mpc}, \\
a_L &= 1100, &  R_L &= 0.6234, &  \tau_{\rm r} &=  0.1115.
\end{align}
While these results are outdated because Planck nowadays implies different values~\cite{Aghanim:2018eyx} for these quantities, the deviations on observable quantities are of the same order of accuracy than those of the fitting functions to the transfer functions~\eqref{T-matter-transfer},~\eqref{S-transfer},~\eqref{Del-transfer} (typically of order $\leq 5\%$). Therefore, we may proceed to give preliminary constraints from 1-point statistics in this simplified approach consistently, and compare with current constraints on non-Gaussianity.

Using the transfer function $T$, it follows that $\langle \Theta_{{\bf k}_1 ... {\bf k}_n}^n \rangle_c =   T(k_1 , \mu_1) \cdot \cdot \cdot T(k_n , \mu_n) \langle \zeta_{{\bf k}_1 ... {\bf k}_n}^n \rangle_c$, from which we write the connected $n$th moment:
\be
 \langle \Theta_L^n \rangle_c =  (2\pi)^3  h_n \left( \sigma_\Theta / \sigma_\zeta \right)^n  I^T_n (\xi ) , \label{n-point-temp-x-space}
\ee
where $h_n$ is defined in terms of the potential in~\eqref{h-n-V} or~\eqref{h-n-V-2}, and $I^T_n (\xi )$ is given by
\bea
I^T_n(\xi) \! &=& \! \frac{n  }{2 (2 \pi^2)^{n+1}} \! \int_{-1}^{+1} \!\!\!\!\!\!\! d\mu \! \int_0^{\infty} \!\! \frac{dx}{x} G^T_\xi ( x , \mu) \! \left[ F^T_\xi( x , \mu) \right]^{n-1} \!\! , \,\,  \quad \label{I_n-integral-mu} \\
 G^T_\xi(x,\mu) \! &=& \! \frac{\sigma_\zeta}{\sigma_\Theta} \sum_{\ell } (2 \ell + 1) P_\ell ( \mu ) \!\! \int_{\xi^{-1}}^1 \!\!\!\!\!\! d z z^2 x^3  T_\ell ( z k_L ) j_{\ell}(z x) , \qquad  \label{G-T} \\
 F^T_\xi(x,\mu) \! &=& \! \frac{\sigma_\zeta}{\sigma_\Theta} \sum_{\ell } (2 \ell + 1)  P_\ell ( \mu ) \!\! \int_{\xi^{-1}}^1 \!\!\! \frac{d y }{y}  T_\ell ( y k_L ) j_{\ell}(y x) . \qquad  \label{F-T}
\eea
In the previous expressions, $P_\ell(x)$ and $j_\ell (x)$ stand for the $\ell$th Legendre polynomial and $\ell$th spherical Bessel function, respectively.  In addition, $T_\ell$ is the Legendre moment of $T(k, \mu )$. The variance of $\Theta$, in the Gaussian theory, is found to be $\sigma_\Theta^2 =  \frac{1}{4 \pi} \sum_{\ell }  (2\ell + 1) C_\ell $, with 
\be
C_\ell = \frac{4 \pi  \sigma_\zeta^2}{(2\ell + 1)^2 \ln \xi} \int^{k_L}_{k_{\rm IR}} \frac{dk}{k}  |T_\ell (k)|^2.
\ee

\begin{figure}[t!]
\begin{center}
\includegraphics[scale=1]{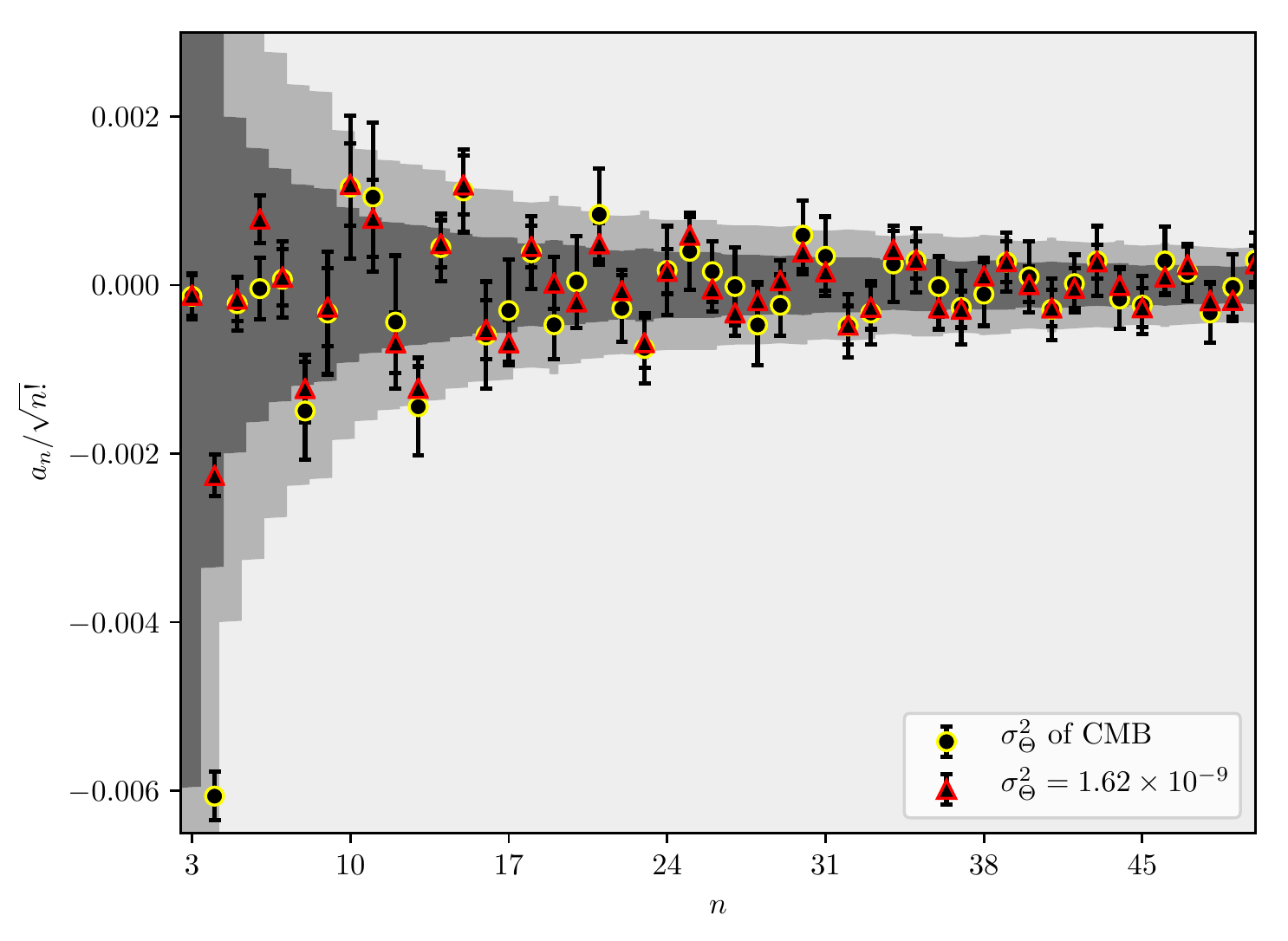}
\caption{The $a_n$ coefficients obtained from Planck. We have limited the data to regions far enough from the galactic plane so that the outcome from SMICA agrees with the other pipelines (and removing each pipeline's masked pixels), effectively considering a third of the sky. The error bars are an estimate of the noise present in the data, computed by comparing half-mission maps. The grey contours represent the intrinsic noise $\sigma(a_n)$ due to Gaussian simulations obtained using full-sky maps generated with CAMB.}
\label{fig:a_n-Planck}
\end{center}
\end{figure}

One can now derive a PDF $\rho (\Theta)$ for $\Theta$ similar to that of Eqs.~(\ref{main-result-1}) and (\ref{main-result-2}). 
\begin{align}
\rho (\zeta) =&\frac{1}{\sqrt{2\pi}\sigma_\zeta } e^{-\frac{\zeta^2}{2\sigma_\zeta^2}}  \left[ 1 + \Delta_T(\zeta) \right],  \label{main-result-1-T} \\
 \Delta_T (\zeta) \equiv& \! \! \int_{-1}^1 d\mu  \int_0^{\infty} \frac{dx}{x} \mathcal{K}_T(x,\mu) \! \int_{-\infty}^{\infty} \!\!\!\! d \bar \zeta \,\, \frac{\exp \Big[{-\frac{ \left(\bar \zeta - \zeta_T(x,\mu) \right)^2}{2\sigma_{\zeta T}^2 (x,\mu )} } \Big] }{\sqrt{2\pi  } \sigma_{\zeta T} (x,\mu)} \frac{\Delta N}{3 H^4}  \left( \sigma_\zeta^2 \frac{\partial^2}{\partial {\bar \zeta}^2} - {\bar \zeta} \frac{\partial}{\partial {\bar \zeta}} \right) \Delta V_{\text{ren}}  \! \left(\psi_{\bar \zeta}\right), \label{main-result-2-T}
\end{align}
where
\bea
\zeta_T(x,\mu) &\equiv&   [ F^T_\xi(x,\mu) /\ln \xi ]  \zeta, \\
\sigma_{\zeta T}^2 (x,\mu) &\equiv&  \sigma_{\zeta}^2 (1 - [ F^T_\xi(x,\mu) /\ln \xi]^2 ), \\
\mathcal{K}_T(x,\mu) &\equiv& 2\pi G^T_\xi(x,\mu) / F^T_\xi(x,\mu), \\
\psi_{\zeta} &\equiv& (\alpha \Delta N/ H)^{-1} \zeta. 
\eea

\begin{figure}[t!]
\begin{center}
\includegraphics[scale=1]{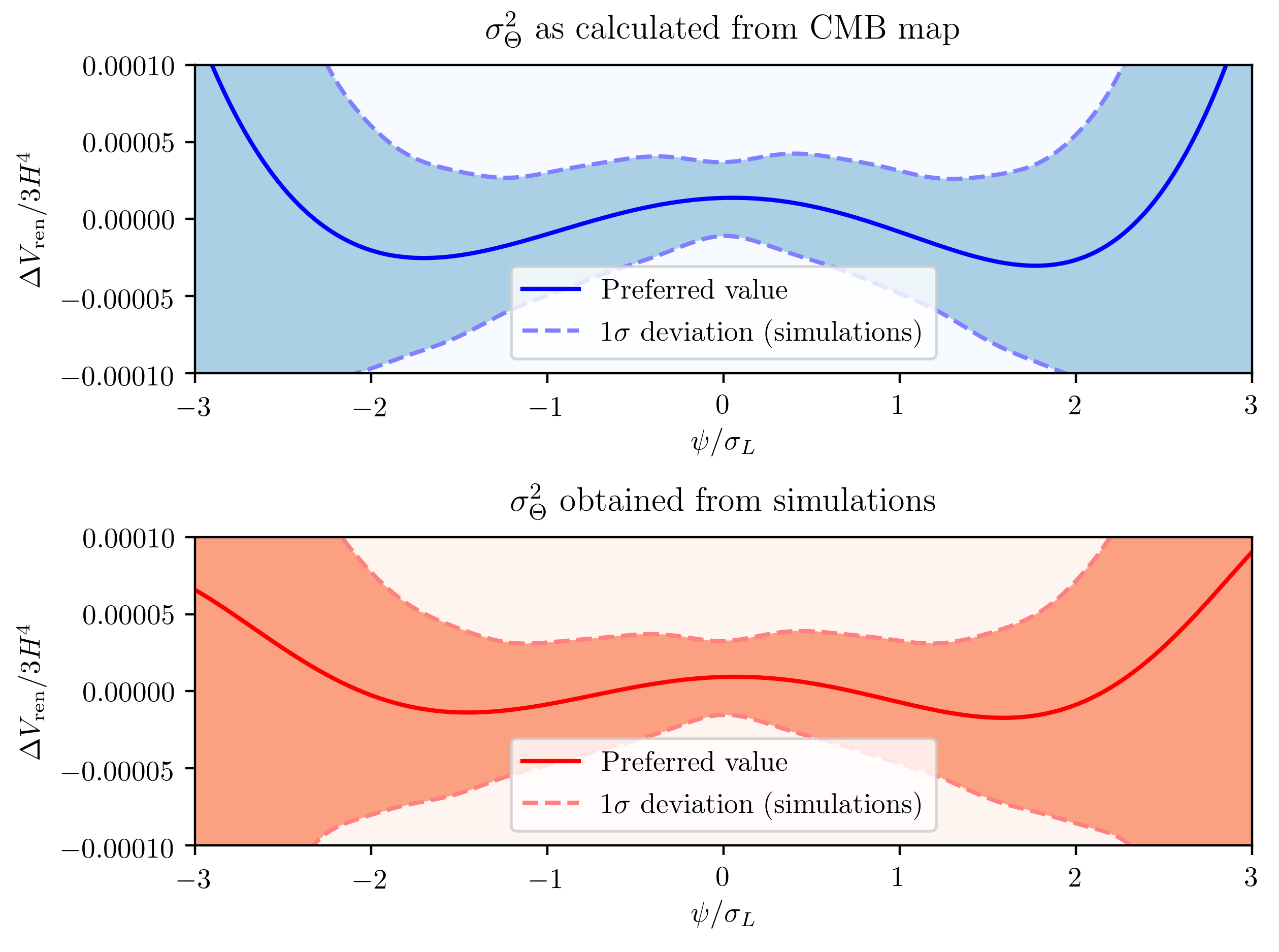}
\caption{The reconstructed potential $\Delta V/3 H^4$ for two different values of $\sigma_\Theta^2$. This reconstruction considers $a_n$ coefficients up to $n=7$. Since $a_2$ is a correction to the 2-point function --and hence, to the propagator-- we do not include this term in the reconstructed potential.}
\label{fig:DeltaV-reconstructed}
\end{center}
\end{figure}

However, the explicit expression for $\rho (\Theta)$ is not necessary to engage in reconstructing $\Delta V$. Equivalently, we may define the following cumulants parametrizing NG:
\be
a_n  \equiv    \int d\Theta \; \rho(\Theta) \, {\rm He}_n\left( \Theta / \sigma_\Theta \right)  . \label{pdf-a_n}
\ee
Independently of the form of $\rho(\Theta) $, these coefficients are directly related to the fully connected moments of $\Theta$ through the relation $\langle \Theta_L^n \rangle_c = \sigma_\Theta^n a_n$. Together with (\ref{n-point-temp-x-space}), this further implies
\be
h_n= a_n \sigma_\zeta^{n}  / [ (2\pi)^3  I^T_n (\xi )] . \label{a-n-h-n}
\ee
Then, by expanding the potential in terms of Hermite polynomials 
\be
\Delta V_{\text{ren}}(\psi)/H^4 = \sum_{m} \frac{b_m}{m!} \, {\rm He}_m \! \left( \psi / \sigma_L \right),
\ee 
one finds that the coefficients $b_n$ determining the shape of the potential are given by
\be
 b_n  = - \frac{3 a_n }{(2\pi)^3 \Delta N I^T_n(\xi)} \left( \frac {\ln \xi}{2 \pi^2}  \right)^n . \label{coeff-b_n}
\ee
The potential $\Delta V_{\text{ren}}(\psi)$ obtained by such a reconstruction has renormalized coefficients $c_{m}^{\rm ren}$ evaluated at the scale $k_{L}$, and so it can be interpreted as the potential generating NG in the range $k_{\rm IR} \leq k \leq k_L$.

Having Eq.~(\ref{coeff-b_n}) at hand, we may proceed to outline the reconstruction process. Figure~\ref{fig:a_n-Planck} shows values of the coefficients $a_n$ acquired from Planck CMB maps (see also Ref.~\cite{Buchert:2017uup} for a similar analysis). The coefficients were obtained by counting the occurrences of $\Theta$ values in Planck's SMICA temperature map. Here we chose two possible values for $\sigma_\Theta^2$: the sample variance computed from the CMB map $\sigma_\Theta^2 = 1.50 \times 10^{-9}$, with which $a_2 = 0$, and the one preferred by simulations $\sigma_\Theta^2 = 1.62 \times 10^{-9}$. The grey contours show the intrinsic noise $\sigma(a_n)$ (1- and 2-$\sigma$ regions) resulting from 500 Gaussian simulations using CAMB~\cite{CAMB-web} with the cosmological parameters reported by Planck~\cite{Ade:2015xua} (with a beam resolution of $5$ arcmin FWHM), and $\sigma_\Theta^2 = 1.62 \times 10^{-9}$, which is the average over simulations of the sample variances. As one might have expected, the observed values are mostly compatible with a Gaussian distribution. To get the $b_n$ coefficients via Eq.~(\ref{coeff-b_n}), we set $\Delta N = 60$ and fix $\ln \xi = 8$, which corresponds to the range of momenta $10^{-4}\;{\rm Mpc}^{-1} \leq k \leq 0.3\; {\rm Mpc}^{-1}$ for the observed modes in the CMB~\cite{Ade:2015lrj, Ade:2015hxq}. Given that Fig.~\ref{fig:a_n-Planck} lacks a conclusive imprint of non-Gaussianity, the potential in Fig.~\ref{fig:DeltaV-reconstructed} serves for illustrative purposes only. However, we must note that this type of analysis is cosmic variance limited, as evidenced by the different results obtained from the two values chosen for $\sigma_\Theta^2$. Additionally, there are a number of anomalies present in the CMB that we disregard herein, such as the statistical differences between the north and south hemispheres~\cite{Ade:2015hxq}. Nevertheless, we encourage the community to keep an eye out for these signatures, as well as to perform more sophisticated analyses with available data sets. For instance, one approach to try and circumvent the aforementioned effects is to compute the transfer functions for a restricted multipole range, which can be done by modifying accordingly the sums in Eqs.~\eqref{G-T} and~\eqref{F-T}, then to consider a filtered CMB map that only contains those contributions, and finally use Eq.~\eqref{coeff-b_n} as before to obtain the reconstructed potential. 

\subsection{2-point PDFs and scale dependence}

It is interesting to write down expressions for quantities that are not typically used directly when characterizing cosmological datasets. For instance, one rarely bothers to write down the full 2-point PDF for the CMB Temperature map, as all of its information (in the Gaussian case) is already specified through the power spectrum. However, the scale dependence of this PDF may be a useful tool to probe for non-Gaussianities, and as such, we deem it worth writing down. The result, analogously to what we had in the simpler case of a spatial window function, is
\be \label{two-point-th}
\begin{split}
\rho_\Theta (\Theta_1,\Theta_2) = &\rho_{G,\Theta}(\Theta_1,\Theta_2)   \Bigg[ 1 - \int_\x \int_{-\infty}^{\infty} \!\!\!\! d \bar \zeta   \frac{\exp \Big[{-\frac{ \left(\bar \zeta - \zeta_\Theta(\x,\n_1,\n_2) \right)^2}{2\sigma_\Theta^2 (\x,\n_1,\n_2)} } \Big] }{\sqrt{2\pi  } \sigma_\Theta (\x,\n_1,\n_2)}  \\ & \;\times   \left\{  \frac{W_\Theta(\x, \n_1)}{s_\Theta^2(\x , \n_1)} \!  \left( G_{11}^{\Theta} \frac{\partial}{\partial {\bar \zeta}} - G_{12}^{\Theta} \right)      +  \frac{W_\Theta(\x, \n_2)}{s_\Theta^2(\x,  \n_2)} \!  \left( G_{21}^{\Theta} \frac{\partial}{\partial {\bar \zeta}} - G_{22}^{\Theta} \right) \right\} F  \! \left({\bar \zeta}\right)  \Bigg].
\end{split}
\ee
As before, let us go through this expression: the first thing to notice is that the two filtering points $\x_1$ and $\x_2$ in~\eqref{two-point} are now replaced by two directions in the sky, $\n_1$ and $\n_2$. Secondly, the Gaussian that is convoluted with $F$ now has a different mean and variance, obtained by conditioning the joint multivariate Gaussian PDF on the values of $\Theta_1$ and $\Theta_2$. 
The starting point to this is a joint Gaussian distribution for $({\bar \zeta}, \Theta_1, \Theta_2)$ with covariance matrix
\be \label{cov-th}
  \pmb{\Sigma}=
  \left[ {\begin{array}{ccc}
   \sigma_\zeta^2 & s_\Theta^2(\x, \n_1) & s_\Theta^2(\x , \n_2) \\
   s_\Theta^2(\x, \n_1) & \sigma_\Theta^2 & \sigma_{\Theta,{\rm ext}}^2(\n_1, \n_2)\\
   s_\Theta^2(\x , \n_2) & \sigma_{\Theta,{\rm ext}}^2(\n_1, \n_2) & \sigma_\Theta^2 \\
  \end{array} } \right],
\ee
where we have written the covariance between the two externally chosen points $\n_1$ and $\n_2$ as
\be
\sigma_{\Theta,{\rm ext}}^2(\n_1, \n_2) = \int_k T(\n_1,\k) T(\n_2,\k)  P_\zeta(k) = \Sigma(\n_1, \n_2),
\ee
and $\sigma_\Theta^2 = \sigma_{\Theta,{\rm ext}}^2(\n, \n)$, which is independent of the direction $\n$. The precise definitions of all the additional functions involved in this section are given in App.~\ref{sec:2-point-det}. It is worth mentioning, as a reminder to the reader, that, as in the 2-point PDF for curvature fluctuations~\eqref{two-point}, both~\eqref{two-point-th} (besides from the temperature variables) and $\Sigma(\n_1,\n_2)$ depend only on the angular distance between $\n_1$ and $\n_2$, or equivalently, on $\n_1 \cdot \n_2$, and that this is a consequence of our universe's homogeneity.

    From this function, \emph{i.e.}, from the 2-point PDF~\eqref{two-point-th}, it is possible to obtain refined constraints on the local ansatz. Given a dataset, one can construct the 2-point PDF as follows: divide the angular distance into $N$ bins of width $\delta\vartheta$ and the temperature in $M \times M$ bins of size $\delta\Theta \times \delta\Theta$, as in the 1-point PDF but now with two axes for the temperature field. Now, for each bin associated to a given angular distance $\vartheta_n = n \cdot \delta\vartheta$, and for each value of $(i,j)$, count how many pairs of pixels separated by that angular distance $\vartheta_n$ have the values $(\Theta_{[i]}, \Theta_{[j]})$ for the temperature in their respective positions. This process would generate $N$ two-dimensional histograms, with two temperature axes, which we label by $(\Theta_1,\Theta_2)$, whose value at coordinate $({\Theta_1}_{[i]}, {\Theta_2}_{[j]})$ would give the number of pairs of pixels with temperatures in the $(i,j)$th bin, separated by angular distance in the $n$th bin. A $1/2$ symmetry factor must be included in the number counts for temperature bins with $i\neq j$, as the bin $({\Theta_1}_{[i]}, {\Theta_2}_{[j]})$ is equivalent to $({\Theta_1}_{[j]}, {\Theta_2}_{[i]})$.
    
    How does this give refined constraints on the local ansatz? Let us appreciate that this set of PDFs contains information on the scale, or more concretely, on the temperature power spectrum and of its expansion in spherical harmonics (the standard $C_\ell$s) through the 2-point correlation $\Sigma(\n_1,\n_2)$. If NG is absent, then at each value of the angular distance the 2-point PDF will be a 2-variable Gaussian probability density with variances $\sigma_\Theta^2$ and covariance $\sigma_{\Theta, {\rm ext} }^2(\n_1,\n_2) = \Sigma(\n_1,\n_2)$. Then, in the presence of NG, each 2-point PDF will undergo a NG deviation induced by the same primordial mechanism $F$, but for each angular separation this deviation will be experienced differently because the covariance matrix implied by the Gaussian part is different. That means that for each angular distance, the kernel that acts upon $F$ in~\eqref{two-point-th} gives a different deviation from Gaussianity, and therefore, each of the 2-point PDFs gives an independent estimator on the primordial NG field. For local NG, all of the $N$ 2-point PDFs at different angular separations in the sky should give consistent\footnote{That is, within the experiment's theoretical and systematic uncertainties included.} constraints/estimations of $F$. Conversely, if NG is measured and it does not adjust to the statistics implied by~\eqref{two-point-th} at different angular scales, then purely local NG would be ruled out. Therefore, looking towards possible future directions to be explored, this type of object (a set of 2-point PDFs) shows promise to disentangle different shapes of NG, such as equilateral or orthogonal templates, and in particular, from the local ansatz.

  In order to search for non-Gaussianity within a 2-point PDF, many approaches are possible. Given a model, \emph{i.e.}, an explicit expression for $F$, and using it as a template with few adjustable parameters is usually the method that will give the best constraints. In the spirit of Eq.~\eqref{F-herm}, however, another one is worth mentioning: one can use bivariate Hermite polynomials on the temperature variables $(\Theta_1, \Theta_2)$, so as to express the PDF in terms of a bivariate Edgeworth expansion~\cite{Sellentin:2017aii,WITHERS2000165}. One then looks for any statistically significant nonzero coefficient in the expansion, in analogy to what was done for the 1-point case in~\cite{Chen:2018brw}. This way, the existence of NG can be tested as a yes/no question, as any nonzero coefficient in an Edgeworth expansion implies a non-Gaussian distribution. This may be particularly useful when searching for NG in the next generation CMB surveys~\cite{Abazajian:2016yjj}.

\section{Prospects for Large-Scale Structure Surveys}

%LSS: $10^7$ more modes than CMB, gained because projection effects onto the sphere make the CMB mix up modes that are in principle independent of each other.

Upcoming cosmological surveys will focus on the statistics of LSS and 21 cm signals, promising to bring precision cosmology to a new era. Indeed, the proliferation of observed modes due to the three dimensional probe offered by LSS will enhance our statistics, albeit with an increased noise level, giving us invaluable information about the fundamental aspects of the early/late universe. 

At small scales several sources of nonlinearity induce NG, like gravitational interactions and galaxy bias, obscuring the primordial contribution to the statistics. However, for relatively long modes, $k\lesssim 10\;$Mpc$^{-1}$ and higher redshifts~\cite{Sefusatti:2007ih}, linear perturbation theory can be trusted, which makes it easier to identify primordial signatures. Perturbative techniques pushing our analytic control towards smaller, weakly nonlinear scales include several schemes like SPT~\cite{Bernardeau:2001qr} and more recently EFTofLSS~\cite{Carrasco:2012cv} and TSPT~\cite{Blas:2015qsi,Blas:2016sfa,Vasudevan:2019ewf}, which are set in a hydrodynamics framework, while going even further requires full Boltzmann solvers via N-body simulations. In this work, we will focus on the purely linear regime, leaving weakly nonlinear evolution with NG initial conditions~\cite{Vasudevan:2019ewf} for future study.

The main probes of non-Gaussianity are the bispectrum and/or trispecturm, number counts and bias. The spectra retain information about the shape of the 3- and 4-point functions in momentum space, which can be linked to the mechanism responsible for generating NG. 
Number counts probe directly the 1-point PDF, which even though loses the shape information, it serves as a complementary and equally powerful estimator of NG. Finally, the halo bias serves as a third independent chanel, which can give clear enhanced signatures of local NG at large scales.
  
The moments of the matter distribution have been computed in several schemes like SPT~\cite{Bernardeau:2001qr} and more recently EFT of LSS~\cite{Carrasco:2012cv} order by order in perturbation theory, starting from a hydrodynamical description of the matter density. More recently, the TSPT method~\cite{Blas:2015qsi,Blas:2016sfa} has focused on constructing a time dependent probability distribution function of the matter perturbation from which moments can be computed. 

In this section, we wish to track how primordial NG, in the form of the generic local ansatz~\eqref{general-asatz-restricted}, gets transmitted to the matter field in the linear regime. We extend our result, the non-Gaussian PDF of curvature fluctuations, in two directions: 1) we deduce a PDF for the matter density constrast $\delta$, and hence, a halo mass distribution; 2) we compute the effect of the tomographic local ansatz on the halo bias. These are complementary probes of the non-Gaussian initial condition via number counts and power spectra, respectively, which should be accessible by surveys such as {\sc Lsst}, {\sc Euclid} and {\sc Ska}.

\subsection{Halo Mass Function} \label{sec:halo}

The matter overdensity $\delta(\x) = \delta \rho(\x)/\bar{\rho}$, with $\bar{\rho} =\Omega_m\rho_{\rm cr}$, is related to the primordial Newtonian potential, $\Phi=\frac{3}{5} \zeta$, as
\be \label{poisson}
\delta(\k) = \alpha(k) \Phi(\k), \qquad \alpha(k)=\frac{2r_H^2k^2T(k)D(z)}{3\Omega_m},
\ee
where $D(z)$ is the linear growth rate, $r_H$ the current Hubble radius and $T(k)$ the transfer function~\cite{Eisenstein:1997ik}. 
We smooth the density field over a radius $R_M=\left(3M/4\pi\bar{\rho}\right)^{1/3}$ as in Eq.~\eqref{z-filtered}, 
\be \label{d-filtered}
\delta_W (\x) = \int_\k \int_\y e^{i{\k} \cdot (\x - \y)} W_M(k) \delta(\k),
\ee
using a top-hat filter $W_M(x)=V_M^{-1}H(R_M-r)$, with $H$ the Heaviside function and $V_M$ the volume of the region within the radius $R_M$. 
The probability distribution for the smoothed overdensity, $\rho(\delta_W)$, is then given by the 1-point PDF of Eq.~\eqref{one-point}, upon the replacements
\be \label{z-to-d}
\zeta\to\delta \quad \text{and} \quad W(k)=\frac35W_M(k)\alpha(k).
\ee

Now, in principle, having the matter distribution function, one can compute the halo number density, that is, the number density of halos of mass between $M$ and $M+ dM$ at redshift $z$, and the observables derived from it. One way to do this is via the Press-Schechter (PS) scheme~\cite{Press:1973iz} extended to the NG case~\cite{Lucchin:1987yv,Chiu:1997xb}. Denoting the tail distribution above some threshold value $\delta_c$ as
\be  \label{mu}
\mu^>(M,z)=\int_{\nu_c(z)}^\infty d \nu \; \rho(\nu),
\ee
with $\nu\equiv\delta/\sigma_W$, then the total fraction of mass collapsed into bound structures will be proportional to this cumulative PDF:$$\frac{n(M)M}{\bar\rho}\propto\mu^>(M).$$ However, as mentioned, we are interested in the number density of objects, \emph{i.e.} clusters, with mass between $M$ and $M+d M.$ This is given by $n(M+d M)-n(M)$ and hence the PS mass function reads\footnote{The fudge factor $2$ corrects for the \emph{cloud-in-cloud} problem, that is, a collapsed object of mass $M_1$ tracing a volume $V_1$ can be part of an object of mass $M_2>M_1$ tracing a larger volume $V_2$ and thus should not be counted as a separate object. The Gaussian value $2$ is a good approximation in case of a small NG deformation~\cite{Matarrese:2000iz}.}
\be  \label{F_PS}
\frac{d n_{\rm PS} }{d M}(M,z)=-2\frac{\bar\rho}{M} \frac{d \mu^>}{d M}(M,z).
\ee
The collapse threshold, through which the $z$ dependence arises, is given by the spherical model as $\delta_c(z)\simeq 1.686\;D(0)/D(z)$. The Gaussian PS function can be evaluated exactly by replacing $\rho\to\rho_{\rm G}$ in Eq.~\eqref{mu}:
\be  \label{F_PS_G}
\frac{d n_{\rm PS}}{d M}\Big|_{\rm G}(M,z)%=-\sqrt\frac2\pi  e^{-\frac{\delta_c^2(z)}{2\sigma_{W}^2}} \frac{\bar\rho}M \frac{\delta_c(z)}{\sigma_W} ({\ln \sigma_W})'
= -2\frac{\bar\rho}M\frac{e^{-\frac{\nu_c^2(z)}{2}}}{\sqrt{2\pi}}  \nu_c(z) (\ln \sigma_W)',
\ee
where a prime stands for the derivative with respect to the mass. For the NG case, the PS mass function is easy to compute from Eq.~\eqref{one-point}:
\be \label{mass-function-PS-F}
\begin{split}
\frac{d n_{\rm PS}}{d M}(M,z) = \frac{d n_{\rm PS}}{d M}\Big|_G (M,z)\left[1-\frac1{\delta_c}\left(1-\nu_c^2-\frac1{(\ln\sigma_W)'} \frac{d }{d M}\right)\int_\x W(x)F\left( {\rm b}\delta_c;{\rm b}^2\sigma_W^2 \right) \right] ,
\end{split}
\ee
where $F\left( {\rm b}\delta_c;{\rm b}^2\sigma_W^2 \right)$ is the Weierstrass transform of the local ansatz [see Eq.~\eqref{W-effect}] evaluated at the threshold $\delta_c(z)$. 
Upon using the Hermite expansion~\eqref{F-herm} of the function $F$, we may obtain a series representation of the mass function~\eqref{mass-function-PS-F}:
\be \label{mass-function-PS-H}
\begin{split}
\frac{d n}{d M}(M,z) = \frac{d n_{\rm PS}}{d M}\Big|_G(M,z)  \left[1+\Delta\left(\nu_c(z)\right)  - \frac{1}{d\nu_c(z)/d M} \sum_{n=2}^\infty \frac{{\rm He}_{n}\left(\nu_c(z)\right)}{(n+1)!} \kappa_{n+1}' \right] ,
\end{split}
\ee
where we have defined the reduced cumulants $\kappa_{n}(M)\equiv \frac{\langle \delta_W^{n} \rangle_c}{\sigma_W^{n}}$. 
%The Gaussian PS function can be evaluated exactly by replacing $\rho\to\rho_{\rm G}$ in Eq.~\eqref{mu}:
%%
%\be  \label{F_PS_G}
%\frac{d \mu^>}{d M}\Big|_{\rm G}(M,z)
%= \frac{e^{-\frac{\nu_c^2(z)}{2}}}{\sqrt{2\pi}}  \nu_c(z) (\ln \sigma_W)',
%\ee
%% 
%where a prime stands for the derivative with respect to the mass.
%where we have used the relation ${\rm b}(R_M)-1=M(\ln \sigma_W)'$, with a prime standing for the derivative with respect to the mass. The bias ${\rm b}(x)$ is defined in Eq.~\eqref{b-1-defs-t}, with the replacements~\eqref{z-to-d} imposed. In this context it is the ratio of the matter-halo to halo-halo correlation functions evaluated at the smoothing scale. 
%Equation~\eqref{mass-function-PS} offers a generalisation of the $f_{\rm NL}$, $g_{\rm NL}$ truncation of the local ansatz (see \emph{e.g.} Refs.~\cite{LoVerde:2007ri,LoVerde:2011iz}) to arbitrary functions $F$. Evidently, positive moments of the PDF ($\kappa_n,\kappa'_n>0$) lead to overabundance of collapsed objects at the high mass end, where $\nu,\nu'\gg1$, as long as $(\ln \kappa_{n+1})'<\nu\nu'$, which is satisfied since the cumulants $\kappa$ depend weakly on the mass~\cite{LoVerde:2011iz}.
Equation~\eqref{mass-function-PS-F} offers a generalisation of the $f_{\rm NL}$, $g_{\rm NL}$ truncation\footnote{For example, when truncated to $n=2$, the expansion~\eqref{mass-function-PS-H} agrees with Eq.~(4.19) of Ref.~\cite{LoVerde:2007ri}.} of the local ansatz (see \emph{e.g.} Refs.~\cite{LoVerde:2007ri,LoVerde:2011iz}) to arbitrary functions $F$. From its moment expansion~\eqref{mass-function-PS-H}, we can see that positive moments of the PDF ($\kappa_n,\kappa'_n>0$) lead to overabundance of collapsed objects at the high mass end, where $\nu,\nu'\gg1$, as long as $(\ln \kappa_{n+1})'<\nu\nu'$, which is satisfied since the cumulants $\kappa$ depend weakly on the mass~\cite{LoVerde:2011iz}.

However, due to the highly nonlinear character of the collapse, one cannot fully parametrize collapsed objects by a single threshold number $\delta_c$. Indeed, it has been shown that the PS prescription does not accurately estimate the halo abundance even in the Gaussian case (for example a small non-spherical perturbation can have a considerable impact)~\cite{Sheth:1999mn}. Hence, the extension to the NG case is guaranteed to also have errors with respect to simulations. What can be done though is to characterise the deviation from Gaussianity by comparing the ratio of G-to-NG densities with that of the PS scheme~\cite{LoVerde:2007ri,Robinson:1998dx}, since the latter is expected to fail equally in both cases: 
\be \label{mass-function}
\frac{d n }{d M} (M,z) = r_G(M,z)\frac{d n_{\rm PS}}{d M}(M,z),\quad\text{with}\quad r_G(M,z)=\dfrac{\frac{d n }{d M}\big|_{\rm G}(M,z)}{\frac{d n_{\rm PS}}{d M}\big|_{\rm G}(M,z)}.
\ee
For the Gaussian mass function, $\frac{d n}{d M}\big|_{\rm G}$, we can adopt a Sheth-Tormen (ST) ansatz~\cite{Sheth:1999mn}, which is better fitted to simulations than the Press-Schechter one, in which case the Gaussian ratio reads
\be \label{gauss-ratio-ST}
r_G[\nu_c(z)]=\sqrt{a}A\left(1+(a\nu_c^2)^{-p}\right)e^{\frac{\nu^2_c}2(1-a)},
\ee
with the ST parameters $a = 0.707, \;A = 0.322184,\; p = 0.3$.

With the mass function~\eqref{mass-function} at hand, which is just Eq.~\eqref{mass-function-PS-H} with the replacement $n_{\rm PS}|_{\rm G} \to  n_{\rm ST}$, we may compute the number of clusters per redshift bin above some mass $M$ as~\cite{LoVerde:2007ri}
\be \label{cluster-counts}
\frac{d N}{d z}(M,z) = \frac{4\pi}{H(z)}\left(\int\frac{d z}{H(z)}\right)^2 f_{sky} \int_{M}^\infty d m \frac{d n}{d m}(m,z),
\ee
where $f_{sky}$ is the fraction of the sky covered by each survey. 

This result, together with the mass function~\eqref{mass-function}, can serve as a template for the number density of clusters. For example, in the context of light isocurvature axions one can compute the non-Gaussian PDF deformation $\Delta$~\cite{Chen:2018uul} and thus the mass function $\frac{d n}{d M}$ and create mock data via simulations. One can then pick an estimator and devise an overlap between the template and the data, in exactly the same manner that one uses the cosine estimator to measure the overlap of the local template with the actual bispectrum. The simplest thing to do is to take the number counts as an estimator. For example, given a model, one can count how many clusters of mass $M$ exist at redshift $z$ in the mock data and compare this number to the real data.

We may also try to reconstruct the PDF from data using other statistical estimators. This can be done by simply solving Eq.~\eqref{mass-function}, together with \eqref{F_PS}, as an ODE for the cumulative PDF, $\mu^>$, to get 
\be \label{mu-dn}
\mu^>[\delta_c(z),\sigma_W]= \frac{1 }{2\bar\rho} \int_M^\infty d m\; \frac{m}{r_G[\nu_c(z)]}  \frac{d n_{\rm }}{d m}(m,z),
\ee
with $r_G$ given by Eq.~\eqref{gauss-ratio-ST}. The left hand side now gives the tail distribution of $\delta$, that is, the probability of $\delta>\delta_c$ at redshift $z$. On the right hand side we have the total mass of collapsed objects with $m>M$ at redshift $z$, accounting for corrections to the spherical collapse model via assigning a mass value $m(z)= m_{\rm obs}/r_G[\nu_c(z)]$ to an object of observed mass $m_{\rm obs}$ and multiplicity $\frac{d n_{\rm }}{d m}(m_{obs},z)$; in principle, this can be deduced from number counts. One can now apply statistical estimators  like Minkowski functionals to the dataset $\left\{\frac{m_{\rm obs}}{r_G[\nu_c(z)]},\frac{d n_{\rm }}{d m}(m_{obs},z)\right\}$, whose difference from their Gaussian estimate, as in the CMB~\cite{Ade:2015hxq,Buchert:2017uup}, will be a direct probe the NG deformation of the primordial PDF.  Let us note, however, that measuring accurately the mass and redshift of clusters is a hard task, which might complicate a direct connection between cluster counts and primordial NG in this manner.

To summarise, in the context of tomographic NG, one can 1) use Eqs.~\eqref{mass-function-PS-H} and~\eqref{mass-function} as a template for the mass function and thus the number density up to arbitrary order in the Edgeworth expansion; 2) assume a microphysical model, compute the PDF (as in \emph{e.g.} Ref.~\cite{Chen:2018uul}) and use Eqs.~\eqref{mass-function-PS-F} and~\eqref{mass-function} as a template without having to refer to moments; and finally, 3) consider statistical estimators on the LSS dataset probing the primordial PDF via the counting scheme implied by Eq.~\eqref{mu-dn}.

\subsection{Halo bias}

Another powerful probe of primordial non-Gaussianity is the halo bias, which enters in the late-time power spectra. As already argued, NG initial conditions alter the halo abundance in a nontrivial way by increasing the number of rare density peaks that collapse to halos. This is easier to understand in the case of the quadratic local ansatz $\delta=\delta_G+f_{\rm NL}  \delta_G^2$~\cite{Dalal:2007cu}: a positive $f_{\rm NL}$ adds positive skewness to the density distribution; thus, the same probability now corresponds to higher values of $\delta$ with respect to the Gaussian field, leading to more probable enhanced peaks. In~\cite{Dalal:2007cu}, it was shown how this is encoded in a scale dependent correction to the halo bias given by
\be \label{bias-fnl-shift}
\Delta b(k) = 2(b-1)\delta_c\frac{f_{\rm NL}}{\alpha(k)},
\ee
where $b$ is the Eulerian Gaussian bias.
The $k^2$ factor in $\alpha$ implies that the effects of non-Gaussianity will be accentuated in large scales (the transfer fucntion goes to 1 for $k\to 0$), which brings surveys like {\sc Ska} to the frontline of NG searches, promising $\sigma(f_{\rm NL})\sim 1$~\cite{Camera:2014bwa}.
This result was rederived and generalised in~\cite{Matarrese:2008nc,Slosar:2008hx} using different approaches and has been confirmed with N-body simulations (see e.g. Refs.~\cite{Baldauf:2015vio,PhysRevD.81.063530,Scoccimarro:2011pz,Desjacques:2008vf,Grossi:2008vf,Pillepich:2008ka}), while in Refs.~\cite{Smith:2011ub,LoVerde:2011iz}, the scale dependent halo bias was computed for the case of cubic $g_{\rm NL},\tau_{\rm NL}$-type local NG ---see Ref.~\cite{Desjacques:2016bnm} for a review. In what follows we extend it to the generalised local ansatz \eqref{general-asatz-restricted} and show how in the case of an isocurvature source, future surveys can probe the landscape potential via measurements of the bias factor.

To begin, let us rewrite the NG ansatz~\eqref{general-asatz-restricted} for the Newtonian potential $\Phi=\frac35 \zeta$ as 
\be \label{general-loc-Phi}
\Phi(\x) = \phi(\x) + \frac{3}{5} \F[\phi](\x),
\ee
where $\phi$ is a Gaussian random field with standard deviation $\sigma_0=\sqrt{\langle \phi^2 \rangle}$. The $3/5$ factor is put explicitly to relate the primordial curvature perturbation $\zeta$ to the late-time (matter-dominated era) gravitational potential $\Phi$.
Following the peak-background split method~\cite{Kaiser:1984sw,Bardeen:1985tr}, we now separate the Gaussian gravitational potential into long and short modes with respect to some characteristic halo scale $R_\star\sim R(M)$ as
\be \label{l/s-phi-split}
\phi = \phi_L + \phi_S,
\ee
which induces a similar split in the variance, $\sigma_0^2=\sigma_L^2+\sigma_S^2 = \langle \phi_L(\x)^2 \rangle + \langle \phi_S(\x)^2 \rangle$. With the help of the expansion in Hermite polynomials~\eqref{F-herm}, and using known identities of these polynomials, we obtain
\be \label{F-L-S-expansion-0}
F(\phi_L,\phi_S)  = \sum_{m=0}^\infty \frac{\beta_{m}\left(\phi_L \right)}{m!} \sigma_S^m \,  {\rm He}_m \!\left( \frac{\phi_S}{\sigma_S} \right) ,
\ee
where
\be \label{betas}
\beta_{m}\left(\phi_L \right) \equiv - \sum_{s=0}^\infty \frac{ f_{m+s}}{s!} \sigma_L^s \,  {\rm He}_s \!\left( \frac{\phi_L}{\sigma_L} \right),  %\;{\rm He}^{\sigma_L^2}_{s-m}\left(\phi_L \right) = \frac{\F^{(m)}(\phi_L)}{m!},
\ee
with $f_0 = f_1 = 0$. How is this expansion of $F$ useful? If one is mainly interested in the short-wavelength dynamics (for instance, to study the gravitational collapse of the matter distribution into galaxies), this expansion is useful to identify the functions $\beta_m$ of long-wavelength fluctuations as effective nonlinearity parameters for the short modes, as may be seen directly from~\eqref{F-L-S-expansion-0}: $m=1$ is a correction to the amplitude of the short modes, $m=2$ a nonlinearity of $f_{\rm NL}$ type, $m=3$ a nonlinearity of $g_{\rm NL}$ type, etc.
%where $\F^{(m)}\left(\phi_L \right) \equiv \frac{\partial^{m}}{\partial \phi^{m}} \F\left(\phi \right)\Big|_{\phi=\phi_L}$ and 
%with the binomial coefficient defined such that ${s \choose m}=0$, for $m>s.$

We now have to understand quantitatively how long wavelength density fluctuations affect the statistics of short modes and thus the halo number density per unit of halo mass, $d n/ d M$, which from now on we denote by $n_L({\x})$, through each term in the expansion~\eqref{F-L-S-expansion-0}. Generically, the halo mass function is a function of the matter contrast and the amplitude of the short modes:
\be
n_L[\rho({\x}),\Delta_{\phi_S}].
\ee
Firstly, because of the dependence on $\rho$, irrespective of the presence of non-Gaussianity, a long perturbation $\delta_L$ will induce a linear background shift in $n$ as 
\be 
n_L[\bar\rho(1+\delta_L({\x}))] = \bar n \left(1 + \frac{\partial \log n_L}{\partial \delta_L} \delta_L({\x}) \right),
\ee
where $\bar n = n_L[\bar\rho]$.
Moving to the NG part, the purely long wavelenght contributions ($m=0$ coefficient in the expansion \eqref{F-L-S-expansion-0}), by definition, will not affect the power spectra and hence the bias to first order in the nonlinearity parameters $f_m$, so we may disregard them. From a short-wavelength modes' perspective, they are constant numbers that will only affect the background density through $\delta_L$ in the previous expression. 

Next, we may observe that the coefficient of the term linear in $\phi_S$ ($m=1$) depends on $\phi_L$. Therefore, a long mode will induce a shift in $n_L({\x})$ through this term, since short modes feel a background perturbed by the local amplitude of the long wavelength perturbation as $\Delta_{\phi_S} \to \left(1+\beta_{1} \! \left(\phi_L \right)\right)\Delta_{\phi_S}$, leading to
\be 
n_L({\x})= \bar n \left(1 + \frac{\partial \log n_L}{\partial \delta_L} \delta_L({\x}) + \beta_{1}\left(\phi_L \right)\frac{\partial \log n_L}{\partial \log\Delta_{\phi_S} }  \right).
\ee

Finally, we need to take into account that the halo density will depend on the local amplitude of the long modes through all the functions $\beta_{m}\left(\phi_L \right)$ for all $m \geq 2$, since each $\beta_{m}$ acts as a nonlinearity parameter assuming a local value set by the long wavelength fluctuation $\phi_L$~\cite{Slosar:2008hx,Smith:2011ub}. That is, we should consider the halo density as a function of the form
\be 
n_L({\x}) = n_L[\rho({\x}); \{\beta_m\}_m].
\ee
where $\beta_1$ controls the amplitude of the short modes, and the rest control the amplitude of their nonlinearities. Thus, we can write
\be \label{n-shift}
n_L({\x}) = 
\bar n \left(1 + \frac{\partial \log n_L}{\partial \delta_L} \delta_L({\x}) + \sum_{m=1} \frac{\partial \log n_L}{\partial \beta_m} \beta_m \! \left(\phi_L (\x) \right)  \right).
\ee

From this expression we can now compute the linear bias, defined as
\be \label{bias-def}
b(k) = \frac{P_{mh}(k)}{P_{mm}(k)},
\ee
where the matter-halo power spectrum is defined as 
\be
P_{mh}(k)=\mathbb{F}_{\x - \y}[ \langle \delta_L(\x) n_L(\y) \rangle] (k),
\ee
and the matter-matter one as 
\be
P_{mm}(k)= \mathbb{F}_{\x - \y}[ \langle \delta_L(\x) \delta_L(\y) \rangle] (k)  =|\delta_k|^2. 
\ee 
We will thus need to compute the matter-halo correlator and Fourier transform it ($\mathbb{ F}$). The matter contrast $\delta$ is related ---in subhorizon scales--- to the Newtonian potential through the Poisson equation~\eqref{poisson},
so that $P_{\phi\delta}=P_{\delta\delta}/\alpha$.
The first term in Eq.~\eqref{n-shift} is trivial since it just yields the $\delta$ propagator, which in Fourier space cancels the denominator in Eq.~\eqref{bias-def}, resulting in the standard constant bias, while the second yields the scale dependent correction. Putting everything together, we get
%\frac{\mathbb{F}_{\x - \y}[ \langle \delta_L(\x) n_L(\y) \rangle] (k) }{\mathbb{F}_{\x - \y}[ \langle \delta_L(\x) \delta_L(\y) \rangle] (k) }
\be \label{linear-b-1}
b(k)  = b_G + \frac{1}{\alpha (k)} \sum_{m=1}^{\infty} \frac{\partial \log n_L}{\partial \beta_m} \langle \beta_m'(\phi_L) \rangle = b_G - \frac{1}{\alpha (k)} \sum_{m=1}^{\infty} \frac{\partial \log n_L}{\partial \beta_m} f_{m+1} .
\ee
%which is the same prediction as with $f_{NL}$ or $g_{NL}$. 
Evidently, the coefficient of the scale dependent correction contains a summation over all the nonlinearity parameters, or better put, the whole function $F$. In order to see this, we can write it down, equivalently, as
\be \label{linear-b-2}
 b(k) = b_G + \frac{1}{\alpha(k)} \int_{-\infty}^{\infty} \mathcal{N}(\phi_L;\sigma_L) \frac{d F(\phi_L;\sigma_L)}{d \phi_L} d \phi_L.
\ee
where $\mathcal{N} \equiv \frac{e^{-\phi_L^2/2\sigma_L^2}}{\sqrt{2\pi} \sigma_L} \sum_{m=1}^\infty \frac{\partial \log n_L}{\partial \beta_m} \frac{1}{\sigma_L^m} {\rm He}_m \! \left( \frac{\phi_L}{\sigma_L} \right) $. Consequently, a scale-dependent halo bias can only signal the presence of \textit{some} form of local NG but not of a specific parametrization of it.

%The functional derivative of $n_L$ w.r.t. the departure from Gaussianity associated to the long modes $F(\phi_L;\sigma_L)$ in Eq.~\eqref{linear-b-2}, or its derivatives w.r.t. $\beta_m$ in Eq.~\eqref{linear-b-1}, would have to be computed from simulations if one is to obtain information about $F$.
%Indeed, in this case one may compute the derivatives of the halo mass function, since one can run simulations with varying initial conditions. However, the coefficients~\eqref{betas} of the scale dependent bias cannot be observed in this form in real data, since they only come once and with the same initial condition. Therefore, the next step is to connect the effective bias coefficient $\int \frac{\delta n_L}{\delta F} \frac{\de F}{\de \phi} \de\phi$, to observable quantities by modeling the halo mass functional $n_L[F]$ in some way. 

A downside of this is that the derivatives of $n_L$ w.r.t $\beta_m$ in Eq.~\eqref{linear-b-2} would have to be computed from simulations if one is to obtain information about $F$, or alternatively, assume a model of collapse into haloes. In the first case, one may compute the derivatives of the halo mass function by varying the initial conditions of the simulations. However, the coefficients~\eqref{betas} of the scale dependent bias cannot be observed in real data in this manner, i.e, by repeating the collapse process as one would do in simulations, because they only come once and with the same initial condition for the curvature perturbation field $\zeta$. Therefore, it becomes necessary that the next step be a connection between the effective bias coefficient $\int \mathcal{N} \frac{d F}{d \phi} d\phi $ and observable quantities by modeling the halo mass functional $n_L$ in some way. 

It turns out that the previous section~\ref{sec:halo} provides enough tools to accomplish this. In particular, if one models local NG at the short scales $\phi_S$ using the PS scheme, the variation with respect to the effective nonlinearity parameter $d \log n_L/ d \beta_m$ may be computed directly as $d \log n_L/ d f_m^S$, where $f_m^S$ is the local nonlinearity parameter of the short-scale theory (and, to first order in the perturbation $F$, it also defines the corresponding nonlinearity parameter of the full theory because of Eq.~\eqref{F-herm}). This can be done systematically using Eqs.~\eqref{Edgw},~\eqref{mass-function-PS-H} and~\eqref{linear-b-1}.

The first scale dependent NG correction $(m=1)$, associated to cubic NG, was computed in Ref.~\cite{Slosar:2008hx} by assuming a universal mass function, in agreement with the result of Ref.~\cite{Dalal:2007cu}:
\be 
\frac{\partial \log n_L}{\partial f_1^S} = 2\delta_c(b_G-1), 
\ee
where $b_G\equiv\frac{\partial \log n_L}{\partial \delta_L}$ is the Gaussian bias. The second term $(m=2)$, corresponding to $g_{\rm NL}$-type NG, was computed in Ref.~\cite{LoVerde:2011iz} using an Edgeworth expansion of the halo mass function and found to be given by
\be 
\frac{\partial \log n_L}{ \partial \log f_2^S} = \frac{\kappa_3(M)}{6}{\rm He}_3\left[\nu_c(M)\right] - \frac{\kappa'_3(M)}{6\nu_c'(M)}{\rm He}_2\left[\nu_c(M)\right], 
\ee
where $\kappa_n(M)$, is the reduced cumulant defined below Eq.~\eqref{mass-function-PS-H}. When expanded up to $f_3=g_{\rm NL}$, the relation~\eqref{linear-b-1} has been shown to be in very good agreement with simulations. Derivatives of the halo mass function with respect to the higher NG nonlinearity parameters $f_4,f_5,\ldots,$ may be easily computed from Eqs.~\eqref{Edgw} and~\eqref{mass-function-PS-H}. Concretely, they are given by
\be
f_m^S \frac{\partial n_L}{\partial \beta_m} = \frac{\partial \log n_L}{\partial \log f_m^S} =  \frac{\kappa_{m+1}(M)}{(m+1)!}{\rm He}_{m+1}\left[\nu_c(M)\right] - \frac{\kappa'_{m+1}(M)}{\nu_c'(M) (m+1)!}{\rm He}_{m}\left[\nu_c(M)\right]
\ee
where primes denote derivatives w.r.t the mass of the halo $M$. These terms should also be tested with N-body simulations for $f_{m>3}^S$ NG initial conditions, but such a computation is beyond the scope of this work. 

From Eq.~\eqref{linear-b-1} it is clear that the bias alone cannot differentiate between $f_{\rm NL},\;g_{\rm NL}$ or any higher order NG but it can give an answer to the question of Gaussianity or not. However, given a specific model for primordial NG, this can be a powerful probe in the sense of matching a template to data. For example, in the previous Chapter~\ref{chap:field}, we showed that one particular realization of this situation is the presence of an isocurvature mode with a potential $\Delta V(\psi)$. In this case, the function $F$ of the generalised local ansatz~\eqref{general-loc-NG-2}, is related to the potential as $F \propto \Delta V'$.  
Thus, a measurement of or a constraint on the bias would translate into a constraint on the parameters of the landscape potential. In such a context, one can choose a well motivated potential $\Delta V$, which fixes the function $F$, depending on few parameters (two or three cover most physically motivated potentials). Then, when proceeding to a simulation, one can use the PDF~\eqref{PDF-windowed} with $W(k) = \frac{3\alpha(k)}{5}$ to draw appropriate NG initial conditions for the density fluctuation field, and then look for a bias of the form~\eqref{linear-b-1}. For example, within the axion parameter space, there are regions that lead to such cases~\cite{Chen:2018uul} and may be probed in the near future with LSS surveys.

\cleardoublepage

\begin{conclusion}

We have studied the generation of a novel class of non-Gaussianity, in multi-field inflation with the help of a well-motivated model involving an axion like field coupled to the co-moving curvature perturbation, and then generalized it to any potential that is well-posed to be treated perturbatively, for instance, a bounded function. We showed that the isocurvature fluctuations can traverse the potential by fluctuating across the barriers at the time of horizon crossing. As a result, the isocurvature field acquires rich statistics that reflect the landscape structure. These statistics may be transferred to the adiabatic mode via a derivative coupling, which is the lowest-order possible coupling that arises from an effective field theory point of view, appearing already at the quadratic level of the action. Other (nonlinear) mechanisms may ultimately accomplish the same effect, but have a natural constraint in the observed value for the amplitude of the non-Gaussian features of the CMB, whereas a quadratic mixing preserves Gaussian statistics with a power spectrum consistent with the data available nowadays, and therefore need not be small. 

The non-Gaussianity in this model is calculated as the outcome of non-perturbative re-summations, at the level of the inner structure of the isocurvature potential, exemplified by the axion decay constant $f$. The result is expressed in terms of a probability distribution function for the curvature perturbation, which consists of a Gaussian part corrected by a non-Gaussian term involving complete information about the landscape potential. In such a case, the traditional perturbative $f_{\rm NL}$ parametrization cannot probe the full non-Gaussian structure and new estimators probing the whole PDF of temperature anisotropies should be developed. In this direction, we have amply pointed out that probability distribution functions, in its various forms, show potential to put tight constraints on nontrivial types of nonlinearities, such as a sinusoidal potential, which are expected to be present in typical scenarios of multi-field inflation. 

Moreover, we established that the perturbative series for the curvature perturbation implied by a quantum-field theory calculation in a de Sitter background is convergent for a given class of potentials, which contains the case of the sinusoidal potential, at least within the well-motivated regime of approximations we considered for super-horizon modes. We also explored connections with the stochastic inflation framework and extended out results from a QFT point of view to a broader class of backgrounds. In the presence of a coupling that allows for it, the transfer of statistics should take place in the same manner within most inflationary setups, as all of them involve a growth of perturbations on super-horizon scales, making the results qualitatively robust.

Finally, we extended and applied our results from the late-time standpoint, where we assume there was a mechanism that seeded local non-Gaussianities and study the resulting statistics and observable effects, guided by our previous results on the Landscape Tomography.  We have focused on the object that contains the full information about the distribution of anisotropies of the temperature and density fields, that is, the probability density function. Our starting point has been the bottom-up parametrization of the curvature perturbation as a Gaussian random field plus an arbitrary analytic function thereof. Instead of truncating the series expansion of this function to the first few terms, corresponding to the standard $f_{\rm NL}$, $g_{\rm NL}$ parametrization, we have kept the entire series enabling us to derive a probability functional that encodes the full local ansatz. We then derived predictions that will be testable with upcoming cosmological surveys, with special focus on two large-scale structure observables that are the halo mass function and halo bias, and constrained the primordial landscape potential with the data available nowadays from the CMB. The next batch of data provided by LSS surveys, and by the next generation CMB surveys, will thus have the potential to reveal new signals in the form of small departures from Gaussianity at the primordial level, and with that constrain our possibilities for a description of the primordial universe.

Let us finish by pointing out that even though this work is focused on local primordial non-Gaussianity, it should be possible to extend the discussion to other cases. Certainly, a different momentum dependence in the correlations would lead to a different scale dependence in the 2-point PDF, which could in turn serve as a NG estimator. Constraining such an object has the potential to be less costly than that of n-point functions using the standard templates. Finally, one may be able to further generalize this by including combinations of spatial derivatives acting on the curvature fluctuation, and in this more general case, one would be able to fully describe any number of momentum dependencies of the correlation functions, employed at the level of the PDF.

Our study gives an example where a stringy landscape may have calculable effects on the CMB and large-scale structure that have not yet been thoroughly searched for in the current data, and even if the theory that dominated our primordial universe is not string theory, the novel type of non-Gaussianity studied herein provides tools to constrain and reconstruct any fundamental theory with analogous setups, which at the level of a low-energy effective field theory cannot differ considerably, in light that all of them have to be able to explain the universe we observe.

\end{conclusion}

\cleardoublepage

\phantomsection 
\addcontentsline{toc}{chapter}{Bibliography} 
\bibliographystyle{plain}
\bibliography{bibliografia}

\begin{appendices}

\mbox{}
\newpage

\chapter{Details on the development of Tomographic non-Gaussianity} \label{chap:details}

\section{Details of $n$-point functions' computations}

Here we provide details of the intermediate steps of the computation discussed in Sec.~\ref{sec:npoints} leading to Eq.~(\ref{n-point-final-res-zeta}). Our starting point consists of plugging the field $u$, expanded in terms of the interaction picture field $u_I$ shown in Eq.~(\ref{u-u_I}), back into $\langle \hat u (\k_1, \tau)  \cdots \hat  u(\k_n , \tau) \rangle $. We obtain
\begin{align}
\langle & u(\k_1, \tau)\cdot \cdot \cdot u(\k_n,\tau) \rangle_{\rm c} =
\sum_{j = 1}^n \sum_{p = 0}^{n-1} \sum_{I_p} i^{n+1} \bra{0} \left(\prod_{\substack{l = 1 \\ l \notin I_p}}^{j -1} \int_{-\infty}^{\tau} d\tau_l [H_I^{\lambda}(\tau_l),\hat{u}_I ( k_l,\tau) ] \right) \left(\prod_{\substack{l = 1 \\ l \in I_p}}^{j -1} \hat{u}_I( k_l,\tau) \right)  \nn \\ 
& \times \left( \int_{-\infty}^{\tau} \!\!\!\! d\tau_j \!  \int_{-\infty}^{\tau_j} \!\!\!\! d\tau' \!  \int_{-\infty}^{\tau'} \!\!\!\! d\tau_1' \!  \ldots \int_{-\infty}^{\tau_{p-1}'} \!\!\!\! d\tau_p' [H_I^{\lambda}(\tau_p'), \ldots [H_I^{\lambda}(\tau_1'),[H_I^{\Lambda}(\tau') ,[H_I^{\lambda}(\tau_j),\hat{u}_I( k_l,\tau)]]] \ldots] \right) \nn  \\ 
& \times \left(\prod_{\substack{l = j+1 \\ l \in I_p}}^{n} \hat{u}_I( k_l,\tau) \right) \left( \prod_{\substack{l = j+1 \\ l \notin I_p}}^{n} \int_{-\infty}^{\tau} d\tau_l [H_I^{\lambda}(\tau_l),\hat{u}_I(k_l,\tau)] \right) \ket{0}_{\rm c} . \label{n-point-comp-1}
\end{align}
The subscript ${\rm c}$ reminds us that we are keeping fully connected contributions only. As discussed after Eq.~(\ref{structure-u}), to obtain these contributions, we must keep only contractions including at least one operator $a_{+}$ (or $a_{+}^{\dag}$) appearing in $H^{\Lambda}_I$. To track these contractions, we have introduced the sum $\sum_{I_p}$ over the sets $I_p = 1_p, 2_p , \cdots$, consisting of all possible arrangements (of dimension $p$) of momenta labels. For example, we could write: $1_2 = (1,2)$, $2_2 = (1,3)$, $3_2 = (2,3)$, etc.

To perform the contractions appearing in (\ref{n-point-comp-1}), it is useful to define the following two quantities:
\bea
\Delta (\tau_a,\tau_b,k) = u_k(\tau_a) u_k^*(\tau_b) = v_k(\tau_a) v_k^*(\tau_b) , \\
D (\tau_a,\tau_b,k) = \Delta (\tau_a,\tau_b,k) - \Delta (\tau_b,\tau_a,k) .
\eea
In terms of these quantities, one can write the following field commutators:
\bea
\left[ \hat{u}_I( \k', \tau'),  \hat{u}_I(\k, \tau) \right] &=& (2 \pi)^3 \delta(\k' + \k) D(\tau', \tau, k) , \quad \\
\left[ \hat{v}_I( \k', \tau'),  \hat{v}_I(\k, \tau) \right] &=& (2 \pi)^3 \delta(\k' + \k) D(\tau', \tau, k)  . \quad
\eea
These relations allow us to further infer the form of the commutators involving $H_I^{\lambda}$ appearing in (\ref{n-point-comp-1}). These are found to be given by
\bea
\left[ H_I^{\lambda}(\tau'),\hat{u}_I(\k,\tau) \right] = -\dfrac{\lambda}{(\tau')^2} \dfrac{\partial [\tau' D(\tau', \tau, k)]}{\partial \tau'} \hat{v}_I(\k,\tau') ,  \qquad \\
\left[ H_I^{\lambda}(\tau'),\hat{v}_I(\k,\tau) \right] = -\dfrac{\lambda}{(\tau')^2}   D(\tau', \tau, k)  \dfrac{\partial [ \tau' \hat{u}_I(\k,\tau') ] }{\partial \tau'} . \qquad
\eea
Then, using these back into (\ref{n-point-comp-1}), and performing the relevant contractions, we find
\bea
\langle u( \k_1,\tau) \cdots u( \k_n,\tau) \rangle_{\rm c} = i^{n+1} (2\pi)^3 \delta^{(3)}\left(\sum_{i = 1}^{n} \k_{i}\right) \dfrac{-\Lambda^4 (-1)^{n/2}}{H^4} \left(\dfrac{-\lambda H}{f} \right)^n \sum_{j = 1}^n \sum_{p = 0}^{n-1} \sum_{I_p}   \nn \\ 
  \times \sum_{{\rm perm}~q_l} \int_{-\infty}^{\tau} \!\!\!\!\!\! d\tau' \tau'^{n-4}
  \int_{\tau'}^{\tau} \!\! d\tau_j \int_{-\infty}^{\tau'} \!\!\!\!\!\! d\tau_1'  \ldots \int_{-\infty}^{\tau_{p-1}'} \!\!\!\!\!\! d\tau_p' \nn \\
\times  \left(\prod_{\substack{l = 1 \\ l \notin I_p}}^{j -1} \int_{-\infty}^{\tau} \!\!\!\! d\tau_l \dfrac{ \Delta(\tau_l, \tau', k_l)}{\tau_l^2} \dfrac{\partial [\tau_l D(\tau_l, \tau, k_l)]}{\partial \tau_l} \right)
 \left(\prod_{\substack{l = 1 \\ l \in I_p}}^{j -1} D(\tau_{q_l}',\tau',k_l) \dfrac{1}{\tau_{q_l}'^2} \dfrac{\partial[\tau_{q_l}' \Delta(\tau, \tau_{q_l}',k_l)]}{\partial \tau_{q_l}'} \right) \nn \\
\times \left( \dfrac{ D(\tau', \tau_j, k_j)}{\tau_j^2} \dfrac{\partial [\tau_j D(\tau_j, \tau, k_j)]}{\partial \tau_j} \right)   \sum_{m=n/2}^{\infty} \frac{1}{(m-n/2) !} \left[ - \frac{1}{2}\left( \frac{H \tau'}{f} \right)^{2} \! \int_k \Delta( \tau' , \tau ' , k) \right]^{m-n/2}  \nn \\
\times \left(\prod_{\substack{l = j+1 \\ l \in I_p}}^{n} \! D(\tau_{q_l}',\tau',k_l) \dfrac{1}{\tau_{q_l}'^2} \dfrac{\partial[\tau_{q_l}' \Delta( \tau_{q_l}', \tau,k_l)]}{\partial \tau_{q_l}'} \right) \!\!
\left( \prod_{\substack{l = j+1 \\ l \notin I_p}}^n \! \int_{-\infty}^{\tau} \!\!\!\! d\tau_l \dfrac{\Delta(\tau', \tau_l, k_l)}{\tau_l^2} \dfrac{\partial [\tau_l D(\tau_l, \tau, k_l)]}{\partial \tau_l} \right) , \quad  \label{n-point-comp-2}
\eea
if $n$ is even, and zero otherwise because the expectation value of an odd number of fields in the interaction picture vanishes identically. This is a consequence of the potential's being an even function of the isocurvature field.

In Eq.~\eqref{n-point-comp-2}, the sum over ``perm $q_l$'' refers to the sum over all possible permutations $l \to q_l$ of labels belonging to the set $I_p$. These permutations affect the arguments $\tau_{q_l}'$ appearing in some functions in the second and fourth lines of this equation.

The previous result can be simplified with the help of the following two steps: first, the sum appearing in the third line comes from the infinite loop contributions shown in Fig.~(\ref{fig:FIG_diagram_res}). They are the consequence of contractions between pairs of creation and annihilation operators $a_+^\dag$ and $a_+$ appearing in $H_I^{\Lambda}$. These terms can be resummed back into the following exponential:
\be \label{re-summation}
\sum_{m'} \frac{1}{m' !} \left[ - \frac{1}{2}\left( \frac{H \tau'}{f} \right)^{2} \! \int_k \Delta( \tau' , \tau ' , k) \right]^{m'} = e^{ - \frac{\sigma_0^2}{2 f^2} } ,
\ee
where $\sigma_0$ is nothing but the variance of $\psi$ already defined in Eq.~(\ref{sigma_0-def}). Notice that $\sigma_0$ is independent of time, and so we may factorize $\exp( - \sigma_0^2 / 2 f^2 )$ out of the $\tau'$ integral.

Secondly, in Eq.~(\ref{n-point-comp-2}) we may relabel every integration variable of the form $\tau'_i$ by a new variable $\tau_l$ (with $l \in I_p$) in such a way that, in the  functions' arguments, the $k_l$'s are always accompanied by a $\tau_l$. This relabeling allows us to recognize that the sum over all possible permutations $q_l$ becomes a sum over all domains of integration for the variables $\{\tau_l\}_{l \in I_p}$. As a result, the nested integrals of Eq.~(\ref{n-point-comp-2}) unravel, and we are led to
\bea
\langle u(\k_1,\tau)\cdot \cdot \cdot u(\k_n,\tau) \rangle_{\rm c} = i^{n+1} (2\pi)^3 \delta^{(3)}\left(\sum_{i = 1}^{n} \k_{i}\right) \dfrac{-\Lambda^4 (-1)^{n/2}}{H^4} \left(\dfrac{-\lambda H}{f} \right)^n e^{-\frac{\sigma_0^2}{2f^2}} \sum_{j = 1}^n \sum_{p = 0}^{n-1} \sum_{I_p} \nn \\ 
\times \int_{-\infty}^{\tau} \!\!\!\! d\tau' \tau'^{n-4}  \left(\prod_{\substack{l = 1 \\ l \in I_p}}^{j -1} \int_{-\infty}^{\tau'} \!\!\!\! d\tau_l  D(\tau_{l},\tau',k_l) \dfrac{1}{\tau_{l}^2} \dfrac{\partial[\tau_{l} \Delta(\tau, \tau_{l},k_l)]}{\partial \tau_{l}} \right) \nn \\
\times  \left(\prod_{\substack{l = 1 \\ l \notin I_p}}^{j -1} \int_{-\infty}^{\tau} \!\!\!\! d\tau_l \dfrac{ \Delta(\tau_l, \tau', k_l)}{\tau_l^2} \dfrac{\partial [\tau_l D(\tau_l, \tau, k_l)]}{\partial \tau_l} \right) 
 \left(\int_{\tau'}^{\tau} d\tau_j \dfrac{ D(\tau', \tau_j, k_j)}{\tau_j^2} \dfrac{\partial [\tau_j D(\tau_j, \tau, k_j)]}{\partial \tau_j} \right) \nn \\
\times \left(\prod_{\substack{l = j+1 \\ l \in I_p}}^{n}  \int_{-\infty}^{\tau'} \!\!\!\! d\tau_l  D(\tau_{l},\tau',k_l) \dfrac{1}{\tau_{l}^2} \dfrac{\partial[\tau_{l} \Delta( \tau_{l}, \tau,k_l)]}{\partial \tau_{l}} \right) \,\,\,\,\, \nn \\
\times \left( \prod_{\substack{l = j+1 \\ l \notin I_p}}^n  \int_{-\infty}^{\tau} \!\!\!\! d\tau_l \dfrac{\Delta(\tau', \tau_l, k_l)}{\tau_l^2} \dfrac{\partial [\tau_l D(\tau_l, \tau, k_l)]}{\partial \tau_l} \right). \,\,\,\,\,\,\,\,\, \label{n-point-comp-3}
\eea
The previous expression can be simplified further by performing the summation over the index $p$ (including the sum over the sets $I_p$ introduced earlier). Notice that these sums allow one to rewrite the second and fourth lines of Eq.~(\ref{n-point-comp-3}) as the multiplication of pairs of terms
\bea
&& \langle  u(\k_1,\tau)  \cdot \cdot \cdot u(\k_n,\tau) \rangle_{\rm c} = \nn \\
&& \,\,\,\,\, i^{n+1} (2\pi)^3 \delta^{(3)}\left(\sum_{i = 1}^{n} \k_{i}\right) \dfrac{-\Lambda^4 (-1)^{n/2}}{H^4} \left(\dfrac{-\lambda H}{f} \right)^n e^{-\frac{\sigma_0^2}{2f^2}} \sum_{j = 1}^n  \int_{-\infty}^{\tau} d\tau' \tau'^{n-4} \nn \\
&& \,\,\,\,\, \times  \prod_{\substack{l = 1}}^{j-1} \left(\int_{-\infty}^{\tau'} \dfrac{d\tau_l}{\tau_{l}^2}  D(\tau_{l},\tau',k_l) \dfrac{\partial[\tau_{l} \Delta(\tau, \tau_{l},k_l)]}{\partial \tau_{l}} + \int_{-\infty}^{\tau} \dfrac{d\tau_l}{\tau_{l}^2}  \Delta(\tau_l, \tau', k_l) \dfrac{\partial [\tau_l D(\tau_l, \tau, k_l)]}{\partial \tau_l} \right) \nn \\
&& \,\,\,\,\, \times  \prod_{\substack{l = j+1}}^{n} \left(\int_{-\infty}^{\tau'} \dfrac{d\tau_l}{\tau_{l}^2}  D(\tau_{l},\tau',k_l) \dfrac{\partial[\tau_{l} \Delta( \tau_{l},\tau,k_l)]}{\partial \tau_{l}} + \int_{-\infty}^{\tau} \dfrac{d\tau_l}{\tau_{l}^2} \Delta(\tau',\tau_l, k_l) \dfrac{\partial [\tau_l D(\tau_l, \tau, k_l)]}{\partial \tau_l} \right) \nn \\
&& \,\,\,\,\, \times  \left(\int_{\tau'}^{\tau} \dfrac{d\tau_j}{\tau_{j}^2} D(\tau', \tau_j, k_j) \dfrac{\partial [\tau_j D(\tau_j, \tau, k_j)]}{\partial \tau_j} \right) \!. \label{n-point-comp-4}
\eea
We can now rewrite this expression in a more compact way by noticing that the propagators of the $\psi$ and $\zeta$ fields are given by (up to a \(\sqrt{2 \epsilon}\) factor for $\zeta$)
\be
\bar{\Delta}(\tau',\tau_l, k_l) = H^2 \tau'\tau_{l}\Delta(\tau',\tau_l, k_l).
\ee
By switching to $\zeta,\psi$ variables the $\tau'^n$ factor from the first line of \eqref{n-point-comp-4} and the $\tau^n$ factor, coming from the conversion of the external fields, combine with the $\Delta$ propagators to form the $\bar{\Delta}$ ones. Then we may conveniently introduce the following set of functions
\begin{align}
\mathcal{G}_{+}(k_l,\tau',\tau) \equiv \frac{(-i\lambda)}{H^3} \int_{-\infty}^{\tau} \dfrac{d\tau_l}{\tau_{l}^3}  & \left(\bar{\Delta}(\tau_l, \tau', k_l) \dfrac{\partial [\bar{D}(\tau_l, \tau, k_l)]}{\partial \tau_l}   \right. \nn \\
& \left. +  \bar{D}(\tau_{l},\tau',k_l) \dfrac{\partial[\bar{\Delta}(\tau, \tau_{l},k_l)]}{\partial \tau_{l}} \Theta(\tau'-\tau_l)  \right), \\
\mathcal{G}_{-}(k_l,\tau',\tau) \equiv \frac{(-i\lambda)}{H^3} \int_{-\infty}^{\tau} \dfrac{d\tau_l}{\tau_{l}^3}  & \left( \bar{\Delta}(\tau',\tau_l, k_l) \dfrac{\partial [\bar{D}(\tau_l, \tau, k_l)]}{\partial \tau_l} \right. \nn \\
& \left. + \bar{D}(\tau_{l},\tau',k_l) \dfrac{\partial[\bar{\Delta}( \tau_{l},\tau,k_l)]}{\partial \tau_{l}} \Theta(\tau'-\tau_l)  \right),
\end{align}
where $\Theta(\tau)$ is the usual Heaviside function. The functions $\mathcal{G}_{\pm}(k_l,\tau',\tau)$ are nothing but the \emph{mixed propagators} defined in Ref.~\cite{Chen:2017ryl}, with the external -- or boundary, in the language of \cite{Chen:2017ryl} -- time, denoted by $\tau$ here, left explicit as an argument. Moreover, subtracting them we find
\be
\mathcal{G}_{+}(k_j,\tau',\tau) - \mathcal{G}_{-}(k_j,\tau',\tau) =
\frac{(+i\lambda)}{H^3} \int_{\tau'}^{\tau} \dfrac{d\tau_j}{\tau_{j}^3} \bar{D}(\tau', \tau_j, k_j) \dfrac{\partial [ \bar{D}(\tau_j, \tau, k_j)]}{\partial \tau_j}.
\ee
Replacing these definitions back into Eq.~(\ref{n-point-comp-4}) and noting that consecutive terms with alternating signs cancel out in the sum over $j$, we end up with the following simple expression for the $n$-point correlation function
\be
\begin{split}
\langle \zeta(\k_1,\tau)\cdots & \zeta(\k_n,\tau) \rangle_{\rm c} = \dfrac{i\Lambda^4}{H^4} (2\pi)^3 \delta^{(3)}\left(\sum_{i = 1}^{n} \k_{i}\right)  e^{-\frac{\sigma_0^2}{2f^2}} \left(-\dfrac{H^2}{2 \epsilon f^2} \right)^{n/2}  \\
& \times \int_{-\infty}^{\tau} \frac{d\tau'}{\tau'^{4}} \Big[ \mathcal{G}_{+}(k_1,\tau',\tau) \ldots \mathcal{G}_{+}(k_n,\tau',\tau)
  - \mathcal{G}_{-}(k_1,\tau',\tau) \ldots \mathcal{G}_{-}(k_n,\tau',\tau) \Big].
\label{n-point-comp-G}
\end{split}
\ee
It is rewarding to notice that this result is exactly what we would have obtained had we used the diagrammatic rules of \cite{Chen:2017ryl}, after adequately treating the loop contributions.

Let us now perform the integrals of Eq.~(\ref{n-point-comp-G}) [or, equivalently, Eq.~(\ref{n-point-comp-4})] to obtain a simple and useful expression for the $n$-point correlation function in momentum space.  As discussed in Sec.~\ref{sec:linearth}, recall that the effect of the $H_I^{\lambda}$ is to source the evolution of the amplitude of $\zeta$ (or, equivalently, $u$) on superhorizon scales. Thus, for a given fixed set of $k_l$'s, the integration domain of each integral appearing in Eq.~(\ref{n-point-comp-G}) can be split into two parts. Before horizon crossing, where $|\tau_l| k_l > 1$ (and $|\tau'| k_{l} > 1$) and after horizon crossing, where the opposite is true: $|\tau_l| k_l < 1$ (and $|\tau'| k_{l} < 1$). The integrants are highly oscillatory in the domain $|\tau_l| k_l > 1$ (and $|\tau'| k_{l} > 1$). These oscillatory contributions would have vanished had we kept track of the $\epsilon$ prescription. And so, we may simply disregard the contributions to (\ref{n-point-comp-4}) coming from these domains.

On the other hand, the integration of these functions over the domain $|\tau_l| k_l < 1$ (and $|\tau'| k_{l} < 1$)  gives us expressions that dominate if the upper limit $\tau$ is such that $|\tau| k_l \ll 1$ (in fact, the integration over the domain  $|\tau_l| k_l < 1$ diverges as $\tau \to 0$). Thus, instead of obtaining exact expressions for these integrals, we may seek the infrared divergent contributions that dominate on superhorizon scales. To do this explicitly for each integral, we introduce an arbitrary time $\tau_0$ such that $|\tau_0| k_l < 1$ for all $k_l$'s. We will use $\tau_0$ to cut off the lower limit of every time integral. The upper limit may be either the end of inflation or some other value, depending on the exact mass of the scalar field. In either case, we assume that the amount of e-folds that different modes spend outside the horizon is approximately the same, as discussed in the main text.

Then, within the integration domains $\tau_l' \in (\tau_0, \tau')$, one may simplify a few of the integrated functions as
\bea
 \dfrac{D(\tau_{l},\tau',k_l)}{\tau_{l}^2} \dfrac{\partial[\tau_{l} \Delta( \tau_{l}, \tau,k_l)]}{\partial \tau_{l}}
\simeq  \dfrac{i e^{-i k_l \tau_l}}{6 k_l \tau_l \tau} \!\! \left[\dfrac{\tau'^2}{\tau_l} - \dfrac{\tau_l^2}{\tau'} \right] , \,\,\, \qquad  \\
 \dfrac{D(\tau_{l},\tau',k_l)}{\tau_{l}^2} \dfrac{\partial[\tau_{l} \Delta( \tau, \tau_{l},k_l)]}{\partial \tau_{l}} \simeq \dfrac{i e^{+i k_l \tau_l}}{6 k_l \tau_l \tau}  \!\! \left[\dfrac{\tau'^2}{\tau_l} - \dfrac{\tau_l^2}{\tau'} \right] . \,\,\, \qquad
\eea
Then, it is direct to find
\be
\int_{\tau_0}^{\tau'} \!\!\!\!\! d\tau_l \, D(\tau_{l},\tau',k_l) \dfrac{1}{\tau_{l}^2} \dfrac{\partial[\tau_{l} \Delta( \tau_{l}, \tau,k_l)]}{\partial \tau_{l}}
 \simeq \dfrac{i e^{-i k_l \tau'}}{6 k_l \tau} \left(\dfrac{\tau'^2}{\tau_0} - \dfrac{3 \tau'}{2} + \dfrac{\tau_0^2}{2 \tau'} \right).
\ee
Similarly, the other relevant integrals may be evaluated as
\bea
\int_{\tau_0}^{\tau'} \!\!\!\!\! d\tau_l \, D(\tau_{l},\tau',k_l) \dfrac{1}{\tau_{l}^2} \dfrac{\partial[\tau_{l} \Delta(  \tau,\tau_{l},k_l)]}{\partial \tau_{l}}
 &\simeq& \dfrac{i e^{+i k_l \tau'}}{6 k_l \tau} \left(\dfrac{\tau'^2}{\tau_0} - \dfrac{3 \tau'}{2} + \dfrac{\tau_0^2}{2 \tau'} \right), \qquad  \\
\int_{\tau_0}^{\tau} \!\!\!\! d\tau_l \,  \dfrac{u_{k_l}(\tau_l)}{\tau_{l}^2} \dfrac{\partial[\tau_{l} D( \tau_{l}, \tau,k_l)]}{\partial \tau_{l}}  &\simeq& \dfrac{2\ln(\tau_0/\tau)}{(2 k_l)^{3/2} \tau} , \,\,\, \qquad  \\
\int_{\tau_0}^{\tau} \!\!\!\! d\tau_l \,  \dfrac{u_{k_l}^*(\tau_l)}{\tau_{l}^2} \dfrac{\partial[\tau_{l} D( \tau_{l}, \tau,k_l)]}{\partial \tau_{l}}  &\simeq& - \dfrac{2\ln(\tau_0/\tau)}{(2 k_l)^{3/2} \tau} , \,\,\, \qquad \\
\int_{\tau'}^{\tau} \!\!\!\! d\tau_j \,  \dfrac{D(\tau',\tau_j,k_j)}{\tau_{j}^2} \dfrac{\partial[\tau_{j} D( \tau_{j}, \tau,k_l)]}{\partial \tau_{j}} 
 &\simeq&  \dfrac{2\ln(\tau'/\tau)}{(2 k_j)^{3/2} \tau} \sqrt{\dfrac{2}{k_j}} \dfrac{k_j^2 \tau'^2}{3} .\,\,\,  \qquad
\eea
Using these results back into (\ref{n-point-comp-4}), and performing the final integral over \(\tau'\), we finally obtain
\bea
 \langle u(\k_1,\tau)\cdot \cdot \cdot u(\k_n,\tau) \rangle_{\rm c} &\simeq& (-1)^{n/2} (2\pi)^3 \delta^{(3)}\left(\sum_{i = 1}^{n} \k_{i}\right)  \quad \nn \\
&& \,\, \times \dfrac{\Lambda^4}{H^4} e^{-\frac{\sigma_0^2}{2f^2}} \left(\dfrac{\lambda H}{2 f \tau} \right)^n \dfrac{1}{3}  \dfrac{k_1^3 + \ldots + k_n^3}{k_1^3 \cdot \cdot \cdot k_n^3} \text{ln}\left(\dfrac{\tau_0}{\tau}\right)^{n+1}. \quad \label{n-point-comp-5}
\eea
Let us remind the reader that the previous equation holds as written for even $n$ only. If $n$ is odd, the correlator vanishes identically because the potential is an even function of the isocurvature field. This result differs by a factor $1/2$ from the Gaussian relation shown in Eq.~(\ref{zeta-n-psi-n}) based on the linear evolution of $\zeta$ at superhorizon scales. This factor appears as a consequence of the integration
\be
\int_{\tau_0}^{\tau} \!\! d\tau' \, \dfrac{\text{ln}(\tau'/\tau)}{\tau'} = - \dfrac{1}{2}\text{ln}^2 (\tau_0/\tau) ,
\ee
and may be understood as a nonlinear effect due to two classes of vertices involved in the computation. The final expression offered in (\ref{n-point-comp-5}) is the main result of this section, and leads directly to Eq.~(\ref{n-point-final-res-zeta}).

Given that $\tau_0$ is chosen in such a way that $|\tau_0| k_l < 1$ (for all $k_l$'s), then we may roughly identify the quantity $\ln (\tau_0 /\tau)$ as the number of $e$-folds after all the modes left the horizon:
\be
\Delta N \simeq \ln (\tau_0/\tau) .
\ee
Then, it is possible to verify that the infrared contributions to the integrals leading to (\ref{n-point-comp-5}) dominate when the condition
\be
 \Delta N \gg  1
\ee
is satisfied. Recall that, in order to deal with the system perturbatively, our computation assumed $\lambda \ll 1$. This means that only after a time $\Delta N \sim 1/\lambda$ the effects (both linear and nonlinear) induced by $\psi$ start to take over, as evidenced by the computations of Sec.~\ref{sec:linearth}.

\section{Structure of $I_n$} \label{app_about_I_n}

Here we derive a few properties of $I_n$ defined in Eq.~(\ref{full-I-n}). To start with, notice that the sum $k_1^3 + \cdots + k_n^3$ appearing in Eq.~(\ref{full-I-n}) leads to $n$ identical integrals. Then, by writing the Dirac delta function as $\delta (\p) = (2 \pi)^{-3} \int_r e^{- i \p \cdot \r}$, the integral $I_n$ becomes
\be
I_n = \frac{n}{(2 \pi)^3}   \int_r  \left[ \int_{k < k_L} \!\!\!\!\! \frac{e^{- i \k \cdot \r} }{k^3 } \right]^{n-1} \int_{q < k_L}   \!\!\!\!\!  e^{- i \q \cdot \r} .
\ee
Let us recall that we are also dealing with an IR cutoff $k_{\rm IR}$. The previous expression may be simplified by first redefining the integration variables as
\be
\r = \x / k_L, \qquad   \k = \y k_L , \qquad \q = \z k_L  ,
\ee
and then by solving the angular parts of the $\x$, $\y$ and $\z$ integrals. These two steps lead to
\be
I_n (\xi) = \frac{n}{(2 \pi^2)^{n+1}}  \int_0^{\infty} \frac{dx}{x} \left[ \int^{1}_{\xi^{-1}} \frac{d y}{y} \frac{\sin (y x)}{y x} \right]^{n-1}  \int^{1}_{\xi^{-1}} \!\!\! d z \, z \,  x^2 \sin (z x)  , \label{I-n-x-beta-0}
\ee
where $\xi =  k_L/ k_{\rm IR}$ is the ratio of scales available to observers introduced in Sec.~\ref{subsec:n-point}. In Eq.~(\ref{I-n-x-beta-0}), we have emphasized that $I_n$ is a function of $\xi$. It may be noticed that this function satisfies
\be
I_n(1) = 0 .
\ee
Next, the $y$ and $z$ integrals may be solved to give
\be
I_n (\xi) = \frac{n }{(2 \pi^2)^{n+1}} \int_0^{\infty} \!\! \frac{dx}{x} G(\xi,x) \left[ F(\xi , x) \right]^{n-1}   ,  \quad \label{app-full-I_n} 
\ee
where we have introduced the functions
\bea
G(\xi , x) &=&  \sin (x) - x \cos (x) - \sin (x / \xi)  + (x / \xi) \cos (x / \xi)  , \qquad \\
F(\xi , x) &=& {\rm Ci} (x) - \frac{\sin (x)}{x}  - {\rm Ci} ( x / \xi ) + \frac{\sin ( x / \xi )}{ x / \xi} . \qquad
\eea
Here, $ {\rm Ci} (x)$ is the cosine integral function.  It is hard to obtain a simple analytical representation of $I_n$ (for arbitrary $n$) beyond that shown in Eq.~(\ref{app-full-I_n}). In the particular case $n=2$, one finds (for $ \xi \geq 1$)
\be
I_2 (\xi)  = \frac{\pi}{(2 \pi^2)^3} \ln \xi .
\ee
In the cases $n=3$ and $n=4$, we are able to obtain exact analytical expressions but it is more useful to write down the results in the limit $\xi \gg 1$:
\bea
I_3 (\xi)  &=& \frac{3 \pi}{2 (2 \pi^2)^{4}}  \left[  \left( \ln \xi \right)^2 + 1 -  \frac{\pi^2}{6}  \right] + \mathcal O(\xi^{-1}), \qquad  \\
I_4 (\xi)  &=& \frac{4 \pi}{2 (2 \pi^2)^{5}} \Bigg[  \left( \ln \xi \right)^3 - \frac{\pi^2 - 3}{4}   \ln \xi  + \frac{11}{4} \zeta (3) - \frac{43}{24} \Bigg] + \mathcal O(\xi^{-1}) . \label{I_4-direct}
\eea
While for $n \geq 5$ we are not able to obtain simple general expressions in the large $\xi$ limit, we can still derive a useful property about $I_n$. Indeed, it is possible to show that $I_n$ has the following asymptotic form for $n \geq 4$:
\be
I_n =  I_n^0 + \Delta I_n , \label{I-n-expansion-0}
\ee
where 
\bea
I_n^0 &=& \frac{n \pi}{2 (2 \pi^2)^{n+1}}   \left( \ln \xi \right)^{n-1} ,  \label{I-n-expansion-1} \\
\Delta I_n &=& - I_n^0 \times C \frac{(n-1) (n-2)}{2} \left( \ln \xi \right)^{-2} + \mathcal O \left[ (\ln \xi )^{-3} \right] ,  \label{I-n-expansion-2}
\eea
with $C =(\pi^2-3)/12 \simeq 0.6$. 

Before deriving Eqs.~(\ref{I-n-expansion-0})-(\ref{I-n-expansion-2})  let us briefly point out their importance: even for very large values of $\xi$, the difference between the leading term $I_n^0$ and the correction $\Delta I_n$ is given by a factor $\left( \ln \xi \right)^2$, which is not necessarily large enough. In fact, the result shown in Eq.~(\ref{I-n-expansion-2}) implies that $\Delta I_n$ will quickly become of order $I_n^0$ as $n$ grows. It is a simple matter to verify that if $\ln \xi = 60$, then $\Delta I_n$ will reach one-tenth of $I_n^0$ around $n\simeq 35$. Nevertheless, the important question here is to what extent the correction $\Delta I_n$ modifies the PDF derived in Sec.~\ref{subsec:PDF-derivation-from-n-points}, obtained under the assumption that $I_n$ can be taken as $I_n^0$. The answer turns out to be rather simple: by performing the same reconstruction carried out in Sec.~\ref{subsec:PDF-derivation-from-n-points}, using the ansatz~(\ref{PDF-ansatz}), we find that $\Delta I_n$ implies an extra contribution to~(\ref{PDF-1}) that has an oscillatory behavior set by the scale $f_{\zeta}$, as expected. This time, however, the oscillations come with different factors in their amplitudes. Among these terms we find a factor of $\sigma_L^4/f^4$ (or, equivalently, $\sigma_{\zeta}^4/f_{\zeta}^4$), and thus comparing with the terms in the uncorrected PDF, the leading of which is $\sigma_L^2/f^2$, we get an estimate of how small $f$ can be so that Eq.~(\ref{PDF-1}) remains accurate. The final answer is that in order to be able to neglect the correction to the PDF, at the very least we must have
\be
f \gtrsim \frac{\sigma_L}{\sqrt{\ln \xi}}    .  \label{f-log-cutoffs}
\ee
This result severely limits the applicability of the PDF of Eq.~(\ref{PDF-1}), and forces us to consider the more general reconstruction carried out in Sec.~\ref{subsec:PDF-derivation-from-n-points-full}.

\subsection{Derivation of $\Delta I_n$}

To show Eqs.~(\ref{I-n-expansion-0})-(\ref{I-n-expansion-2}), we may split the integral (\ref{I-n-x-beta-0}) into two parts, one containing the term $\sin(x) - x \cos (x)$ and the other containing the term $- \sin(x / \xi) + (x / \xi) \cos ( x / \xi)$. Then by changing the integration variable of the second part as $x \to x / \xi$, we end up with the following expression
\be
I_n (\xi) = \frac{ n \pi}{2 (2 \pi^2)^{n+1}}  \left[ \bar I_n (\xi)  + (-1)^n  \bar I_n (1 / \xi)  \right], \label{I-n-x-beta-v2}
\ee
where we have defined $ \bar I_n (\xi)$ as
\be
\bar I_n (\xi) \equiv  \frac{2}{\pi} \int_0^{\infty} \frac{dx}{x} \left[ \sin(x) - x \cos (x)  \right]   \left[ F (\xi, x) \right]^{n-1}   .  \qquad
\ee
Notice that for large $\xi$ the second contribution to (\ref{I-n-x-beta-v2}), given by $\bar I_n(1/\xi)$, is very suppressed compared to the first one, given by $\bar I_n(\xi)$. To see this, it is enough to see that $F(1/\xi , x)$ is a function that becomes quickly independent of $\xi$ for values $x > 1/ \xi$, and so one finds that in the relevant integration domain the function is essentially given by
\be
F(1 / \xi, x) \simeq  {\rm Ci} (x) - \frac{\sin (x)}{x} .
\ee
This implies that the contribution to (\ref{I-n-x-beta-v3}) coming from $\bar I_n(1/\xi)$ is at most of order 1. Then, we are left with
\be
I_n (\xi) = \frac{n \pi}{2 (2 \pi^2)^{n+1}}  \bar I_n (\xi) + \mathcal O (1)  . \label{I-n-x-beta-v3}
\ee
Next, notice that the function $F(\xi,x)$ that appears inside the integral $\bar I_n (\xi) $ satisfies
\be
F(\xi, 0) = \ln \frac{1}{\xi} .
\ee
In addition, in the range $1 < x  \leq \xi$, it is well approximated by $F_\xi(x)  \simeq 1 - \gamma +  \ln \frac{\xi}{ x } $, whereas, for values $x > \xi$, the function $F_\xi(x) $ vanishes quickly. We can therefore write
\be
F(\xi,x)  \simeq 1 - \gamma +  \ln \frac{\xi}{ x }  + \epsilon (\xi x), \label{approx-F}
\ee
where $\gamma$ is the Euler-Mascheroni constant and $\epsilon (y)$ is a slowly varying function that remains small in the relevant interval $1 < x  \leq \xi$. In fact, this function is suppressed everywhere $1 < x  \leq \xi$ and its largest value is of order $0.1$ when $x / \xi  = 1$. It is enough to take $\epsilon (y) = y^2 / 12$. Then, we can write
\be
\bar I_n (\xi)  =  \frac{2}{\pi} \int_0^{x_*} \frac{dx}{x} \left[ \sin(x) - x \cos (x)  \right] 
\left[ 1 - \gamma +  \ln \frac{\xi }{x }  + \epsilon (\xi x) \right]^{n-1}  ,
\ee
where $x^* \gtrsim \xi$ has been introduced to cut off the integral, since for values larger than $x^*$ the function $F(\xi , x)$ is highly suppressed. The introduction of $x_*$ has the benefit of allowing us to use approximation (\ref{approx-F}) inside the integral.  Now, taking a derivative with respect to $\xi$, we find
\be
\bar I_n ' (\xi)  =   (n-1)  \frac{2}{\pi} \int_0^{x_*} \frac{dx}{x} \left[ \sin(x) - x \cos (x)  \right]  \times
\left[  \frac{1}{\xi} + \epsilon ' (\xi x) x \right]
\left[ 1 - \gamma +  \ln \frac{1}{\xi x }  + \epsilon (\xi x) \right]^{n-2}  . 
\ee
This equation leads to
\bea
\bar I_n ' (\xi)  =  \frac{(n - 1)}{\xi} \bar I_{n-1}  (\xi) &+&  (n-1)  \frac{2}{\pi} \int_0^{x_*} \frac{dx}{x} \left[ \sin(x) - x \cos (x)  \right]
  \nn \\
&& \times \,  \epsilon ' (\xi x) \, x \left[ 1 - \gamma +  \ln \frac{1}{\xi x }  + \epsilon (\xi x) \right]^{n-2} .
\eea
Because $\xi x \epsilon ' (\xi x)$ is of order $\epsilon$ in the entire domain $1 < x  \xi$, the second term is clearly subleading with respect to the first one. Then, we can simply disregard the second term, and write
\be
\bar I_n ' (\xi)  = \frac{(n - 1)}{\xi} \bar I_{n-1}  (\xi) .
\ee
This relation allows us to obtain $\bar I_{n} (\xi)$ from $ \bar I_{n-1}  (\xi)$. For instance, by direct computation, we are able to deduce that in the large $\xi$ limit, $\bar I_3(\xi)$ and $\bar I_3(1/\xi)$ are given by
\bea
\bar I_{3} (\xi) &=& ( \ln \xi )^2 - \frac{\pi^2 - 3}{12} + \mathcal O(\xi^{-1}) , \\
\bar I_{3} (1/\xi) &=&  \frac{\pi^2 - 9}{12} + \mathcal O(\xi^{-1}) .
\eea
From this result, we identify $C = (\pi^2 - 3) / 12$. Then, we have
\be
\bar I_4 ' (\xi)  =  \frac{3}{\xi} \left( [ \ln (1/ \xi) ]^2 - C \right) .
\ee
Solving this relation, we end up with
\be
\bar I_4  (\xi)  = ( \ln   \xi )^3 - 3 C \ln \xi + \mathcal O(1) .
\ee
Notice that this result correctly accounts for the form of $I_4$ previously shown in Eq.~(\ref{I_4-direct}). Repeating this step many times, we end up with the following general expression for $\bar I_n  (\xi)$:
\be
\bar I_n  (\xi)  =  ( \ln \xi )^{n-1}  -  C \frac{(n-1)(n-2)}{2} ( \ln \xi )^{n-3} ,
\ee
from where Eqs.~(\ref{I-n-expansion-0})-(\ref{I-n-expansion-2}) follow.

%%%%%%%%%%%%%%%%%%%%%%%%%%%%%%%%%%%%
%%%%%%%%%%%%%%%%%%%%%%%%%%%%%%%%%%%%

\cleardoublepage

\chapter{PDFs in time-dependent Quantum Field Theory} \label{chap:ininPDF}

In what follows we will study the dynamics of a self-interacting real scalar field \(\varphi\), which for concreteness we take to live inside a 4-dimensional spacetime (although this is not essential). The theory is described by a Lagrangian density 
\be \label{lagrangian}
\L= \L_{\text{free}} - V(\varphi, x).
\ee
Here \(x\) is the spacetime coordinate, and \(\L_{\text{free}}\) defines a free theory, of which we assume its solutions to be known in terms of mode functions \(\varphi_k(t)\):
\bea
\varphi(\x , t) &=& \int_k \, \hat \varphi (\k , t) \, e^{- i \k \cdot \x } \\  \hat \varphi (\k , t) &=& \varphi_k(t) a(\k) + \varphi_k^*(t) a^\dag(-\k) 
\eea
where we have introduced $\int_k \equiv (2\pi)^{-3} \int d^3 k$ as shorthand notation.
The operators $a(\k)$ and $a^\dag(\k)$ are creation and annihilation operators that satisfy the following commutation relations:
\be
\left[ a (\k) , a^{\dag} (\k') \right] = (2 \pi)^3 \delta^{(3)} (\k - \k').
\ee 
Equivalently, we may describe the complete theory through a hamiltonian density $\H= \H_{\rm {free}} + V(\varphi, x)$. Either using this starting point or the lagrangian density~\eqref{lagrangian}, this setup is capable of describing a vast number of theories, including non-autonomous systems where there are explicit time dependences in the parameters of the theory. 
Having established the foundations, our attention will now be focused on describing the machinery we will use to solve the theory perturbatively.

We now proceed to quantize the system adopting the interaction picture framework. That is, the quantum field $\varphi$ is written as $\varphi (\x , t) = U^\dag (t)  \varphi_I (\x , t)  U (t)$,
where $\varphi_I (\x , t)$ is the interaction picture field and evolves as a field of the free theory. Explicitly, it is given by
\bea
\varphi_I(\x , t) &=& \int_k \, \hat \varphi_I (\k , t) \, e^{- i \k \cdot \x }  \\
\hat{\varphi}_I (\k , t) &=& \varphi_k(t) a(\k) + \varphi_k^*(t) a^{\dag}(-\k).
\eea

On the other hand, $U (t)$ is the time evolution operator in the interaction picture (sometimes dubbed as propagator), which is given by
\be \label{propagator}
U (t) = \text{T} \exp \left\{ - i \int^{t}_{t_0 + i\epsilon |t_0|} \!\!\!\!\!\!\!\!\! d t' H_I (t')  \right\} 
\ee
where $\text{T}$ is the time ordering symbol, instructing to place operators evaluated at later times at the left of the expression, and operators evaluated at earlier times at the right. Here we have incorporated a prescription to evaluate the integral, in the form of a positive infinitesimal quantity \(\epsilon\), which takes care of selecting the proper \textit{in} state when \(t_0 \to -\infty\)~\cite{Weinberg:1995mt}. This could also be implemented by adding an imaginary part to the argument of the interaction Hamiltonian $H_I$~\cite{Adshead:2009cb}. 

Clearly, the object that determines how temporal evolution takes place is $H_I$, which is given by the potential evaluated at the interaction picture fields,
\be \label{inter}
H_I(\tau) =  \int_{\x} V( \varphi_I(\x,t), \x, t),
\ee
where $\int_{\x} \equiv \int d^3 \x$. In order to deal with $H_I$, one of the methods we will consider is to make a Fourier expansion of the potential over field space and expand the exponential through its power series
\be \label{expansion}
\begin{split}
V(\varphi, \x, t) =& \, \frac{1}{2\pi} \int_{-\infty}^{\infty} d \gamma \tilde{V}(\gamma, \x, t) e^{-i \gamma \varphi} \\
 =& \, \frac{1}{2\pi} \int_{-\infty}^{\infty} d \gamma \tilde{V}(\gamma, \x, t) \sum_{m=0}^{\infty} \frac{(-i\gamma \varphi)^m}{m!},
\end{split}
\ee
which is formally possible since $\varphi$ is a hermitian field with real eigenvalues. This representation of the potential has also been utilized in other recent QFT results~\cite{Chebotarev:2018udi,Alexandrov:2017xmr} (referred therein as part of the construction of the $\mathcal{S}$-matrix in the Efimov representation, introduced earlier on in~\cite{Efimov:1977tw} and other previous works of the same author). In principle, either starting from here or from a Taylor expansion we will have to deal with an infinite number of vertices, with an arbitrary number of external legs. Although this expansion in Fourier modes is not essential to the final result, it turns out to be helpful in some derivations, being particularly convenient within a diagrammatic approach.  

Our first matter of interest will be to compute \(n\)-point functions for the \(\varphi\) field
\be
\langle \varphi(\x_1,t) \ldots \varphi(\x_n,t) \rangle = \langle U^\dag (t) \varphi_I(\x_1,t) \ldots \varphi_I(\x_n,t) U (t) \rangle.
\ee
Expanding the interaction picture propagator using the Dyson series one can readily write down
\be \label{dyson}
\begin{split}
\langle \varphi(\x_1,t) \ldots \varphi(\x_n,t) \rangle =& \, \langle \sum_{N=0}^{\infty} (-i)^N \sum_{l=0}^{N} (-1)^l \int_{t_0 - i\epsilon |t_0|}^t \!\!\!\!\!\!\!\!\! dt_l \ldots
\int_{t_0 - i\epsilon |t_0|}^{t_2} \!\!\!\!\!\!\!\!\! dt_1 H_I(t_1) \ldots H_I(t_l) \\ & \times \varphi_I(\x_1,t) \ldots \varphi_I(\x_n,t) \int_{t_0 + i\epsilon |t_0|}^t \!\!\!\!\!\!\!\!\! dt_{l+1} \ldots \int_{t_0 + i\epsilon |t_0|}^{t_{N-1}} \!\!\!\!\!\!\!\!\! dt_N H_I(t_{l+1}) \ldots H_I(t_N) \rangle,
\end{split}
\ee
and thus we only need to evaluate expectation values of the form
\be
\langle H_I(t_1) \ldots H_I(t_l)   \varphi_I(\x_1,t) \ldots \varphi_I(\x_n,t) H_I(t_{l+1}) \ldots H_I(t_N) \rangle
\ee
where we have yet to specify the \textit{in} state. 

We construct these states using the creation and annihilation operators of the free theory, which we can write in terms of the interaction picture field operator and its conjugate canonical momentum field operator. This may be accomplished by inverting the relations that define the field observables in momentum space
\bea
 \hat \varphi_I (\k , t) &=& \varphi_k(t) a(\k) + \varphi_k^*(t) a^\dag(-\k) \\ 
 \Pi^\varphi_I (\k , t) &=& \dot{\varphi}_k(t) a(\k) + \dot{\varphi}_k^*(t) a^\dag(-\k), 
\eea
and then write down the operator of interest that specifies the \textit{in} state. For instance, one could write a superposition of one-particle states as
\be
\ket{\Phi} = \int_\x W(\x) \hat \varphi_I(\x) \! \ket{0},
\ee
or, more generally, for a superposition of multi-particle states,
\be
\ket{\Phi} = \sum_{n}  \left( \int_{\x_1} \ldots \int_{\x_n}  W_n(\x_1,\ldots,\x_n) \hat \varphi_I(\x_1) \ldots \hat \varphi_I(\x_n) \right) \! \ket{0},
\ee
where the functions $\{W_n(\x_1,\ldots,\x_n)\}_n$ characterize the state, with $n$ labelling the number of ``particles'' in each term of the sum. Here we have introduced the Fock vacuum $\ket{0}$, which is annihilated by the $a(\k)$ operators, i.e.,  $a(\k) \ket{0} = 0$, and supports a ladder of states upon acting on it with the $a^\dagger (\k)$ operators.

Finally, we note that because \(\Pi^{\varphi}_I(x) = d\varphi_I(x)/dt\), and the temporal derivative only affects the mode functions, we only need to compute
\be \label{npoint}
\begin{split}
\bra{0} \! \varphi_I(\y_1, t_0) \ldots \varphi_I(\y_J, t_0) & H_I(t_1) \ldots H_I(t_l)   \varphi_I(\x_1,t) \ldots
 \varphi_I(\x_n,t) \\
& \times  H_I(t_{l+1}) \ldots H_I(t_N) \varphi_I(\y_1, t_0) \ldots \varphi_I(\y_J, t_0) \! \ket{0}
\end{split}
\ee
while keeping track of which field is to be differentiated with respect to time at \(t_0\), and finally summing the necessary terms to reconstruct the desired \textit{in} state, which will lie within the Fock space of the free theory vacuum. In this process, it is neater to let the positions $\y_i$ be different at each side of the inner product; thus we will first compute the correlation with no repeated positions in the fields, and take the corresponding limits at the end of the computation. Note that we may reconstruct the expectation value of other operators of interest by taking derivatives and linear combinations of the \(n\)-point correlation functions~\eqref{npoint} at time $t$. 
 
In the following subsections we proceed to outline the steps leading to the main result, while leaving most of the details to Appendices~\ref{sec:loops} and~\ref{sec:arbitrary-initial}. With the benefit of hindsight, we will appreciate that the results we will obtain are a direct consequence of Wick's theorem~\cite{Wick:1950ee}, and thus they do not rely on the particular representation of the expansion chosen to compute perturbations with the potential (Taylor series, Fourier series, etc.).

\section{Main result}

Since usually all of the relevant information to describe the theory can be stored within the path integral, it is natural to expect that we may write down an explicit functional that generates the $n$-point functions at any order in perturbation theory.

In this section we derive such a functional, which allows us to compute $n$-point functions at a given time slice $t$. We refer the interested reader to Appendix~\ref{sec:loops} for some of the technical details.

It will prove useful to write down the propagator of the free theory as a fundamental object, both in momentum and position space
\bea
\Delta(t,t',p) &\equiv& \varphi^{}_p(t)\varphi^*_p(t') \\  \sigma^2(t,t',r) &\equiv& \frac{1}{(2\pi)^3} \int d^3p \; \Delta(t,t',p) e^{i\p \cdot \r} \nonumber \\ &=& \braket{0|\varphi_I(t,\x) \varphi_I(t',\y) |0},
\eea
where $r = |\x - \y|$.

\subsection{The loop contributions}

Ultimately, it is the interacting term $V(\varphi, \r, t)$ that which will lead to non-trivial signatures, if any, in the spectrum of the field $\varphi$. Given that we are taking a perturbative approach, these signatures must be reflected solely through correlations, such as equation~\eqref{npoint}. If we expand the potential $V$ inside $H_I(\tau) =  \int_{\x} V( \varphi_I(\x,t), \x, t)$ as a power series on $\varphi_I$, each term in the expansion will ``interact'' with up to as many other spacetime positions as the power of the particular term. These interactions are usually represented with diagrams, joining ``outer legs'', which represent the fields that are used to construct the observed states (\textit{in} or \textit{out} states), and ``internal legs'' that arise from the interaction terms. Each of these connections gives a contribution of $\sigma^2(t,t',r)$ to the process under consideration, which we will call propagators or covariances depending on the context. Within all of these connections one can find contractions between two vertices (in this context, vertex is short for a spacetime position where an interaction potential is evaluated and the number of fields associated to it). As a result, it is possible to encounter ``closed circuits'', such as $\sigma^2(t_1,t_2,r_{12}) \sigma^2(t_2,t_3,r_{23}) \sigma^2(t_3,t_1,r_{31}) $. Diagrammatically, each propagator is represented with a line; hence these kinds of contributions are represented by a line segment that closes in itself. Therefore one calls them ``loops''.

To evaluate the correlation in equation~\eqref{dyson}, we will first address the fully interacting contribution to equation~\eqref{npoint}, i.e. that which connects all coordinates of the fields that define the \textit{in} state with interaction vertices. Put simply, for the moment we will not be interested in the contributions that arise directly from the free theory and can be factored out. Note that this is not equivalent to a fully connected contribution because this would require all external legs to interact with each other through the vertices, which would be reflected through an overall Dirac delta in momentum space (provided that the system is translationally invariant).

As is shown in Appendix~\ref{sec:loops}, the fully interacting $n$-point correlator contains the following loop structure as a factor:
\be \label{npoint6}
\begin{split}
\bra{0} \! \varphi_I(\z_1, t_0) \ldots \varphi_I(\z_J, t_0) & H_I(t_1) \ldots H_I(t_l)   \varphi_I(\x_1,t) \ldots \varphi_I(\x_n,t)
\\ & H_I(t_{l+1}) \ldots H_I(t_N) \varphi_I(\y_1, t_0) \ldots \varphi_I(\y_J, t_0) \! \ket{0}_{FI} \\
& \propto
 \int_{\r_1} \int_{\varphi_1} \ldots \int_{\r_N} \int_{\varphi_N} \dfrac{\partial^{n_1} V}{\partial \varphi_1^{n_1}} \ldots \dfrac{\partial^{n_N} V}{\partial \varphi_N^{n_N}}  \frac{\exp \left( -\frac{1}{2} \varphi_i \left(\Sigma_I^{-1}\right)_{ij} \varphi_j \right)}{\sqrt{(2\pi)^N |\text{det} \Sigma_I|}},
\end{split}
\ee
where we have omitted the propagators that connect the vertices with the outer legs. Here $n_l$ is defined as the number of ``legs'' at vertex $l$ of the perturbative expansion that are connected to the fields defining the \textit{in} state, and $\Sigma_I$ is a (complex) symmetric matrix that has the position space propagators connecting the vertices $\r_l$ as entries. This matrix plays the role of a covariance matrix, and so we will treat as such.

Let us appreciate an important aspect of this last result: it is an expectation value over a (multivariate) gaussian probability density function. With this in mind (and Wick's thoerem), we claim that
\be \label{npointn}
\begin{split}
\bra{0} \! \varphi_I(\z_1, t_0) & \ldots \varphi_I(\z_J, t_0)  H_I(t_1) \ldots H_I(t_l)   \varphi_I(\x_1,t) \ldots \varphi_I(\x_n,t) \\ & \,\,\,\,\,\,\,\,\,\,\,\,\,\,\,\,\,\,\,\,\,\,\,\,\,\,\,\,\,\,\, \times H_I(t_{l+1}) \ldots H_I(t_N) \varphi_I(\y_1, t_0) \ldots \varphi_I(\y_J, t_0) \! \ket{0} \\  = & \int_{\varphi_{\z_1}} \!\!\!\!\!\! \ldots \! \int_{\varphi_{\z_J}}\!\!\!\!\!\!  \ldots 
\int_{\varphi_{\x_1}} \!\!\!\!\!\!  \ldots \! \int_{\varphi_{\x_N}} \!  \int_{\varphi_{\y_1}}  \!\!\!\!\!\! \ldots \! \int_{\varphi_{\y_J}} \! \int_{\r_1} \int_{\varphi_1} \ldots \int_{\r_N} \int_{\varphi_N} \\
& \times  V(\varphi_1, \r_1, t_1) \ldots  V(\varphi_N, \r_N, t_N)  \frac{\exp \left( -\frac{1}{2} \pmb{\varphi}^T \cdot \left(\pmb{\Sigma}^{-1}\right) \cdot \pmb{\varphi} \right)}{\sqrt{(2\pi)^{N+n+2J} |\text{det} \pmb{\Sigma} |}} \\ & \times \varphi_{\z_1} \ldots \varphi_{\z_J} \varphi_{\x_1} \ldots \varphi_{\x_n} \varphi_{\y_1} \ldots \varphi_{\y_J} 
\end{split}
\ee
where \(\pmb{\varphi}^T \equiv (\varphi_{\z_1} \,\,\, ... \,\,\, \varphi_{\z_J}  \,\,\, \varphi_{1} \,\,\, ... \,\,\, \varphi_l \,\,\, \varphi_{\x_1} \,\,\, ... \,\,\, \varphi_{\x_n} \,\,\, \varphi_{l+1} \,\,\, ... \,\,\, \varphi_N \,\,\, \varphi_{\y_1} \,\,\, ... \,\,\, \varphi_{\y_J}  )\), and \(\pmb{\Sigma}\) is the corresponding $(N+n+2J) \times (N+n+2J)$ covariance matrix.  
The covariances in this matrix are the propagators between the fields' corresponding spacetime positions, and they have their respective temporal arguments ordered within the propagators' arguments as the fields are in the definition of $\pmb{\varphi}^T$. For example, the covariance relating $\varphi_{\z_a}$ and $\varphi_{b}$ is $\sigma^2(t_0,t_b,|\z_a-\r_b|)$, and the one relating $\varphi_{\x_i}$ with $\varphi_{b}$ would be $\sigma^2(t,t_b,|\x_i-\r_b|)$ if $b\geq l+1$, while it would be $\sigma^2(t_b,t,|\x_i-\r_b|)$ if $b \leq l$. We omit the dependence of \(\pmb{\Sigma}\) on $(N,l)$ to ease the notation. Also, the integrals $\int_{\varphi}$ are shorthand for $\int_{-\infty}^{\infty} d\varphi$.

Equation~\eqref{npointn} describes the full $n$-point function, including both the free theory contributions and the interacting ones. 
Note that the free theory pairings are given precisely by a Gaussian distribution as in~\eqref{npointn}, only without the $\varphi_i$ terms. Instances of these would be $\sigma^2(t_0,t_0,|\z_a-\y_b|)$ or $\sigma^2(t,t_0,|\x_i-\y_b|)$. Since it is fairly easy to check that these contractions also arise from this expression, we have obtained an indication that we are on the right track.

\subsection{A corollary of Wick's theorem}

We now proceed to prove the claim introduced in the previous section: let the potential $V$ be given by its Taylor expansion about $\varphi = 0$, with spacetime-dependent coefficients $c_m(\r,t)$
\be \label{taylorV}
V(\varphi,\r,t) = \sum_{m=0}^{\infty} \frac{c_m(\r,t)}{m!} \varphi^m,
\ee
and let us use this in~\eqref{npointn}. The right-hand side of the equation now reads
\be
\begin{split} \label{proof1}
& \int_{\r_1} \ldots \int_{\r_N} \sum_{m_1=0}^{\infty} \ldots \sum_{m_N=0}^{\infty} \frac{c_{m_1}(\r_1,t_1) \ldots c_{m_N}(\r_N,t_N)}{m_1! \ldots m_N!} \\ 
& \times
 \int_{\varphi_{\z_1}} \!\!\!\!\!\! \ldots \! \int_{\varphi_{\z_J}} \int_{\varphi_{\x_1}} \!\!\!\!\!\!  \ldots \! \int_{\varphi_{\x_N}} \!\int_{\varphi_{\y_1}}  \!\!\!\!\!\! \ldots \! \int_{\varphi_{\y_J}} \!  \int_{\varphi_1} \int_{\varphi_N} \frac{\exp \left( -\frac{1}{2} \pmb{\varphi}^T \cdot \left(\pmb{\Sigma}^{-1}\right) \cdot \pmb{\varphi} \right)}{\sqrt{(2\pi)^{N+n+2J} |\text{det} \pmb{\Sigma} |}} \\
& \times \varphi_{\z_1} \ldots \varphi_{\z_J} \varphi_{\x_1} \ldots \varphi_{\x_n} \varphi_{\y_1} \ldots \varphi_{\y_J} \varphi_1^{m_1} \ldots \varphi_N^{m_N}.
\end{split}
\ee
The second line in this last expression is nothing more than a moment of a multivariate gaussian distribution. Therefore, the result is the sum over all pairings of fields of the product of the corresponding covariances. On the other hand, if we go back to the starting point~\eqref{npoint}, we have
\be \label{proof2}
\begin{split}
 \int_{\r_1} \ldots \int_{\r_N} \sum_{m_1=0}^{\infty} \ldots \sum_{m_N=0}^{\infty} & \frac{c_{m_1}(\r_1,t_1) \ldots c_{m_N}(\r_N,t_N)}{m_1! \ldots m_N!}
 \bra{0} \! \varphi_I(\z_1, t_0) \ldots \varphi_I(\z_J, t_0) \varphi_I(\r_1, t_1)^{m_1} \\ 
 & \ldots \times \varphi_I(\r_l, t_l)^{m_l} \varphi_I(\x_1,t) \ldots \varphi_I(\x_n,t) \varphi_I(\r_{l+1}, t_{l+1})^{m_{l+1}} \\ & \ldots  \times \varphi_I(\r_{N}, t_N)^{m_N} \varphi_I(\y_1, t_0) \ldots \varphi_I(\y_J, t_0) \! \ket{0}, 
\end{split}
\ee 
where the vacuum expectation value, per Wick's theorem, is exactly the sum, over all the possible pairings of fields, of the product of the free-theory two-point functions associated to the pairings, which in turn are exactly the covariances we have defined earlier. Hence the last two expressions are equal and therefore~\eqref{npointn} holds as written.

In order to connect this with the usual diagrammatic approach, note that in this last step the sum over all possible pairings is exactly what gives rise to propagators connecting vertices, and as may be seen from Appendix~\ref{sec:loops}, the flow of momenta through the diagrams appears by taking the Fourier transform to momentum space of each propagator. In this sense, we have only rewritten a known statement in an apparently more complicated manner. However, in this way it is possible to appreciate some aspects of perturbation theory that usually remain obscure in a diagrammatic approach. 

\subsection{An explicit result for the PDF of a self-interacting scalar field at every order} \label{subsec:main-PDF}

Now we return to equation~\eqref{dyson}, which is the object we want to characterize through a probability distribution. This PDF must be able to generate $n$-point functions to any order in perturbation theory, and should deliver a path integral in the formal limit $N \to \infty$. Furthermore, it should always be positive when reduced to a finite number of points at which to evaluate the field. The breakdown of this property would be a clear indicator that higher-order terms are required to give a meaningful result.

In what follows, we will write down explicit results taking the free theory vacuum $\ket{0}$ as the \textit{in} state (thus omitting $\varphi_\z$ and $\varphi_\y$ in $\pmb \varphi$), but it is straightforward to get a more general result, which we list in Appendix~\ref{sec:arbitrary-initial}. The reason for doing this is that the structure of the computation we wish to emphasize, namely the interacting terms, is already contained within the correlations that come out of this choice.

Using equation~\eqref{npointn}, the $n$-point function for the $\varphi$ field is given by
\be
\begin{split}
\langle \varphi(\x_1,t) ... \varphi(\x_n,t) \rangle =&  \sum_{N=0}^{\infty} (-i)^N \sum_{l=0}^{N} (-1)^l \int_{t_0 - i\epsilon |t_0|}^t \!\!\!\!\!\!\!\!\! dt_l ...
\int_{t_0 - i\epsilon |t_0|}^{t_2} \!\!\!\!\!\!\!\!\! dt_1 \int_{t_0 + i\epsilon |t_0|}^t \!\!\!\!\!\!\!\!\! dt_{l+1} ... \int_{t_0 + i\epsilon |t_0|}^{t_{N-1}} \!\!\!\!\!\!\!\!\! dt_N \int_{\r_1} ... \int_{\r_N}  \\ & \times \int_{\varphi_{\x_1}} \!\!\!\!\!\!  ...  \int_{\varphi_{\x_N}} \!\int_{\varphi_1} \!\!\!  ... \int_{\varphi_N} V(\varphi_1, \r_1, t_1) ...  V(\varphi_N, \r_N, t_N) \\ & \times \frac{\exp \left( -\frac{1}{2} \pmb{\varphi}^T \cdot \left(\pmb{\Sigma}^{-1}\right) \cdot \pmb{\varphi} \right)}{\sqrt{(2\pi)^{N+n} |\text{det} \pmb{\Sigma} |}} \varphi_{\x_1} ... \varphi_{\x_n}.
\end{split}
\ee
This correlation is a moment of the distribution
\be \label{PDF}
\begin{split}
\rho_{\varphi} = \sum_{N=0}^{\infty} (-i)^N \sum_{l=0}^{N} (-1)^l \int_{t_0 - i\epsilon |t_0|}^t \!\!\!\!\!\!\!\!\! dt_l ...
\int_{t_0 - i\epsilon |t_0|}^{t_2} \!\!\!\!\!\!\!\!\! dt_1 \int_{t_0 + i\epsilon |t_0|}^t \!\!\!\!\!\!\!\!\! dt_{l+1} ... \int_{t_0 + i\epsilon |t_0|}^{t_{N-1}} \!\!\!\!\!\!\!\!\! dt_N \int_{\r_1} ... \int_{\r_N}  \\ \times \!\int_{\varphi_1} \!\!\!  ... \int_{\varphi_N} V(\varphi_1, \r_1, t_1) ...  V(\varphi_N, \r_N, t_N)  \frac{\exp \left( -\frac{1}{2} \pmb{\varphi}^T \cdot \left(\pmb{\Sigma}^{-1}\right) \cdot \pmb{\varphi} \right)}{\sqrt{(2\pi)^{N+n} |\text{det} \pmb{\Sigma} |}},
\end{split}
\ee
which is, in its own right, a probability density function for the field $\varphi$. It is important to keep in mind that, even though it is not explicitly stated, the covariance matrix $\Sigma$ is different for each pair $(N,l)$, in the manner discussed between~\eqref{npointn} and~\eqref{taylorV}. We can summarize this by stating that the times that are integrated over a contour shifted by $-i\epsilon |t_0|$ always go to the left in the covariances, and that those with $+i\epsilon |t_0|$ always go to the right. When two times have the same imaginary component, the covariance has its arguments time-ordered if the integration is with $+i\epsilon |t_0|$, and anti-time-ordered if the integration goes with $-i\epsilon |t_0|$. Keeping this in mind, we may formally factor the distribution defined by the exponential out of the spacetime integrations, write it as a functional integral, and then further rearrange~\eqref{PDF} to get
\be \label{Path-integral1}
\begin{split}
\rho_{\varphi} =& \int D\varphi \frac{\exp \left( - \frac{1}{2} \varphi \cdot \left( {\Sigma}^{-1} \right) \cdot \varphi \right)}{\sqrt{|\text{det} (2\pi \Sigma) |}} \\ & \times \sum_{l=0}^{\infty} (+i)^l \int_{t_0 - i\epsilon |t_0|}^t \!\!\!\!\!\!\!\!\! dt_l \int_{\r_l} V(\varphi(\r_l, t_l), \r_l, t_l) \,  ...
\int_{t_0 - i\epsilon |t_0|}^{t_2} \!\!\!\!\!\!\!\!\! dt_1 \int_{\r_1} V(\varphi(\r_1, t_1), \r_1, t_1) \\ & \times \sum_{N=0}^{\infty} (-i)^N   \int_{t_0 + i\epsilon |t_0|}^t \!\!\!\!\!\!\!\!\! dt_{l+1} \int_{\r_{l+1}} V(\varphi(\r_{l+1}, t_{l+1}), \r_{l+1}, t_{l+1})  \\ & \,\,\,\,\,\,\,\,\,\,\,\,\,\,\,\,  \ldots  \times \int_{t_0 + i\epsilon |t_0|}^{t_{N+l-1}} \!\!\!\!\!\!\!\!\! dt_{N+l} \int_{\r_{N+l}}  V(\varphi(\r_{N+l}, t_{N+l}), \r_{N+l}, t_{N+l}),  
\end{split}
\ee
where we have to stress that the arguments of the fields inside the potential are there merely as a label for the functional integral to read; they are \textit{not} to be integrated over by the spacetime integrals right away.

In this last expression, $\varphi$ contains both ``internal'' (those in the arguments of the interaction $V$) and ``external'' fields (those that appear in the observable, characterized by the positions $\x_i$). 
One additional formal step gives the resummation of the Dyson series
\be \label{Path-integral2}
\begin{split}
\rho_{\varphi} =& \int \! D\varphi_+ D\varphi_-  \exp \left\{ +i\int^{t}_{t_0 - i\epsilon |t_0|} \!\!\!\!\!\!\!\!\! d t' \int_{\r} V(\varphi_-(\r,t'),\r,t') \right\}  \! \\ & \times \frac{\exp \left( - \frac{1}{2} \varphi \cdot \left( {\Sigma}^{-1} \right) \cdot \varphi \right)}{\sqrt{|\text{det} (2\pi \Sigma) |}}   \exp \left\{ - i \int^{t}_{t_0 + i\epsilon |t_0|} \!\!\!\!\!\!\!\!\! d t' \int_{\r} V(\varphi_+(\r,t'),\r,t') \right\}.
\end{split}
\ee
Here we have made a distinction between the fields $\varphi_+$ and $\varphi_-$ (as is usually done in the Schwinger-Keldysh formalism) in order to be unambiguous regarding the order in which the covariances have their temporal arguments arranged: the $\varphi_+$ field always has its corresponding times to the right and time-ordered among themselves, while those of $\varphi_-$ always go to the left and anti-time-ordered among themselves. It is no longer necessary to write down the time ordering symbols, because the ordering prescription is already implemented through the definition of $\Sigma$. 
From this point one can derive an expression with more resemblance to the usual path integral formulation~\cite{Weinberg:2005vy}. It is also relevant to keep in mind that the matrix $\Sigma$ also has entries for external $\varphi$ fields, which are neither $\varphi_+$ nor $\varphi_-$: the Schwinger-Keldysh fields only account for the inner structure of the theory. To emphasize this, note that an expectation value for an observable quantity is computed by integrating over the corresponding external field variables 
\be
\langle f \rangle (\r_1, ..., \r_n ; t) = \int D\varphi \, \rho_{\varphi} \, f(\varphi(\r_1,t), ... ,\varphi(\r_n,t);t),
\ee
as one would expect.

A remark is in order here. Throughout these derivations, when we write $\int D\varphi$ we mean to integrate over the range of eigenvalues of the field operator (heretofore the real line) for each spacetime position relevant in the integration, and reduce the corresponding (usually Gaussian) distribution to the relevant coordinates. For instance, in this last equation $\int D\varphi = \int_{\varphi(\r_1,t)} ... \int_{\varphi(\r_n,t)} $: the RHS is composed of a product of $n$ real integrals from $-\infty$ to $\infty$ for each field, thus sweeping over all of their possible eigenvalues.

The distribution $\rho_{\varphi}$ is normalized, in the sense that $\int D\varphi \rho_{\varphi} = 1$, as can be readily seen order by order from the perturbative expansion~\eqref{PDF}: if we disregard the $\epsilon$ prescription, then only $N=0$ gives a nonzero contribution, since by having integrated out the external fields, the argument of the time integrations is the same for every $l$, and what remains is a sum (with signs) over integration domains, which cancel out identically. Hence, for finite $t_0$ we have $\int D\varphi \rho_{\varphi} = 1$, and then $\lim_{t_0 \to -\infty} \int D\varphi \rho_{\varphi} = 1$, which is what is usually meant by $t_0 = -\infty$. Diagrammatically, this implies the cancellation of loop diagrams in the interacting theory that are disconnected from the external legs.

Me must note, though, that for computational purposes, equation~\eqref{Path-integral2} may be as useful as the starting point~\eqref{propagator} because ultimately both are formal expressions. However, it makes manifest one of the fundamental aspects of perturbation theory: the result is expressed only in terms of the quantities of the free theory, modulated by the perturbation $V$; no new propagators are introduced at a basic level.

As this section's final comment, we note that including self-interactions involving derivatives of the field is also feasible. However, in order to represent all the contractions through a Gaussian distribution in an unambiguous manner, it is desirable that the field and its conjugate momenta be ordered in a definite and uniform way within the Hamiltonian. If this is not the case, then one would probably be forced to add extra labels to the integrations so as to implement the different ordering prescriptions.

\section{The loop structure within the $n$-point functions} \label{sec:loops}

For the sake of familiarity with traditional approaches to QFT, we will proceed with the computation mostly in momentum space, even though the final result will reveal this step as unnecessary.

Let $\int_{\gamma} = (2\pi)^{-1} \int_{-\infty}^\infty d\gamma$ and $\int_{\varphi} = \int_{-\infty}^\infty d\varphi$. Then replacing equations~\eqref{inter} and~\eqref{expansion} into~\eqref{npoint}, we get
\be \label{npoint2}
\begin{split}
& \int_{\r_1} \int_{\varphi_1} \int_{\gamma_1} \ldots \int_{\r_N} \int_{\varphi_N} \int_{\gamma_N} V(\varphi_1, \r_1, t_1) e^{i \gamma_1 \varphi_1} \ldots V(\varphi_N, \r_N, t_N) e^{i \gamma_N \varphi_N} \\
& \times \sum_{m_1 = 0}^{\infty} \ldots \sum_{m_N = 0}^{\infty} \frac{(-i \gamma_1)^{m_1} \ldots (-i \gamma_N)^{m_N}}{m_1! \ldots m_N!} \int_{k_{11}} \ldots \int_{k_{1m_1}} 
\ldots \int_{k_{N1}} \ldots \int_{k_{Nm_N}} \\ 
& \times  \bra{0} \! \hat{\varphi}_I(\q_1, t_0) \ldots \hat{\varphi}_I(\q_J, t_0) \hat{\varphi}_I(\k_{11}, t_1) \ldots \hat{\varphi}_I(\k_{1m_1}, t_1) \ldots \hat{\varphi}_I(\k_{l1}, t_l) \ldots \hat{\varphi}_I(\k_{lm_l}, t_l)  \\ 
& \times \hat{\varphi}_I(\k_1,t) \ldots \hat{\varphi}_I(\k_n,t) 
\hat{\varphi}_I(\k_{(l+1)1}, t_{l+1}) \ldots \hat{\varphi}_I(\k_{(l+1)m_{l+1}}, t_{l+1}) \ldots \hat{\varphi}_I(\k_{N1}, t_N) \ldots \hat{\varphi}_I(\k_{Nm_N}, t_N) \\ 
& \times \hat{\varphi}_I(\p_1, t_0) \ldots \hat{\varphi}_I(\p_J, t_0) \! \ket{0} \prod_{j=1}^N e^{-i \sum_{a=1}^{m_j} \k_{ja} \cdot \r_j}.
\end{split}
\ee
Let us not get distracted by the size of the previous equation and instead focus on how to deal with it. The previous vacuum expectation value can be evaluated by moving all annihilation operators to the right, giving rise to contractions between pairs of field operators in all possible ways. Hence, we need only distinguish the non-equivalent pairings and count the number of equivalent contractions for each pairing. 

Now we take a diagrammatic approach and try to obtain the fully interacting contributions. That is, in what follows we will only keep track of the terms where all fields at times \(t_0\) or \(t\) are contracted with fields arising from the interaction-picture hamiltonian. We regard two contractions as equivalent if they are connected to the same pair of spacetime positions (or vertices), indexed by the letter \(l\). Additionally, we define $n_{ij}$ as the number of field contractions between vertices $i$ and $j$, thus making $n_{ii}$ the number of closed loops formed from vertex $i$ alone. Also, we define $n_i$ as the number of contractions from vertex $i$ to the outer fields in the correlation (i.e., the ones evaluated at time $t$ or $t_0$). 

Firstly, let us count the number of possible ways of assigning roles to each field in the correlation: at vertex $i$, we have that the following multinomial coefficient
\be
\binom{m_i}{n_i, 2n_{ii}, n_{i1}, \ldots, n_{iN}}
\ee
is the number of possible ways of assigning the fields of vertex $i$ to the different roles they can undertake, with 
\be
m_i =  n_i + 2n_{ii} + \sum_{j \neq i} n_{ij} 
\ee
referring to the indices in the power series of~\eqref{npoint2} that represents $e^{-i\gamma_i \hat \varphi}$. Once these roles have been assigned, we may count the number of equivalent ways to achieve a certain configuration of contractions: given $n_{ij}$, if $i \neq j$, there are $n_{ij}!$ ways of forming contractions between vertices $i$ and $j$, and similarly there are
\be
\frac{(2n_{ii})!}{2^{n_{ii}} n_{ii}!}
\ee
ways of arranging the contractions of vertex $i$ with itself. Since the ``outer legs'' of~\eqref{npoint2} (the fields with momenta $\p_i$ or $\k_i$) are distinguishable, the only remaining combinatorial factor to account for is $n_i!$, which is the number of possible ways of assigning the $n_i$ ``outer legs'' to the $n_i$ fields of the vertex available for these contractions.

With all the previous statements considered,~\eqref{npoint2} is equal to
\be \label{npoint3}
\begin{split}
& \int_{\r_1} \! \int_{\varphi_1} \! \int_{\gamma_1} \! \ldots \! \int_{\r_N} \! \int_{\varphi_N} \! \int_{\gamma_N} \! V(\varphi_1, \r_1, t_1) e^{i \gamma_1 \varphi_1} \ldots V(\varphi_N, \r_N, t_N) e^{i \gamma_N \varphi_N } \\ 
& \times \underbrace{\sum_{n_1 = 0}^{n+J} \ldots \sum_{n_N = 0}^{n+J}}_{n_1 + \ldots + n_N = n + J}  \sum_{n_{11} = 0}^{\infty} \sum_{n_{12} = 0}^{\infty} \ldots \! \sum_{n_{1N} = 0}^{\infty}  \sum_{n_{22} = 0}^{\infty} \sum_{n_{23} = 0}^{\infty} \ldots \! \sum_{n_{NN} = 0}^{\infty} \\
& \times \left[ \frac{(-i \gamma_1)^{ n_1 + 2n_{11} + \sum_{j \neq 1} n_{1j} } \ldots (-i \gamma_N)^{n_N  + 2n_{NN} + \sum_{j \neq N} n_{Nj}  }}{\left( n_1 + 2n_{11} + \sum_{j \neq 1} n_{1j}  \right)! \, \ldots \, \left(n_N + 2n_{NN} + \sum_{j \neq N} n_{Nj}  \right)!} \binom{ n_1 + 2n_{11} + \sum_{j \neq 1} n_{1j}  }{n_1, 2n_{11}, n_{11}, \ldots, n_{1N}} \right. \\ 
& \,\,\,\,\,\, \left. \ldots \times \binom{ n_N + 2n_{NN} + \sum_{j \neq N} n_{Nj}  }{n_N, 2n_{NN}, n_{N1}, \ldots, n_{NN}} n_1! \ldots n_N! \right. \\
& \,\,\,\,\,\, \left. \times \frac{(2n_{11})!}{2^{n_{11}} n_{11}!} \left( \int_k \Delta(t_1, t_1, k) \right)^{n_{11}} \ldots \,  \frac{(2n_{NN})!}{2^{n_{NN}} n_{NN}!} \left( \int_k \Delta(t_N, t_N, k) \right)^{n_{NN}} \right. \\
& \,\,\,\,\,\, \left. \times n_{12}! \left( \int_k \Delta(t_1, t_2, k) e^{i \k \cdot (\r_1 - \r_2)}\right)^{n_{12}}  \cdots n_{(N-1)N}! \left( \int_k \Delta(t_{(N-1)}, t_N, k) e^{i \k \cdot (\r_{N-1} - \r_N)}\right)^{n_{(N-1)N}}   \right] \\
& \times \sum_{\{i_b\}} \exp\left(-i \sum_{b=1}^{n+2J} \q_{b} \cdot \r_{i_b}\right) \left( \Delta(t_0, t_{i_1}, q_1) \ldots \Delta(t_0, t_{i_{J}}, q_J) \Delta(t_{i_{J+1}}, t, k_1) \right. \\ 
& \left.  \ldots \times \Delta(t, t_{i_{J+n}}, k_n) \Delta(t_{i_{J+n+1}}, t_0, p_1) \ldots \Delta(t_{i_{n+2J}}, t_0, p_J) \right)
\end{split}
\ee
where we have written $\q_{b} = \k_{b-J}$ for $b = J+1, \ldots, J+n$ and $\q_b = \p_{b-(J+n)}$ for $b=J+n+1, \ldots, 2J+N$ in the exponential of the last line. This last sum (over $\{i_b\}$) accounts for all possible ways of connecting the outer legs to the vertices, with the restriction that $i_b$ must take the value $a$ for $n_a$ values of $b$. On the other hand, a myriad of cancellations occur inside the square bracket. After carrying them out, we obtain:
\be \label{npoint4}
\begin{split}
& \int_{\r_1} \! \int_{\varphi_1} \! \int_{\gamma_1} \! \ldots \! \int_{\r_N} \! \int_{\varphi_N} \! \int_{\gamma_N} \! V(\varphi_1, \r_1, t_1) e^{i \gamma_1 \varphi_1} \ldots V(\varphi_N, \r_N, t_N) e^{i \gamma_N \varphi_N } \\
& \times \underbrace{\sum_{n_1 = 0}^{n+J} \ldots \sum_{n_N = 0}^{n+J}}_{n_1 + \ldots + n_N = n + J}  \sum_{n_{11} = 0}^{\infty} \sum_{n_{12} = 0}^{\infty} \ldots \! \sum_{n_{1N} = 0}^{\infty}  \sum_{n_{22} = 0}^{\infty} \sum_{n_{23} = 0}^{\infty} \ldots \! \sum_{n_{NN} = 0}^{\infty} \\ 
& \times (-i \gamma_1)^{n_1} \ldots (-i \gamma_N)^{n_N}  \left( \prod_{i<j}^N \frac{1}{n_{ij}!} \left( - \gamma_i \gamma_j \int_k \Delta(t_i, t_j, k) e^{i \k \cdot (\r_i - \r_j)}\right)^{n_{ij}} \right) \\ 
& \times \left(\prod_{i=1}^N \frac{1}{n_{ii}!} \left( - \frac{\gamma_i^2}{2} \int_k \Delta(t_i, t_i, k) \right)^{n_{ii}}  \right) \\
& \times \sum_{\{i_b\}} \exp\left(-i \sum_{b=1}^{n+2J} \q_{b} \cdot \r_{i_b}\right) \left( \Delta(t_0, t_{i_1}, q_1) \ldots \Delta(t_0, t_{i_{J}}, q_J) \Delta(t_{i_{J+1}}, t, k_1) \right. \\ 
& \left.  \ldots \times \Delta(t, t_{i_{J+n}}, k_n) \Delta(t_{i_{J+n+1}}, t_0, p_1) \ldots \Delta(t_{i_{n+2J}}, t_0, p_J) \right)
\end{split}
\ee
which can be recast as
\be \label{npoint5}
\begin{split}
& \underbrace{\sum_{n_1 = 0}^{n+2J} \ldots \sum_{n_N = 0}^{n+2J}}_{n_1 + \ldots + n_N = n + 2J} \!\! \int_{\r_1} \!\! \int_{\varphi_1} \!\! \int_{\gamma_1} \!\! \ldots \!\! \int_{\r_N} \!\! \int_{\varphi_N} \!\! \int_{\gamma_N} \! \dfrac{\partial^{n_1} V}{\partial \varphi_1^{n_1}} \ldots \dfrac{\partial^{n_N} V}{\partial \varphi_N^{n_N}} e^{i \sum_{j} \gamma_j \varphi_j} \\
& \times \exp \left( \sum_{i=1}^N \left( -\frac{1}{2} \gamma_i^2 \sigma_0^2(t_i) \right) + \sum_{\substack{i<j \\ i,j=1}}^N \left( - \gamma_i \gamma_j \sigma^2(t_i, t_j, |\r_i - \r_j|) \right) \right) \\
& \times \sum_{\{i_b\}} \exp\left(-i \sum_{b=1}^{n+2J} \q_{b} \cdot \r_{i_b}\right) \left( \Delta(t_0, t_{i_1}, q_1) \ldots \Delta(t_0, t_{i_{J}}, q_J) \Delta(t_{i_{J+1}}, t, k_1) \right. \\ 
& \left. \ldots \times \Delta(t, t_{i_{J+n}}, k_n) \Delta(t_{i_{J+n+1}}, t_0, p_1) \ldots \Delta(t_{i_{n+2J}}, t_0, p_J) \right)
\end{split}
\ee
where we have omitted the arguments of the potential $V(\varphi, \r, t)$. Now performing the integrations over the \(\gamma\) variables, we end up with
\be \label{npoint55}
\begin{split}
& \underbrace{\sum_{n_1 = 0}^{n+2J} \ldots \sum_{n_N = 0}^{n+2J}}_{n_1 + \ldots + n_N = n + 2J} \int_{\r_1} \int_{\varphi_1} \ldots \int_{\r_N} \int_{\varphi_N} \dfrac{\partial^{n_1} V}{\partial \varphi_1^{n_1}} \ldots \dfrac{\partial^{n_N} V}{\partial \varphi_N^{n_N}}  \frac{\exp \left( -\frac{1}{2} \varphi_i \left(\Sigma_I^{-1}\right)_{ij} \varphi_j \right)}{\sqrt{(2\pi)^N |\text{det} \Sigma_I|}} \\
& \times \sum_{\{i_b\}} \exp\left(-i \sum_{b=1}^{n+2J} \q_{b} \cdot \r_{i_b}\right) \left( \Delta(t_0, t_{i_1}, q_1) \ldots \Delta(t_0, t_{i_{J}}, q_J) \Delta(t_{i_{J+1}}, t, k_1) \right. \\ 
& \left. \ldots \times \Delta(t, t_{i_{J+n}}, k_n) \Delta(t_{i_{J+n+1}}, t_0, p_1) \ldots \Delta(t_{i_{n+2J}}, t_0, p_J) \right)
\end{split}
\ee
where the matrix elements of $\Sigma_I$ are given by $(\Sigma_I)_{ij} = \sigma^2(t_{\text{min}\{i,j\}},t_{\text{max}\{i,j\}}, |\r_i -\r_j|)$, and we sum over repeated indices at this instance. Taking Fourier transform to position space over $\q_i$ (with conjugate variables $\z_i$), $\k_i$ (with conjugate variables $\x_i$) and $\p_i$ (with conjugate variables $\y_i$), we arrive at
\be \label{npoint6a}
\begin{split}
\bra{0} \! \varphi_I(\z_1, t_0) & \ldots \varphi_I(\z_J, t_0)  H_I(t_1) \ldots H_I(t_l)   \varphi_I(\x_1,t) \ldots \varphi_I(\x_n,t) \\ & \,\,\,\,\,\,\,\,\,\,\,\,\,\,\,\,\,\,\,\,\,\,\,\,\,\,\,\,\,\,\, \times H_I(t_{l+1}) \ldots H_I(t_N) \varphi_I(\y_1, t_0) \ldots \varphi_I(\y_J, t_0) \! \ket{0}_{FI} \\
 = & \underbrace{\sum_{n_1 = 0}^{n+2J} \ldots \sum_{n_N = 0}^{n+2J}}_{n_1 + \ldots + n_N = n + 2J} \int_{\r_1} \int_{\varphi_1} \ldots \int_{\r_N} \int_{\varphi_N} \dfrac{\partial^{n_1} V}{\partial \varphi_1^{n_1}} \ldots \dfrac{\partial^{n_N} V}{\partial \varphi_N^{n_N}}  \frac{\exp \left( -\frac{1}{2} \varphi_i \left(\Sigma_I^{-1}\right)_{ij} \varphi_j \right)}{\sqrt{(2\pi)^N |\text{det} \Sigma_I|}} \\
& \times \sum_{\{i_b\}} \left( \sigma^2(t_0, t_{i_1}, |\z_1 - \r_{i_1}|) \ldots \sigma^2(t_0, t_{i_{J}}, |\z_J - \r_{i_J}|) \sigma^2(t_{i_{J+1}}, t, |\x_1 - \r_{i_{J+1}}|) \right. \\
&  \left. \ldots \times \sigma^2(t, t_{i_{J+n}}, |\x_n - \r_{i_{J+n}}|) \sigma^2(t_{i_{J+n+1}}, t_0, |\y_1 - \r_{i_{J+n+1}}|)  \right. \\
& \left. \ldots \times \sigma^2(t_{i_{n+2J}}, t_0, |\y_J - \r_{i_{n+2J}}|) \right).
\end{split}
\ee
So far, we have only dealt with the fully interacting contributions to the correlation. But from this end of the computation, we can appreciate some structure emerging in the result: it is an expectation value over a gaussian probability density function. With this in mind, we claim that we can write down
\be \label{npoint7}
\begin{split}
\bra{0} \! \varphi_I(\z_1, t_0) & \ldots \varphi_I(\z_J, t_0)  H_I(t_1) \ldots H_I(t_l)   \varphi_I(\x_1,t) \ldots \varphi_I(\x_n,t) \\ & \,\,\,\,\,\,\,\,\,\,\,\,\,\,\,\,\,\,\,\,\,\,\,\,\,\,\,\,\,\,\, \times H_I(t_{l+1}) \ldots H_I(t_N) \varphi_I(\y_1, t_0) \ldots \varphi_I(\y_J, t_0) \! \ket{0} \\
 = & \int_{\varphi_{\z_1}} \!\!\!\!\!\! \ldots \! \int_{\varphi_{\z_J}} 
\int_{\varphi_{\x_1}} \!\!\!\!\!\!  \ldots \! \int_{\varphi_{\x_N}} \!\int_{\varphi_{\y_1}}  \!\!\!\!\!\! \ldots \! \int_{\varphi_{\y_J}} \! \int_{\r_1} \int_{\varphi_1} \ldots \int_{\r_N} \int_{\varphi_N}  \\
& \times V(\varphi_1, \r_1, t_1) \ldots  V(\varphi_N, \r_N, t_N)  \frac{\exp \left( -\frac{1}{2} \pmb{\varphi}^T \cdot \left(\pmb{\Sigma}^{-1}\right) \cdot \pmb{\varphi} \right)}{\sqrt{(2\pi)^{N+n+2J} |\text{det} \pmb{\Sigma} |}} \\
& \times \varphi_{\z_1} \ldots \varphi_{\z_J} \varphi_{\x_1} \ldots \varphi_{\x_n} \varphi_{\y_1} \ldots \varphi_{\y_J} 
\end{split}
\ee
with \(\pmb{\varphi}^T \equiv (\varphi_{\z_1} \,\,\, \ldots \,\,\, \varphi_{\z_J}  \,\,\, \varphi_{1} \,\,\, \ldots \,\,\, \varphi_l \,\,\, \varphi_{\x_1} \,\,\, \ldots \,\,\, \varphi_{\x_n} \,\,\, \varphi_{l+1} \,\,\, \ldots \,\,\, \varphi_N \,\,\, \varphi_{\y_1} \,\,\, \ldots \,\,\, \varphi_{\y_J}  )\), and \(\pmb{\Sigma}\) as the corresponding covariance matrix. 

The covariances in the last expression are the propagators between the fields' corresponding spacetime positions, with their respective temporal arguments ordered as the fields are in the definition of $\pmb{\varphi}^T$. For example, the covariance relating $\varphi_{\z_a}$ and $\varphi_{b}$ is $\sigma^2(t_0,t_b,|\z_a-\r_b|)$, and the one relating $\varphi_{\x_i}$ with $\varphi_{b}$ would be $\sigma^2(t,t_b,|\x_i-\r_b|)$ if $b\geq l+1$, while it would be $\sigma^2(t_b,t,|\x_i-\r_b|)$ if $b \leq l$.  

Note that the part of the correlations which we haven't computed explicitly in this appendix is given by terms that are products of free theory pairings between the external legs and a fully interacting contribution involving the remaining fields (it is not essential that the number of fields at $t_0$ is equal at both sides of the interaction), and it is fairly easy to check that the correlators of the free theory are given by a gaussian distribution as in~\eqref{npoint7} without the $\varphi_i$ terms. So, \textit{a posteriori}, the claim doesn't seem unreasonable. The proof is given in the main text.

\vspace{0.5cm}

\section{The in-in PDF for an arbitrary initial state} \label{sec:arbitrary-initial}

Had we kept the fields that define the \textit{in} state within section~\ref{subsec:main-PDF}, we would have arrived to
\be
\begin{split}
\langle \varphi(\x_1,t) \ldots \varphi(\x_n,t) \rangle =  & \sum_{N=0}^{\infty} (-i)^N \sum_{l=0}^{N} (-1)^l \int_{t_0 - i\epsilon |t_0|}^t \!\!\!\!\!\!\!\!\! dt_l \ldots
\int_{t_0 - i\epsilon |t_0|}^{t_2} \!\!\!\!\!\!\!\!\! dt_1 \int_{t_0 + i\epsilon |t_0|}^t \!\!\!\!\!\!\!\!\! dt_{l+1} \ldots \int_{t_0 + i\epsilon |t_0|}^{t_{N-1}} \!\!\!\!\!\!\!\!\! dt_N \\
& \times \int_{\r_1} \ldots \int_{\r_N} \!\int_{\varphi_1} \!\!\!  \ldots \int_{\varphi_N}  \int_{\varphi_{\z_1}} \!\!\!\!\!\! \ldots \! \int_{\varphi_{\z_J}} \! \int_{\varphi_{\x_1}} \!\!\!\!\!\!  \ldots \! \int_{\varphi_{\x_N}} \! \int_{\varphi_{\y_1}}  \!\!\!\!\!\! \ldots \! \int_{\varphi_{\y_J}} \\ 
& \times V(\varphi_1, \r_1, t_1) \ldots  V(\varphi_N, \r_N, t_N)  \frac{\exp \left( -\frac{1}{2} \pmb{\varphi}^T \cdot \left(\pmb{\Sigma}^{-1}\right) \cdot \pmb{\varphi} \right)}{\sqrt{(2\pi)^{N+n} |\text{det} \pmb{\Sigma} |}} \\
& \times \varphi_{\z_1} \ldots \varphi_{\z_J} \varphi_{\x_1} \ldots \varphi_{\x_n} \varphi_{\y_1} \ldots \varphi_{\y_J} 
\end{split}
\ee
where we haven't set $\z_i = \y_i$ yet in order to avoid equivocal statements, since the covariance associated to a contraction of $\varphi_\z$ with a vertex is not equal to that of $\varphi_\y$ (in fact they are complex conjugates). 

The subsequent steps follow in the same way as in the main text, yielding
\be \label{PDFalt}
\begin{split}
\rho_{\varphi} =& \sum_{N=0}^{\infty} (-i)^N \sum_{l=0}^{N} (-1)^l \int_{t_0 - i\epsilon |t_0|}^t \!\!\!\!\!\!\!\!\! dt_l \ldots
\int_{t_0 - i\epsilon |t_0|}^{t_2} \!\!\!\!\!\!\!\!\! dt_1 \int_{t_0 + i\epsilon |t_0|}^t \!\!\!\!\!\!\!\!\! dt_{l+1} \ldots \int_{t_0 + i\epsilon |t_0|}^{t_{N-1}} \!\!\!\!\!\!\!\!\! dt_N \\ 
& \times \int_{\r_1} \ldots \int_{\r_N} \!\int_{\varphi_1} \!\!\!  \ldots \int_{\varphi_N}  \int_{\varphi_{\z_1}} \!\!\!\!\!\! \ldots \! \int_{\varphi_{\z_J}}
 \! \int_{\varphi_{\y_1}}  \!\!\!\!\!\! \ldots \! \int_{\varphi_{\y_J}}  \\ 
& \times \frac{\exp \left( -\frac{1}{2} \pmb{\varphi}^T \cdot \left(\pmb{\Sigma}^{-1}\right) \cdot \pmb{\varphi} \right)}{\sqrt{(2\pi)^{N+n} |\text{det} \pmb{\Sigma} |}} \\
& \times \varphi_{\z_1} \ldots \varphi_{\z_J} V(\varphi_1, \r_1, t_1) \ldots  V(\varphi_N, \r_N, t_N) \varphi_{\y_1} \ldots \varphi_{\y_J} \big|_{\y_i = \z_i} ,
\end{split}
\ee
where we have to stress that when performing the field integrals in this last expression the arguments of the covariances must be taken in the same order as the fields are written, and after doing that, set $\y_i = \z_i$ so that the \textit{in} states match. 

We can also write this as a functional integral
\be \label{Path-integral2a}
\begin{split}
\rho_{\varphi} = \int \!  D\varphi_- D\varphi_+  &  \left( \varphi_-(\y_1,t_0) \ldots \varphi_-(\y_J,t_0) \exp \left\{ +i\int^{t}_{t_0 - i\epsilon |t_0|} \!\!\!\!\!\!\!\!\! d t' \int_{\r} V(\varphi_-(\r,t'),\r,t') \right\} \right) \! 
\\ & \times \frac{\exp \left( - \frac{1}{2} \varphi \cdot \left( {\Sigma}^{-1} \right) \cdot \varphi \right)}{\sqrt{|\text{det} (2\pi \Sigma) |}} \\
& \times \left( \exp \left\{ - i \int^{t}_{t_0 + i\epsilon |t_0|} \!\!\!\!\!\!\!\!\! d t' \int_{\r} V(\varphi_+(\r,t'),\r,t') \right\} \varphi_+(\y_1,t_0) \ldots \varphi_+(\y_J,t_0) \right),
\end{split}
\ee
in which the distinction between $\varphi_+$ and $\varphi_-$ fields, with their respective time orderings, makes the result easier to write down. As in the main text, the resulting propagators/covariances involving a $\varphi_+$ field and another type of field ($\varphi_+$ or $\varphi$) always have the corresponding time in the second temporal argument, and are time-ordered if it is a $\varphi_+ \varphi_+$ contraction. Conversely, the entries of the propagators corresponding to $\varphi_-$ always go to the left when contracted with another type of field and are anti-time-ordered when considering a $\varphi_- \varphi_-$ contraction.

\cleardoublepage

\chapter{Partition Function and 2-point PDFs in the presence of Local NG} 

\section{Partition Function} \label{sec:Partition}

The defining property of a partition function is that upon functional differentiation as in~\eqref{Z-corr}, it should give the $n$-point functions
\be
\langle \zeta({\k}_1 ) \cdots \zeta({\k}_n )\rangle = \frac{\delta^n Z[J]}{(i\delta J(-{\k}_1 )) \cdots (i \delta J(-{\k}_n)) }\bigg|_{J = 0} = \int D\zeta \rho[\zeta] \zeta({\k}_1 ) \cdots \zeta({\k}_n ),
\ee
which is accomplished by taking the functional Fourier transform of the PDF~\eqref{Z-Fourier}
\be \label{Z-Fourier-app}
Z[J] = \int D \zeta \, \rho[\zeta] \, e^{i\int_\k \zeta({\k}) J(-{\k})} = \int D \zeta \, \rho[\zeta] \, e^{i\int_\y \zeta({\y}) J({\y})}.
\ee
To evaluate this expression, let us use that $\rho[\zeta]$ is given by a Gaussian distribution times a correction:
\begin{align}
\rho_F[\zeta] =& \rho_G[\zeta] \times \left[ 1 - \int_\x  \int_{\k} \frac{\partial F}{\partial \zeta} (\zeta(\x)) + \int_\x \int_\y \zeta(\x) \int_{\k} \frac{e^{i \k \cdot (\x - \y)} }{P_\zeta(k)}  \int_\z \int_{\bf q} e^{i {\bf q} \cdot (\y - \z)} F(\zeta(\z))   \right] \nn  \\
=&  \rho_G[\zeta] \times \left[ 1 - \int_\x  \int_{\k} \frac{\partial F}{\partial \zeta} (\zeta(\x)) + \int_\x  \int_{\k} \int_\z \zeta(\x)  \frac{e^{i \k \cdot (\x - \z)} }{P_\zeta(k)} F(\zeta(\z))   \right].
\end{align}

We will now evaluate~\eqref{Z-Fourier-app} by expanding $F$ in a power series and using Wick's theorem. 
%Let
%\be
%F(\zeta) = \sum_{n=0}^\infty \frac{g_n}{n!} \zeta^n.
%\ee
Since~\eqref{Z-Fourier-app} may be read as computing the expectation value of
\begin{align} \label{expectand}
 e^{i\int_\y \zeta({\y}) J({\y})} \left[ 1 - \int_\x  \int_{\k} \frac{\partial F}{\partial \zeta} (\zeta(\x)) + \int_\x  \int_{\k} \int_\z \zeta(\x)  \frac{e^{i \k \cdot (\x - \z)} }{P_\zeta(k)} F(\zeta(\z))   \right]
%=& \sum_{m = 0}^\infty \left( \frac{i^m}{m!} \left(\int_{\k} \zeta(\k) J(-\k)  \right)^m \right) \left[ 1 - \sum_{n=0}^\infty \left\{ \int_\x  \int_{\k} \frac{\partial F}{\partial \zeta} (\zeta(\x)) + \int_\x  \int_{\k} \int_\z \zeta(\x)  \frac{e^{i \k \cdot (\x - \z)} }{P_\zeta(k)} F(\zeta(\z))  \right\}  \right]
\end{align}
over a Gaussian measure, we will do exactly that. Therefore, we will have to compute
\begin{align} \label{expectands}
\left\langle e^{i\int_\y \zeta({\y}) J(\y)} \frac{\partial F}{\partial \zeta} (\zeta(\x))  \right\rangle_G, & & \left\langle e^{i\int_\y \zeta({\y}) J(\y)}  \zeta(\x) F(\zeta(\z))  \right\rangle_G,
\end{align}
where the subscript $G$ instructs to take the expectation value over a Gaussian measure. Note that the second expectation value, per Wick's theorem, can be written as the sum of two expectation values
\be
\begin{split}
\left\langle e^{i\int_\y \zeta({\y}) J({\y})}  \zeta(\x) F(\zeta(\z))  \right\rangle_G =& \, i \int_\w J(\w) \left\langle \zeta(\w) \zeta(\x)  \right\rangle_G \left\langle e^{i\int_\y \zeta({\y}) J({\y})} F(\zeta(\z))  \right\rangle_G \\ 
& + \left\langle  \zeta(\x) \zeta(\z)  \right\rangle_G \left\langle e^{i\int_\y \zeta({\y}) J({\y})}  \frac{\partial F}{\partial \zeta}(\zeta(\z))  \right\rangle_G \\
=& \, i \int_\w J(\w) \Sigma(\w,\x) \left\langle e^{i\int_\y \zeta({\y}) J({\y})} F(\zeta(\z))  \right\rangle_G \\ 
& + \Sigma(\x,\z) \left\langle e^{i\int_\y \zeta({\y}) J({\y})}  \frac{\partial F}{\partial \zeta}(\zeta(\z))  \right\rangle_G.
\end{split}
\ee
Now, using that $\Sigma(\x,\z) = \int_\q P_\zeta(q) e^{i \q \cdot(\x - \z) }$, and replacing the last term into the corresponding term of~\eqref{expectand}, we get
\be
\int_\x  \int_{\k} \int_\z \frac{e^{i \k \cdot (\x - \z)} }{P_\zeta(k)}  \int_\q P_\zeta(q) e^{i \q \cdot(\x - \z) } \left\langle e^{i\int_\y \zeta({\y}) J({\y})}  \frac{\partial F}{\partial \zeta}(\zeta(\z))  \right\rangle_G.
\ee 
Then, integrating over $\x$ gives $|\q| = |\k|$, and thus yielding
\be
\int_\z  \int_{\k} \left\langle e^{i\int_\y \zeta({\y}) J({\y})}  \frac{\partial F}{\partial \zeta}(\zeta(\z))  \right\rangle_G,
\ee
which is equal but opposite in sign to the first term of~\eqref{expectands}. Therefore, those two cancel out, and we only have to compute
\be
\int_\x  \int_{\k} \int_\z \frac{e^{i \k \cdot (\x - \z)} }{P_\zeta(k)} i \int_\w J(\w) \Sigma(\w,\x) \left\langle e^{i\int_\y \zeta({\y}) J({\y})} F(\zeta(\z))  \right\rangle_G.
\ee

If we define
\be
F(\zeta) = \sum_{n=0}^\infty \frac{g_n}{n!} \zeta^n,
\ee
then the object of interest in the computation is
\be
\frac{i^m g_n}{m! n!} \left\langle \left( \int_\y \zeta({\y}) J({\y})\right)^m \zeta(\z)^n  \right\rangle_G.
\ee
There are three type of contractions in this expression: two self-contractions, of fields originating from equivalent expressions, and a mixed contraction. Performing the contractions and combinatorics gives
\be
\begin{split}
\frac{i^m g_n}{m! n!} \sum_{\substack{m',n',\ell' \\ 2m' + \ell' = m \\ 2n' + \ell' = n }} & \frac{m!}{2^{m'} m'! (m - 2m' - \ell')! } \left( \int_x \int_y J(\x) \Sigma(\x,\y) J(\y) \right)^{m'} \\ 
& \times \frac{1}{\ell'!} \left( \int_\y J(\y) \Sigma(\y,\z) \right)^{\ell'} \\ 
& \times \frac{n!}{2^{n'} n'! (n - 2n' - \ell')! } \left( \Sigma(\z,\z) \right)^{n'},
\end{split}
\ee
which one can sum over $n,m$ to eliminate the constraints of the $m',n',\ell'$ sums. This gives
\begin{align}
& \left\langle e^{i\int_\y \zeta({\y}) J({\y})} F(\zeta(\z))  \right\rangle_G \nn \\
&= \sum_{n',m',\ell'=0}^\infty \frac{i^{2m' + \ell'} g_{2n' + \ell}}{2^{m'} m'! 2^{n'} n'! \ell'!} \left( \int_x \int_y J(\x) \Sigma(\x,\y) J(\y) \right)^{m'}  \left( \int_\y J(\y) \Sigma(\y,\z) \right)^{\ell'}  \left( \Sigma(\z,\z) \right)^{n'} \nn \\
&= \exp \left[- \frac{1}{2} \int_x \int_y J(\x) \Sigma(\x,\y) J(\y) \right] \sum_{n',\ell'=0}^\infty g_{2n'+\ell'} \frac{1}{n'!} \left( \frac{1}{2} \Sigma(\z,\z) \right)^{n'} \frac{1}{\ell'!} \left(i \int_\y J(\y) \Sigma(\y,\z) \right)^{\ell'} \nn \\
&=  e^{- \frac{1}{2} \int_x \int_y J(\x) \Sigma(\x,\y) J(\y)} \sum_{n',\ell'=0}^\infty \frac{\left(i \int_\y J(\y) \Sigma(\y,\z) \frac{\partial }{\partial \zeta}  \right)^{\ell'}}{\ell'!}   \frac{\left( \frac{1}{2} \Sigma(\z,\z) \frac{\partial^2 }{\partial \zeta^2} \right)^{n'}}{n'!}  \left( \sum_{n=0}^\infty \frac{ g_{n}}{n!} \zeta^n \right) \bigg|_{\zeta = 0} \nn \\
&=  e^{- \frac{1}{2} \int_x \int_y J(\x) \Sigma(\x,\y) J(\y)}   \exp \left[\frac{\Sigma(\z,\z)}{2} \frac{\partial^2}{\partial \zeta^2} \right]  \left( F(\zeta) \right) \bigg|_{\zeta = i \int_\y J(\y) \Sigma(\y,\z)},
\end{align}
and identifying $\sigma_\zeta^2 = \Sigma(\z,\z)$, if we use the Weierstrass transform we get
\begin{align}
& \int_\x  \int_{\k} \int_\z \frac{e^{i \k \cdot (\x - \z)} }{P_\zeta(k)} i \int_\w J(\w) \Sigma(\w,\x) \left\langle e^{i\int_\y \zeta({\y}) J({\y})} F(\zeta(\z))  \right\rangle_G \nn \\
&= i e^{- \frac{1}{2} \int_x \int_y J(\x) \Sigma(\x,\y) J(\y)} \int_\z J(\z) \!\!  \int_{-\infty}^\infty \!\!\!\! d\zeta \frac{\exp \left\{- \frac{ \left( \zeta - i \int_\y J(\y) \Sigma(\y,\z) \right)^2 }{2 \sigma_\zeta^2} \right\} }{\sqrt{2\pi \sigma_\zeta^2}} F(\zeta) .
\end{align}
Finally, we use that
\begin{align}
 \frac{\partial }{\partial \zeta} e^{- \frac{ \left( \zeta - i \int_\y J(\y) \Sigma(\y,\z) \right)^2 }{2 \sigma_\zeta^2} } &= e^{- \frac{ \left( \zeta - i \int_\y J(\y) \Sigma(\y,\z) \right)^2 }{2 \sigma_\zeta^2} } \left( i \int_\y J(\y) \Sigma(\y,\z) - \zeta \right) \frac{1}{\sigma_\zeta^2} \nn \\
 \implies e^{- \frac{ \left( \zeta - i \int_\y J(\y) \Sigma(\y,\z) \right)^2 }{2 \sigma_\zeta^2} } &= \frac{1}{i \int_\y J(\y) \Sigma(\y,\z)} \left( \zeta + \sigma_\zeta^2 \frac{\partial }{\partial \zeta}  \right) e^{- \frac{ \left( \zeta - i \int_\y J(\y) \Sigma(\y,\z) \right)^2 }{2 \sigma_\zeta^2} },
\end{align}
to get, after integration by parts,
\begin{align}
& \int_\x  \int_{\k} \int_\z \frac{e^{i \k \cdot (\x - \z)} }{P_\zeta(k)} i \int_\w J(\w) \Sigma(\w,\x) \left\langle e^{i\int_\y \zeta({\y}) J({\y})} F(\zeta(\z))  \right\rangle_G \nn \\
&=  e^{- \frac{1}{2} \int_x \int_y J(\x) \Sigma(\x,\y) J(\y)} \!\! \int_\z \frac{J(\z)}{\int_\y J(\y) \Sigma(\y,\z)} \!\!  \int_{-\infty}^\infty \!\!\!\! d\zeta \frac{\exp \left\{- \frac{ \left( \zeta - i \int_\y J(\y) \Sigma(\y,\z) \right)^2 }{2 \sigma_\zeta^2} \right\} }{\sqrt{2\pi \sigma_\zeta^2}} \left( \zeta - \sigma_\zeta^2 \frac{\partial^2}{\partial \zeta^2} \right) F(\zeta) \nn \\
&=  \exp \left[- \frac{1}{2} \int_k \int_y |J(\k)|^2 P_\zeta(k) \right] \nn \\ 
& \,\,\,\, \times \int_\x \frac{\int_\k e^{i\k \cdot \x} J(-\k)}{\int_\k  e^{i\k \cdot \x} J(-\k)  P_\zeta(k) } \!\!  \int_{-\infty}^\infty \!\!\!\! d\zeta \frac{\exp \left\{- \frac{ \left( \zeta - i \int_k e^{i\k \cdot \x} J(-\k) P_\zeta(k) \right)^2 }{2 \sigma_\zeta^2} \right\} }{\sqrt{2\pi \sigma_\zeta^2}} \left( \zeta - \sigma_\zeta^2 \frac{\partial^2}{\partial \zeta^2} \right) F(\zeta),
\end{align}
which finally gives, as shown in the main text,
\be 
\begin{split}
Z[&J]  = \, \exp \left[-\frac{1}{2} \int_\k J(\k) J(-\k) P_\zeta(k) \right] \;\times\\  &  \left( 1 - \int_\x \frac{\int_\k e^{i\k \cdot \x} J(-\k)}{ \int_\k e^{i\k \cdot \x} J(-\k) P_\zeta(k) } \int_{ \zeta} \frac{\exp \left[ - \frac{\left({ \zeta} - i\int_\k e^{i\k \cdot \x} J(-\k) P_\zeta(k) \right)^2}{2 \sigma_\zeta^2} \right] }{\sqrt{2 \pi} \sigma_\zeta} \left(\sigma_\zeta^2 \frac{\partial }{\partial \zeta} - {\zeta}\right)  F(\bar \zeta) \right).
\end{split}
\ee

\section{The 2-point PDFs} \label{sec:2-point-det}

\subsection{2-point PDF: three-dimensional case}

The 2-point PDF, to our knowledge, has not been derived earlier and thus we outline the procedure with more detail. As in the previous case, the PDF must include the non-fully connected contributions. This is a combinatorial mess, for we expect
\be
\begin{split}
\langle \zeta^{n_1}(\x_1) \zeta^{n_2}(\x_2) \rangle = \sum_{m_1, m_2, m_t} & \#_{n_i,n_2,m_1,m_2, m_t} \left(\sigma_{\zeta}^2|_{\x_1}\right)^{m_1} \left(\sigma_{\zeta}^2(x) \right)^{m_t} \left(\sigma_{\zeta}^2|_{\x_2}\right)^{m_2} \\ & \times \langle \zeta^{n_1 - 2m_1 - m_t}(\x_1) \zeta^{n_2 - 2m_2 - m_t}(\x_2) \rangle_c.
\end{split}
\ee
Let us calculate $\#_{n_1,n_2,m_1,m_2,m_t}$: we have an overall factor $n_1! n_2!$ from which we must divide the overcounted terms. In this counting, we have $m_t!$ redundant permutations when connecting $\x_1$ and $\x_2$, plus $2^{m_1} m_1! 2^{m_2} m_2!$ when pairing amongst themselves. Finally, the ones that are assigned to the fully connected contribution undergo no further permutation, thus we must also divide by $(n_1 - 2m_1 - m_t)! (n_2 - 2m_2 - m_t)!$. Thus,
\be \label{bipoint}
\begin{split}
\langle \zeta^{n_1}(\x_1) \zeta^{n_2}(\x_2) \rangle = \sum_{m_1, m_2, m_t} & \frac{n_1! \, n_2! \left(\sigma_{\zeta}^2|_{\x_1}\right)^{m_1} \left(\sigma_{\zeta}^2(x) \right)^{m_t} \left(\sigma_{\zeta}^2|_{\x_2}\right)^{m_2}  }{2^{m_1} m_1! \, (n_1 - 2m_1 - m_t)! \, m_t! \, (n_2 - 2m_2 - m_t)! \,  2^{m_2} m_2!} \\ & \times \langle \zeta^{n_1 - 2m_1 - m_t}(\x_1) \zeta^{n_2 - 2m_2 - m_t}(\x_2) \rangle_c.
\end{split}
\ee
Careful inspection of this result reveals that~\eqref{bipoint} leads to a two-point distribution analogous to what was obtained in~\cite{Chen:2018brw}, but with two points defining the filtering instead of one:
\be
\begin{split} \label{PDF-2-app}
\rho (\zeta_1,\zeta_2) = \rho_G(\zeta_1,\zeta_2) &  \left[ 1 + \int_\x \! \int_{-\infty}^{\infty} \!\!\!\! d \bar \zeta   \frac{\exp \Big[{-\frac{ \left(\bar \zeta - \zeta(\x,\x_1,\x_2) \right)^2}{2\sigma_\zeta^2 (\x,\x_1,\x_2)} } \Big] }{\sqrt{2\pi  } \sigma_\zeta (\x,\x_1,\x_2)} \right. \\ & \left. \times \left\{  \frac{W(|\x - \x_1|)}{s^2(|\x - \x_1|)} \!  \left( G_{11} \frac{\partial}{\partial {\bar \zeta}} - G_{12} \right)  +  \frac{W(|\x - \x_2|)}{s^2(|\x - \x_2|)} \!  \left( G_{21} \frac{\partial}{\partial {\bar \zeta}} - G_{22} \right) \right\} F  \! \left({\bar \zeta}\right)  \right],
\end{split}
\ee
where the coefficients $F_{ij}$ depend on both the field variables and the spacetime positions $\x, \x_1, \x_2$.

In obtaining this expression, we have defined a number of functions that depend uniquely on the structure of the Gaussian theory. The variance $\sigma_{\zeta}^2(\x,\x_1,\x_2)$ and $\zeta(\x,\x_1,\x_2)$ are the regression coefficients obtained by conditioning a Gaussian distribution of $(\bar \zeta, \zeta_1, \zeta_2)$ over $(\zeta_1^W, \zeta_2^W)$, with covariance matrix given by~\eqref{cov}:
\begin{align}
\sigma_{W}^2(\x,\x_1,\x_2) =& \, \sigma_{\zeta}^2 + \frac{  2 s^2(|\x-\x_1|) s^2(|\x-\x_2|)  }{\sigma_{W}^4 - \sigma_{W,{\rm ext}}^4(|\x_1-\x_2|)} \sigma_{W,{\rm ext}}^2(|\x_1-\x_2|) \nn \\ & - \frac{s^4(|\x-\x_1|)  + s^4(|\x-\x_2|)   }{\sigma_{W}^4 - \sigma_{W,{\rm ext}}^4(|\x_1-\x_2|)} \sigma_{W}^2, \\
\zeta(\x,\x_1,\x_2) =& \, \frac{s^2(|\x-\x_1|)  \zeta_1  + s^2(|\x-\x_2|)  \zeta_2 }{\sigma_{W}^4 - \sigma_{W,{\rm ext}}^4(|\x_1-\x_2|)} \sigma_{W}^2 \nn \\ & - \frac{ s^2(|\x-\x_1|) \zeta_2   +  s^2(|\x-\x_2|)  \zeta_1 }{\sigma_{W}^4 - \sigma_{W,{\rm ext}}^4(|\x_1-\x_2|)} \sigma_{W,{\rm ext}}^2(|\x_1-\x_2|).
\end{align}
Furthermore, the coefficients $G_{ij}$ are given by
\begin{align}
G_{11} &\equiv \sigma_\zeta^2 - s^2(|\x-\x_2|) \frac{s^2(|\x-\x_2|) \sigma_W^2 - s^2(|\x-\x_1|)\sigma_{W,{\rm ext}}^2(|\x_1-\x_2|)}{\sigma_{W}^4 - \sigma_{W,{\rm ext}}^4(|\x_1-\x_2|)}, \\
G_{12} &\equiv {\bar \zeta} - s^2(|\x-\x_2|) \frac{ \sigma_W^2  \zeta_2 - \sigma_{W,{\rm ext}}^2(|\x_1-\x_2|) \zeta_1}{\sigma_{W}^4 - \sigma_{W,{\rm ext}}^4(|\x_1-\x_2|)}, \\
G_{21} &\equiv \sigma_\zeta^2 - s^2(|\x-\x_1|) \frac{s^2(|\x-\x_1|) \sigma_W^2 - s^2(|\x-\x_2|)\sigma_{W,{\rm ext}}^2(|\x_1-\x_2|)}{\sigma_{W}^4 - \sigma_{W,{\rm ext}}^4(|\x_1-\x_2|)}, \\
G_{22} &\equiv {\bar \zeta} - s^2(|\x-\x_1|) \frac{ \sigma_W^2  \zeta_1 - \sigma_{W,{\rm ext}}^2(|\x_1-\x_2|) \zeta_2 }{\sigma_{W}^4 - \sigma_{W,{\rm ext}}^4(|\x_1-\x_2|)}.
\end{align}

\subsection{2-point PDF: map to the sphere $S^2$}

The only difference with the previous Appendix is that the window function should be a map onto an angular coordinate $\n$ instead of a three-dimensional (flat) space. In the text, we wrote
\be
\begin{split}
\rho_\Theta (\Theta_1,\Theta_2) =& \, \rho_{G,W}(\Theta_1,\Theta_2)   \Bigg[ 1 - \int_\x \int_{-\infty}^{\infty} \!\!\!\! d \bar \zeta   \frac{\exp \Big[{-\frac{ \left(\bar \zeta - \zeta_\Theta(\x,\n_1,\n_2) \right)^2}{2\sigma_\Theta^2 (\x,\n_1,\n_2)} } \Big] }{\sqrt{2\pi  } \sigma_\Theta (\x,\n_1,\n_2)} \\ &  \;\times   \left\{  \frac{W_\Theta(\x, \n_1)}{s_\Theta^2(\x , \n_1)} \!  \left( G_{11}^{\Theta} \frac{\partial}{\partial {\bar \zeta}} - G_{12}^{\Theta} \right)   +  \frac{W_\Theta(\x, \n_2)}{s_\Theta^2(\x,  \n_2)} \!  \left( G_{21}^{\Theta} \frac{\partial}{\partial {\bar \zeta}} - G_{22}^{\Theta} \right) \right\} F  \! \left({\bar \zeta}\right)  \Bigg].
\end{split}
\ee
Here we have
\begin{align}
W_\Theta(\x,\n) \equiv \int_k e^{i \k \cdot \x} T(k,\n) & & s^2_\Theta(\x,\n) \equiv \int_k e^{i \k \cdot \x} T(k,\n) P_\zeta(k),
\end{align}
while the regression coefficients are given by
\begin{align}
\sigma_{\Theta}^2(\x,\n_1,\n_2) =& \, \sigma_{\zeta}^2 + \frac{  2 s_\Theta^2(\x,\n_1) s_\Theta^2(\x,\n_2)  }{\sigma_{\Theta}^4 - \sigma_{\Theta,{\rm ext}}^4(\n_1,\n_2)} \sigma_{\Theta,{\rm ext}}^2(\n_1,\n_2) \nn \\ & - \frac{s^4(\x,\n_1)  + s^4(\x,\n_2)   }{\sigma_{\Theta}^4 - \sigma_{\Theta,{\rm ext}}^4(\n_1,\n_2)} \sigma_{\Theta}^2, \\
\zeta_\Theta(\x,\n_1,\n_2) =& \, \frac{s_\Theta^2(\x,\n_1)  \Theta_1  + s_\Theta^2(\x,\n_2)  \Theta_2 }{\sigma_{\Theta}^4 - \sigma_{\Theta,{\rm ext}}^4(\n_1,\n_2)} \sigma_{\Theta}^2 \nn \\ & - \frac{ s_\Theta^2(\x,\n_1) \Theta_2   +  s_\Theta^2(\x,\n_2)  \Theta_1 }{\sigma_{\Theta}^4 - \sigma_{\Theta,{\rm ext}}^4(\n_1,\n_2)} \sigma_{\Theta,{\rm ext}}^2(\n_1,\n_2).
\end{align}
Finally, the functions $G_{ij}^\Theta$ are given by
\begin{align}
G_{11}^\Theta &\equiv \sigma_\zeta^2 - s_\Theta^2(\x,\n_2) \frac{s_\Theta^2(\x,\n_2) \sigma_\Theta^2 - s_\Theta^2(\x,\n_1)\sigma_{\Theta,{\rm ext}}^2(\n_1,\n_2)}{\sigma_{\Theta}^4 - \sigma_{\Theta,{\rm ext}}^4(\n_1,\n_2)}, \\
G_{12}^\Theta &\equiv {\bar \zeta} - s_\Theta^2(\x,\n_2) \frac{ \sigma_\Theta^2  \Theta_2 - \sigma_{\Theta,{\rm ext}}^2(\n_1,\n_2) \Theta_1}{\sigma_{\Theta}^4 - \sigma_{\Theta,{\rm ext}}^4(\n_1,\n_2)}, \\
G_{21}^\Theta &\equiv \sigma_\zeta^2 - s_\Theta^2(\x,\n_1) \frac{s_\Theta^2(\x,\n_1) \sigma_\Theta^2 - s_\Theta^2(\x,\n_2)\sigma_{\Theta,{\rm ext}}^2(\n_1,\n_2)}{\sigma_{\Theta}^4 - \sigma_{\Theta,{\rm ext}}^4(\n_1,\n_2)}, \\
G_{22}^\Theta &\equiv {\bar \zeta} - s_\Theta^2(\x,\n_1) \frac{ \sigma_\Theta^2  \Theta_1 - \sigma_{\Theta,{\rm ext}}^2(\n_1,\n_2) \Theta_2 }{\sigma_{\Theta}^4 - \sigma_{\Theta,{\rm ext}}^4(\n_1,\n_2)}.
\end{align}

\newpage
\thispagestyle{empty}
\mbox{}

%\section{A toy model for primordial non-Gaussianity}

%Try a Gauss-law like argument

%\lipsum[50-60]

\end{appendices} 

\end{document}